\documentclass[12pt,fleqn]{article}
\usepackage[margin=1in,footskip=0.25in]{geometry}
\usepackage[utf8]{inputenc}
\usepackage{longtable}
\usepackage{hyperref}
\usepackage{graphicx}
\usepackage{natbib}
\setcitestyle{square,numbers,colon}
\usepackage{amsmath}
\usepackage{listings,xcolor}
\usepackage{python}
\usepackage[draft]{minted}
\usepackage[nottoc,numbib]{tocbibind}

\hypersetup{
  colorlinks=true,
  linkcolor={blue!50!black},
  citecolor={red!50!black},
  urlcolor={blue!80!black}
}

\newcommand{\iu}{{i\mkern1mu}} 

\usepackage{verbatim}
\usepackage{tcolorbox}

\definecolor{Qblue}{HTML}{001C78}
\definecolor{QBackground}{HTML}{EAEAEA}

\newtcolorbox{jsonexamplebox}[2][]
{
  colframe = Qblue,
  colback  = QBackground,
  coltitle = white,
  title    = \texttt{Example JSON} - #2,
  #1,
}

\makeatletter
\def\PYG@reset{\let\PYG@it=\relax \let\PYG@bf=\relax%
    \let\PYG@ul=\relax \let\PYG@tc=\relax%
    \let\PYG@bc=\relax \let\PYG@ff=\relax}
\def\PYG@tok#1{\csname PYG@tok@#1\endcsname}
\def\PYG@toks#1+{\ifx\relax#1\empty\else%
    \PYG@tok{#1}\expandafter\PYG@toks\fi}
\def\PYG@do#1{\PYG@bc{\PYG@tc{\PYG@ul{%
    \PYG@it{\PYG@bf{\PYG@ff{#1}}}}}}}
\def\PYG#1#2{\PYG@reset\PYG@toks#1+\relax+\PYG@do{#2}}

\expandafter\def\csname PYG@tok@w\endcsname{\def\PYG@tc##1{\textcolor[rgb]{0.73,0.73,0.73}{##1}}}
\expandafter\def\csname PYG@tok@c\endcsname{\let\PYG@it=\textit\def\PYG@tc##1{\textcolor[rgb]{0.25,0.50,0.50}{##1}}}
\expandafter\def\csname PYG@tok@cp\endcsname{\def\PYG@tc##1{\textcolor[rgb]{0.74,0.48,0.00}{##1}}}
\expandafter\def\csname PYG@tok@k\endcsname{\let\PYG@bf=\textbf\def\PYG@tc##1{\textcolor[rgb]{0.00,0.50,0.00}{##1}}}
\expandafter\def\csname PYG@tok@kp\endcsname{\def\PYG@tc##1{\textcolor[rgb]{0.00,0.50,0.00}{##1}}}
\expandafter\def\csname PYG@tok@kt\endcsname{\def\PYG@tc##1{\textcolor[rgb]{0.69,0.00,0.25}{##1}}}
\expandafter\def\csname PYG@tok@o\endcsname{\def\PYG@tc##1{\textcolor[rgb]{0.40,0.40,0.40}{##1}}}
\expandafter\def\csname PYG@tok@ow\endcsname{\let\PYG@bf=\textbf\def\PYG@tc##1{\textcolor[rgb]{0.67,0.13,1.00}{##1}}}
\expandafter\def\csname PYG@tok@nb\endcsname{\def\PYG@tc##1{\textcolor[rgb]{0.00,0.50,0.00}{##1}}}
\expandafter\def\csname PYG@tok@nf\endcsname{\def\PYG@tc##1{\textcolor[rgb]{0.00,0.00,1.00}{##1}}}
\expandafter\def\csname PYG@tok@nc\endcsname{\let\PYG@bf=\textbf\def\PYG@tc##1{\textcolor[rgb]{0.00,0.00,1.00}{##1}}}
\expandafter\def\csname PYG@tok@nn\endcsname{\let\PYG@bf=\textbf\def\PYG@tc##1{\textcolor[rgb]{0.00,0.00,1.00}{##1}}}
\expandafter\def\csname PYG@tok@ne\endcsname{\let\PYG@bf=\textbf\def\PYG@tc##1{\textcolor[rgb]{0.82,0.25,0.23}{##1}}}
\expandafter\def\csname PYG@tok@nv\endcsname{\def\PYG@tc##1{\textcolor[rgb]{0.10,0.09,0.49}{##1}}}
\expandafter\def\csname PYG@tok@no\endcsname{\def\PYG@tc##1{\textcolor[rgb]{0.53,0.00,0.00}{##1}}}
\expandafter\def\csname PYG@tok@nl\endcsname{\def\PYG@tc##1{\textcolor[rgb]{0.63,0.63,0.00}{##1}}}
\expandafter\def\csname PYG@tok@ni\endcsname{\let\PYG@bf=\textbf\def\PYG@tc##1{\textcolor[rgb]{0.60,0.60,0.60}{##1}}}
\expandafter\def\csname PYG@tok@na\endcsname{\def\PYG@tc##1{\textcolor[rgb]{0.49,0.56,0.16}{##1}}}
\expandafter\def\csname PYG@tok@nt\endcsname{\let\PYG@bf=\textbf\def\PYG@tc##1{\textcolor[rgb]{0.00,0.50,0.00}{##1}}}
\expandafter\def\csname PYG@tok@nd\endcsname{\def\PYG@tc##1{\textcolor[rgb]{0.67,0.13,1.00}{##1}}}
\expandafter\def\csname PYG@tok@s\endcsname{\def\PYG@tc##1{\textcolor[rgb]{0.73,0.13,0.13}{##1}}}
\expandafter\def\csname PYG@tok@sd\endcsname{\let\PYG@it=\textit\def\PYG@tc##1{\textcolor[rgb]{0.73,0.13,0.13}{##1}}}
\expandafter\def\csname PYG@tok@si\endcsname{\let\PYG@bf=\textbf\def\PYG@tc##1{\textcolor[rgb]{0.73,0.40,0.53}{##1}}}
\expandafter\def\csname PYG@tok@se\endcsname{\let\PYG@bf=\textbf\def\PYG@tc##1{\textcolor[rgb]{0.73,0.40,0.13}{##1}}}
\expandafter\def\csname PYG@tok@sr\endcsname{\def\PYG@tc##1{\textcolor[rgb]{0.73,0.40,0.53}{##1}}}
\expandafter\def\csname PYG@tok@ss\endcsname{\def\PYG@tc##1{\textcolor[rgb]{0.10,0.09,0.49}{##1}}}
\expandafter\def\csname PYG@tok@sx\endcsname{\def\PYG@tc##1{\textcolor[rgb]{0.00,0.50,0.00}{##1}}}
\expandafter\def\csname PYG@tok@m\endcsname{\def\PYG@tc##1{\textcolor[rgb]{0.40,0.40,0.40}{##1}}}
\expandafter\def\csname PYG@tok@gh\endcsname{\let\PYG@bf=\textbf\def\PYG@tc##1{\textcolor[rgb]{0.00,0.00,0.50}{##1}}}
\expandafter\def\csname PYG@tok@gu\endcsname{\let\PYG@bf=\textbf\def\PYG@tc##1{\textcolor[rgb]{0.50,0.00,0.50}{##1}}}
\expandafter\def\csname PYG@tok@gd\endcsname{\def\PYG@tc##1{\textcolor[rgb]{0.63,0.00,0.00}{##1}}}
\expandafter\def\csname PYG@tok@gi\endcsname{\def\PYG@tc##1{\textcolor[rgb]{0.00,0.63,0.00}{##1}}}
\expandafter\def\csname PYG@tok@gr\endcsname{\def\PYG@tc##1{\textcolor[rgb]{1.00,0.00,0.00}{##1}}}
\expandafter\def\csname PYG@tok@ge\endcsname{\let\PYG@it=\textit}
\expandafter\def\csname PYG@tok@gs\endcsname{\let\PYG@bf=\textbf}
\expandafter\def\csname PYG@tok@gp\endcsname{\let\PYG@bf=\textbf\def\PYG@tc##1{\textcolor[rgb]{0.00,0.00,0.50}{##1}}}
\expandafter\def\csname PYG@tok@go\endcsname{\def\PYG@tc##1{\textcolor[rgb]{0.53,0.53,0.53}{##1}}}
\expandafter\def\csname PYG@tok@gt\endcsname{\def\PYG@tc##1{\textcolor[rgb]{0.00,0.27,0.87}{##1}}}
\expandafter\def\csname PYG@tok@err\endcsname{\def\PYG@tc##1{\textcolor[rgb]{0,0.11,0.47}{##1}}}
\expandafter\def\csname PYG@tok@l+m+mi\endcsname{\def\PYG@tc##1{\textcolor[rgb]{0,0.11,0.47}{##1}}}
\expandafter\def\csname PYG@tok@kc\endcsname{\let\PYG@bf=\textbf\def\PYG@tc##1{\textcolor[rgb]{0.00,0.50,0.00}{##1}}}
\expandafter\def\csname PYG@tok@kd\endcsname{\let\PYG@bf=\textbf\def\PYG@tc##1{\textcolor[rgb]{0.00,0.50,0.00}{##1}}}
\expandafter\def\csname PYG@tok@kn\endcsname{\let\PYG@bf=\textbf\def\PYG@tc##1{\textcolor[rgb]{0.00,0.50,0.00}{##1}}}
\expandafter\def\csname PYG@tok@kr\endcsname{\let\PYG@bf=\textbf\def\PYG@tc##1{\textcolor[rgb]{0.00,0.50,0.00}{##1}}}
\expandafter\def\csname PYG@tok@bp\endcsname{\def\PYG@tc##1{\textcolor[rgb]{0.00,0.50,0.00}{##1}}}
\expandafter\def\csname PYG@tok@fm\endcsname{\def\PYG@tc##1{\textcolor[rgb]{0.00,0.00,1.00}{##1}}}
\expandafter\def\csname PYG@tok@vc\endcsname{\def\PYG@tc##1{\textcolor[rgb]{0.10,0.09,0.49}{##1}}}
\expandafter\def\csname PYG@tok@vg\endcsname{\def\PYG@tc##1{\textcolor[rgb]{0.10,0.09,0.49}{##1}}}
\expandafter\def\csname PYG@tok@vi\endcsname{\def\PYG@tc##1{\textcolor[rgb]{0.10,0.09,0.49}{##1}}}
\expandafter\def\csname PYG@tok@vm\endcsname{\def\PYG@tc##1{\textcolor[rgb]{0.10,0.09,0.49}{##1}}}
\expandafter\def\csname PYG@tok@sa\endcsname{\def\PYG@tc##1{\textcolor[rgb]{0.73,0.13,0.13}{##1}}}
\expandafter\def\csname PYG@tok@sb\endcsname{\def\PYG@tc##1{\textcolor[rgb]{0.73,0.13,0.13}{##1}}}
\expandafter\def\csname PYG@tok@sc\endcsname{\def\PYG@tc##1{\textcolor[rgb]{0.73,0.13,0.13}{##1}}}
\expandafter\def\csname PYG@tok@dl\endcsname{\def\PYG@tc##1{\textcolor[rgb]{0.73,0.13,0.13}{##1}}}
\expandafter\def\csname PYG@tok@s2\endcsname{\def\PYG@tc##1{\textcolor[rgb]{0.73,0.13,0.13}{##1}}}
\expandafter\def\csname PYG@tok@sh\endcsname{\def\PYG@tc##1{\textcolor[rgb]{0.73,0.13,0.13}{##1}}}
\expandafter\def\csname PYG@tok@s1\endcsname{\def\PYG@tc##1{\textcolor[rgb]{0.73,0.13,0.13}{##1}}}
\expandafter\def\csname PYG@tok@mb\endcsname{\def\PYG@tc##1{\textcolor[rgb]{0.40,0.40,0.40}{##1}}}
\expandafter\def\csname PYG@tok@mf\endcsname{\def\PYG@tc##1{\textcolor[rgb]{0.40,0.40,0.40}{##1}}}
\expandafter\def\csname PYG@tok@mh\endcsname{\def\PYG@tc##1{\textcolor[rgb]{0.40,0.40,0.40}{##1}}}
\expandafter\def\csname PYG@tok@mi\endcsname{\def\PYG@tc##1{\textcolor[rgb]{0.40,0.40,0.40}{##1}}}
\expandafter\def\csname PYG@tok@il\endcsname{\def\PYG@tc##1{\textcolor[rgb]{0.40,0.40,0.40}{##1}}}
\expandafter\def\csname PYG@tok@mo\endcsname{\def\PYG@tc##1{\textcolor[rgb]{0.40,0.40,0.40}{##1}}}
\expandafter\def\csname PYG@tok@ch\endcsname{\let\PYG@it=\textit\def\PYG@tc##1{\textcolor[rgb]{0.25,0.50,0.50}{##1}}}
\expandafter\def\csname PYG@tok@cm\endcsname{\let\PYG@it=\textit\def\PYG@tc##1{\textcolor[rgb]{0.25,0.50,0.50}{##1}}}
\expandafter\def\csname PYG@tok@cpf\endcsname{\let\PYG@it=\textit\def\PYG@tc##1{\textcolor[rgb]{0.25,0.50,0.50}{##1}}}
\expandafter\def\csname PYG@tok@c1\endcsname{\let\PYG@it=\textit\def\PYG@tc##1{\textcolor[rgb]{0.25,0.50,0.50}{##1}}}
\expandafter\def\csname PYG@tok@cs\endcsname{\let\PYG@it=\textit\def\PYG@tc##1{\textcolor[rgb]{0.25,0.50,0.50}{##1}}}

\def\PYGZbs{\char`\\}
\def\PYGZus{\char`\_}
\def\PYGZob{\char`\{}
\def\PYGZcb{\char`\}}
\def\PYGZca{\char`\^}

\def\PYGZgt{\char`\>}

\def\PYGZhy{\char`\-}

\def\PYGZdq{\char`\"}

\makeatother
\begin{document}
 
\title{Qiskit Backend Specifications for\\OpenQASM and OpenPulse Experiments}
\author{David C. McKay$^1$\footnote{dcmckay@us.ibm.com}, Thomas Alexander$^1$, Luciano Bello$^1$, Michael J. Biercuk$^2$, \\
Lev Bishop$^1$, Jiayin Chen$^2$, Jerry M. Chow$^1$, Antonio D. C\'orcoles$^1$, \\
Daniel Egger$^1$, Stefan Filipp$^1$, Juan Gomez$^1$, Michael Hush$^2$, \\
Ali Javadi-Abhari$^1$, Diego Moreda$^1$, Paul Nation$^1$, Brent Paulovicks$^1$,\\
Erick Winston$^1$, Christopher J. Wood$^1$, James Wootton$^1$ and\\
Jay M. Gambetta$^1$\\ \\
$^1$IBM Research\\
$^2$Q-CTRL Pty Ltd, Sydney NSW Australia}
\date{\today}

\maketitle

\abstract
As interest in quantum computing grows, there is a pressing need for standardized API's so that algorithm designers, circuit designers, and physicists can be provided a common reference frame for designing, executing, and optimizing experiments. There is also a need for a language specification that goes beyond gates and allows users to specify the time dynamics of a quantum experiment and recover the time dynamics of the output. In this document we provide a specification for a common interface to backends (simulators and experiments) and a standarized data structure (Qobj --- quantum object) for sending experiments to those backends via Qiskit. We also introduce OpenPulse, a language for specifying pulse level control (i.e. control of the continuous time dynamics) of a general quantum device independent of the specific hardware implementation.

\newpage
\tableofcontents

\section{Introduction}

In recent years, the emergence of cloud quantum computing platforms (e.g. the IBM Q Experience~\cite{ibmqx}) has generated interest in programming NISQ (noisy intermediate scalable quantum) devices~\cite{preskill:2018}. There are now several software stacks that allow users to explore quantum computing over the cloud~\cite{qiskit,qsharp,forest,cirq} with Qiskit being the most complete and widely used. Qiskit is an open-source framework for quantum computing.  It allows users to create, compile, and execute quantum programs online, either in a simulator or a real quantum processor. To continue progress in this field, it is important to define a specification that meets future improvements in both devices and their classical control systems, and that allows the science required to make quantum computing possible. \\

Qiskit is designed to accommodate three user levels -- the algorithm designer, the circuit design and the quantum physicist. The algorithm designer wants to research and develop quantum algorithms and applications. The circuit designer wants to optimize quantum circuits for a given device and explore topics such as error correction, quantum verification and validation, and circuit optimizations. The quantum physicist wants to optimize and design quantum gates to perform the best circuit on a given device. They must be allowed to explore noise in these systems, apply dynamical decoupling and perform optimal control theory.  This document introduces and focuses on the specification and data structures that enable these three user levels to communicate with real quantum devices and simulators, collectively called `Backends',  to run/simulate experiments and retrieve results. In particular, we specify a self-contained quantum object (Qobj) data structure that defines a complete quantum experiment in one of two languages, OpenQASM or OpenPulse. The Qobj is a JSON file and so can be easily validated against schemas for correctness before executing on a backend. For the algorithm and circuit designer, the targeted language is OpenQASM 2.0~\cite{cross:2017}. Future extensions to OpenQASM are required to enable classical commands. However, the current representation covers the experiments that can be run in the foreseeable future. For the quantum physicists, this document introduces the OpenPulse language which is targeted for any system that obeys the rotating wave approximation (e.g., superconducting transmon qubits, ion trap qubits, NV qubits and quantum dots).   \\

\subsection{Intended Audience}
This document is geared towards three different readers:
\begin{enumerate}
\item Users of quantum devices who want to know the exact time dynamics of the experiment and the measurements that are collected and so will write experiments in OpenPulse.
\item Contributors to Qiskit who want to know the details of Qiskit's data structures and API design.
\item Providers of cloud-access quantum devices and simulators, who want to offer their backends for use through Qiskit.
\end{enumerate}

\subsection{Outline of Document}
This document is organized as follows: \S~\ref{sect:api} consists of an API specification that details the function calls and configuration data structures required by Qiskit-compatible devices. \S~\ref{sect:openqasm} and \S~\ref{sect:exqasm} introduce the specifications for an OpenQASM-type experiment, along with examples. \S~\ref{sect:openpulse} introduces the OpenPulse specifications. In \S~\ref{sect:expulse} and \S~\ref{sect:samples}, we give examples for several types of OpenPulse-type experiments.

\subsection{Outside of the Scope}

There are certain operations on a quantum device that are considered outside of the scope of this specification. These include fully generic classical processing, closed-loop feedback, and reconfiguration of the device (e.g., changing a magnetic field) on timescales longer than a single experiment. These may be added as extensions on a case-by-case basis.  \\

\subsection{Interface Language and Schemas \label{sect:language}}

In this specification, all interfacing between the user and a backend is according to the JSON format with snake case naming convention. Since complex numbers are not natively supported in the JSON specification, lists of complex numbers will be represented by nested lists of the form $[[a,b],[c,d],\ldots]$, corresponding to $[a+b\iu, c+d\iu, \ldots ]$. Date/times are in ISO 8601 notation.\\

All the data structures are defined by schemas which can be found at \url{https://github.com/Qiskit/qiskit-terra/tree/master/qiskit/schemas}. The schemas supersede anything detailed in this document.  \\

Non-integer numeric values are specified as JSON floats (\texttt{type: number}). There is no particular precision implied at the level of the abstract model specified in this document. There are precision constraints on experimental hardware (e.g. from DACs), and specific devices may enforce a precision constraint on the sample values in the pulse library. However, there will generally be an indirect relation between the sample values in the JSON pulse library and outputs from the hardware.

\section{Qiskit API \label{sect:api}}

\begin{figure}
\centering
\includegraphics[width=0.6\textwidth]{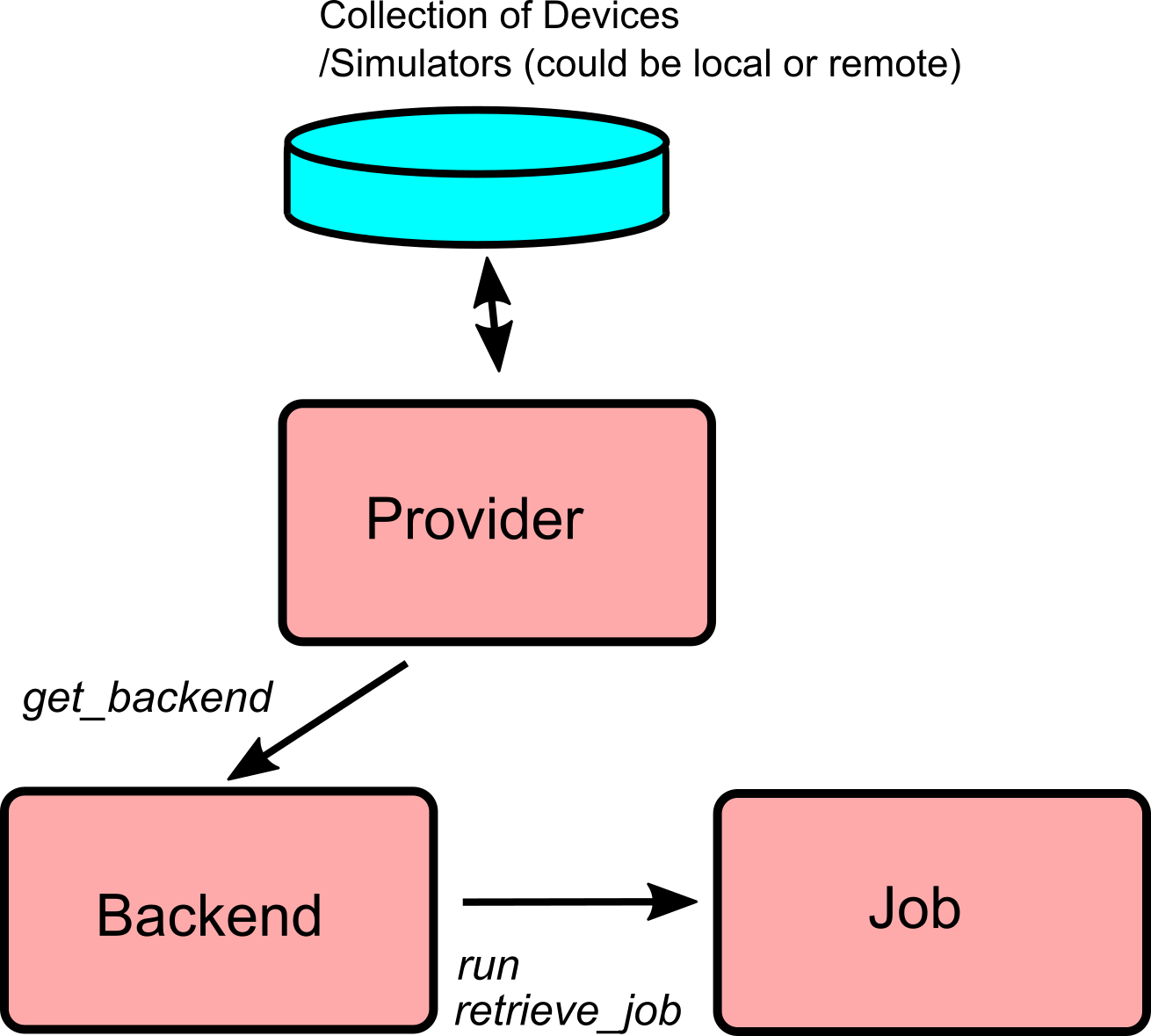}
\caption[captiont]{Interoperability of API objects for a Qiskit compatible system.\\\hspace{\textwidth}See text for details.\label{fig:api}}
\end{figure}

The API is the set of classes, functions and data structures for interfacing with devices and simulators, and running experiments. The classes are summarized in Figure~\ref{fig:api}, the function calls in Table~\ref{table:api}, and the data structures in Table~\ref{table:data}. There is a providers class (\S~\ref{sect:platform}) which manages collections of backends corresponding to a single authentication point and/or single connection point (e.g., a provider would be a set of backends from the same source, such as backends provided by IBM Q). Each backend (device/simulator that can run experiments) has a class (\S~\ref{sect:backend}) which can be used to obtain configuration and status information from the device and to run experiments. Experiments (jobs) are defined by a quantum object (\texttt{Qobj}, \S~\ref{sect:rundict}) data structure that contains configuration information and the experiment sequences. The job class is created via a run call to the backend with a \texttt{Qobj} that is validated against the general \texttt{Qobj} schema as well as any specific schema imposed by the backend. Once created, the job object can be used to get status information about the job, to cancel the job and to retrieve results.

\begin{table}
\centering
    \begin{tabular}{|p{0.3\textwidth}|p{0.3\textwidth}|p{0.3\textwidth}|}
        \hline Provider & Backend & Job \\ \hline
    - \texttt{get\_backend}   \newline
    - \texttt{available\_backends} &
    - \texttt{configuration} \newline
    - \texttt{properties} \newline
    - \texttt{defaults} \newline
    - \texttt{schema} \newline
    - \texttt{status} \newline
    - \texttt{run} \newline
    - \texttt{jobs} \newline
    - \texttt{retrieve\_job} &
    - \texttt{status} \newline
    - \texttt{done} \newline
        - \texttt{running} \newline
        - \texttt{cancelled} \newline
    - \texttt{job\_id} \newline
    - \texttt{backend} \newline
    - \texttt{cancel} \newline
        - \texttt{result}  \\ \hline
    \end{tabular}
\caption{Summary of function calls for the different classes of the Qiskit API. \label{table:api}}
\end{table}

\begin{table}
\centering
    \begin{tabular}{|p{0.32\textwidth}|p{0.32\textwidth}|}
        \hline OpenQASM & OpenPulse \\ \hline
    - \texttt{backend\_config} \newline
    - \texttt{backend\_props} \newline
    - \texttt{backend\_status} \newline
    - \texttt{job\_status} \newline
    - \texttt{Qobj} \newline
    - \texttt{result}   &
    - \texttt{default\_pulse\_config} \newline
    - \texttt{cmd\_def} \newline \\ \hline
    \end{tabular}
\caption[caption]{Summary of data structures for the Qiskit API. OpenPulse inherits all the OpenQASM structures. \label{table:data}}
\end{table}

\subsection{General Overview of a Qiskit Experiment}

\begin{figure}
\centering
\includegraphics[width=0.75\textwidth]{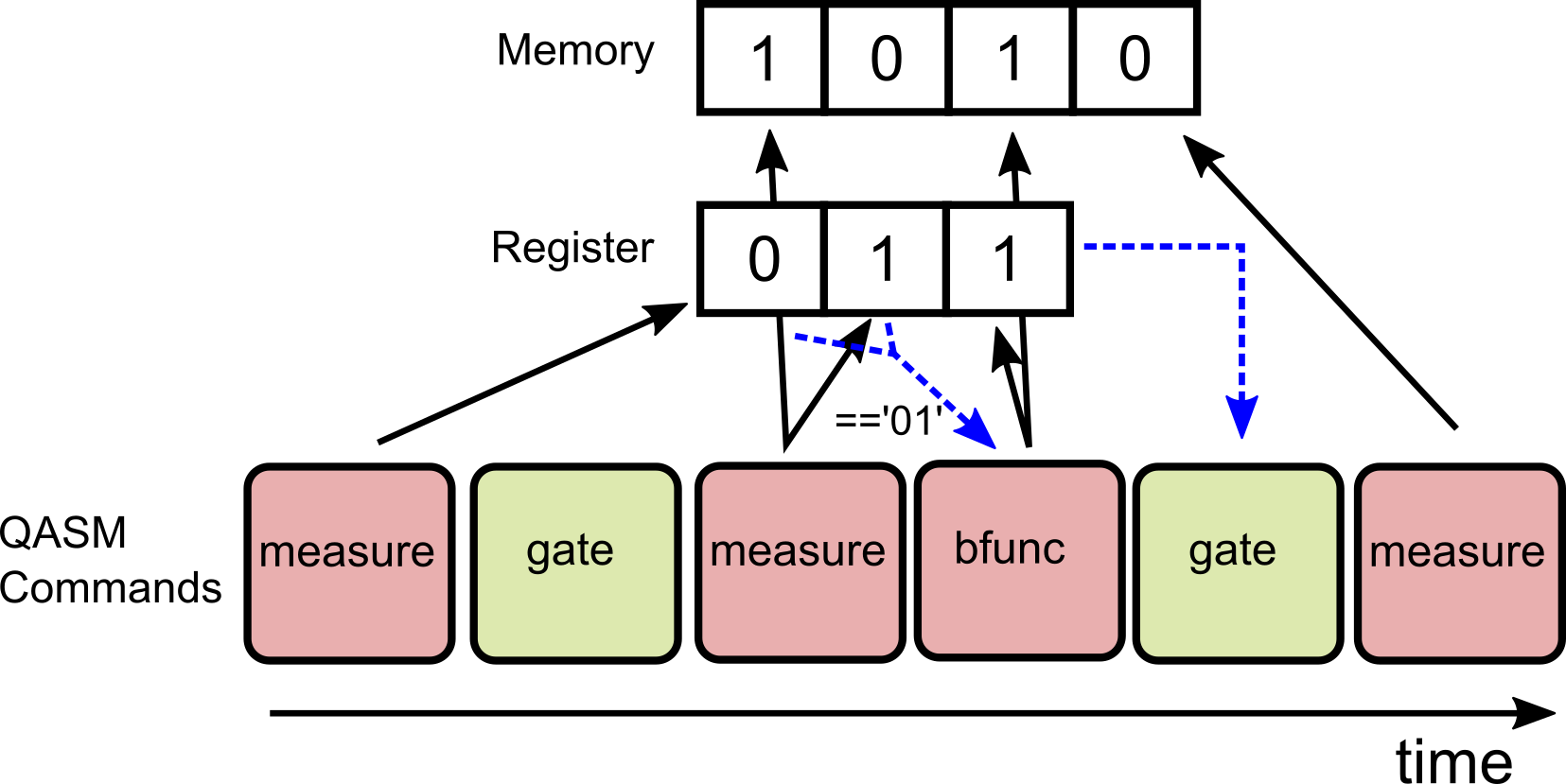}
\caption{Experiment flow between commands and the \texttt{memory} and \texttt{register}. Here the \texttt{memory} contains 4 slots and the \texttt{register} 3 slots. The first measurement is stored in the first register slot. Then an unconditional gate is performed on the qubit. The second measurement is stored in both the \texttt{memory} (first slot) and \texttt{register} (second slot). A boolean function operating on the first two bits of the register writes the result to the third register bit. This \texttt{register} bit is then used to conditionally run a \texttt{gate} command. The next measurement writes only to the \texttt{memory}. \label{fig:memoryflow}}
\end{figure}

The experiment sequences in the \texttt{Qobj} define quantum operations that run on the \texttt{backends}. A single \texttt{Qobj} defines a batch of experiments to run concurrently, i.e., one shot of each experiment in the \texttt{Qobj} will run in the order they are listed, and the entire sequence of experiments is repeated until the specified number of shots is collected. There are two languages to express these sequences, OpenQASM (\S~\ref{sect:openqasm}) and OpenPulse (\S~\ref{sect:openpulse}). OpenQASM is an abstract representation of the operations in terms of gates, measurements, and conditionals. OpenPulse defines the continuous time-dynamics of the operations and measurements, i.e., the pulses applied on the \texttt{backend}.  In experiments, measurements on the qubits are stored on the \texttt{backend} in two types of storage, slow access \texttt{memory} and a fast access \texttt{register}. \texttt{Memory} is dynamically sized by the user and is assumed unlimited in size. It is read out at the end of an experiment through the \texttt{result} data object. In OpenQASM experiments, measurements on the qubit(s) return a bit value for the state of the qubit(s). OpenPulse experiments can support additional measurement data types, for example, the raw demodulated signal or the IQ values after application of the measurement kernel, and these can also be stored in the memory. The \texttt{register} is fast access storage of bit data which stores either the qubit state or the output of a boolean function. If supported, gates can be conditional on whether a specified register is 1 (true). The number of \texttt{register} slots is defined by the backend.  A typical flow of measuring to \texttt{memory} and the \texttt{register} is shown in Figure~\ref{fig:memoryflow}.

\subsection{Provider \label{sect:platform}}

The provider object manages a set of backends offered by a provider, e.g., backends that have the same user authentication and address. The provider methods are listed in Table.~\ref{table:api}. The provider has a method \texttt{Provider.available\_backends} which returns all backends from the provider. The provider class has a function \texttt{Provider.\-get\_backend(backend\_name)} which returns a \texttt{Backend} object for the given backend name (\texttt{backend\_name}).

\subsection{Backend \label{sect:backend}}

The \texttt{Backend} class is an interface to an available device or simulator (something capable of running a quantum experiment). A backend may be online or local and may support OpenPulse (all backends must support OpenQASM). Here we list the required data structures (Table~\ref{table:data}) and function calls (Table~\ref{table:api}) for a general backend. Additional OpenPulse structures will be detailed in Section~\ref{sect:addpulseapi}.
\subsubsection{Configuration \label{sect:apiconfig}}

The backend will have a method \texttt{Backend.configuration()} which returns the required \texttt{backend\_config} data structure. Backends can include additional items to this structure (if the backend supports OpenPulse, then the additional required configuration items are listed in \S~\ref{sect:pulseconfig}).
\begin{jsonexamplebox}{Backend Configuration}
\begin{Verbatim}[commandchars=\\\{\}]
\PYG{p}{\PYGZob{}}
  \PYG{n+nt}{\PYGZdq{}backend\PYGZus{}name\PYGZdq{}}\PYG{p}{:} \PYG{l+s+s2}{\PYGZdq{}ibmqx2\PYGZdq{}}\PYG{p}{,}
  \PYG{n+nt}{\PYGZdq{}backend\PYGZus{}version\PYGZdq{}}\PYG{p}{:} \PYG{l+s+s2}{\PYGZdq{}1.1.1\PYGZdq{}}\PYG{p}{,}
  \PYG{n+nt}{\PYGZdq{}n\PYGZus{}qubits\PYGZdq{}}\PYG{p}{:} \PYG{l+m+mi}{5}\PYG{p}{,}
  \PYG{n+nt}{\PYGZdq{}basis\PYGZus{}gates\PYGZdq{}}\PYG{p}{:} \PYG{p}{[}\PYG{l+s+s2}{\PYGZdq{}u1\PYGZdq{}}\PYG{p}{,}\PYG{l+s+s2}{\PYGZdq{}u2\PYGZdq{}}\PYG{p}{,}\PYG{l+s+s2}{\PYGZdq{}u3\PYGZdq{}}\PYG{p}{,}\PYG{l+s+s2}{\PYGZdq{}cx\PYGZdq{}}\PYG{p}{],}
  \PYG{n+nt}{\PYGZdq{}coupling\PYGZus{}map\PYGZdq{}}\PYG{p}{:} \PYG{p}{[[}\PYG{l+m+mi}{0}\PYG{p}{,}\PYG{l+m+mi}{1}\PYG{p}{],[}\PYG{l+m+mi}{0}\PYG{p}{,}\PYG{l+m+mi}{2}\PYG{p}{],[}\PYG{l+m+mi}{0}\PYG{p}{,}\PYG{l+m+mi}{3}\PYG{p}{],[}\PYG{l+m+mi}{1}\PYG{p}{,}\PYG{l+m+mi}{2}\PYG{p}{],[}\PYG{l+m+mi}{0}\PYG{p}{,}\PYG{l+m+mi}{4}\PYG{p}{]],}
  \PYG{n+nt}{\PYGZdq{}gates\PYGZdq{}}\PYG{p}{:} \PYG{p}{[}\PYG{err}{gate\PYGZus{}config1}\PYG{p}{,}\PYG{err}{gate\PYGZus{}config2}\PYG{p}{,}\PYG{err}{gate\PYGZus{}config3}\PYG{p}{,}\PYG{err}{...}\PYG{p}{],}
  \PYG{n+nt}{\PYGZdq{}local\PYGZdq{}}\PYG{p}{:} \PYG{k+kc}{false}\PYG{p}{,}
  \PYG{n+nt}{\PYGZdq{}simulator\PYGZdq{}}\PYG{p}{:} \PYG{k+kc}{false}\PYG{p}{,}
  \PYG{n+nt}{\PYGZdq{}conditional\PYGZdq{}}\PYG{p}{:} \PYG{k+kc}{true}\PYG{p}{,}
  \PYG{n+nt}{\PYGZdq{}configurable\PYGZdq{}}\PYG{p}{:} \PYG{k+kc}{false}\PYG{p}{,}
  \PYG{n+nt}{\PYGZdq{}n\PYGZus{}registers\PYGZdq{}}\PYG{p}{:} \PYG{l+m+mi}{5}\PYG{p}{,}
  \PYG{n+nt}{\PYGZdq{}register\PYGZus{}map\PYGZdq{}}\PYG{p}{:} \PYG{p}{[[}\PYG{l+m+mi}{1}\PYG{p}{,}\PYG{l+m+mi}{1}\PYG{p}{,}\PYG{l+m+mi}{0}\PYG{p}{,}\PYG{l+m+mi}{0}\PYG{p}{,}\PYG{l+m+mi}{0}\PYG{p}{],[}\PYG{l+m+mi}{1}\PYG{p}{,}\PYG{l+m+mi}{1}\PYG{p}{,}\PYG{l+m+mi}{0}\PYG{p}{,}\PYG{l+m+mi}{0}\PYG{p}{,}\PYG{l+m+mi}{0}\PYG{p}{],[}\PYG{l+m+mi}{0}\PYG{p}{,}\PYG{l+m+mi}{0}\PYG{p}{,}\PYG{l+m+mi}{1}\PYG{p}{,}\PYG{l+m+mi}{1}\PYG{p}{,}\PYG{l+m+mi}{0}\PYG{p}{],}
                   \PYG{p}{[}\PYG{l+m+mi}{0}\PYG{p}{,}\PYG{l+m+mi}{0}\PYG{p}{,}\PYG{l+m+mi}{0}\PYG{p}{,}\PYG{l+m+mi}{1}\PYG{p}{,}\PYG{l+m+mi}{1}\PYG{p}{],[}\PYG{l+m+mi}{0}\PYG{p}{,}\PYG{l+m+mi}{0}\PYG{p}{,}\PYG{l+m+mi}{0}\PYG{p}{,}\PYG{l+m+mi}{1}\PYG{p}{,}\PYG{l+m+mi}{1}\PYG{p}{]],}
  \PYG{n+nt}{\PYGZdq{}open\PYGZus{}pulse\PYGZdq{}}\PYG{p}{:} \PYG{k+kc}{false}
\PYG{p}{\PYGZcb{}}
\end{Verbatim}
\end{jsonexamplebox}

where
\begin{itemize}
\item \texttt{backend\_name}: Unique (to provider) backend identifier name. This could describe a setup that goes through several changes, but retains common elements (e.g., for a physical device backend this could include the same coupling\_map and the physical location, etc.)
\item \texttt{backend\_version}: Backend version string in the form ``X.X.X''.  Versions could indicate, e.g., code changes, equipement upgrades, different cooldowns, new optimizations, etc.
\item \texttt{n\_qubits}: Number of qubits in the backend. Simulator backends return ``-1".
\item \texttt{basis\_gates}: List of the available gates on the backend as an array of gate names (these should match the entries in \texttt{gates}).
\item \texttt{coupling\_map}: Representation of the physical coupling map on the device (the coupling maps for each gate are defined in \texttt{gates}).
\item \texttt{gates}: List of the available gates on the backend as a \texttt{gate\_config} data structure (defined below in \S~\ref{sect:gatedict}).
\item \texttt{local}: Backend runs locally (true) or online (false).
\item \texttt{simulator}: Backend is a simulator (true) or an experimental device (false).
\item \texttt{conditional}: Backend supports conditional gates (true) or does not (false). Individual gates may also support or not support conditionals (see the \texttt{gate\_config} data structure).
\item \texttt{configurable}: Backend (if simulator) is configurable (true). If true then there are user specified configuration parameters (e.g., the topology, noise parameters, etc.). The data structure for these settings is set by the specific backend.
\item \texttt{n\_registers} (required if \texttt{conditional} is true): Specifies the number of registers slots (i.e. the number of register bits) that are available for conditional operations. Each register can hold a bit value.
\item \texttt{register\_map} (required if \texttt{conditional} is true): Specifies the registers that each qubit can store measurements. For this example, qubits 0 and 1 can store in registers 0 and 1, qubit 2 in registers 2 and 3, and qubits 3 and 4 in registers 3 and 4.
\item \texttt{open\_pulse}: OpenPulse experiments are accepted on this backend (bool).
\end{itemize}

The configuration structure may also have the following optional fields,
\begin{jsonexamplebox}{Optional Backend Configurations}
\begin{Verbatim}[commandchars=\\\{\}]
\PYG{p}{\PYGZob{}}
  \PYG{n+nt}{\PYGZdq{}online\PYGZus{}date\PYGZdq{}}\PYG{p}{:} \PYG{l+s+s2}{\PYGZdq{}2018\PYGZhy{}04\PYGZhy{}02 15:00:00Z\PYGZdq{}}\PYG{p}{,}
  \PYG{n+nt}{\PYGZdq{}display\PYGZus{}name\PYGZdq{}}\PYG{p}{:} \PYG{l+s+s2}{\PYGZdq{}IBM Q 5 Yorktown\PYGZdq{}}\PYG{p}{,}
  \PYG{n+nt}{\PYGZdq{}sample\PYGZus{}name\PYGZdq{}}\PYG{p}{:} \PYG{l+s+s2}{\PYGZdq{}sparrow\PYGZdq{}}\PYG{p}{,}
  \PYG{n+nt}{\PYGZdq{}description\PYGZdq{}}\PYG{p}{:} \PYG{l+s+s2}{\PYGZdq{}A 3 qubit superconducting processor}
\PYG{l+s+s2}{                  with fixed\PYGZhy{}frequency qubits\PYGZdq{}}\PYG{p}{,}
  \PYG{n+nt}{\PYGZdq{}url\PYGZdq{}}\PYG{p}{:} \PYG{l+s+s2}{\PYGZdq{}https://ibm.biz/qiskit\PYGZhy{}yorktown\PYGZdq{}}\PYG{p}{,}
  \PYG{n+nt}{\PYGZdq{}tags\PYGZdq{}}\PYG{p}{:} \PYG{p}{[}\PYG{l+s+s2}{\PYGZdq{}credits\PYGZus{}required\PYGZdq{}}\PYG{p}{]}
\PYG{p}{\PYGZcb{}}
\end{Verbatim}
\end{jsonexamplebox}

where
\begin{itemize}
\item \texttt{online\_date}: Date that the backend was put online.
\item \texttt{display\_name}: Alternate name for the backend that is more descriptive that can be used for display purposes.
\item \texttt{sample\_name}: Name of the sample for this given backend (likely blank for a simulator).
\item \texttt{description}: String to describe the backend.
\item \texttt{url}: Internet address to the backend (if applicable).
\item \texttt{tags}: List of tag strings for the backend that indicate true/false properties, e.g., ``credits\_required'' (backend requires credits to run). Any absent tag means that the property is false and new tags can be added.
\end{itemize}

\subsubsection{gate\_config Data Structure \label{sect:gatedict}}

The \texttt{gate\_config} data structure has the following keys,
\begin{jsonexamplebox}{Gate Config}
\begin{Verbatim}[commandchars=\\\{\}]
\PYG{p}{\PYGZob{}}
  \PYG{n+nt}{\PYGZdq{}name\PYGZdq{}}\PYG{p}{:} \PYG{l+s+s2}{\PYGZdq{}u3\PYGZdq{}}\PYG{p}{,}
  \PYG{n+nt}{\PYGZdq{}parameters\PYGZdq{}}\PYG{p}{:} \PYG{p}{[}\PYG{err}{theta}\PYG{p}{,}\PYG{err}{phi}\PYG{p}{,}\PYG{err}{lambda}\PYG{p}{],}
  \PYG{n+nt}{\PYGZdq{}coupling\PYGZus{}map\PYGZdq{}}\PYG{p}{:} \PYG{p}{[[}\PYG{l+m+mi}{0}\PYG{p}{],[}\PYG{l+m+mi}{1}\PYG{p}{],[}\PYG{l+m+mi}{2}\PYG{p}{],[}\PYG{l+m+mi}{3}\PYG{p}{]],}
  \PYG{n+nt}{\PYGZdq{}qasm\PYGZus{}def\PYGZdq{}}\PYG{p}{:} \PYG{l+s+s2}{\PYGZdq{}gate u3(theta,phi,lambda) q}
\PYG{l+s+s2}{               \PYGZob{} U(theta,phi,lambda) q; \PYGZcb{}\PYGZdq{}}\PYG{p}{,}
  \PYG{n+nt}{\PYGZdq{}conditional\PYGZdq{}}\PYG{p}{:} \PYG{k+kc}{true}\PYG{p}{,}
  \PYG{n+nt}{\PYGZdq{}latency\PYGZus{}map\PYGZdq{}}\PYG{p}{:} \PYG{p}{[[}\PYG{l+m+mi}{1}\PYG{p}{,}\PYG{l+m+mi}{0}\PYG{p}{,}\PYG{l+m+mi}{0}\PYG{p}{,}\PYG{l+m+mi}{0}\PYG{p}{,}\PYG{l+m+mi}{0}\PYG{p}{],[}\PYG{l+m+mi}{0}\PYG{p}{,}\PYG{l+m+mi}{1}\PYG{p}{,}\PYG{l+m+mi}{1}\PYG{p}{,}\PYG{l+m+mi}{0}\PYG{p}{,}\PYG{l+m+mi}{0}\PYG{p}{],[}\PYG{l+m+mi}{0}\PYG{p}{,}\PYG{l+m+mi}{1}\PYG{p}{,}\PYG{l+m+mi}{1}\PYG{p}{,}\PYG{l+m+mi}{0}\PYG{p}{,}\PYG{l+m+mi}{0}\PYG{p}{],}
                  \PYG{p}{[}\PYG{l+m+mi}{0}\PYG{p}{,}\PYG{l+m+mi}{0}\PYG{p}{,}\PYG{l+m+mi}{0}\PYG{p}{,}\PYG{l+m+mi}{1}\PYG{p}{,}\PYG{l+m+mi}{0}\PYG{p}{],[}\PYG{l+m+mi}{0}\PYG{p}{,}\PYG{l+m+mi}{0}\PYG{p}{,}\PYG{l+m+mi}{0}\PYG{p}{,}\PYG{l+m+mi}{0}\PYG{p}{,}\PYG{l+m+mi}{1}\PYG{p}{]],}
  \PYG{n+nt}{\PYGZdq{}description\PYGZdq{}}\PYG{p}{:} \PYG{l+s+s2}{\PYGZdq{}SU(2) gate with three rotation angles\PYGZdq{}}
\PYG{p}{\PYGZcb{}}
\end{Verbatim}
\end{jsonexamplebox}

where
\begin{itemize}
\item \texttt{name}: Gate name, as it will be referred to in the OpenQASM circuit.
\item \texttt{parameters}: List of parameters for the gate (empty if no parameters).
\item \texttt{coupling\_map}: List of qubits that the gate applies to, each element of the list is an n-qubit list where n is the size of the gate (e.g. 1-qubit gate, 2-qubit gate).
\item \texttt{qasm\_def}: OpenQASM definition of the gate in terms of the basis gates [U,CX]. Each unitary gate has an efficient representation in this basis.
\item \texttt{conditional} (optional): Gate supports conditional operation (true/false). If not listed then defaults to the backend setting.
\item \texttt{latency\_map} (optional): List for each gate of length \texttt{n\_registers} that indicates if the feedback speed to the register is fast (1) or slow (0). In the above example the $u3$ gate for qubit 0 has low latency for conditionals to \texttt{register} 0, but qubits 1 and 2 have low latency to both registers 1 and 2.
\item \texttt{description} (optional): Description of the gate.
\end{itemize}

Here is an example for \texttt{cx}
\begin{jsonexamplebox}{\texttt}
\begin{Verbatim}[commandchars=\\\{\}]
\PYG{p}{\PYGZob{}}
  \PYG{n+nt}{\PYGZdq{}name\PYGZdq{}}\PYG{p}{:} \PYG{l+s+s2}{\PYGZdq{}cx\PYGZdq{}}\PYG{p}{,}
  \PYG{n+nt}{\PYGZdq{}parameters\PYGZdq{}}\PYG{p}{:} \PYG{p}{[],}
  \PYG{n+nt}{\PYGZdq{}coupling\PYGZus{}map\PYGZdq{}}\PYG{p}{:} \PYG{p}{[[}\PYG{l+m+mi}{0}\PYG{p}{,}\PYG{l+m+mi}{1}\PYG{p}{],[}\PYG{l+m+mi}{0}\PYG{p}{,}\PYG{l+m+mi}{2}\PYG{p}{],[}\PYG{l+m+mi}{0}\PYG{p}{,}\PYG{l+m+mi}{3}\PYG{p}{],[}\PYG{l+m+mi}{1}\PYG{p}{,}\PYG{l+m+mi}{2}\PYG{p}{],[}\PYG{l+m+mi}{0}\PYG{p}{,}\PYG{l+m+mi}{4}\PYG{p}{]],}
  \PYG{n+nt}{\PYGZdq{}qasm\PYGZus{}def\PYGZdq{}}\PYG{p}{:} \PYG{l+s+s2}{\PYGZdq{}gate cx q1,q2 \PYGZob{} CX q1, q2; \PYGZcb{}\PYGZdq{}}\PYG{p}{,}
  \PYG{n+nt}{\PYGZdq{}conditional\PYGZdq{}}\PYG{p}{:} \PYG{k+kc}{true}\PYG{p}{,}
  \PYG{n+nt}{\PYGZdq{}latency\PYGZus{}map\PYGZdq{}}\PYG{p}{:} \PYG{p}{[[}\PYG{l+m+mi}{1}\PYG{p}{,}\PYG{l+m+mi}{0}\PYG{p}{,}\PYG{l+m+mi}{0}\PYG{p}{,}\PYG{l+m+mi}{0}\PYG{p}{,}\PYG{l+m+mi}{0}\PYG{p}{],[}\PYG{l+m+mi}{1}\PYG{p}{,}\PYG{l+m+mi}{0}\PYG{p}{,}\PYG{l+m+mi}{0}\PYG{p}{,}\PYG{l+m+mi}{0}\PYG{p}{,}\PYG{l+m+mi}{0}\PYG{p}{],}
                  \PYG{p}{[}\PYG{l+m+mi}{1}\PYG{p}{,}\PYG{l+m+mi}{0}\PYG{p}{,}\PYG{l+m+mi}{0}\PYG{p}{,}\PYG{l+m+mi}{0}\PYG{p}{,}\PYG{l+m+mi}{0}\PYG{p}{],[}\PYG{l+m+mi}{0}\PYG{p}{,}\PYG{l+m+mi}{1}\PYG{p}{,}\PYG{l+m+mi}{0}\PYG{p}{,}\PYG{l+m+mi}{0}\PYG{p}{,}\PYG{l+m+mi}{0}\PYG{p}{],}
                  \PYG{p}{[}\PYG{l+m+mi}{0}\PYG{p}{,}\PYG{l+m+mi}{0}\PYG{p}{,}\PYG{l+m+mi}{0}\PYG{p}{,}\PYG{l+m+mi}{0}\PYG{p}{,}\PYG{l+m+mi}{1}\PYG{p}{]],}
  \PYG{n+nt}{\PYGZdq{}description\PYGZdq{}}\PYG{p}{:} \PYG{l+s+s2}{\PYGZdq{}CNOT gate\PYGZdq{}}
\PYG{p}{\PYGZcb{}}
\end{Verbatim}
\end{jsonexamplebox}

and for \texttt{swap}
\begin{jsonexamplebox}{\texttt}
\begin{Verbatim}[commandchars=\\\{\}]
\PYG{p}{\PYGZob{}}
  \PYG{n+nt}{\PYGZdq{}name\PYGZdq{}}\PYG{p}{:} \PYG{l+s+s2}{\PYGZdq{}swap\PYGZdq{}}\PYG{p}{,}
  \PYG{n+nt}{\PYGZdq{}parameters\PYGZdq{}}\PYG{p}{:} \PYG{p}{[],}
  \PYG{n+nt}{\PYGZdq{}coupling\PYGZus{}map\PYGZdq{}}\PYG{p}{:} \PYG{p}{[[}\PYG{l+m+mi}{2}\PYG{p}{,}\PYG{l+m+mi}{3}\PYG{p}{]],}
  \PYG{n+nt}{\PYGZdq{}qasm\PYGZus{}def\PYGZdq{}}\PYG{p}{:} \PYG{l+s+s2}{\PYGZdq{}gate swap a,b \PYGZob{} CX a,b; CX b,a; CX a,b; \PYGZcb{}\PYGZdq{}}\PYG{p}{,}
  \PYG{n+nt}{\PYGZdq{}conditional\PYGZdq{}}\PYG{p}{:} \PYG{k+kc}{false}\PYG{p}{,}
  \PYG{n+nt}{\PYGZdq{}description\PYGZdq{}}\PYG{p}{:} \PYG{l+s+s2}{\PYGZdq{}SWAP gate\PYGZdq{}}
\PYG{p}{\PYGZcb{}}
\end{Verbatim}
\end{jsonexamplebox}

\subsubsection{Properties}

The backend will have a call \texttt{Backend.properties()} which will return a \texttt{backend\_props} data structure with backend properties (e.g. calibrations and coherences). Note that this information is optionally provided by the backend, which will set how often and/or under what conditions calibrations and characterizations need to be updated.
\begin{jsonexamplebox}{Backend Properties}
\begin{Verbatim}[commandchars=\\\{\}]
\PYG{p}{\PYGZob{}}
  \PYG{n+nt}{\PYGZdq{}backend\PYGZus{}name\PYGZdq{}}\PYG{p}{:} \PYG{l+s+s2}{\PYGZdq{}ibmqx2\PYGZdq{}}\PYG{p}{,}
  \PYG{n+nt}{\PYGZdq{}backend\PYGZus{}version\PYGZdq{}}\PYG{p}{:} \PYG{l+s+s2}{\PYGZdq{}1.1.1\PYGZdq{}}\PYG{p}{,}
  \PYG{n+nt}{\PYGZdq{}last\PYGZus{}update\PYGZus{}date\PYGZdq{}}\PYG{p}{:} \PYG{l+s+s2}{\PYGZdq{}2018\PYGZhy{}04\PYGZhy{}02 15:00:00Z\PYGZdq{}}\PYG{p}{,}
  \PYG{n+nt}{\PYGZdq{}gates\PYGZdq{}}\PYG{p}{:} \PYG{p}{[}\PYG{err}{gate\PYGZus{}prop1}\PYG{p}{,}\PYG{err}{gate\PYGZus{}prop2}\PYG{p}{,}\PYG{err}{...}\PYG{p}{],}
  \PYG{n+nt}{\PYGZdq{}qubits\PYGZdq{}}\PYG{p}{:} \PYG{p}{[[}\PYG{err}{nduv\PYGZus{}struct}\PYG{p}{,}\PYG{err}{...}\PYG{p}{],[}\PYG{err}{nduv\PYGZus{}struct}\PYG{p}{,}\PYG{err}{...}\PYG{p}{],}\PYG{err}{...}\PYG{p}{],}
  \PYG{n+nt}{\PYGZdq{}general\PYGZdq{}}\PYG{p}{:} \PYG{p}{[}\PYG{err}{nduv\PYGZus{}struct}\PYG{p}{,}\PYG{err}{nduv\PYGZus{}struct}\PYG{p}{,}\PYG{err}{...}\PYG{p}{]}
\PYG{p}{\PYGZcb{}}
\end{Verbatim}
\end{jsonexamplebox}

where
\begin{itemize}
\item \texttt{backend\_name}, \texttt{backend\_version}: Backend identifiers (from \texttt{Backend.configuration()}) that specify what backend these results were obtained from.
\item \texttt{last\_update\_date}: Date/time of the last run calibration.
\item \texttt{gates}: List of the qubit gate parameters (as a \texttt{gate\_prop} structure, see below).
\item \texttt{qubits}:  List of list of qubit parameters (e.g. coherences) which is in order of the qubits. The qubit parameters could generically include ``T1'', ``T2'', ``readoutErr''  and ``frequency''.
\item \texttt{general}:  List of general backend parameters (as a \texttt{nduv\_struct}, see below).
\end{itemize}
Each of the parameters is expressed as a \texttt{nduv\_struct} (name-date-unit-value structure) as below,
\begin{jsonexamplebox}{NDUV Struct}
\begin{Verbatim}[commandchars=\\\{\}]
\PYG{p}{\PYGZob{}}
  \PYG{n+nt}{\PYGZdq{}name\PYGZdq{}}\PYG{p}{:} \PYG{l+s+s2}{\PYGZdq{}T1\PYGZdq{}}\PYG{p}{,}
  \PYG{n+nt}{\PYGZdq{}date\PYGZdq{}}\PYG{p}{:} \PYG{l+s+s2}{\PYGZdq{}2018\PYGZhy{}04\PYGZhy{}02 15:00:00Z\PYGZdq{}}\PYG{p}{,}
  \PYG{n+nt}{\PYGZdq{}unit\PYGZdq{}}\PYG{p}{:} \PYG{l+s+s2}{\PYGZdq{}us\PYGZdq{}}\PYG{p}{,}
  \PYG{n+nt}{\PYGZdq{}value\PYGZdq{}}\PYG{p}{:} \PYG{l+m+mi}{60}
\PYG{p}{\PYGZcb{}}
\end{Verbatim}
\end{jsonexamplebox}

where
\begin{itemize}
\item \texttt{name}: Name of the parameter.
\item \texttt{date}: Date the parameter was measured.
\item \texttt{unit}: Unit (as a string) for the value.
\item \texttt{value}: Parameter value.
\end{itemize}

For the gate parameters the form of the \texttt{gate\_prop} structure is,
\begin{jsonexamplebox}{Gate Properties}
\begin{Verbatim}[commandchars=\\\{\}]
\PYG{p}{\PYGZob{}}
  \PYG{n+nt}{\PYGZdq{}qubits\PYGZdq{}}\PYG{p}{:} \PYG{p}{[}\PYG{l+m+mi}{1}\PYG{p}{],}
  \PYG{n+nt}{\PYGZdq{}gate\PYGZdq{}}\PYG{p}{:} \PYG{l+s+s2}{\PYGZdq{}u1\PYGZdq{}}\PYG{p}{,}
  \PYG{n+nt}{\PYGZdq{}parameters\PYGZdq{}}\PYG{p}{:} \PYG{p}{[}\PYG{err}{nduv\PYGZus{}struct1}\PYG{p}{,} \PYG{err}{nduv\PYGZus{}struct2}\PYG{p}{,}\PYG{err}{...}\PYG{p}{]}
\PYG{p}{\PYGZcb{}}
\end{Verbatim}
\end{jsonexamplebox}

where
\begin{itemize}
\item \texttt{qubits}: Qubits involved in the gate.
\item \texttt{gate}: Gate name, must be one of the gates from ``gates'' in the backend configuration structure.
\item \texttt{parameters}: List of parameter structures which could generically include ``gate\_err'' (by gate error we mean the $1-F_{\mathrm{avg}}$ for the particular gate) and ``gate\_time''. Note that each backend may measure gate error using different methodologies, this will have to be conveyed by the backend over separate channels (e.g. at the URL).
\end{itemize}

For the \texttt{general} parameters in backend properties, the form of each structure is,
\begin{jsonexamplebox}{General Parameters}
\begin{Verbatim}[commandchars=\\\{\}]
\PYG{p}{\PYGZob{}}
  \PYG{n+nt}{\PYGZdq{}general\PYGZus{}parameters\PYGZdq{}}\PYG{p}{:} \PYG{p}{[}
    \PYG{p}{\PYGZob{}}\PYG{n+nt}{\PYGZdq{}name\PYGZdq{}}\PYG{p}{:} \PYG{l+s+s2}{\PYGZdq{}fridge\PYGZus{}temperature\PYGZdq{}}\PYG{p}{,} \PYG{err}{...}\PYG{p}{\PYGZcb{},}
    \PYG{p}{\PYGZob{}}\PYG{n+nt}{\PYGZdq{}name\PYGZdq{}}\PYG{p}{:} \PYG{l+s+s2}{\PYGZdq{}cooldown\PYGZus{}date\PYGZdq{}}\PYG{p}{,} \PYG{err}{...}\PYG{p}{\PYGZcb{}}
  \PYG{p}{]}
\PYG{p}{\PYGZcb{}}
\end{Verbatim}
\end{jsonexamplebox}

where
\begin{itemize}
\item \texttt{parameters}: List of parameter structures for general system properties which may include \texttt{fridge\_temperature} and \texttt{cooldown\_date} (e.g., for a specific type of backend).
\end{itemize}

\subsubsection{Defaults}

The backend will have a call \texttt{Backend.defaults()} which gives a data structure of typical default parameters for the \texttt{Qobj} job (\S~\ref{sect:rundict}). This is particularly important for OpenPulse, see \S~\ref{sect:pulsedefaults}.

\subsubsection{Schema}

The backend will have a call \texttt{Backend.schema()} which will return a JSON schema for the \texttt{Qobj} data structure used for running jobs on the backend (\S~\ref{sect:rundict}). This allows users to validate their \texttt{Qobj} file. These are typically the extra constraints imposed by a particular backend, in addition to those imposed by the generic Qobj schema.

\subsubsection{Status}

The backend will have a call \texttt{Backend.status()} which returns status information on the backend in the \texttt{backend\_status} structure (only the \texttt{operational} and \texttt{status\_msg} fields are required),
\begin{jsonexamplebox}{Backend Status}
\begin{Verbatim}[commandchars=\\\{\}]
\PYG{p}{\PYGZob{}}
  \PYG{n+nt}{\PYGZdq{}backend\PYGZus{}name\PYGZdq{}}\PYG{p}{:} \PYG{l+s+s2}{\PYGZdq{}ibmqx2\PYGZdq{}}\PYG{p}{,}
  \PYG{n+nt}{\PYGZdq{}backend\PYGZus{}version\PYGZdq{}}\PYG{p}{:} \PYG{l+s+s2}{\PYGZdq{}1.1.1\PYGZdq{}}\PYG{p}{,}
  \PYG{n+nt}{\PYGZdq{}operational\PYGZdq{}}\PYG{p}{:} \PYG{k+kc}{true}\PYG{p}{,}
  \PYG{n+nt}{\PYGZdq{}pending\PYGZus{}jobs\PYGZdq{}}\PYG{p}{:} \PYG{l+m+mi}{20}\PYG{p}{,}
  \PYG{n+nt}{\PYGZdq{}status\PYGZus{}msg\PYGZdq{}}\PYG{p}{:} \PYG{l+s+s2}{\PYGZdq{}This is a status message\PYGZdq{}}
\PYG{p}{\PYGZcb{}}
\end{Verbatim}
\end{jsonexamplebox}

where
\begin{itemize}
\item \texttt{backend\_name}, \texttt{backend\_version}: Backend identifiers that specify the backend.
\item \texttt{operational}: Backend is operational (true/false), i.e., currently running jobs.
\item \texttt{pending\_jobs}: Number of jobs in the queue for the backend (if no queue return 0).
\item \texttt{status\_msg}: Status message for the backend. For example, ``The backend is down for calibration, will be back at 19:00''.
\end{itemize}

\subsubsection{Run}

The backend will have a function call to run jobs, \texttt{Backend.run(Qobj)}, which makes a job object based on a \texttt{Qobj} data structure (see \S~\ref{sect:rundict}) and returns the job object.

\subsubsection{Jobs}

The backend may also keep track of jobs that have been previously run, and so there is a method to list all jobs run on the backend by the user \texttt{Backend.jobs()} which will return a list of job objects. The backend may elect to only show jobs for a given period of time, e.g., jobs run in the last week. The backend may also allow a filtering mechanism on this call (e.g. by job id or job date).

\subsection{Job}

The job object is a submitted and accepted experiment on the backend (i.e. the backend has assigned it a provider-unique ID ``job\_id'', but the job may be in the queue, running or completed). The job methods are listed in Table~\ref{table:api}. The job\_id is obtained by calling the method \texttt{Job.job\_id()}. The backend associated with the job is obtained by calling the method \texttt{Job.backend()}. The job object is used to get status updates on the job, to cancel a job and to retrieve experiment results. The job itself is created from a \texttt{Qobj} structure that is defined in the next section.

\subsubsection{Qobj Data Structure \label{sect:rundict}}

Experiments are loaded in through the backend using a Qobj data structure which encapsulates the user configuration settings and experiment sequences. The basic form of the \texttt{Qobj} is,
\begin{jsonexamplebox}{Qobj Data Structure}
\begin{Verbatim}[commandchars=\\\{\}]
\PYG{p}{\PYGZob{}}
  \PYG{n+nt}{\PYGZdq{}qobj\PYGZus{}id\PYGZdq{}}\PYG{p}{:} \PYG{l+s+s2}{\PYGZdq{}qobj\PYGZus{}id\PYGZdq{}}\PYG{p}{,}
  \PYG{n+nt}{\PYGZdq{}type\PYGZdq{}}\PYG{p}{:} \PYG{l+s+s2}{\PYGZdq{}PULSE\PYGZdq{}}\PYG{p}{,}
  \PYG{n+nt}{\PYGZdq{}schema\PYGZus{}version\PYGZdq{}}\PYG{p}{:} \PYG{l+s+s2}{\PYGZdq{}1.0\PYGZdq{}}\PYG{p}{,}
  \PYG{n+nt}{\PYGZdq{}experiments\PYGZdq{}}\PYG{p}{:} \PYG{p}{[}\PYG{err}{exp\PYGZus{}struct1}\PYG{p}{,}\PYG{err}{exp\PYGZus{}struct2}\PYG{p}{,}\PYG{err}{...}\PYG{p}{],}
  \PYG{n+nt}{\PYGZdq{}header\PYGZdq{}}\PYG{p}{:} \PYG{p}{\PYGZob{}\PYGZcb{},}
  \PYG{n+nt}{\PYGZdq{}config\PYGZdq{}}\PYG{p}{:} \PYG{err}{user\PYGZus{}config}
\PYG{p}{\PYGZcb{}}
\end{Verbatim}
\end{jsonexamplebox}
where
\begin{itemize}
\item \texttt{qobj\_id}: User generated run identifier.
\item \texttt{type}: Type of experiment, can be either ``QASM'' for openQASM experiments or ``PULSE'' for OpenPulse experiments.
\item \texttt{schema\_version}: Version of the schema that was used to generate and validate this Qobj.
\item \texttt{experiments}: List of $m$ experiment sequences to run. Each experiment is an experiment data structure. Each experiment is run once in the order that they are specified in this list and then the sequence is repeated until the specified number of shots has been performed.
\item \texttt{header}: User-defined structure that contains metadata on the job and \emph{is not} used by the backend. The \texttt{header} will be passed through to the \text{result} data structure unchanged. For example, this may contain a description of the full job and/or the backend that the experiments were compiled for.
\item \texttt{config}: Configuration settings structure as defined below.
\end{itemize}
The experiments are defined in the following experiment data structure (\texttt{exp\_struct}):
\begin{jsonexamplebox}{Experiment Data Structure}
\begin{Verbatim}[commandchars=\\\{\}]
\PYG{p}{\PYGZob{}}
  \PYG{n+nt}{\PYGZdq{}header\PYGZdq{}}\PYG{p}{:} \PYG{p}{\PYGZob{}\PYGZcb{},}
  \PYG{n+nt}{\PYGZdq{}config\PYGZdq{}}\PYG{p}{:} \PYG{err}{user\PYGZus{}config}\PYG{p}{,}
  \PYG{n+nt}{\PYGZdq{}instructions\PYGZdq{}}\PYG{p}{:} \PYG{p}{[}\PYG{err}{cmd1}\PYG{p}{,} \PYG{err}{cmd2}\PYG{p}{,}\PYG{err}{...}\PYG{p}{]}
\PYG{p}{\PYGZcb{}}
\end{Verbatim}
\end{jsonexamplebox}
where
\begin{itemize}
\item \texttt{header}: User-defined structure that contains metadata on each experiment and \emph{is not} used by the backend. The \texttt{header} will be passed through to the \text{result} data structure unchanged. For example, this may contain a fitting parameters  for the experiment. In addition, this header can contain a mapping of backend memory and backend qubits to OpenQASM registers. This is because an OpenQASM circuit may contain multiple classical and quantum registers, but Qobj flattens them into a single memory and single set of qubits. 
\item \texttt{config}: Configuration structure for user settings that can be different in each experiment. These will override the configuration settings of the whole job.
\item \texttt{instructions}: List of sequence commands that define the experiment. For OpenQASM these commands are defined in \S~\ref{sect:qasmcmds} and for OpenPulse these are defined in \S~\ref{sect:actionitems}.
\end{itemize}

The structure \texttt{user\_config} passed in through the \texttt{Qobj} is a structure for backends and experiment types to be configured by the user (see \S~\ref{sect:pulseQObjconfig} for additional OpenPulse config settings). There are only a few required fields and the rest of the specification for the \texttt{user\_config} would be conveyed by each individual backend and the schemas obtained from \texttt{Backend.schema()}. The structure (with some possible additional fields) is,
\begin{jsonexamplebox}{User Config}
\begin{Verbatim}[commandchars=\\\{\}]
\PYG{p}{\PYGZob{}}
  \PYG{n+nt}{\PYGZdq{}shots\PYGZdq{}}\PYG{p}{:} \PYG{l+m+mi}{1024}\PYG{p}{,}
  \PYG{n+nt}{\PYGZdq{}memory\PYGZus{}slots\PYGZdq{}}\PYG{p}{:} \PYG{l+m+mi}{5}\PYG{p}{,}
  \PYG{n+nt}{\PYGZdq{}seed\PYGZdq{}}\PYG{p}{:} \PYG{l+m+mi}{1}\PYG{p}{,}
  \PYG{n+nt}{\PYGZdq{}max\PYGZus{}credits\PYGZdq{}}\PYG{p}{:} \PYG{l+m+mi}{3}
\PYG{p}{\PYGZcb{}}
\end{Verbatim}
\end{jsonexamplebox}
where
\begin{itemize}
\item \texttt{shots}: Number of times to repeat the experiment (for some simulators this may be limited to 1, e.g., a unitary simulator).
\item \texttt{memory\_slots}: Number of classical memory slots used in this job. Memory slots are used to record the results of qubit measurements and read out at the end of an experiment. They cannot be used for feedback (those are the registers).
\item \texttt{seed} (optional): Randomization seed for simulators.
\item \texttt{max\_credits} (optional): For credit-based backends, the maximum number of credits that a user is willing to spend on this run (an error will be thrown if the run required more than max\_credits).

\end{itemize}

\subsubsection{Job Status \label{sect:statusdict}}

A call to \texttt{Job.status()} returns back a \texttt{job\_status} data structure of the following form,
\begin{jsonexamplebox}{Job Status}
\begin{Verbatim}[commandchars=\\\{\}]
\PYG{p}{\PYGZob{}}
  \PYG{n+nt}{\PYGZdq{}job\PYGZus{}id\PYGZdq{}}\PYG{p}{:} \PYG{l+s+s2}{\PYGZdq{}job\PYGZus{}id\PYGZdq{}}\PYG{p}{,}
  \PYG{n+nt}{\PYGZdq{}status\PYGZdq{}}\PYG{p}{:} \PYG{l+s+s2}{\PYGZdq{}QUEUED\PYGZdq{}}\PYG{p}{,}
  \PYG{n+nt}{\PYGZdq{}status\PYGZus{}msg\PYGZdq{}}\PYG{p}{:} \PYG{l+s+s2}{\PYGZdq{}In the queue. Expected to run in 43 minutes.\PYGZdq{}}
\PYG{p}{\PYGZcb{}}
\end{Verbatim}
\end{jsonexamplebox}
where
\begin{itemize}
\item \texttt{job\_id}: Backend generated id corresponding to this job (this will only be nonzero if the job has been successfully initialized and accepted to run on the backend).
\item \texttt{status}: String value corresponding to the job status (``ERROR'',``QUEUED'', ``INITIALIZING'', ``RUNNING'', ``CANCELLED'' and ``DONE'').
\item \texttt{status\_msg}: Backend defined status message.
\end{itemize}
The job also has additional calls \texttt{Job.done()}, \texttt{Job.running()} and \texttt{Job.cancelled()} which returns a boolean (true/false) to indicate whether the job is done, running, or cancelled.

\subsubsection{Job Cancel}

The job object will have a method to cancel a job. The call to \texttt{Job.cancel()} removes the job from the queue (if it is in the queue) or stops a job that is running. On credit-based backends, this prevents credits from being deducted.

\subsubsection{Job Result (Measurement Retrieval) \label{sect:jobresult}}

The job will have a method to retrieve measurement results \texttt{Job.result()}. The exact behavior of this function may be backend specific. For example, the backend may require the user to call this function immediately after the call to \texttt{Backend.run()} and then this is a blocking function that only returns with all the measurement results (e.g., if the backend does not want to cache all the data). Alternatively, the backend may want to store the data for some amount of time and with each call to \texttt{Job.result()} some amount is returned and subsequently removed from the backend storage. The form of the \texttt{result} data structure is defined in \S~\ref{sect:results}.

\subsection{Result Data Structure \label{sect:results}}

The \texttt{results} data structure from \texttt{Job.result()} has the form:
\begin{jsonexamplebox}{Results}
\begin{Verbatim}[commandchars=\\\{\}]
\PYG{p}{\PYGZob{}}
  \PYG{n+nt}{\PYGZdq{}backend\PYGZus{}name\PYGZdq{}}\PYG{p}{:} \PYG{l+s+s2}{\PYGZdq{}ibmqx2\PYGZdq{}}\PYG{p}{,}
  \PYG{n+nt}{\PYGZdq{}backend\PYGZus{}version\PYGZdq{}}\PYG{p}{:} \PYG{l+s+s2}{\PYGZdq{}2.1.2\PYGZdq{}}\PYG{p}{,}
  \PYG{n+nt}{\PYGZdq{}qobj\PYGZus{}id\PYGZdq{}}\PYG{p}{:} \PYG{l+s+s2}{\PYGZdq{}qobj\PYGZus{}id\PYGZdq{}}\PYG{p}{,}
  \PYG{n+nt}{\PYGZdq{}job\PYGZus{}id\PYGZdq{}}\PYG{p}{:} \PYG{l+s+s2}{\PYGZdq{}job\PYGZus{}id\PYGZdq{}}\PYG{p}{,}
  \PYG{n+nt}{\PYGZdq{}date\PYGZdq{}}\PYG{p}{:} \PYG{l+s+s2}{\PYGZdq{}2018\PYGZhy{}04\PYGZhy{}02 15:00:00Z\PYGZdq{}}\PYG{p}{,}
  \PYG{n+nt}{\PYGZdq{}header\PYGZdq{}}\PYG{p}{:} \PYG{p}{\PYGZob{}\PYGZcb{},}
  \PYG{n+nt}{\PYGZdq{}status\PYGZdq{}}\PYG{p}{:} \PYG{l+s+s2}{\PYGZdq{}COMPLETED\PYGZdq{}}\PYG{p}{,}
  \PYG{n+nt}{\PYGZdq{}success\PYGZdq{}}\PYG{p}{:} \PYG{k+kc}{true}\PYG{p}{,}
  \PYG{n+nt}{\PYGZdq{}results\PYGZdq{}}\PYG{p}{:} \PYG{p}{[}\PYG{err}{exp\PYGZus{}result1}\PYG{p}{,}\PYG{err}{exp\PYGZus{}result2}\PYG{p}{,}\PYG{err}{...}\PYG{p}{]}
\PYG{p}{\PYGZcb{}}
\end{Verbatim}
\end{jsonexamplebox}

where
\begin{itemize}
\item \texttt{backend\_name}, \texttt{backend\_version}: Backend identifiers (from \texttt{Backend.configuration()}) that specify what backend these results were obtained from.
\item \texttt{qobj\_id}: User generated id corresponding to the \texttt{qobj\_id} in the \texttt{Qobj}.
\item \texttt{job\_id}: Unique backend job identifier corresponding to these results.
\item \texttt{date}: Date when the job was run.
\item \texttt{header}: Header structure for the job that was passed in with the \texttt{Qobj}.
\item \texttt{results}: List of $m$ (number of experiments) \texttt{exp\_result} data structures (defined below).
\end{itemize}

Each individual experiment returns an \texttt{exp\_result} data structure,
\begin{jsonexamplebox}{Experiment Result}
\begin{Verbatim}[commandchars=\\\{\}]
\PYG{p}{\PYGZob{}}
  \PYG{n+nt}{\PYGZdq{}shots\PYGZdq{}}\PYG{p}{:} \PYG{err}{n2}\PYG{p}{,}
  \PYG{n+nt}{\PYGZdq{}shots\PYGZdq{}}\PYG{p}{:} \PYG{p}{[}\PYG{err}{n1}\PYG{p}{,}\PYG{err}{n2}\PYG{p}{],}
  \PYG{n+nt}{\PYGZdq{}status\PYGZdq{}}\PYG{p}{:} \PYG{l+s+s2}{\PYGZdq{}status string\PYGZdq{}}\PYG{p}{,}
  \PYG{n+nt}{\PYGZdq{}success\PYGZdq{}}\PYG{p}{:} \PYG{k+kc}{true}\PYG{p}{,}
  \PYG{n+nt}{\PYGZdq{}header\PYGZdq{}}\PYG{p}{:} \PYG{p}{\PYGZob{}\PYGZcb{},}
  \PYG{n+nt}{\PYGZdq{}seed\PYGZdq{}}\PYG{p}{:} \PYG{l+m+mi}{1}\PYG{p}{,}
  \PYG{n+nt}{\PYGZdq{}meas\PYGZus{}return\PYGZdq{}}\PYG{p}{:} \PYG{l+s+s2}{\PYGZdq{}single\PYGZdq{}}\PYG{p}{,}
  \PYG{n+nt}{\PYGZdq{}data\PYGZdq{}}\PYG{p}{:} \PYG{err}{exp\PYGZus{}data}
\PYG{p}{\PYGZcb{}}
\end{Verbatim}
\end{jsonexamplebox}

where
\begin{itemize}
\item \texttt{shots}: If a single integer, then this is the number of shots taken to obtain this data ($s=n1$). If the backend allows asynchronous calls to measurement, the value of $n2$ will increase as more data is taken. For backends that return the data in sections (e.g. for bandwidth reasons) shots is given as a two-element list where the data is from shot $n1$ to shot $n2$ ($s=n2-n1$). The next call will give the data starting at $n2+1$.
\item \texttt{status}: Status message for this particular experiment.
\item \texttt{success}: Success of the experiment (bool).
\item \texttt{header}: Header structure for the experiment that was passed in with the \texttt{Qobj}.
\item \texttt{seed} (optional): Experiment seed (for simulator backends).
\item \texttt{meas\_return} (optional): String which determines whether the returned data is averaged over the shots \texttt{avg} or contains each shot \texttt{single}. This is an OpenPulse option, but could also apply to snapshots.
\item \texttt{data}: Generic return experiment data structure \texttt{exp\_data} that will depend on the type of experiment (``QASM'' or ``PULSE'') and/or the type of backend (e.g. simulator data). See below.
\end{itemize}

\subsubsection{Measurement Data \label{sect:measdata}}

The measurement data that is returned in \texttt{exp\_data} has one of several possible forms: as a histogram of counts of the memory states, the memory, or (for simulators) the statevector or unitary matrix. These different forms are as follows,
\begin{jsonexamplebox}{Measurement Data}
\begin{Verbatim}[commandchars=\\\{\}]
\PYG{p}{\PYGZob{}}
  \PYG{n+nt}{\PYGZdq{}counts\PYGZdq{}}\PYG{p}{:} \PYG{p}{\PYGZob{}}\PYG{n+nt}{\PYGZdq{}0x0\PYGZdq{}}\PYG{p}{:} \PYG{l+m+mi}{10}\PYG{p}{,} \PYG{n+nt}{\PYGZdq{}0x1\PYGZdq{}}\PYG{p}{:} \PYG{l+m+mi}{100}\PYG{p}{,} \PYG{n+nt}{\PYGZdq{}0x2\PYGZdq{}}\PYG{p}{:} \PYG{l+m+mi}{100}\PYG{p}{,} \PYG{err}{...}\PYG{p}{\PYGZcb{},}
  \PYG{n+nt}{\PYGZdq{}memory\PYGZdq{}}\PYG{p}{:} \PYG{p}{[}\PYG{l+s+s2}{\PYGZdq{}0x1\PYGZdq{}}\PYG{p}{,}\PYG{l+s+s2}{\PYGZdq{}0x2\PYGZdq{}}\PYG{p}{,}\PYG{l+s+s2}{\PYGZdq{}0x2\PYGZdq{}}\PYG{p}{,}\PYG{l+s+s2}{\PYGZdq{}0x1\PYGZdq{}}\PYG{p}{,}\PYG{l+s+s2}{\PYGZdq{}0x0\PYGZdq{}}\PYG{p}{,}\PYG{err}{...}\PYG{p}{],}
  \PYG{n+nt}{\PYGZdq{}statevector\PYGZdq{}}\PYG{p}{:} \PYG{p}{[[[}\PYG{l+m+mf}{1.0}\PYG{p}{,}\PYG{l+m+mi}{0}\PYG{p}{],[}\PYG{l+m+mi}{0}\PYG{p}{,}\PYG{l+m+mi}{0}\PYG{p}{],[}\PYG{l+m+mi}{0}\PYG{p}{,}\PYG{l+m+mi}{0}\PYG{p}{],[}\PYG{l+m+mi}{0}\PYG{p}{,}\PYG{l+m+mi}{0}\PYG{p}{]],}\PYG{err}{...}\PYG{p}{],}
  \PYG{n+nt}{\PYGZdq{}unitary\PYGZdq{}}\PYG{p}{:} \PYG{p}{[[[}\PYG{l+m+mf}{1.0}\PYG{p}{,}\PYG{l+m+mi}{0}\PYG{p}{],[}\PYG{l+m+mi}{0}\PYG{p}{,}\PYG{l+m+mi}{0}\PYG{p}{],[}\PYG{l+m+mi}{0}\PYG{p}{,}\PYG{l+m+mi}{0}\PYG{p}{],[}\PYG{l+m+mi}{0}\PYG{p}{,}\PYG{l+m+mi}{0}\PYG{p}{]],}\PYG{err}{...}\PYG{p}{],}
  \PYG{n+nt}{\PYGZdq{}snapshots\PYGZdq{}}\PYG{p}{:} \PYG{p}{\PYGZob{}}\PYG{err}{snapshot\PYGZus{}data}\PYG{p}{\PYGZcb{}}
\PYG{p}{\PYGZcb{}}
\end{Verbatim}
\end{jsonexamplebox}
where
\begin{itemize}
\item \texttt{counts}: Histogram of counts in the different memory states. Only states with non-zero counts are listed as keys. The states are labeled in \emph{hex} (e.g., a 4 slot memory ``1010'' (bit string) is decimal 10 and hex ``0xA'').
\item \texttt{memory}: State of the classical memory. For OpenQASM (or OpenPulse Level 2) this is a list of hex strings indicating the state for each shot. For OpenPulse Level 0 and 1 see \S~\ref{sect:meas}.
\item \texttt{statevector}: Final statevector corresponding to evolution of the zero state.
\item \texttt{unitary}: Final unitary matrix corresponding to the quantum circuit.
\item \texttt{snapshots}: Snapshots data structure that returns data as dictated by the \texttt{snapshot} command used by simulators (\S~\ref{sect:snapshot}), the form of the data structure is described in \S~\ref{sect:snapshot_results}.
\end{itemize}

\subsubsection{Snapshots \label{sect:snapshot_results}}

The snapshots are returned as a data structure with a key for each snapshot type, and subkeys are given by the names specified in the snapshot command (\S~\ref{sect:snapshot}). Each of these returns information about the state of the simulator at the point where the snapshot was set in the command sequence. Details on the different snapshot types and schemas will be specified in a separate document for the simulators. The data structure is,
\begin{jsonexamplebox}{Snapshot Data}
\begin{Verbatim}[commandchars=\\\{\}]
\PYG{p}{\PYGZob{}}
   \PYG{n+nt}{\PYGZdq{}snap\PYGZus{}type\PYGZus{}1\PYGZdq{}}\PYG{p}{:\PYGZob{}}
    \PYG{n+nt}{\PYGZdq{}snap\PYGZus{}label1\PYGZdq{}}\PYG{p}{:} \PYG{p}{\PYGZob{}\PYGZcb{},}
    \PYG{n+nt}{\PYGZdq{}snap\PYGZus{}label2\PYGZdq{}}\PYG{p}{:} \PYG{p}{\PYGZob{}\PYGZcb{}}
    \PYG{p}{\PYGZcb{}}
\PYG{p}{\PYGZcb{}}
\end{Verbatim}
\end{jsonexamplebox}
where
\begin{itemize}
\item \texttt{snap\_type}: All snapshots of the type ``snap\_type''.
\item \texttt{snap\_label}: Data for the snapshot with name ``snap\_label'.
\end{itemize}

\section{OpenQASM Representation as a Quantum Object (\texttt{Qobj}) Data Structure \label{sect:openqasm}}

As previously mentioned, and defined elsewhere~\cite{cross:2017}, OpenQASM is a language for specifying quantum circuits in terms of gates, measurements, and conditionals. An OpenQASM circuit is a string of OpenQASM commands with a header to define the number and size of quantum registers (groups of qubits) and classical registers (groups of classical bits). This section specifies how to represent a group of experiments (circuits) defined in OpenQASM as a \texttt{Qobj} data structure (\S~\ref{sect:rundict}) that can be sent to a device or simulator for execution. This is not a redefinition of OpenQASM, but rather a specification for how to parse the OpenQASM instructions into a more appropriate data format for execution. For an OpenQASM circuit to be represented in the \texttt{Qobj} format, the multiple quantum/classical registers in an OpenQASM circuit must be mapped to a single quantum/classical register and the classical register separated into slow memory (``memory'') and fast memory (``register'').

\subsection{Experiment Sequence Commands \label{sect:qasmcmds}}

OpenQASM in \texttt{Qobj} will have the following sequence commands to define an experiment. In general the only required field of the data structure is \texttt{name} and there are several reserved keywords, ``\texttt{bfunc}'', ``\texttt{copy}'', ``\texttt{reset}'', ``\texttt{barrier}'', ``\texttt{measure}'', and ``\texttt{snapshot}''.

\subsubsection{Boolean Function \label{sect:qasmclassicalfunc}}

A boolean function is a command that takes in a register as an argument and computes a boolean value that is written back into one of the register slots. These are defined by the following structure:
\begin{jsonexamplebox}{Boolean Function}
\begin{Verbatim}[commandchars=\\\{\}]
\PYG{p}{\PYGZob{}}
  \PYG{n+nt}{\PYGZdq{}name\PYGZdq{}}\PYG{p}{:} \PYG{l+s+s2}{\PYGZdq{}bfunc\PYGZdq{}}\PYG{p}{,}
  \PYG{n+nt}{\PYGZdq{}mask\PYGZdq{}}\PYG{p}{:} \PYG{l+s+s2}{\PYGZdq{}0xF\PYGZdq{}}\PYG{p}{,}
  \PYG{n+nt}{\PYGZdq{}relation\PYGZdq{}}\PYG{p}{:} \PYG{l+s+s2}{\PYGZdq{}==\PYGZdq{}}\PYG{p}{,}
  \PYG{n+nt}{\PYGZdq{}val\PYGZdq{}}\PYG{p}{:} \PYG{l+s+s2}{\PYGZdq{}0x5\PYGZdq{}}\PYG{p}{,}
  \PYG{n+nt}{\PYGZdq{}register\PYGZdq{}}\PYG{p}{:} \PYG{l+m+mi}{0}\PYG{p}{,}
  \PYG{n+nt}{\PYGZdq{}memory\PYGZdq{}}\PYG{p}{:} \PYG{l+m+mi}{0}
\PYG{p}{\PYGZcb{}}
\end{Verbatim}
\end{jsonexamplebox}
where
\begin{itemize}
\item \texttt{name}: ``bfunc''.
\item \texttt{mask}: Hex value which is applied as an \texttt{AND} to the register bits. In the given example ,``0xF'' uses the first 4 bits of the register. The backend may put constraints on the number of register bits that can be used in this function.
\item \texttt{relation}: Relational operator for comparing the masked register to the \texttt{val} (``=='': equals, ``!='' not equals).
\item \texttt{val}: Value to which to compare the masked register. In other words, the output of the function is (\texttt{register} \texttt{AND} \texttt{mask}) \texttt{relation} \texttt{val}. In the above example this is true when the first 4 register bits are ``0101''
\item \texttt{register}: Register slot in which to store the boolean function result. This register value can then be used to apply conditional commands (see the following sections).
\item \texttt{memory} (optional): Memory slot in which to store the boolean function result.
\end{itemize}

\subsubsection{Copy Function \label{sect:qasmcopy}}

A copy function is a command that copies a register slot. This is defined by the following structure:
\begin{jsonexamplebox}{Copy Function}
\begin{Verbatim}[commandchars=\\\{\}]
\PYG{p}{\PYGZob{}}
  \PYG{n+nt}{\PYGZdq{}name\PYGZdq{}}\PYG{p}{:} \PYG{l+s+s2}{\PYGZdq{}copy\PYGZdq{}}\PYG{p}{,}
  \PYG{n+nt}{\PYGZdq{}register\PYGZus{}orig\PYGZdq{}}\PYG{p}{:} \PYG{l+m+mi}{0}\PYG{p}{,}
  \PYG{n+nt}{\PYGZdq{}register\PYGZus{}copy\PYGZdq{}}\PYG{p}{:} \PYG{p}{[}\PYG{l+m+mi}{1}\PYG{p}{,}\PYG{l+m+mi}{2}\PYG{p}{,}\PYG{l+m+mi}{5}\PYG{p}{]}
\PYG{p}{\PYGZcb{}}
\end{Verbatim}
\end{jsonexamplebox}

where
\begin{itemize}
\item \texttt{name}: ``copy''.
\item \texttt{register\_orig}: Register slot to copy.
\item \texttt{register\_copy}: Register slot(s) to copy to.
\end{itemize}

\subsubsection{Gate \label{sect:qasmgate}}

Each OpenQASM gate is defined by the following structure:
\begin{jsonexamplebox}{OpenQASM Gate}
\begin{Verbatim}[commandchars=\\\{\}]
\PYG{p}{\PYGZob{}}
  \PYG{n+nt}{\PYGZdq{}name\PYGZdq{}}\PYG{p}{:} \PYG{l+s+s2}{\PYGZdq{}u2\PYGZdq{}}\PYG{p}{,}
  \PYG{n+nt}{\PYGZdq{}qubits\PYGZdq{}}\PYG{p}{:} \PYG{p}{[}\PYG{l+m+mi}{1}\PYG{p}{],}
  \PYG{n+nt}{\PYGZdq{}params\PYGZdq{}}\PYG{p}{:} \PYG{p}{[}\PYG{l+m+mf}{0.0}\PYG{p}{,} \PYG{l+m+mf}{3.141592653589793}\PYG{p}{],}
  \PYG{n+nt}{\PYGZdq{}texparams\PYGZdq{}}\PYG{p}{:} \PYG{p}{[}\PYG{l+s+s2}{\PYGZdq{}0\PYGZdq{}}\PYG{p}{,}\PYG{l+s+s2}{\PYGZdq{}\PYGZbs{}\PYGZbs{}pi\PYGZdq{}}\PYG{p}{],}
  \PYG{n+nt}{\PYGZdq{}conditional\PYGZdq{}}\PYG{p}{:} \PYG{l+m+mi}{3}
\PYG{p}{\PYGZcb{}}
\end{Verbatim}
\end{jsonexamplebox}
where
\begin{itemize}
\item \texttt{name}: Name of the gate.
\item \texttt{qubits}: List of qubits to apply the gate.
\item \texttt{params} (optional): List of parameters for the gate (if the gate has parameters, such as \texttt{u1}, \texttt{u2}, \texttt{u3}).
\item \texttt{texparams} (optional): List of parameters for the gate in latex notation.
\item \texttt{conditional} (optional): Apply the gate if the given register (in this example register 3) is 1 (true) and conditionals are supported. If left blank then the gate has no conditional element (i.e. no feedback). By default this is blank.
\end{itemize}

\subsubsection{Barrier}

If the instruction name is ``barrier'' then the command ensures that all instructions listed after (for the specified qubits) occur (in time) after all instructions listed before the ``barrier''. The barrier instructions is defined by the following structure:
\begin{jsonexamplebox}{Barrier Gate}
\begin{Verbatim}[commandchars=\\\{\}]
\PYG{p}{\PYGZob{}}
  \PYG{n+nt}{\PYGZdq{}name\PYGZdq{}}\PYG{p}{:} \PYG{l+s+s2}{\PYGZdq{}barrier\PYGZdq{}}\PYG{p}{,}
  \PYG{n+nt}{\PYGZdq{}qubits\PYGZdq{}}\PYG{p}{:} \PYG{p}{[}\PYG{l+m+mi}{1}\PYG{p}{]}
\PYG{p}{\PYGZcb{}}
\end{Verbatim}
\end{jsonexamplebox}

where
\begin{itemize}
\item \texttt{name}: ``barrier''.
\item \texttt{qubits}: List of qubits on which to apply the barrier.
\end{itemize}

\subsubsection{Reset}

If the instruction name is ``reset'' then the listed qubits are reset back to the ground state. The reset instruction is defined by the following structure:
\begin{jsonexamplebox}{Reset Gate}
\begin{Verbatim}[commandchars=\\\{\}]
\PYG{p}{\PYGZob{}}
  \PYG{n+nt}{\PYGZdq{}name\PYGZdq{}}\PYG{p}{:} \PYG{l+s+s2}{\PYGZdq{}reset\PYGZdq{}}\PYG{p}{,}
  \PYG{n+nt}{\PYGZdq{}qubits\PYGZdq{}}\PYG{p}{:} \PYG{p}{[}\PYG{l+m+mi}{1}\PYG{p}{]}
\PYG{p}{\PYGZcb{}}
\end{Verbatim}
\end{jsonexamplebox}
where
\begin{itemize}
\item \texttt{name}: ``reset''.
\item \texttt{qubits}: List of qubits to reset.
\end{itemize}

\subsubsection{Measure}

An OpenQASM measurement projects the state of the specified qubits based on the principles of quantum measurement. The resulting measurement is recorded in the specified memory/register slots. The OpenQASM measurement is defined by the following structure:
\begin{jsonexamplebox}{Measurement Instructiond}
\begin{Verbatim}[commandchars=\\\{\}]
\PYG{p}{\PYGZob{}}
  \PYG{n+nt}{\PYGZdq{}name\PYGZdq{}}\PYG{p}{:} \PYG{l+s+s2}{\PYGZdq{}measure\PYGZdq{}}\PYG{p}{,}
  \PYG{n+nt}{\PYGZdq{}qubits\PYGZdq{}}\PYG{p}{:} \PYG{p}{[}\PYG{l+m+mi}{1}\PYG{p}{,}\PYG{l+m+mi}{2}\PYG{p}{],}
  \PYG{n+nt}{\PYGZdq{}memory\PYGZdq{}}\PYG{p}{:} \PYG{p}{[}\PYG{l+m+mi}{1}\PYG{p}{,}\PYG{l+m+mi}{2}\PYG{p}{],}
  \PYG{n+nt}{\PYGZdq{}register\PYGZdq{}}\PYG{p}{:} \PYG{p}{[}\PYG{l+m+mi}{1}\PYG{p}{,}\PYG{l+m+mi}{2}\PYG{p}{]}
\PYG{p}{\PYGZcb{}}
\end{Verbatim}
\end{jsonexamplebox}
where
\begin{itemize}
\item \texttt{name}: ``measure''.
\item \texttt{qubits}: List of qubits to measure.
\item \texttt{memory}: List of memory slots in which to store the measurement results (must be the same length as qubits). Subsequent measurements that write to the same memory slot will overwrite the previous measurement.
\item \texttt{register} (optional): List of register slots in which to store the measurement results (must be the same length as qubits). These can be used for fast feedback (if allowed). The allowed slots for a qubit may be constrained by the backend \texttt{register\_map}.
\end{itemize}

\subsubsection{Snapshot \label{sect:snapshot}}

The snapshot is a special command reserved for simulators which allows a ``snapshot'' of the simulator state to be recorded. The OpenQASM command for the snapshot is,
\begin{jsonexamplebox}{Snapshot Instruction}
\begin{Verbatim}[commandchars=\\\{\}]
\PYG{p}{\PYGZob{}}
  \PYG{n+nt}{\PYGZdq{}name\PYGZdq{}}\PYG{p}{:} \PYG{l+s+s2}{\PYGZdq{}snapshot\PYGZdq{}}\PYG{p}{,}
  \PYG{n+nt}{\PYGZdq{}label\PYGZdq{}}\PYG{p}{:} \PYG{l+s+s2}{\PYGZdq{}snap1\PYGZdq{}}\PYG{p}{,}
  \PYG{n+nt}{\PYGZdq{}type\PYGZdq{}}\PYG{p}{:} \PYG{l+s+s2}{\PYGZdq{}state\PYGZdq{}}
\PYG{p}{\PYGZcb{}}
\end{Verbatim}
\end{jsonexamplebox}

where
\begin{itemize}
\item \texttt{name}: ``snapshot''.
\item \texttt{label}: Snapshot label which is used to identify the snapshot in the output.
\item \texttt{type}: Type of snapshot, e.g., ``state'' (take a snapshot of the quantum state). The types of snapshots offered are defined in a separate specification document for simulators.
\end{itemize}
There may be additional configuration fields which will be defined in a separate specification document for simulators.

\subsection{Measurement Results \label{sect:gateresults}}

See \S~\ref{sect:measdata}.

\section{OpenQASM \texttt{Qobj} Examples\label{sect:exqasm}}

Here are several examples of \texttt{qoj} OpenQASM data objects and their corresponding result data objects for some common experiments -- a Bell state, teleportation and the three-qubit repetition code.

\subsection{Bell State}

This \texttt{Qobj} defines two experiments which create two different Bell states. The \texttt{Qobj} is,

\begin{jsonexamplebox}{Bell State - Qobj}
\begin{Verbatim}[commandchars=\\\{\}]
\PYG{p}{\PYGZob{}}
  \PYG{n+nt}{\PYGZdq{}qobj\PYGZus{}id\PYGZdq{}}\PYG{p}{:} \PYG{l+s+s2}{\PYGZdq{}bell\PYGZus{}Qobj\PYGZus{}07272018\PYGZdq{}}\PYG{p}{,}
  \PYG{n+nt}{\PYGZdq{}type\PYGZdq{}}\PYG{p}{:} \PYG{l+s+s2}{\PYGZdq{}QASM\PYGZdq{}}\PYG{p}{,}
  \PYG{n+nt}{\PYGZdq{}schema\PYGZus{}version\PYGZdq{}}\PYG{p}{:} \PYG{l+s+s2}{\PYGZdq{}1.0\PYGZdq{}}\PYG{p}{,}
  \PYG{n+nt}{\PYGZdq{}experiments\PYGZdq{}}\PYG{p}{:} \PYG{p}{[}\PYG{err}{bell\PYGZus{}exp1}\PYG{p}{,}\PYG{err}{bell\PYGZus{}exp2}\PYG{p}{],}
  \PYG{n+nt}{\PYGZdq{}header\PYGZdq{}}\PYG{p}{:} \PYG{p}{\PYGZob{}}\PYG{n+nt}{\PYGZdq{}description\PYGZdq{}}\PYG{p}{:} \PYG{l+s+s2}{\PYGZdq{}Bell states\PYGZdq{}}\PYG{p}{\PYGZcb{},}
  \PYG{n+nt}{\PYGZdq{}config\PYGZdq{}}\PYG{p}{:} \PYG{p}{\PYGZob{}}\PYG{n+nt}{\PYGZdq{}shots\PYGZdq{}}\PYG{p}{:} \PYG{l+m+mi}{1000}\PYG{p}{,} \PYG{n+nt}{\PYGZdq{}memory\PYGZus{}slots\PYGZdq{}}\PYG{p}{:} \PYG{l+m+mi}{2}\PYG{p}{\PYGZcb{}}
\PYG{p}{\PYGZcb{}}
\end{Verbatim}
\end{jsonexamplebox}

where the first experiment \texttt{bell\_exp1} is,
\begin{jsonexamplebox}{Bell State - Qobj Experiment 1}
\begin{Verbatim}[commandchars=\\\{\}]
\PYG{p}{\PYGZob{}}
  \PYG{n+nt}{\PYGZdq{}header\PYGZdq{}}\PYG{p}{:} \PYG{p}{\PYGZob{}}\PYG{n+nt}{\PYGZdq{}description\PYGZdq{}}\PYG{p}{:} \PYG{l+s+s2}{\PYGZdq{}|11\PYGZgt{}+|00\PYGZgt{} Bell\PYGZdq{}}\PYG{p}{\PYGZcb{},}
  \PYG{n+nt}{\PYGZdq{}config\PYGZdq{}}\PYG{p}{:} \PYG{p}{\PYGZob{}\PYGZcb{},}
  \PYG{n+nt}{\PYGZdq{}instructions\PYGZdq{}}\PYG{p}{:} \PYG{p}{[}
  \PYG{p}{\PYGZob{}}\PYG{n+nt}{\PYGZdq{}name\PYGZdq{}}\PYG{p}{:} \PYG{l+s+s2}{\PYGZdq{}u2\PYGZdq{}}\PYG{p}{,}
   \PYG{n+nt}{\PYGZdq{}qubits\PYGZdq{}}\PYG{p}{:} \PYG{p}{[}\PYG{l+m+mi}{0}\PYG{p}{],}
   \PYG{n+nt}{\PYGZdq{}params\PYGZdq{}}\PYG{p}{:} \PYG{p}{[}\PYG{l+m+mf}{0.0}\PYG{p}{,}\PYG{l+m+mf}{3.14159}\PYG{p}{]\PYGZcb{},}
  \PYG{p}{\PYGZob{}}\PYG{n+nt}{\PYGZdq{}name\PYGZdq{}}\PYG{p}{:} \PYG{l+s+s2}{\PYGZdq{}cx\PYGZdq{}}\PYG{p}{,}
   \PYG{n+nt}{\PYGZdq{}qubits\PYGZdq{}}\PYG{p}{:} \PYG{p}{[}\PYG{l+m+mi}{0}\PYG{p}{,}\PYG{l+m+mi}{1}\PYG{p}{]\PYGZcb{},}
  \PYG{p}{\PYGZob{}}\PYG{n+nt}{\PYGZdq{}name\PYGZdq{}}\PYG{p}{:} \PYG{l+s+s2}{\PYGZdq{}measure\PYGZdq{}}\PYG{p}{,}
   \PYG{n+nt}{\PYGZdq{}qubits\PYGZdq{}}\PYG{p}{:} \PYG{p}{[}\PYG{l+m+mi}{0}\PYG{p}{,}\PYG{l+m+mi}{1}\PYG{p}{],}
   \PYG{n+nt}{\PYGZdq{}memory\PYGZdq{}}\PYG{p}{:} \PYG{p}{[}\PYG{l+m+mi}{0}\PYG{p}{,}\PYG{l+m+mi}{1}\PYG{p}{]\PYGZcb{}]}
\PYG{p}{\PYGZcb{}}
\end{Verbatim}
\end{jsonexamplebox}
and \texttt{bell\_exp2} is,
\begin{jsonexamplebox}{Bell State - Qobj Experiment 2}
\begin{Verbatim}[commandchars=\\\{\}]
\PYG{p}{\PYGZob{}}
  \PYG{n+nt}{\PYGZdq{}header\PYGZdq{}}\PYG{p}{:} \PYG{p}{\PYGZob{}}\PYG{n+nt}{\PYGZdq{}description\PYGZdq{}}\PYG{p}{:} \PYG{l+s+s2}{\PYGZdq{}|01\PYGZgt{}+|10\PYGZgt{} Bell\PYGZdq{}}\PYG{p}{\PYGZcb{},}
  \PYG{n+nt}{\PYGZdq{}config\PYGZdq{}}\PYG{p}{:} \PYG{p}{\PYGZob{}\PYGZcb{},}
  \PYG{n+nt}{\PYGZdq{}instructions\PYGZdq{}}\PYG{p}{:} \PYG{p}{[\PYGZob{}}\PYG{n+nt}{\PYGZdq{}name\PYGZdq{}}\PYG{p}{:} \PYG{l+s+s2}{\PYGZdq{}u2\PYGZdq{}}\PYG{p}{,} \PYG{n+nt}{\PYGZdq{}qubits\PYGZdq{}}\PYG{p}{:} \PYG{p}{[}\PYG{l+m+mi}{0}\PYG{p}{],}
                    \PYG{n+nt}{\PYGZdq{}params\PYGZdq{}}\PYG{p}{:} \PYG{p}{[}\PYG{l+m+mf}{0.0}\PYG{p}{,}\PYG{l+m+mf}{3.14159}\PYG{p}{]\PYGZcb{},}
  \PYG{p}{\PYGZob{}}\PYG{n+nt}{\PYGZdq{}name\PYGZdq{}}\PYG{p}{:} \PYG{l+s+s2}{\PYGZdq{}cx\PYGZdq{}}\PYG{p}{,} \PYG{n+nt}{\PYGZdq{}qubits\PYGZdq{}}\PYG{p}{:} \PYG{p}{[}\PYG{l+m+mi}{0}\PYG{p}{,}\PYG{l+m+mi}{1}\PYG{p}{]\PYGZcb{},}
  \PYG{p}{\PYGZob{}}\PYG{n+nt}{\PYGZdq{}name\PYGZdq{}}\PYG{p}{:} \PYG{l+s+s2}{\PYGZdq{}u3\PYGZdq{}}\PYG{p}{,} \PYG{n+nt}{\PYGZdq{}qubits\PYGZdq{}}\PYG{p}{:} \PYG{p}{[}\PYG{l+m+mi}{0}\PYG{p}{],}
   \PYG{n+nt}{\PYGZdq{}params\PYGZdq{}}\PYG{p}{:} \PYG{p}{[}\PYG{l+m+mf}{3.14159}\PYG{p}{,}\PYG{l+m+mf}{0.0}\PYG{p}{,}\PYG{l+m+mf}{3.14159}\PYG{p}{]\PYGZcb{},}
  \PYG{p}{\PYGZob{}}\PYG{n+nt}{\PYGZdq{}name\PYGZdq{}}\PYG{p}{:} \PYG{l+s+s2}{\PYGZdq{}measure\PYGZdq{}}\PYG{p}{,} \PYG{n+nt}{\PYGZdq{}qubits\PYGZdq{}}\PYG{p}{:} \PYG{p}{[}\PYG{l+m+mi}{0}\PYG{p}{,}\PYG{l+m+mi}{1}\PYG{p}{],} \PYG{n+nt}{\PYGZdq{}memory\PYGZdq{}}\PYG{p}{:} \PYG{p}{[}\PYG{l+m+mi}{0}\PYG{p}{,}\PYG{l+m+mi}{1}\PYG{p}{]\PYGZcb{}]}
\PYG{p}{\PYGZcb{}}
\end{Verbatim}
\end{jsonexamplebox}

The result data object containing the results from these experiments would be the following (the exact counts will change slightly each time this experiment is run),
\begin{jsonexamplebox}{Bell State - Result}
\begin{Verbatim}[commandchars=\\\{\}]
\PYG{p}{\PYGZob{}}
  \PYG{n+nt}{\PYGZdq{}backend\PYGZus{}name\PYGZdq{}}\PYG{p}{:} \PYG{l+s+s2}{\PYGZdq{}ibmqx2\PYGZdq{}}\PYG{p}{,}
  \PYG{n+nt}{\PYGZdq{}backend\PYGZus{}version\PYGZdq{}}\PYG{p}{:} \PYG{l+s+s2}{\PYGZdq{}2.1.2\PYGZdq{}}\PYG{p}{,}
  \PYG{n+nt}{\PYGZdq{}qobj\PYGZus{}id\PYGZdq{}}\PYG{p}{:} \PYG{l+s+s2}{\PYGZdq{}bell\PYGZus{}Qobj\PYGZus{}07272018\PYGZdq{}}\PYG{p}{,}
  \PYG{n+nt}{\PYGZdq{}job\PYGZus{}id\PYGZdq{}}\PYG{p}{:} \PYG{l+s+s2}{\PYGZdq{}XY1253GSEF\PYGZdq{}}\PYG{p}{,}
  \PYG{n+nt}{\PYGZdq{}date\PYGZdq{}}\PYG{p}{:} \PYG{l+s+s2}{\PYGZdq{}2018\PYGZhy{}04\PYGZhy{}02 15:00:00Z\PYGZdq{}}\PYG{p}{,}
  \PYG{n+nt}{\PYGZdq{}header\PYGZdq{}}\PYG{p}{:} \PYG{p}{\PYGZob{}}\PYG{n+nt}{\PYGZdq{}description\PYGZdq{}}\PYG{p}{:} \PYG{l+s+s2}{\PYGZdq{}Bell states\PYGZdq{}}\PYG{p}{\PYGZcb{},}
  \PYG{n+nt}{\PYGZdq{}success\PYGZdq{}}\PYG{p}{:} \PYG{k+kc}{true}\PYG{p}{,}
  \PYG{n+nt}{\PYGZdq{}results\PYGZdq{}}\PYG{p}{:} \PYG{p}{[}\PYG{err}{expResult1}\PYG{p}{,}\PYG{err}{expResult2}\PYG{p}{]}
\PYG{p}{\PYGZcb{}}
\end{Verbatim}
\end{jsonexamplebox}
where \texttt{exp\_result1} is
\begin{jsonexamplebox}{Bell State - Experiment 1 Result}
\begin{Verbatim}[commandchars=\\\{\}]
\PYG{p}{\PYGZob{}}
  \PYG{n+nt}{\PYGZdq{}shots\PYGZdq{}}\PYG{p}{:} \PYG{l+m+mi}{1000}\PYG{p}{,}
  \PYG{n+nt}{\PYGZdq{}status\PYGZdq{}}\PYG{p}{:} \PYG{l+s+s2}{\PYGZdq{}DONE\PYGZdq{}}\PYG{p}{,}
  \PYG{n+nt}{\PYGZdq{}success\PYGZdq{}}\PYG{p}{:} \PYG{k+kc}{true}\PYG{p}{,}
  \PYG{n+nt}{\PYGZdq{}header\PYGZdq{}}\PYG{p}{:} \PYG{p}{\PYGZob{}}\PYG{n+nt}{\PYGZdq{}description\PYGZdq{}}\PYG{p}{:} \PYG{l+s+s2}{\PYGZdq{}|11\PYGZgt{}+|00\PYGZgt{} Bell\PYGZdq{}}\PYG{p}{\PYGZcb{},}
  \PYG{n+nt}{\PYGZdq{}data\PYGZdq{}}\PYG{p}{:} \PYG{p}{\PYGZob{}}\PYG{n+nt}{\PYGZdq{}counts\PYGZdq{}}\PYG{p}{:} \PYG{p}{\PYGZob{}}\PYG{n+nt}{\PYGZdq{}0x0\PYGZdq{}}\PYG{p}{:} \PYG{l+m+mi}{450}\PYG{p}{,} \PYG{n+nt}{\PYGZdq{}0x1\PYGZdq{}}\PYG{p}{:} \PYG{l+m+mi}{10}\PYG{p}{,}
                      \PYG{n+nt}{\PYGZdq{}0x2\PYGZdq{}}\PYG{p}{:} \PYG{l+m+mi}{20}\PYG{p}{,} \PYG{n+nt}{\PYGZdq{}0x3\PYGZdq{}}\PYG{p}{:} \PYG{l+m+mi}{520}\PYG{p}{\PYGZcb{},}
           \PYG{n+nt}{\PYGZdq{}memory\PYGZdq{}}\PYG{p}{:} \PYG{p}{[}\PYG{l+s+s2}{\PYGZdq{}0x0\PYGZdq{}}\PYG{p}{,}\PYG{l+s+s2}{\PYGZdq{}0x0\PYGZdq{}}\PYG{p}{,}\PYG{l+s+s2}{\PYGZdq{}0x3\PYGZdq{}}\PYG{p}{,}\PYG{l+s+s2}{\PYGZdq{}0x2\PYGZdq{}}\PYG{p}{,}\PYG{l+s+s2}{\PYGZdq{}0x3\PYGZdq{}}\PYG{p}{,}\PYG{l+s+s2}{\PYGZdq{}0x0\PYGZdq{}}\PYG{p}{,}\PYG{err}{...}\PYG{p}{]\PYGZcb{}}
\PYG{p}{\PYGZcb{}}
\end{Verbatim}
\end{jsonexamplebox}
and \texttt{exp\_result2} is
\begin{jsonexamplebox}{Bell State - Experiment 2 Result}
\begin{Verbatim}[commandchars=\\\{\}]
\PYG{p}{\PYGZob{}}
  \PYG{n+nt}{\PYGZdq{}shots\PYGZdq{}}\PYG{p}{:} \PYG{l+m+mi}{1000}\PYG{p}{,}
  \PYG{n+nt}{\PYGZdq{}status\PYGZdq{}}\PYG{p}{:} \PYG{l+s+s2}{\PYGZdq{}DONE\PYGZdq{}}\PYG{p}{,}
  \PYG{n+nt}{\PYGZdq{}success\PYGZdq{}}\PYG{p}{:} \PYG{k+kc}{true}\PYG{p}{,}
  \PYG{n+nt}{\PYGZdq{}header\PYGZdq{}}\PYG{p}{:} \PYG{p}{\PYGZob{}}\PYG{n+nt}{\PYGZdq{}description\PYGZdq{}}\PYG{p}{:} \PYG{l+s+s2}{\PYGZdq{}|01\PYGZgt{}+|10\PYGZgt{} Bell\PYGZdq{}}\PYG{p}{\PYGZcb{},}
  \PYG{n+nt}{\PYGZdq{}data\PYGZdq{}}\PYG{p}{:} \PYG{p}{\PYGZob{}}\PYG{n+nt}{\PYGZdq{}counts\PYGZdq{}}\PYG{p}{:} \PYG{p}{\PYGZob{}}\PYG{n+nt}{\PYGZdq{}0x0\PYGZdq{}}\PYG{p}{:} \PYG{l+m+mi}{5}\PYG{p}{,} \PYG{n+nt}{\PYGZdq{}0x1\PYGZdq{}}\PYG{p}{:} \PYG{l+m+mi}{510}\PYG{p}{,}
                      \PYG{n+nt}{\PYGZdq{}0x2\PYGZdq{}}\PYG{p}{:} \PYG{l+m+mi}{480}\PYG{p}{,} \PYG{n+nt}{\PYGZdq{}0x3\PYGZdq{}}\PYG{p}{:} \PYG{l+m+mi}{5}\PYG{p}{\PYGZcb{},}
           \PYG{n+nt}{\PYGZdq{}memory\PYGZdq{}}\PYG{p}{:} \PYG{p}{[}\PYG{l+s+s2}{\PYGZdq{}0x2\PYGZdq{}}\PYG{p}{,}\PYG{l+s+s2}{\PYGZdq{}0x1\PYGZdq{}}\PYG{p}{,}\PYG{l+s+s2}{\PYGZdq{}0x1\PYGZdq{}}\PYG{p}{,}\PYG{l+s+s2}{\PYGZdq{}0x2\PYGZdq{}}\PYG{p}{,}\PYG{l+s+s2}{\PYGZdq{}0x2\PYGZdq{}}\PYG{p}{,}\PYG{l+s+s2}{\PYGZdq{}0x2\PYGZdq{}}\PYG{p}{,}\PYG{err}{...}\PYG{p}{]\PYGZcb{}}
\PYG{p}{\PYGZcb{}}
\end{Verbatim}
\end{jsonexamplebox}

\subsection{Teleportation}

This \texttt{Qobj} defines an experiment to ``teleport'' a quantum state. This circuit takes the state of Q0 (which starts in the superposition state) and ``teleports'' it to the state of Q2 using a Bell state between Q1 and Q2 and conditional operations on Q2 based on joint measurement of Q0 and Q1. Assume the device has two register slots available for feedback and that there are no conditions on feedback or measurement to these registers. The \texttt{Qobj} is,

\begin{jsonexamplebox}{Teleportation - Qobj}
\begin{Verbatim}[commandchars=\\\{\}]
\PYG{p}{\PYGZob{}}
  \PYG{n+nt}{\PYGZdq{}qobj\PYGZus{}id\PYGZdq{}}\PYG{p}{:} \PYG{l+s+s2}{\PYGZdq{}teleport\PYGZus{}07272018\PYGZdq{}}\PYG{p}{,}
  \PYG{n+nt}{\PYGZdq{}type\PYGZdq{}}\PYG{p}{:} \PYG{l+s+s2}{\PYGZdq{}QASM\PYGZdq{}}\PYG{p}{,}
  \PYG{n+nt}{\PYGZdq{}schema\PYGZus{}version\PYGZdq{}}\PYG{p}{:} \PYG{l+s+s2}{\PYGZdq{}1.0\PYGZdq{}}\PYG{p}{,}
  \PYG{n+nt}{\PYGZdq{}experiments\PYGZdq{}}\PYG{p}{:} \PYG{p}{[}\PYG{err}{teleport\PYGZus{}exp1}\PYG{p}{],}
  \PYG{n+nt}{\PYGZdq{}header\PYGZdq{}}\PYG{p}{:} \PYG{p}{\PYGZob{}}\PYG{n+nt}{\PYGZdq{}description\PYGZdq{}}\PYG{p}{:} \PYG{l+s+s2}{\PYGZdq{}Teleport circuit\PYGZdq{}}\PYG{p}{\PYGZcb{},}
  \PYG{n+nt}{\PYGZdq{}config\PYGZdq{}}\PYG{p}{:} \PYG{p}{\PYGZob{}}\PYG{n+nt}{\PYGZdq{}shots\PYGZdq{}}\PYG{p}{:} \PYG{l+m+mi}{1000}\PYG{p}{,} \PYG{n+nt}{\PYGZdq{}memory\PYGZus{}slots\PYGZdq{}}\PYG{p}{:} \PYG{l+m+mi}{2}\PYG{p}{\PYGZcb{}}
\PYG{p}{\PYGZcb{}}
\end{Verbatim}
\end{jsonexamplebox}
where \texttt{teleport\_exp1} is,
\begin{jsonexamplebox}{Teleportation - Qobj Experiment}
\begin{Verbatim}[commandchars=\\\{\}]
\PYG{p}{\PYGZob{}}
  \PYG{n+nt}{\PYGZdq{}header\PYGZdq{}}\PYG{p}{:} \PYG{p}{\PYGZob{}\PYGZcb{},}
  \PYG{n+nt}{\PYGZdq{}config\PYGZdq{}}\PYG{p}{:} \PYG{p}{\PYGZob{}\PYGZcb{},}
  \PYG{n+nt}{\PYGZdq{}instructions\PYGZdq{}}\PYG{p}{:} \PYG{p}{[}
  \PYG{p}{\PYGZob{}}\PYG{n+nt}{\PYGZdq{}name\PYGZdq{}}\PYG{p}{:} \PYG{l+s+s2}{\PYGZdq{}u2\PYGZdq{}}\PYG{p}{,} \PYG{n+nt}{\PYGZdq{}qubits\PYGZdq{}}\PYG{p}{:} \PYG{p}{[}\PYG{l+m+mi}{0}\PYG{p}{],} \PYG{n+nt}{\PYGZdq{}params\PYGZdq{}}\PYG{p}{:} \PYG{p}{[}\PYG{l+m+mf}{0.0}\PYG{p}{,}\PYG{l+m+mf}{3.14159}\PYG{p}{]\PYGZcb{},}
  \PYG{p}{\PYGZob{}}\PYG{n+nt}{\PYGZdq{}name\PYGZdq{}}\PYG{p}{:} \PYG{l+s+s2}{\PYGZdq{}u2\PYGZdq{}}\PYG{p}{,} \PYG{n+nt}{\PYGZdq{}qubits\PYGZdq{}}\PYG{p}{:} \PYG{p}{[}\PYG{l+m+mi}{1}\PYG{p}{],} \PYG{n+nt}{\PYGZdq{}params\PYGZdq{}}\PYG{p}{:} \PYG{p}{[}\PYG{l+m+mf}{0.0}\PYG{p}{,}\PYG{l+m+mf}{3.14159}\PYG{p}{]\PYGZcb{},}
  \PYG{p}{\PYGZob{}}\PYG{n+nt}{\PYGZdq{}name\PYGZdq{}}\PYG{p}{:} \PYG{l+s+s2}{\PYGZdq{}cx\PYGZdq{}}\PYG{p}{,} \PYG{n+nt}{\PYGZdq{}qubits\PYGZdq{}}\PYG{p}{:} \PYG{p}{[}\PYG{l+m+mi}{1}\PYG{p}{,}\PYG{l+m+mi}{2}\PYG{p}{]\PYGZcb{},}
  \PYG{p}{\PYGZob{}}\PYG{n+nt}{\PYGZdq{}name\PYGZdq{}}\PYG{p}{:} \PYG{l+s+s2}{\PYGZdq{}cx\PYGZdq{}}\PYG{p}{,} \PYG{n+nt}{\PYGZdq{}qubits\PYGZdq{}}\PYG{p}{:} \PYG{p}{[}\PYG{l+m+mi}{0}\PYG{p}{,}\PYG{l+m+mi}{1}\PYG{p}{]\PYGZcb{},}
  \PYG{p}{\PYGZob{}}\PYG{n+nt}{\PYGZdq{}name\PYGZdq{}}\PYG{p}{:} \PYG{l+s+s2}{\PYGZdq{}measure\PYGZdq{}}\PYG{p}{,} \PYG{n+nt}{\PYGZdq{}qubits\PYGZdq{}}\PYG{p}{:} \PYG{p}{[}\PYG{l+m+mi}{1}\PYG{p}{],}
   \PYG{n+nt}{\PYGZdq{}memory\PYGZdq{}}\PYG{p}{:} \PYG{p}{[}\PYG{l+m+mi}{1}\PYG{p}{],} \PYG{n+nt}{\PYGZdq{}register\PYGZdq{}}\PYG{p}{:} \PYG{p}{[}\PYG{l+m+mi}{1}\PYG{p}{]\PYGZcb{},}
  \PYG{p}{\PYGZob{}}\PYG{n+nt}{\PYGZdq{}name\PYGZdq{}}\PYG{p}{:} \PYG{l+s+s2}{\PYGZdq{}u2\PYGZdq{}}\PYG{p}{,} \PYG{n+nt}{\PYGZdq{}qubits\PYGZdq{}}\PYG{p}{:} \PYG{p}{[}\PYG{l+m+mi}{0}\PYG{p}{],} \PYG{n+nt}{\PYGZdq{}params\PYGZdq{}}\PYG{p}{:} \PYG{p}{[}\PYG{l+m+mf}{0.0}\PYG{p}{,}\PYG{l+m+mf}{3.14159}\PYG{p}{]\PYGZcb{},}
  \PYG{p}{\PYGZob{}}\PYG{n+nt}{\PYGZdq{}name\PYGZdq{}}\PYG{p}{:} \PYG{l+s+s2}{\PYGZdq{}measure\PYGZdq{}}\PYG{p}{,} \PYG{n+nt}{\PYGZdq{}qubits\PYGZdq{}}\PYG{p}{:} \PYG{p}{[}\PYG{l+m+mi}{0}\PYG{p}{],}
   \PYG{n+nt}{\PYGZdq{}memory\PYGZdq{}}\PYG{p}{:} \PYG{p}{[}\PYG{l+m+mi}{0}\PYG{p}{],} \PYG{n+nt}{\PYGZdq{}register\PYGZdq{}}\PYG{p}{:} \PYG{p}{[}\PYG{l+m+mi}{0}\PYG{p}{]\PYGZcb{},}
  \PYG{p}{\PYGZob{}}\PYG{n+nt}{\PYGZdq{}name\PYGZdq{}}\PYG{p}{:} \PYG{l+s+s2}{\PYGZdq{}u1\PYGZdq{}}\PYG{p}{,} \PYG{n+nt}{\PYGZdq{}qubits\PYGZdq{}}\PYG{p}{:} \PYG{p}{[}\PYG{l+m+mi}{2}\PYG{p}{],} \PYG{n+nt}{\PYGZdq{}params\PYGZdq{}}\PYG{p}{:} \PYG{p}{[}\PYG{l+m+mf}{3.14159}\PYG{p}{],}
   \PYG{n+nt}{\PYGZdq{}conditional\PYGZdq{}}\PYG{p}{:} \PYG{l+m+mi}{0}\PYG{p}{\PYGZcb{},}
  \PYG{p}{\PYGZob{}}\PYG{n+nt}{\PYGZdq{}name\PYGZdq{}}\PYG{p}{:} \PYG{l+s+s2}{\PYGZdq{}u3\PYGZdq{}}\PYG{p}{,} \PYG{n+nt}{\PYGZdq{}qubits\PYGZdq{}}\PYG{p}{:} \PYG{p}{[}\PYG{l+m+mi}{2}\PYG{p}{],}
   \PYG{n+nt}{\PYGZdq{}params\PYGZdq{}}\PYG{p}{:} \PYG{p}{[}\PYG{l+m+mf}{3.14159}\PYG{p}{,}\PYG{l+m+mf}{0.0}\PYG{p}{,}\PYG{l+m+mf}{3.14159}\PYG{p}{],}\PYG{n+nt}{\PYGZdq{}conditional\PYGZdq{}}\PYG{p}{:} \PYG{l+m+mi}{1}\PYG{p}{\PYGZcb{}]}
\PYG{p}{\PYGZcb{}}
\end{Verbatim}
\end{jsonexamplebox}

In this particular example, the only measurements stored in memory are the first measurements on Q0 and Q1, which will be equally distributed since the starting state of Q0 is the superposition state, The result data object containing the results would be the following (the exact counts will change slightly each time this experiment is run),
\begin{jsonexamplebox}{Teleportation - Result}
\begin{Verbatim}[commandchars=\\\{\}]
\PYG{p}{\PYGZob{}}
  \PYG{n+nt}{\PYGZdq{}backend\PYGZus{}name\PYGZdq{}}\PYG{p}{:} \PYG{l+s+s2}{\PYGZdq{}ibmqx2\PYGZdq{}}\PYG{p}{,}
  \PYG{n+nt}{\PYGZdq{}backend\PYGZus{}version\PYGZdq{}}\PYG{p}{:} \PYG{l+s+s2}{\PYGZdq{}2.1.2\PYGZdq{}}\PYG{p}{,}
  \PYG{n+nt}{\PYGZdq{}qobj\PYGZus{}id\PYGZdq{}}\PYG{p}{:} \PYG{l+s+s2}{\PYGZdq{}teleport\PYGZus{}07272018\PYGZdq{}}\PYG{p}{,}
  \PYG{n+nt}{\PYGZdq{}job\PYGZus{}id\PYGZdq{}}\PYG{p}{:} \PYG{l+s+s2}{\PYGZdq{}XX12353ISEL\PYGZdq{}}\PYG{p}{,}
  \PYG{n+nt}{\PYGZdq{}date\PYGZdq{}}\PYG{p}{:} \PYG{l+s+s2}{\PYGZdq{}2018\PYGZhy{}04\PYGZhy{}02 15:00:00Z\PYGZdq{}}\PYG{p}{,}
  \PYG{n+nt}{\PYGZdq{}header\PYGZdq{}}\PYG{p}{:} \PYG{p}{\PYGZob{}}\PYG{n+nt}{\PYGZdq{}description\PYGZdq{}}\PYG{p}{:} \PYG{l+s+s2}{\PYGZdq{}Teleport circuit\PYGZdq{}}\PYG{p}{\PYGZcb{},}
  \PYG{n+nt}{\PYGZdq{}success\PYGZdq{}}\PYG{p}{:} \PYG{k+kc}{true}\PYG{p}{,}
  \PYG{n+nt}{\PYGZdq{}results\PYGZdq{}}\PYG{p}{:} \PYG{p}{[}\PYG{err}{exp\PYGZus{}result1}\PYG{p}{]}
\PYG{p}{\PYGZcb{}}
\end{Verbatim}
\end{jsonexamplebox}
where \texttt{expResults1} is
\begin{jsonexamplebox}{Teleportation - Experiment Result}
\begin{Verbatim}[commandchars=\\\{\}]
\PYG{p}{\PYGZob{}}
  \PYG{n+nt}{\PYGZdq{}shots\PYGZdq{}}\PYG{p}{:} \PYG{l+m+mi}{1000}\PYG{p}{,}
  \PYG{n+nt}{\PYGZdq{}status\PYGZdq{}}\PYG{p}{:} \PYG{l+s+s2}{\PYGZdq{}DONE\PYGZdq{}}\PYG{p}{,}
  \PYG{n+nt}{\PYGZdq{}success\PYGZdq{}}\PYG{p}{:} \PYG{k+kc}{true}\PYG{p}{,}
  \PYG{n+nt}{\PYGZdq{}header\PYGZdq{}}\PYG{p}{:} \PYG{p}{\PYGZob{}\PYGZcb{},}
  \PYG{n+nt}{\PYGZdq{}data\PYGZdq{}}\PYG{p}{:} \PYG{p}{\PYGZob{}}\PYG{n+nt}{\PYGZdq{}counts\PYGZdq{}}\PYG{p}{:} \PYG{p}{\PYGZob{}}\PYG{n+nt}{\PYGZdq{}0x0\PYGZdq{}}\PYG{p}{:} \PYG{l+m+mi}{250}\PYG{p}{,} \PYG{n+nt}{\PYGZdq{}0x1\PYGZdq{}}\PYG{p}{:} \PYG{l+m+mi}{220}\PYG{p}{,}
                      \PYG{n+nt}{\PYGZdq{}0x2\PYGZdq{}}\PYG{p}{:} \PYG{l+m+mi}{260}\PYG{p}{,} \PYG{n+nt}{\PYGZdq{}0x3\PYGZdq{}}\PYG{p}{:} \PYG{l+m+mi}{270}\PYG{p}{\PYGZcb{},}
          \PYG{n+nt}{\PYGZdq{}memory\PYGZdq{}}\PYG{p}{:} \PYG{p}{[}\PYG{l+s+s2}{\PYGZdq{}0x0\PYGZdq{}}\PYG{p}{,}\PYG{l+s+s2}{\PYGZdq{}0x1\PYGZdq{}}\PYG{p}{,}\PYG{l+s+s2}{\PYGZdq{}0x2\PYGZdq{}}\PYG{p}{,}\PYG{l+s+s2}{\PYGZdq{}0x1\PYGZdq{}}\PYG{p}{,}\PYG{l+s+s2}{\PYGZdq{}0x3\PYGZdq{}}\PYG{p}{,}\PYG{l+s+s2}{\PYGZdq{}0x0\PYGZdq{}}\PYG{p}{,}\PYG{err}{...}\PYG{p}{]\PYGZcb{}}
\PYG{p}{\PYGZcb{}}
\end{Verbatim}
\end{jsonexamplebox}

\subsection{Three-Qubit Repetition Code}

This \texttt{Qobj} defines an experiment to perform bit-flip error correction on a quantum state encoded in three qubits in the states $|000\rangle$ and $|111\rangle$. Parity measurements to two ancillas are used to detect if a bit flip has occurred on one of the data qubits. If a bit flip is detected, then a conditional pi-pulse is performed to undo the error. The \texttt{Qobj} is,
\begin{jsonexamplebox}{Three-Qubit Repetition Code - Qobj}
\begin{Verbatim}[commandchars=\\\{\}]
\PYG{p}{\PYGZob{}}
  \PYG{n+nt}{\PYGZdq{}qobj\PYGZus{}id\PYGZdq{}}\PYG{p}{:} \PYG{l+s+s2}{\PYGZdq{}repcode\PYGZus{}07272018\PYGZdq{}}\PYG{p}{,}
  \PYG{n+nt}{\PYGZdq{}type\PYGZdq{}}\PYG{p}{:} \PYG{l+s+s2}{\PYGZdq{}QASM\PYGZdq{}}\PYG{p}{,}
  \PYG{n+nt}{\PYGZdq{}schema\PYGZus{}version\PYGZdq{}}\PYG{p}{:} \PYG{l+s+s2}{\PYGZdq{}1.0\PYGZdq{}}\PYG{p}{,}
  \PYG{n+nt}{\PYGZdq{}experiments\PYGZdq{}}\PYG{p}{:} \PYG{p}{[}\PYG{err}{repcode\PYGZus{}exp1}\PYG{p}{],}
  \PYG{n+nt}{\PYGZdq{}header\PYGZdq{}}\PYG{p}{:} \PYG{p}{\PYGZob{}}\PYG{n+nt}{\PYGZdq{}description\PYGZdq{}}\PYG{p}{:} \PYG{l+s+s2}{\PYGZdq{}Three qubit repetition code\PYGZdq{}}\PYG{p}{\PYGZcb{},}
  \PYG{n+nt}{\PYGZdq{}config\PYGZdq{}}\PYG{p}{:} \PYG{p}{\PYGZob{}}\PYG{n+nt}{\PYGZdq{}shots\PYGZdq{}}\PYG{p}{:} \PYG{l+m+mi}{1000}\PYG{p}{,} \PYG{n+nt}{\PYGZdq{}memory\PYGZus{}slots\PYGZdq{}}\PYG{p}{:} \PYG{l+m+mi}{5}\PYG{p}{\PYGZcb{}}
\PYG{p}{\PYGZcb{}}
\end{Verbatim}
\end{jsonexamplebox}

where the circuit commands are,
\begin{jsonexamplebox}{Three-Qubit Reptition Code - Experiment}
\begin{Verbatim}[commandchars=\\\{\}]
\PYG{p}{\PYGZob{}}
  \PYG{n+nt}{\PYGZdq{}header\PYGZdq{}}\PYG{p}{:} \PYG{p}{\PYGZob{}}\PYG{n+nt}{\PYGZdq{}data\PYGZus{}qubits\PYGZdq{}}\PYG{p}{:} \PYG{p}{[}\PYG{l+m+mi}{0}\PYG{p}{,}\PYG{l+m+mi}{1}\PYG{p}{,}\PYG{l+m+mi}{2}\PYG{p}{],}
             \PYG{n+nt}{\PYGZdq{}ancilla\PYGZus{}qubits\PYGZdq{}}\PYG{p}{:} \PYG{p}{[}\PYG{l+m+mi}{3}\PYG{p}{,}\PYG{l+m+mi}{4}\PYG{p}{]\PYGZcb{},}
  \PYG{n+nt}{\PYGZdq{}config\PYGZdq{}}\PYG{p}{:} \PYG{p}{\PYGZob{}\PYGZcb{},}
  \PYG{n+nt}{\PYGZdq{}instructions\PYGZdq{}}\PYG{p}{:} \PYG{p}{[}
  \PYG{p}{\PYGZob{}}\PYG{n+nt}{\PYGZdq{}name\PYGZdq{}}\PYG{p}{:} \PYG{l+s+s2}{\PYGZdq{}u2\PYGZdq{}}\PYG{p}{,} \PYG{n+nt}{\PYGZdq{}qubits\PYGZdq{}}\PYG{p}{:} \PYG{p}{[}\PYG{l+m+mi}{0}\PYG{p}{],} \PYG{n+nt}{\PYGZdq{}params\PYGZdq{}}\PYG{p}{:} \PYG{p}{[}\PYG{l+m+mf}{0.0}\PYG{p}{,}\PYG{l+m+mf}{3.14159}\PYG{p}{]\PYGZcb{},}
  \PYG{p}{\PYGZob{}}\PYG{n+nt}{\PYGZdq{}name\PYGZdq{}}\PYG{p}{:} \PYG{l+s+s2}{\PYGZdq{}cx\PYGZdq{}}\PYG{p}{,} \PYG{n+nt}{\PYGZdq{}qubits\PYGZdq{}}\PYG{p}{:} \PYG{p}{[}\PYG{l+m+mi}{0}\PYG{p}{,}\PYG{l+m+mi}{1}\PYG{p}{]\PYGZcb{},}
  \PYG{p}{\PYGZob{}}\PYG{n+nt}{\PYGZdq{}name\PYGZdq{}}\PYG{p}{:} \PYG{l+s+s2}{\PYGZdq{}cx\PYGZdq{}}\PYG{p}{,} \PYG{n+nt}{\PYGZdq{}qubits\PYGZdq{}}\PYG{p}{:} \PYG{p}{[}\PYG{l+m+mi}{0}\PYG{p}{,}\PYG{l+m+mi}{2}\PYG{p}{]\PYGZcb{},}
  \PYG{p}{\PYGZob{}}\PYG{n+nt}{\PYGZdq{}name\PYGZdq{}}\PYG{p}{:} \PYG{l+s+s2}{\PYGZdq{}cx\PYGZdq{}}\PYG{p}{,} \PYG{n+nt}{\PYGZdq{}qubits\PYGZdq{}}\PYG{p}{:} \PYG{p}{[}\PYG{l+m+mi}{0}\PYG{p}{,}\PYG{l+m+mi}{3}\PYG{p}{]\PYGZcb{},}
  \PYG{p}{\PYGZob{}}\PYG{n+nt}{\PYGZdq{}name\PYGZdq{}}\PYG{p}{:} \PYG{l+s+s2}{\PYGZdq{}cx\PYGZdq{}}\PYG{p}{,} \PYG{n+nt}{\PYGZdq{}qubits\PYGZdq{}}\PYG{p}{:} \PYG{p}{[}\PYG{l+m+mi}{1}\PYG{p}{,}\PYG{l+m+mi}{3}\PYG{p}{]\PYGZcb{},}
  \PYG{p}{\PYGZob{}}\PYG{n+nt}{\PYGZdq{}name\PYGZdq{}}\PYG{p}{:} \PYG{l+s+s2}{\PYGZdq{}cx\PYGZdq{}}\PYG{p}{,} \PYG{n+nt}{\PYGZdq{}qubits\PYGZdq{}}\PYG{p}{:} \PYG{p}{[}\PYG{l+m+mi}{1}\PYG{p}{,}\PYG{l+m+mi}{4}\PYG{p}{]\PYGZcb{},}
  \PYG{p}{\PYGZob{}}\PYG{n+nt}{\PYGZdq{}name\PYGZdq{}}\PYG{p}{:} \PYG{l+s+s2}{\PYGZdq{}cx\PYGZdq{}}\PYG{p}{,} \PYG{n+nt}{\PYGZdq{}qubits\PYGZdq{}}\PYG{p}{:} \PYG{p}{[}\PYG{l+m+mi}{2}\PYG{p}{,}\PYG{l+m+mi}{4}\PYG{p}{]\PYGZcb{},}
  \PYG{p}{\PYGZob{}}\PYG{n+nt}{\PYGZdq{}name\PYGZdq{}}\PYG{p}{:} \PYG{l+s+s2}{\PYGZdq{}measure\PYGZdq{}}\PYG{p}{,} \PYG{n+nt}{\PYGZdq{}qubits\PYGZdq{}}\PYG{p}{:} \PYG{p}{[}\PYG{l+m+mi}{3}\PYG{p}{],} \PYG{n+nt}{\PYGZdq{}memory\PYGZdq{}}\PYG{p}{:} \PYG{p}{[}\PYG{l+m+mi}{0}\PYG{p}{],}
   \PYG{n+nt}{\PYGZdq{}register\PYGZdq{}}\PYG{p}{:} \PYG{p}{[}\PYG{l+m+mi}{0}\PYG{p}{]\PYGZcb{},}
  \PYG{p}{\PYGZob{}}\PYG{n+nt}{\PYGZdq{}name\PYGZdq{}}\PYG{p}{:} \PYG{l+s+s2}{\PYGZdq{}measure\PYGZdq{}}\PYG{p}{,} \PYG{n+nt}{\PYGZdq{}qubits\PYGZdq{}}\PYG{p}{:} \PYG{p}{[}\PYG{l+m+mi}{4}\PYG{p}{],} \PYG{n+nt}{\PYGZdq{}memory\PYGZdq{}}\PYG{p}{:} \PYG{p}{[}\PYG{l+m+mi}{1}\PYG{p}{],}
   \PYG{n+nt}{\PYGZdq{}register\PYGZdq{}}\PYG{p}{:} \PYG{p}{[}\PYG{l+m+mi}{1}\PYG{p}{]\PYGZcb{},}
  \PYG{p}{\PYGZob{}}\PYG{n+nt}{\PYGZdq{}name\PYGZdq{}}\PYG{p}{:} \PYG{l+s+s2}{\PYGZdq{}bfunc\PYGZdq{}}\PYG{p}{,} \PYG{n+nt}{\PYGZdq{}mask\PYGZdq{}}\PYG{p}{:} \PYG{l+s+s2}{\PYGZdq{}0x3\PYGZdq{}}\PYG{p}{,} \PYG{n+nt}{\PYGZdq{}relation\PYGZdq{}}\PYG{p}{:} \PYG{l+s+s2}{\PYGZdq{}==\PYGZdq{}}\PYG{p}{,}
   \PYG{n+nt}{\PYGZdq{}val\PYGZdq{}}\PYG{p}{:} \PYG{l+s+s2}{\PYGZdq{}0x1\PYGZdq{}}\PYG{p}{,} \PYG{n+nt}{\PYGZdq{}register\PYGZdq{}}\PYG{p}{:} \PYG{p}{[}\PYG{l+m+mi}{2}\PYG{p}{]\PYGZcb{},}
  \PYG{p}{\PYGZob{}}\PYG{n+nt}{\PYGZdq{}name\PYGZdq{}}\PYG{p}{:} \PYG{l+s+s2}{\PYGZdq{}bfunc\PYGZdq{}}\PYG{p}{,} \PYG{n+nt}{\PYGZdq{}mask\PYGZdq{}}\PYG{p}{:} \PYG{l+s+s2}{\PYGZdq{}0x3\PYGZdq{}}\PYG{p}{,} \PYG{n+nt}{\PYGZdq{}relation\PYGZdq{}}\PYG{p}{:} \PYG{l+s+s2}{\PYGZdq{}==\PYGZdq{}}\PYG{p}{,}
   \PYG{n+nt}{\PYGZdq{}val\PYGZdq{}}\PYG{p}{:} \PYG{l+s+s2}{\PYGZdq{}0x2\PYGZdq{}}\PYG{p}{,} \PYG{n+nt}{\PYGZdq{}register\PYGZdq{}}\PYG{p}{:} \PYG{p}{[}\PYG{l+m+mi}{3}\PYG{p}{]\PYGZcb{},}
  \PYG{p}{\PYGZob{}}\PYG{n+nt}{\PYGZdq{}name\PYGZdq{}}\PYG{p}{:} \PYG{l+s+s2}{\PYGZdq{}bfunc\PYGZdq{}}\PYG{p}{,} \PYG{n+nt}{\PYGZdq{}mask\PYGZdq{}}\PYG{p}{:} \PYG{l+s+s2}{\PYGZdq{}0x3\PYGZdq{}}\PYG{p}{,} \PYG{n+nt}{\PYGZdq{}relation\PYGZdq{}}\PYG{p}{:} \PYG{l+s+s2}{\PYGZdq{}==\PYGZdq{}}\PYG{p}{,}
   \PYG{n+nt}{\PYGZdq{}val\PYGZdq{}}\PYG{p}{:} \PYG{l+s+s2}{\PYGZdq{}0x3\PYGZdq{}}\PYG{p}{,} \PYG{n+nt}{\PYGZdq{}register\PYGZdq{}}\PYG{p}{:} \PYG{p}{[}\PYG{l+m+mi}{4}\PYG{p}{]\PYGZcb{},}
  \PYG{p}{\PYGZob{}}\PYG{n+nt}{\PYGZdq{}name\PYGZdq{}}\PYG{p}{:} \PYG{l+s+s2}{\PYGZdq{}u3\PYGZdq{}}\PYG{p}{,} \PYG{n+nt}{\PYGZdq{}qubits\PYGZdq{}}\PYG{p}{:} \PYG{p}{[}\PYG{l+m+mi}{0}\PYG{p}{],}
   \PYG{n+nt}{\PYGZdq{}params\PYGZdq{}}\PYG{p}{:} \PYG{p}{[}\PYG{l+m+mf}{3.14159}\PYG{p}{,}\PYG{l+m+mf}{0.0}\PYG{p}{,}\PYG{l+m+mf}{3.14159}\PYG{p}{],} \PYG{n+nt}{\PYGZdq{}conditional\PYGZdq{}}\PYG{p}{:} \PYG{l+m+mi}{2}\PYG{p}{\PYGZcb{},}
  \PYG{p}{\PYGZob{}}\PYG{n+nt}{\PYGZdq{}name\PYGZdq{}}\PYG{p}{:} \PYG{l+s+s2}{\PYGZdq{}u3\PYGZdq{}}\PYG{p}{,} \PYG{n+nt}{\PYGZdq{}qubits\PYGZdq{}}\PYG{p}{:} \PYG{p}{[}\PYG{l+m+mi}{1}\PYG{p}{],}
   \PYG{n+nt}{\PYGZdq{}params\PYGZdq{}}\PYG{p}{:} \PYG{p}{[}\PYG{l+m+mf}{3.14159}\PYG{p}{,}\PYG{l+m+mf}{0.0}\PYG{p}{,}\PYG{l+m+mf}{3.14159}\PYG{p}{],} \PYG{n+nt}{\PYGZdq{}conditional\PYGZdq{}}\PYG{p}{:} \PYG{l+m+mi}{4}\PYG{p}{\PYGZcb{},}
  \PYG{p}{\PYGZob{}}\PYG{n+nt}{\PYGZdq{}name\PYGZdq{}}\PYG{p}{:} \PYG{l+s+s2}{\PYGZdq{}u3\PYGZdq{}}\PYG{p}{,} \PYG{n+nt}{\PYGZdq{}qubits\PYGZdq{}}\PYG{p}{:} \PYG{p}{[}\PYG{l+m+mi}{2}\PYG{p}{],}
   \PYG{n+nt}{\PYGZdq{}params\PYGZdq{}}\PYG{p}{:} \PYG{p}{[}\PYG{l+m+mf}{3.14159}\PYG{p}{,}\PYG{l+m+mf}{0.0}\PYG{p}{,}\PYG{l+m+mf}{3.14159}\PYG{p}{],} \PYG{n+nt}{\PYGZdq{}conditional\PYGZdq{}}\PYG{p}{:} \PYG{l+m+mi}{3}\PYG{p}{\PYGZcb{},}
  \PYG{p}{\PYGZob{}}\PYG{n+nt}{\PYGZdq{}name\PYGZdq{}}\PYG{p}{:} \PYG{l+s+s2}{\PYGZdq{}measure\PYGZdq{}}\PYG{p}{,} \PYG{n+nt}{\PYGZdq{}qubits\PYGZdq{}}\PYG{p}{:} \PYG{p}{[}\PYG{l+m+mi}{0}\PYG{p}{],} \PYG{n+nt}{\PYGZdq{}memory\PYGZdq{}}\PYG{p}{:} \PYG{p}{[}\PYG{l+m+mi}{2}\PYG{p}{]\PYGZcb{},}
  \PYG{p}{\PYGZob{}}\PYG{n+nt}{\PYGZdq{}name\PYGZdq{}}\PYG{p}{:} \PYG{l+s+s2}{\PYGZdq{}measure\PYGZdq{}}\PYG{p}{,} \PYG{n+nt}{\PYGZdq{}qubits\PYGZdq{}}\PYG{p}{:} \PYG{p}{[}\PYG{l+m+mi}{1}\PYG{p}{],} \PYG{n+nt}{\PYGZdq{}memory\PYGZdq{}}\PYG{p}{:} \PYG{p}{[}\PYG{l+m+mi}{3}\PYG{p}{]\PYGZcb{},}
  \PYG{p}{\PYGZob{}}\PYG{n+nt}{\PYGZdq{}name\PYGZdq{}}\PYG{p}{:} \PYG{l+s+s2}{\PYGZdq{}measure\PYGZdq{}}\PYG{p}{,} \PYG{n+nt}{\PYGZdq{}qubits\PYGZdq{}}\PYG{p}{:} \PYG{p}{[}\PYG{l+m+mi}{2}\PYG{p}{],} \PYG{n+nt}{\PYGZdq{}memory\PYGZdq{}}\PYG{p}{:} \PYG{p}{[}\PYG{l+m+mi}{4}\PYG{p}{]\PYGZcb{}]}
\PYG{p}{\PYGZcb{}}
\end{Verbatim}
\end{jsonexamplebox}

The measurement results, assuming perfect feedback and a 3\% bit flip rate, are
\begin{jsonexamplebox}{Three-Qubit Reptition Code - Result}
\begin{Verbatim}[commandchars=\\\{\}]
\PYG{p}{\PYGZob{}}
  \PYG{n+nt}{\PYGZdq{}backend\PYGZus{}name\PYGZdq{}}\PYG{p}{:} \PYG{l+s+s2}{\PYGZdq{}ibmqx2\PYGZdq{}}\PYG{p}{,}
  \PYG{n+nt}{\PYGZdq{}backend\PYGZus{}version\PYGZdq{}}\PYG{p}{:} \PYG{l+s+s2}{\PYGZdq{}2.1.2\PYGZdq{}}\PYG{p}{,}
  \PYG{n+nt}{\PYGZdq{}qobj\PYGZus{}id\PYGZdq{}}\PYG{p}{:} \PYG{l+s+s2}{\PYGZdq{}repcode\PYGZus{}07272018\PYGZdq{}}\PYG{p}{,}
  \PYG{n+nt}{\PYGZdq{}job\PYGZus{}id\PYGZdq{}}\PYG{p}{:} \PYG{l+s+s2}{\PYGZdq{}ZZY1353JSIF\PYGZdq{}}\PYG{p}{,}
  \PYG{n+nt}{\PYGZdq{}date\PYGZdq{}}\PYG{p}{:} \PYG{l+s+s2}{\PYGZdq{}2018\PYGZhy{}04\PYGZhy{}02 15:00:00Z\PYGZdq{}}\PYG{p}{,}
  \PYG{n+nt}{\PYGZdq{}header\PYGZdq{}}\PYG{p}{:} \PYG{p}{\PYGZob{}}\PYG{n+nt}{\PYGZdq{}description\PYGZdq{}}\PYG{p}{:} \PYG{l+s+s2}{\PYGZdq{}Three qubit repetition code\PYGZdq{}}\PYG{p}{\PYGZcb{},}
  \PYG{n+nt}{\PYGZdq{}success\PYGZdq{}}\PYG{p}{:} \PYG{k+kc}{true}\PYG{p}{,}
  \PYG{n+nt}{\PYGZdq{}results\PYGZdq{}}\PYG{p}{:} \PYG{p}{[}\PYG{err}{exp\PYGZus{}result1}\PYG{p}{]}
\PYG{p}{\PYGZcb{}}
\end{Verbatim}
\end{jsonexamplebox}
where \texttt{exp\_result1} is
\begin{jsonexamplebox}{Three-Qubit Reptition Code - Experiment Result}
\begin{Verbatim}[commandchars=\\\{\}]
\PYG{p}{\PYGZob{}}
  \PYG{n+nt}{\PYGZdq{}shots\PYGZdq{}}\PYG{p}{:} \PYG{l+m+mi}{1000}\PYG{p}{,}
  \PYG{n+nt}{\PYGZdq{}status\PYGZdq{}}\PYG{p}{:} \PYG{l+s+s2}{\PYGZdq{}DONE\PYGZdq{}}\PYG{p}{,}
  \PYG{n+nt}{\PYGZdq{}success\PYGZdq{}}\PYG{p}{:} \PYG{k+kc}{true}\PYG{p}{,}
  \PYG{n+nt}{\PYGZdq{}header\PYGZdq{}}\PYG{p}{:} \PYG{p}{\PYGZob{}}
    \PYG{n+nt}{\PYGZdq{}data\PYGZus{}qubits\PYGZdq{}}\PYG{p}{:} \PYG{p}{[}\PYG{l+m+mi}{0}\PYG{p}{,}\PYG{l+m+mi}{1}\PYG{p}{,}\PYG{l+m+mi}{2}\PYG{p}{],}
    \PYG{n+nt}{\PYGZdq{}ancilla\PYGZus{}qubits\PYGZdq{}}\PYG{p}{:} \PYG{p}{[}\PYG{l+m+mi}{3}\PYG{p}{,}\PYG{l+m+mi}{4}\PYG{p}{]\PYGZcb{},}
  \PYG{n+nt}{\PYGZdq{}data\PYGZdq{}}\PYG{p}{:} \PYG{p}{\PYGZob{}}\PYG{n+nt}{\PYGZdq{}counts\PYGZdq{}}\PYG{p}{:} \PYG{p}{\PYGZob{}}\PYG{n+nt}{\PYGZdq{}0x00\PYGZdq{}}\PYG{p}{:} \PYG{l+m+mi}{455}\PYG{p}{,} \PYG{n+nt}{\PYGZdq{}0x1C\PYGZdq{}}\PYG{p}{:} \PYG{l+m+mi}{455}\PYG{p}{,} \PYG{n+nt}{\PYGZdq{}0x1E\PYGZdq{}}\PYG{p}{:} \PYG{l+m+mi}{15}\PYG{p}{,}
                      \PYG{n+nt}{\PYGZdq{}0x02\PYGZdq{}}\PYG{p}{:} \PYG{l+m+mi}{15}\PYG{p}{,} \PYG{n+nt}{\PYGZdq{}0x01\PYGZdq{}}\PYG{p}{:} \PYG{l+m+mi}{15}\PYG{p}{,} \PYG{n+nt}{\PYGZdq{}0x1D\PYGZdq{}}\PYG{p}{:} \PYG{l+m+mi}{15}\PYG{p}{,}
                      \PYG{n+nt}{\PYGZdq{}0x03\PYGZdq{}}\PYG{p}{:} \PYG{l+m+mi}{15}\PYG{p}{,} \PYG{n+nt}{\PYGZdq{}0x1F\PYGZdq{}}\PYG{p}{:} \PYG{l+m+mi}{15}\PYG{p}{\PYGZcb{},}
           \PYG{n+nt}{\PYGZdq{}memory\PYGZdq{}}\PYG{p}{:} \PYG{p}{[}\PYG{l+s+s2}{\PYGZdq{}0x00\PYGZdq{}}\PYG{p}{,}\PYG{l+s+s2}{\PYGZdq{}0x1F\PYGZdq{}}\PYG{p}{,}\PYG{l+s+s2}{\PYGZdq{}0x1C\PYGZdq{}}\PYG{p}{,}\PYG{l+s+s2}{\PYGZdq{}0x1C\PYGZdq{}}\PYG{p}{,}\PYG{l+s+s2}{\PYGZdq{}0x00\PYGZdq{}}\PYG{p}{,}\PYG{err}{...}\PYG{p}{]\PYGZcb{}}
\PYG{p}{\PYGZcb{}}
\end{Verbatim}
\end{jsonexamplebox}

\section{OpenPulse Specification \label{sect:openpulse}}

The goal of the OpenPulse specification is to allow a user to have pulse-level control (i.e. control of the continuous time dynamics) of a general quantum device. The specification is designed to be hardware and implementation agnostic, thus supporting a broad class of quantum devices (e.g., superconducting qubits, solid-state spins, trapped ions, NV centers, etc.). Pulses in the specification are defined in terms of a complex-valued envelope, relative to global clocks that the quantum device will provide, that the user has some ability to set. This specification is tailored to qubits with single-meter measurements (e.g., each qubit is coupled to a single readout cavity, fluorescence detection, etc.).\\

The motivation for this specification is to enable experiments that are beyond the scope of OpenQASM. For example, the OpenPulse specification enables user experimentation of improved decoupling schemes, calibrations, pulse-shaping, and optimal control. However, the idea is not for users to start from scratch, and as such, the device can provide a library of calibrated pulses and estimated system parameters as a starting point for user experimentation. The device may also provide a mapping from gates to sequences of these pulses, thus allowing the user to run gate based circuits through OpenPulse, and experiment with their own mappings. \\

This specification also enables experimentation with readout that is not possible in OpenQASM. For example, users may obtain raw measurement outcomes (after downconversion, i.e., the raw output pulse envelope) to better construct measurement kernels and discriminators. OpenPulse is designed for simple classical feedback; pulses may be run conditionally on measurement results. For example, this allows active qubit reset that may be required for certain high-fidelity experiments. \\

\begin{figure}
\includegraphics[width=\textwidth]{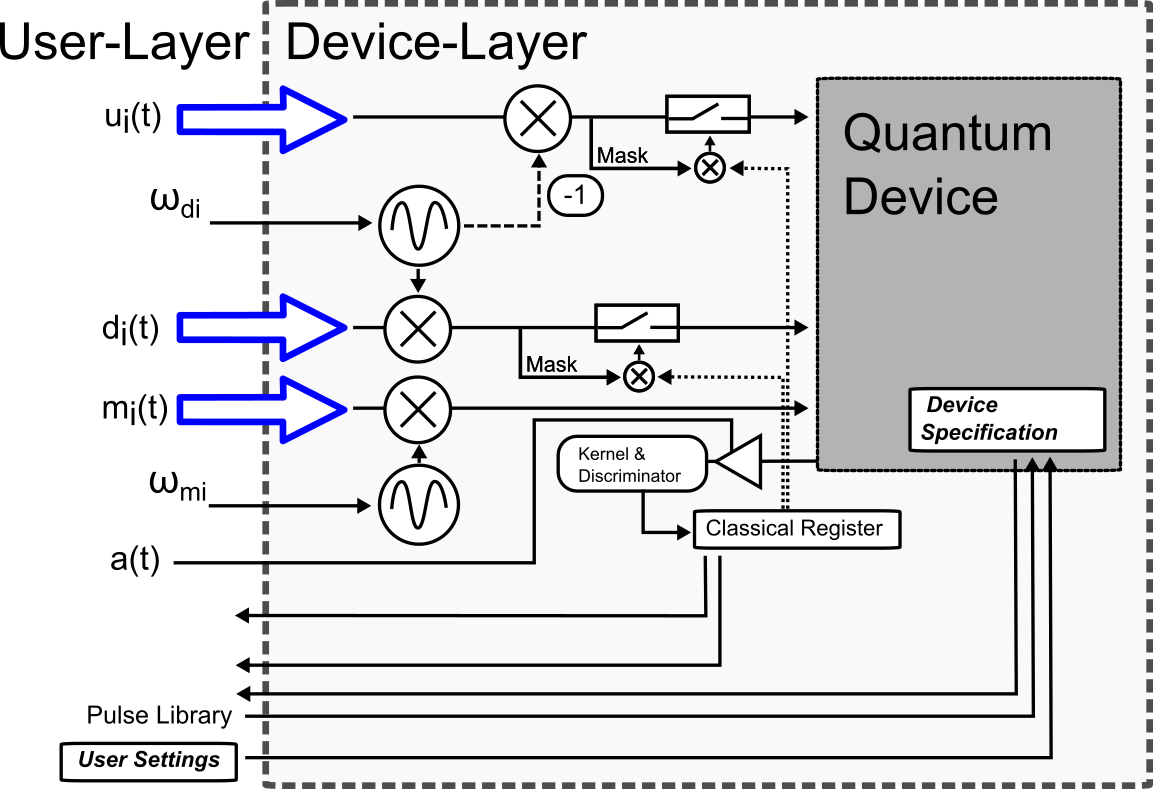}
\caption{Interface to the device for OpenPulse. Here are the inputs for a single qubit $i$. The device will provide certain specifications when queried. The user can supply a pulse library and modify certain settings. Measurement results are loaded into a classical memory on the device when indicated by an acquisition command. In Level 2 output the measurement first passes through a kernel and discriminator and can then be used as a mask on drive pulses. \label{fig:device}}
\end{figure}

The interface to the device is given in Figure~\ref{fig:device}. Each qubit has two channels, a drive channel and a measurement input (stimulus) channel. The signal on the drive channel for qubit $i$, given as $d_i(t)$, is mixed up with a user-specified LO (local oscillator) at frequency $\omega_{d_i}$ as $D_i(t) = Re[d_i(t)e^{\iu\omega_{d_i} t}]$ and then interacts with the qubit as $\hat{H} = D_i(t)\sigma_{i}^{X}$. Upon request the device must provide a general estimate of the qubit frequency good to within the bandwidth of the drive channel. The signal on the measurement stimulus channel for qubit $i$, given as $m_i(t)$, is mixed up with a user-specified LO at frequency $\omega_{m_i}$ as $M_i(t) = Re[m_i(t) e^{\iu\omega_{m_i} t}]$. This signal is applied to the qubit for measurement (see below).  \\

Additional control channels (``U Channels'') may be present depending on the device. The action of these channels is described by the Hamiltonian returned by the device in enough detail to allow for their operation (\S~\ref{sect:ham}). The bandwidth of these channels is also specified by the device. The signal on U channel $i$ is assumed to be mixed up by the some combination of qubit LO's (e.g. $U_{i}(t)=\mathrm{Re}\left[u_{i}(t)e^{\iu\left(\omega_{d_i}-\omega_{d_j}\right)t}\right]$) as specified by the device. DC control channels, i.e., control channels that cannot be set for each experiment, are beyond the scope of this specification. Special care would be required for these channels in order to coordinate across several users and to lock out the experiment while the channel is being set.  \\

An experiment is defined by a single time sequence of pulse commands over the various channels(\S~\ref{sect:seq}). These pulse commands are given in terms of pre-defined pulses (short sequences), and the start time of these pulses. Times are specified in steps of a device defined time unit $dt$. These pre-defined pulses (the pulse library) are sent to the device via the API. Upon request, the device can provide a default pulse library, and the typical sequences for OpenQASM pulses (OpenQASM command definition), so that an OpenQASM circuit can be programmed via OpenPulse.\\

Measurement outputs from the device (\S~\ref{sect:meas}) are stored in a series of measurement memory slots (the number of memory slots is set by the user). Acquisition commands on the experiment time sequence, $a(t)$, specify when to record measurements, and into which memory slot to store the measurements. Typically, this would correspond to a pulse command on the measurement stimulus channel. The type of information stored depends on the measurement level selected by the user. For Level 0, measurements of the raw signal (after mixing back down, i.e., the measurement output pulse envelope) are stored. For a Level 1 measurement, a measurement kernel is selected (from a list of available kernels), and the measurement returns a complex number obtained after applying the measurement kernel to the measurement output signal. For Level 2 measurements, a discriminator is selected (from a list of available discriminators), and the qubit state is stored (0 or 1). If Level 2 measurement is selected, then measurements can also be pushed to registers from which feedback is possible. Drive pulses can be conditional (applied or not applied) based on the value of a register bit. \\

There is a backend method to load the experiment, specified as a \texttt{Qobj} (\S~\ref{sect:rundict}), into the run queue (\S~\ref{sect:backend}), and a status query to return the experiment status. Once the experiment is running, the measurement memory can be streamed, although it is not guaranteed that measurements can be streamed in real time (\S~\ref{sect:jobresult}). Multiple experiments loaded in a single \texttt{Qobj} will run concurrently (one shot of each experiment is run, followed by the next experiment and then the whole set is repeated for as many shots as specified). No other Qobj Experiment can be run in this period.

\subsection{Additional API Calls}\label{sect:addpulseapi}

In addition to the general API description in \S~\ref{sect:api}, the data structures reported by an OpenPulse backend will have additional fields.

\subsubsection{Backend Configuration \label{sect:pulseconfig}}

In OpenPulse the call to \texttt{Backend.configuration()} will have the following additional fields for the \texttt{configuration} data structure:
\begin{jsonexamplebox}{OpenPulse Backend Configuration}
\begin{Verbatim}[commandchars=\\\{\}]
\PYG{p}{\PYGZob{}}
  \PYG{n+nt}{\PYGZdq{}n\PYGZus{}uchannels\PYGZdq{}}\PYG{p}{:} \PYG{l+m+mi}{2}\PYG{p}{,}
  \PYG{n+nt}{\PYGZdq{}hamiltonian\PYGZdq{}}\PYG{p}{:} \PYG{err}{ham\PYGZus{}dict}\PYG{p}{,}
  \PYG{n+nt}{\PYGZdq{}u\PYGZus{}channel\PYGZus{}lo\PYGZdq{}}\PYG{p}{:} \PYG{p}{[[\PYGZob{}}\PYG{n+nt}{\PYGZdq{}q\PYGZdq{}}\PYG{p}{:} \PYG{l+m+mi}{0}\PYG{p}{,} \PYG{n+nt}{\PYGZdq{}scale\PYGZdq{}}\PYG{p}{:} \PYG{p}{[}\PYG{l+m+mi}{1}\PYG{p}{,}\PYG{l+m+mi}{0}\PYG{p}{]\PYGZcb{}],}
  \PYG{p}{[\PYGZob{}}\PYG{n+nt}{\PYGZdq{}q\PYGZdq{}}\PYG{p}{:} \PYG{l+m+mi}{0}\PYG{p}{,} \PYG{n+nt}{\PYGZdq{}scale\PYGZdq{}}\PYG{p}{:} \PYG{p}{[}\PYG{l+m+mi}{\PYGZhy{}1}\PYG{p}{,}\PYG{l+m+mi}{0}\PYG{p}{]\PYGZcb{},\PYGZob{}}\PYG{n+nt}{\PYGZdq{}q\PYGZdq{}}\PYG{p}{:} \PYG{l+m+mi}{1}\PYG{p}{,} \PYG{n+nt}{\PYGZdq{}scale\PYGZdq{}}\PYG{p}{:} \PYG{p}{[}\PYG{l+m+mi}{1}\PYG{p}{,}\PYG{l+m+mi}{0}\PYG{p}{]\PYGZcb{}]],}
  \PYG{n+nt}{\PYGZdq{}meas\PYGZus{}levels\PYGZdq{}}\PYG{p}{:} \PYG{p}{[}\PYG{l+m+mi}{1}\PYG{p}{,}\PYG{l+m+mi}{2}\PYG{p}{],}
  \PYG{n+nt}{\PYGZdq{}qubit\PYGZus{}lo\PYGZus{}range\PYGZdq{}}\PYG{p}{:} \PYG{p}{[[}\PYG{l+m+mf}{4.5}\PYG{p}{,}\PYG{l+m+mf}{5.5}\PYG{p}{],[}\PYG{l+m+mf}{4.5}\PYG{p}{,}\PYG{l+m+mf}{5.5}\PYG{p}{]],}
  \PYG{n+nt}{\PYGZdq{}meas\PYGZus{}lo\PYGZus{}range\PYGZdq{}}\PYG{p}{:} \PYG{p}{[[}\PYG{l+m+mf}{6.0}\PYG{p}{,}\PYG{l+m+mf}{7.0}\PYG{p}{],[}\PYG{l+m+mf}{6.0}\PYG{p}{,}\PYG{l+m+mf}{7.0}\PYG{p}{]],}
  \PYG{n+nt}{\PYGZdq{}dt\PYGZdq{}}\PYG{p}{:} \PYG{l+m+mf}{1.3333}\PYG{p}{,}
  \PYG{n+nt}{\PYGZdq{}dtm\PYGZdq{}}\PYG{p}{:} \PYG{l+m+mf}{10.5}\PYG{p}{,}
  \PYG{n+nt}{\PYGZdq{}rep\PYGZus{}times\PYGZdq{}}\PYG{p}{:} \PYG{p}{[}\PYG{l+m+mi}{100}\PYG{p}{,}\PYG{l+m+mi}{250}\PYG{p}{,}\PYG{l+m+mi}{500}\PYG{p}{,}\PYG{l+m+mi}{1000}\PYG{p}{],}
  \PYG{n+nt}{\PYGZdq{}meas\PYGZus{}map\PYGZdq{}}\PYG{p}{:} \PYG{p}{[[}\PYG{l+m+mi}{0}\PYG{p}{],[}\PYG{l+m+mi}{1}\PYG{p}{,}\PYG{l+m+mi}{2}\PYG{p}{]],}
  \PYG{n+nt}{\PYGZdq{}channel\PYGZus{}bandwidth\PYGZdq{}}\PYG{p}{:} \PYG{p}{[[}\PYG{l+m+mf}{\PYGZhy{}0.2}\PYG{p}{,}\PYG{l+m+mf}{0.4}\PYG{p}{],[}\PYG{l+m+mf}{\PYGZhy{}0.3}\PYG{p}{,}\PYG{l+m+mf}{0.3}\PYG{p}{],}
         \PYG{p}{[}\PYG{l+m+mf}{\PYGZhy{}0.3}\PYG{p}{,}\PYG{l+m+mf}{0.3}\PYG{p}{],[}\PYG{l+m+mf}{\PYGZhy{}0.02}\PYG{p}{,}\PYG{l+m+mf}{0.02}\PYG{p}{],[}\PYG{l+m+mf}{\PYGZhy{}0.02}\PYG{p}{,}\PYG{l+m+mf}{0.02}\PYG{p}{],}
         \PYG{p}{[}\PYG{l+m+mf}{\PYGZhy{}0.02}\PYG{p}{,}\PYG{l+m+mf}{0.02}\PYG{p}{]],}
  \PYG{n+nt}{\PYGZdq{}meas\PYGZus{}kernels\PYGZdq{}}\PYG{p}{:} \PYG{p}{[}\PYG{l+s+s2}{\PYGZdq{}kernel1\PYGZdq{}}\PYG{p}{,}\PYG{l+s+s2}{\PYGZdq{}kernel2\PYGZdq{}}\PYG{p}{,}\PYG{err}{...}\PYG{p}{],}
  \PYG{n+nt}{\PYGZdq{}discriminators\PYGZdq{}}\PYG{p}{:} \PYG{p}{[}\PYG{l+s+s2}{\PYGZdq{}disc1\PYGZdq{}}\PYG{p}{,}\PYG{l+s+s2}{\PYGZdq{}disc2\PYGZdq{}}\PYG{p}{,}\PYG{err}{...}\PYG{p}{],}
  \PYG{n+nt}{\PYGZdq{}acquisition\PYGZus{}latency\PYGZdq{}}\PYG{p}{:} \PYG{p}{[[}\PYG{l+m+mi}{100}\PYG{p}{,}\PYG{l+m+mi}{100}\PYG{p}{],} \PYG{p}{[}\PYG{l+m+mi}{100}\PYG{p}{,}\PYG{l+m+mi}{100}\PYG{p}{]],}
  \PYG{n+nt}{\PYGZdq{}conditional\PYGZus{}latency\PYGZdq{}}\PYG{p}{:} \PYG{p}{[[}\PYG{l+m+mi}{100}\PYG{p}{,}\PYG{l+m+mi}{1000}\PYG{p}{],[}\PYG{l+m+mi}{1000}\PYG{p}{,}\PYG{l+m+mi}{100}\PYG{p}{],[}\PYG{l+m+mi}{100}\PYG{p}{,}\PYG{l+m+mi}{1000}\PYG{p}{],}
                          \PYG{p}{[}\PYG{l+m+mi}{1000}\PYG{p}{,}\PYG{l+m+mi}{100}\PYG{p}{],[}\PYG{l+m+mi}{100}\PYG{p}{,}\PYG{l+m+mi}{1000}\PYG{p}{],[}\PYG{l+m+mi}{1000}\PYG{p}{,}\PYG{l+m+mi}{100}\PYG{p}{]]}
\PYG{p}{\PYGZcb{}}
\end{Verbatim}
\end{jsonexamplebox}
where
\begin{itemize}
\item \texttt{n\_uchannels}: Number of `U' channels (control channels) in the device.
\item \texttt{hamiltonian}: Hamiltonian for the device (as a \texttt{ham\_dict}) which gives the approximate qubit frequencies and describes the operation of each `u' channel. Details are given in \S~\ref{sect:ham}.
\item \texttt{u\_channel\_lo}: List of length \texttt{n\_uchannels} that specifies each U channel LO in terms of qubit LO's. See example above for the data structure.  The LO scales as $\omega_{U_i} = \sum_{j} \alpha_{ij} \omega_{qj}$ where $\alpha_{ij}$ are the coefficients specified by the complex value of the \texttt{scale} item in the data structure.
\item \texttt{meas\_levels}: List of device allowable measurement output levels (0,1,2). In this example, level 0 is not allowed.
\item \texttt{qubit\_lo\_range}: Range of allowable frequencies for each qubit LO.
\item \texttt{meas\_lo\_range}: Range of allowable frequencies for each measurement LO.
\item \texttt{dt}: Discretization of the \emph{input} time sequences in ns.
\item \texttt{dtm}: Discretization of the \emph{output} time sequences in ns (only needed for measurement level 0).
\item \texttt{rep\_times}: List of possible experiment repetition times in $\mu$s.
\item \texttt{meas\_map} (optional): List of lists indicating all measurements which are multiplexed together. In the above example Q0 is measured independently, but Q1 and Q2 are multiplexed together.
\item \texttt{channel\_bandwidth} (optional): List of tuples that specifies the lower and upper bandwidth for each input channel (ordered as the qubit channels, then measurement channels then U channels). This bandwidth is specified assuming the LOs are at the default values.
\item \texttt{meas\_kernels}: List of available measurement kernels (list of names). The kernels are described in more detail in \S~\ref{sect:kernel}.
\item \texttt{discriminators}: List of available measurement discriminators (list of names). The discriminators are described in more detail in \S~\ref{sect:kernel}.
\item \texttt{acquisition\_latency} (optional): List of latencies (in units of $dt$) for the acquisition to be set into the registers. This is a list of length \texttt{n$\times$m} where \texttt{n} is \texttt{n\_qubits} and \texttt{m} is \texttt{n\_registers}. It supersedes \texttt{register\_map} as it provides more accurate latency information.
\item \texttt{conditional\_latency} (optional): List of latencies (in units of $dt$) for each channel (drive, then $u$, then $m$) to the registers if these registers are used for a conditional operation. This supersedes the \texttt{latency\_map} in the \texttt{gate\_config} for OpenQASM.
\end{itemize}

\subsubsection{Backend Defaults \label{sect:pulsedefaults}}

In OpenPulse the Backend method \texttt{Backend.defaults()} will return a \texttt{default\_pulse\_config} data structure giving the default device settings:
\begin{jsonexamplebox}{Backend Defaults}
\begin{Verbatim}[commandchars=\\\{\}]
\PYG{p}{\PYGZob{}}
  \PYG{n+nt}{\PYGZdq{}qubit\PYGZus{}freq\PYGZus{}est\PYGZdq{}}\PYG{p}{:} \PYG{p}{[}\PYG{l+m+mf}{4.9}\PYG{p}{,}\PYG{l+m+mf}{5.0}\PYG{p}{,}\PYG{l+m+mf}{5.1}\PYG{p}{],}
  \PYG{n+nt}{\PYGZdq{}meas\PYGZus{}freq\PYGZus{}est\PYGZdq{}}\PYG{p}{:} \PYG{p}{[}\PYG{l+m+mf}{6.5}\PYG{p}{,}\PYG{l+m+mf}{6.6}\PYG{p}{,}\PYG{l+m+mf}{6.7}\PYG{p}{],}
  \PYG{n+nt}{\PYGZdq{}buffer\PYGZdq{}}\PYG{p}{:} \PYG{l+m+mi}{10}\PYG{p}{,}
  \PYG{n+nt}{\PYGZdq{}pulse\PYGZus{}library\PYGZdq{}}\PYG{p}{:} \PYG{err}{pulse\PYGZus{}lib}\PYG{p}{,}
  \PYG{n+nt}{\PYGZdq{}cmd\PYGZus{}def\PYGZdq{}}\PYG{p}{:} \PYG{err}{cmd\PYGZus{}def}\PYG{p}{,}
  \PYG{n+nt}{\PYGZdq{}meas\PYGZus{}kernel\PYGZdq{}}\PYG{p}{:} \PYG{p}{\PYGZob{}}\PYG{n+nt}{\PYGZdq{}name\PYGZdq{}}\PYG{p}{:} \PYG{l+s+s2}{\PYGZdq{}kernel1\PYGZdq{}}\PYG{p}{,} \PYG{n+nt}{\PYGZdq{}params\PYGZdq{}}\PYG{p}{:} \PYG{p}{[]\PYGZcb{},}
  \PYG{n+nt}{\PYGZdq{}discriminator\PYGZdq{}}\PYG{p}{:} \PYG{p}{\PYGZob{}}\PYG{n+nt}{\PYGZdq{}name\PYGZdq{}}\PYG{p}{:} \PYG{l+s+s2}{\PYGZdq{}max\PYGZus{}1Q\PYGZus{}fidelity\PYGZdq{}}\PYG{p}{,}
                    \PYG{n+nt}{\PYGZdq{}params\PYGZdq{}}\PYG{p}{:} \PYG{p}{[}\PYG{l+m+mi}{0}\PYG{p}{,}\PYG{l+m+mi}{0}\PYG{p}{]\PYGZcb{}}
\PYG{p}{\PYGZcb{}}
\end{Verbatim}
\end{jsonexamplebox}

where
\begin{itemize}
\item \texttt{qubit\_freq\_est}: List of \emph{estimated} qubit frequencies in GHz. These correspond to the $\nu$ terms in the Hamiltonian (see \S~\ref{sect:ham}).
\item \texttt{meas\_freq\_est}: List of \emph{estimated} measurement cavity frequencies in GHz.
\item \texttt{buffer}: The default buffer time (in units of $dt$) between pulses.
\item \texttt{pulse\_library}: This is a default pulse library (see \S~\ref{sect:seq} given as a \texttt{pulse\_lib} data structure as described in \S~\ref{sect:pulselib}). This may be the latest calibration or a default estimation. This should be specified to the user offline, but the main motivation of providing a default pulse library (and \texttt{cmd\_def}) is to prevent bottlenecking the device with calibration runs.
\item \texttt{cmd\_def}: This is a default OpenQASM command to OpenPulse command definition as a \texttt{cmd\_def} data structure (see \S~\ref{sect:qasmgatedef} for the format) for converting OpenQASM to OpenPulse. This is relatively static because the pulses are defined in terms of the pulse library.
\item \texttt{meas\_kernel}: The default measurement kernel from the list given in the call to \texttt{Backend.configuration()}.
\item \texttt{discriminator}: The default discriminator from the list given in the call to \texttt{Backend.configuration()}.
\end{itemize}

\subsubsection{Hamiltonian Specification \label{sect:ham}}

The Hamiltonian is specified as a \texttt{ham\_struct} data structure (Hamiltonian data structure) with the following form,
\begin{jsonexamplebox}{Hamiltonian Specification}
\begin{Verbatim}[commandchars=\\\{\}]
\PYG{p}{\PYGZob{}}
  \PYG{n+nt}{\PYGZdq{}h\PYGZus{}latex\PYGZdq{}}\PYG{p}{:} \PYG{l+s+s2}{\PYGZdq{}H = \PYGZbs{}\PYGZbs{}sum\PYGZus{}\PYGZob{}i\PYGZcb{}\PYGZca{}\PYGZob{}N\PYGZcb{} D\PYGZus{}i(t) \PYGZbs{}\PYGZbs{}sigma\PYGZus{}i\PYGZca{}\PYGZob{}X\PYGZcb{} +}
\PYG{l+s+s2}{    \PYGZbs{}\PYGZbs{}sum\PYGZus{}\PYGZob{}i\PYGZcb{}\PYGZca{}\PYGZob{}N\PYGZcb{} 2\PYGZbs{}\PYGZbs{}pi \PYGZbs{}nu\PYGZus{}i \PYGZbs{}\PYGZbs{}sigma\PYGZus{}i\PYGZca{}\PYGZob{}+\PYGZcb{}\PYGZbs{}\PYGZbs{}sigma\PYGZus{}i\PYGZca{}\PYGZob{}\PYGZhy{}\PYGZcb{}\PYGZdq{}}\PYG{p}{,}
  \PYG{n+nt}{\PYGZdq{}h\PYGZus{}str\PYGZdq{}}\PYG{p}{:}  \PYG{p}{[}\PYG{l+s+s2}{\PYGZdq{}\PYGZus{}\PYGZus{}SUM[i,0,N,\PYGZus{}X\PYGZob{}i\PYGZcb{}\PYGZus{}||\PYGZus{}D\PYGZob{}i\PYGZcb{}\PYGZus{}]\PYGZdq{}}\PYG{p}{,}
             \PYG{l+s+s2}{\PYGZdq{}\PYGZus{}\PYGZus{}SUM[i,0,N,2*pi*\PYGZus{}v\PYGZob{}i\PYGZcb{}\PYGZus{}*\PYGZus{}O\PYGZob{}i\PYGZcb{}\PYGZus{}]\PYGZdq{}}\PYG{p}{],}
  \PYG{n+nt}{\PYGZdq{}vars\PYGZdq{}} \PYG{p}{:} \PYG{p}{\PYGZob{}}\PYG{n+nt}{\PYGZdq{}v0\PYGZdq{}}\PYG{p}{:} \PYG{l+m+mf}{5.0}\PYG{p}{,} \PYG{n+nt}{\PYGZdq{}v1\PYGZdq{}} \PYG{p}{:} \PYG{l+m+mf}{5.25}\PYG{p}{\PYGZcb{},}
  \PYG{n+nt}{\PYGZdq{}osc\PYGZdq{}} \PYG{p}{:} \PYG{p}{\PYGZob{}\PYGZcb{}}
\PYG{p}{\PYGZcb{}}
\end{Verbatim}
\end{jsonexamplebox}
where
\begin{itemize}
\item \texttt{h\_latex}: Latex string describing the Hamiltonian (required).
\item \texttt{h\_str}: List of parsable terms in the Hamiltonian (required for machine readability of the Hamiltonian, e.g., for interoperability with OpenPulse simulators).
\item \texttt{vars}: Structure of variable values in the \texttt{h\_str}.
\item \texttt{osc}: Structure of oscillator values in the \texttt{h\_str}.
\end{itemize}
The minimum requirement is that the Hamiltonian structure return a latex readable string that can specify the coupling map and control channel operation. The latex string must specify the Hamiltonian in terms of Pauli ($\sigma^X,\sigma^Y,\sigma^Z$) or ladder operators ($a,a^{\dagger}$). At minimum, this string must have,
\begin{equation}
    H = \sum_{i}^{N} D_i(t) \sigma_i^{X} + \sum_{i}^{N} 2\pi \nu_i (1-\sigma_i^{Z})/2
\end{equation}
specifying a set of independent qubits. The value of $\nu_i$ is given by the \texttt{qubit\_freq\_est} spec. All other terms in the Hamiltonian can be returned as symbols \emph{or} as values in units of frequency (e.g. GHz for superconducting devices). In \S~\ref{sect:expulse} we will give more specific Hamiltonian examples. \\

To specify the higher levels of the system being used as a qubit (e.g. a transmon), the qubit terms can be written in terms of ladder operators as well, e.g. for a transmon represented as a Duffing oscillator (where $\delta_i$ is the anharmonicity),
\begin{equation}
    H = \sum_{i}^{N} D_i(t) (a_i+a_i^{\dagger}) + \sum_{i}^{N} 2\pi \nu_i a_i^{\dagger}a_i + \frac{\delta_i}{2} (1- a_i^{\dagger} a_i) a_i^{\dagger}a_i
\end{equation}

\subsubsection{Kernel and Discriminator Specification \label{sect:kernel}}

The device will specify the kernel and discriminator as a list of available kernels/discriminators. A kernel/discriminator data structure has the form:
\begin{jsonexamplebox}{Kernel}
\begin{Verbatim}[commandchars=\\\{\}]
\PYG{p}{\PYGZob{}}
  \PYG{n+nt}{\PYGZdq{}name\PYGZdq{}}\PYG{p}{:} \PYG{l+s+s2}{\PYGZdq{}boxcar\PYGZdq{}}\PYG{p}{,}
  \PYG{n+nt}{\PYGZdq{}params\PYGZdq{}}\PYG{p}{:} \PYG{p}{[]}
\PYG{p}{\PYGZcb{}}
\end{Verbatim}
\end{jsonexamplebox}

which would be a boxcar averaging kernel. No parameters are needed because this function averages the entire measurement output. A sample discriminator is
\begin{jsonexamplebox}{Discriminator}
\begin{Verbatim}[commandchars=\\\{\}]
\PYG{p}{\PYGZob{}}
  \PYG{n+nt}{\PYGZdq{}name\PYGZdq{}}\PYG{p}{:} \PYG{l+s+s2}{\PYGZdq{}max\PYGZus{}1Q\PYGZus{}fidelity\PYGZdq{}}\PYG{p}{,}
  \PYG{n+nt}{\PYGZdq{}params\PYGZdq{}}\PYG{p}{:} \PYG{p}{[}\PYG{l+m+mf}{\PYGZhy{}0.05}\PYG{p}{,}\PYG{l+m+mf}{0.1}\PYG{p}{]}
\PYG{p}{\PYGZcb{}}
\end{Verbatim}
\end{jsonexamplebox}

that is a discriminator that maximizes the one-qubit readout fidelity by thresholding 0 and 1 to be two halves of the IQ plane. The parameters indicate the line to use for the thresholding. \\

Each kernel/discriminator has a name and a list of default parameters (if any). There is no formal specification for how the action of these kernels/discriminators behaves, and the device is not obliged to reveal that information. If the device wishes to convey that information, it would have to be outside this API, and would likely need to include language-specific code. Since discriminators act on level 1 measurement output, users are free to construct their own discriminators \emph{locally}. If users construct kernels or discriminators they would like included on this list then they need to contact the device administrators for it to be added. \\

There may be kernels/discriminators that take multi-qubit measurements as inputs (this would be conveyed by the device). If these are to be used, then the corresponding measurement acquisition must occur with the proper number of qubits, and that kernel/discriminator listed as the only kernel/discriminator, see \S~\ref{sect:acq}.

\subsubsection{Pulse Library \label{sect:pulselib}}

In OpenPulse, experiments are specified as a sequence of predefined pulses, which are defined in the pulse library. The pulse library is a list of \texttt{pulse\_lib} data structures of the form,
\begin{jsonexamplebox}{Pulse}
\begin{Verbatim}[commandchars=\\\{\}]
\PYG{p}{\PYGZob{}}
  \PYG{n+nt}{\PYGZdq{}name\PYGZdq{}}\PYG{p}{:} \PYG{l+s+s2}{\PYGZdq{}pulse\PYGZus{}name\PYGZdq{}}\PYG{p}{,}
  \PYG{n+nt}{\PYGZdq{}samples\PYGZdq{}}\PYG{p}{:} \PYG{p}{[[}\PYG{l+m+mf}{0.1}\PYG{p}{,}\PYG{l+m+mf}{0.0}\PYG{p}{],[}\PYG{l+m+mf}{0.2}\PYG{p}{,}\PYG{l+m+mf}{0.0}\PYG{p}{],} \PYG{p}{[}\PYG{l+m+mf}{0.1}\PYG{p}{,}\PYG{l+m+mf}{0.0}\PYG{p}{],[}\PYG{l+m+mf}{0.0}\PYG{p}{,}\PYG{l+m+mf}{0.0}\PYG{p}{]]}
\PYG{p}{\PYGZcb{}}
\end{Verbatim}
\end{jsonexamplebox}
where
\begin{itemize}
  \item \texttt{name}: Name of the pulse. This is a unique string identifier used to refer to the pulse in the command sequence for the experiment (see \S~\ref{sect:drivepulse}).
  \item \texttt{samples}: List of complex values (specified using the convention discussed in \S~\ref{sect:language}) which define the amplitude points for the pulse envelope. The time between the amplitude points is specified by the device time unit $dt$ (see \S~\ref{sect:pulseconfig}). These amplitudes have an absolute value less than or equal to 1 (for a tuple $[a,b]$, $\sqrt{a^2+b^2} \leq 1$).
\end{itemize}
If the user would like parameterized symbolic pulses, these can be added as another layer on top of this pulse library \emph{locally}, however, in order to be language agnostic, only lists of points can be sent through to the device. A symbolic pulse layer will be part of Qiskit. An example pulse library is shown below,
\begin{jsonexamplebox}{Pulse Library}
\begin{Verbatim}[commandchars=\\\{\}]
\PYG{p}{[}
 \PYG{p}{\PYGZob{}}\PYG{n+nt}{\PYGZdq{}name\PYGZdq{}}\PYG{p}{:} \PYG{l+s+s2}{\PYGZdq{}pulse1\PYGZdq{}}\PYG{p}{,}  \PYG{n+nt}{\PYGZdq{}samples\PYGZdq{}}\PYG{p}{:} \PYG{p}{[[}\PYG{l+m+mf}{0.1}\PYG{p}{,}\PYG{l+m+mf}{0.0}\PYG{p}{],[}\PYG{l+m+mf}{0.2}\PYG{p}{,}\PYG{l+m+mf}{0.0}\PYG{p}{],}
  \PYG{p}{[}\PYG{l+m+mf}{0.1}\PYG{p}{,}\PYG{l+m+mf}{0.0}\PYG{p}{],[}\PYG{l+m+mf}{0.0}\PYG{p}{,}\PYG{l+m+mf}{0.0}\PYG{p}{],[}\PYG{l+m+mf}{\PYGZhy{}0.1}\PYG{p}{,}\PYG{l+m+mf}{0.0}\PYG{p}{],[}\PYG{l+m+mf}{\PYGZhy{}0.2}\PYG{p}{,}\PYG{l+m+mf}{0.0}\PYG{p}{],}
  \PYG{p}{[}\PYG{l+m+mf}{0.1}\PYG{p}{,}\PYG{l+m+mf}{0.0}\PYG{p}{],[}\PYG{l+m+mf}{0.1}\PYG{p}{,}\PYG{l+m+mf}{0.0}\PYG{p}{],[}\PYG{l+m+mf}{0.05}\PYG{p}{,}\PYG{l+m+mf}{0.0}\PYG{p}{]]\PYGZcb{},}
 \PYG{p}{\PYGZob{}}\PYG{n+nt}{\PYGZdq{}name\PYGZdq{}}\PYG{p}{:} \PYG{l+s+s2}{\PYGZdq{}drag\PYGZus{}pulse\PYGZdq{}}\PYG{p}{,} \PYG{n+nt}{\PYGZdq{}samples\PYGZdq{}}\PYG{p}{:} \PYG{p}{[[}\PYG{l+m+mf}{0.004}\PYG{p}{,}\PYG{l+m+mf}{0.009}\PYG{p}{],}
  \PYG{p}{[}\PYG{l+m+mf}{0.029}\PYG{p}{,}\PYG{l+m+mf}{0.05}\PYG{p}{],[}\PYG{l+m+mf}{0.135}\PYG{p}{,}\PYG{l+m+mf}{0.18}\PYG{p}{],[}\PYG{l+m+mf}{0.41}\PYG{p}{,}\PYG{l+m+mf}{0.365}\PYG{p}{],[}\PYG{l+m+mf}{0.8}\PYG{p}{,}\PYG{l+m+mf}{0.355}\PYG{p}{],}
  \PYG{p}{[}\PYG{l+m+mf}{1.0}\PYG{p}{,}\PYG{l+m+mf}{0.0}\PYG{p}{],[}\PYG{l+m+mf}{0.8}\PYG{p}{,}\PYG{l+m+mf}{\PYGZhy{}0.355}\PYG{p}{],[}\PYG{l+m+mf}{0.41}\PYG{p}{,}\PYG{l+m+mf}{\PYGZhy{}0.365}\PYG{p}{],[}\PYG{l+m+mf}{0.135}\PYG{p}{,}\PYG{l+m+mf}{\PYGZhy{}0.18}\PYG{p}{],}
  \PYG{p}{[}\PYG{l+m+mf}{0.029}\PYG{p}{,}\PYG{l+m+mf}{\PYGZhy{}0.05}\PYG{p}{],[}\PYG{l+m+mf}{0.004}\PYG{p}{,}\PYG{l+m+mf}{\PYGZhy{}0.009}\PYG{p}{]]\PYGZcb{},}
 \PYG{p}{\PYGZob{}}\PYG{n+nt}{\PYGZdq{}name\PYGZdq{}}\PYG{p}{:} \PYG{l+s+s2}{\PYGZdq{}square\PYGZus{}pulse\PYGZdq{}}\PYG{p}{,} \PYG{n+nt}{\PYGZdq{}samples\PYGZdq{}}\PYG{p}{:} \PYG{p}{[[}\PYG{l+m+mf}{0.1}\PYG{p}{,}\PYG{l+m+mf}{0.0}\PYG{p}{],}
  \PYG{p}{[}\PYG{l+m+mf}{0.1}\PYG{p}{,}\PYG{l+m+mf}{0.0}\PYG{p}{],} \PYG{p}{[}\PYG{l+m+mf}{0.1}\PYG{p}{,}\PYG{l+m+mf}{0.0}\PYG{p}{],[}\PYG{l+m+mf}{0.1}\PYG{p}{,}\PYG{l+m+mf}{0.0}\PYG{p}{],[}\PYG{l+m+mf}{0.1}\PYG{p}{,}\PYG{l+m+mf}{0.0}\PYG{p}{],[}\PYG{l+m+mf}{0.1}\PYG{p}{,}\PYG{l+m+mf}{0.0}\PYG{p}{]]\PYGZcb{}}
\PYG{p}{]}
\end{Verbatim}
\end{jsonexamplebox}

The pulse library is passed to the device via the \texttt{Qobj} (see \S~\ref{sect:pulseQObjconfig}), and a default pulse library may be requested from the device also through the backend object (\S~\ref{sect:qasmgatedef}). Pulses from the user supplied pulse library will supersede pulses of the same name from the default pulse library. We give sample pulse libraries in \S~\ref{sect:samplelibs}.  

\subsubsection{OpenQASM Command Definition \label{sect:qasmgatedef}}

The device will return a default pulse library and OpenQASM command to OpenPulse command definition (\texttt{cmd\_def}) that defines the low-level OpenQASM commands in terms of this pulse library. The \texttt{cmd\_def} is returned for each OpenQASM gate returned in the set of basis gates (see \S~\ref{sect:apiconfig}) plus the measurement command. \texttt{cmd\_def} is a list, where each entry has the form,
\begin{jsonexamplebox}{Command Definition Entry}
\begin{Verbatim}[commandchars=\\\{\}]
\PYG{p}{\PYGZob{}}
  \PYG{n+nt}{\PYGZdq{}name\PYGZdq{}}\PYG{p}{:} \PYG{l+s+s2}{\PYGZdq{}u1\PYGZdq{}}\PYG{p}{,}
  \PYG{n+nt}{\PYGZdq{}qubits\PYGZdq{}}\PYG{p}{:} \PYG{p}{[}\PYG{l+m+mi}{0}\PYG{p}{],}
  \PYG{n+nt}{\PYGZdq{}sequence\PYGZdq{}}\PYG{p}{:} \PYG{p}{[\PYGZob{}}\PYG{n+nt}{\PYGZdq{}name\PYGZdq{}}\PYG{p}{:} \PYG{l+s+s2}{\PYGZdq{}fc\PYGZdq{}}\PYG{p}{,} \PYG{n+nt}{\PYGZdq{}phase\PYGZdq{}}\PYG{p}{:} \PYG{l+s+s2}{\PYGZdq{}p0\PYGZdq{}}\PYG{p}{,}
                \PYG{n+nt}{\PYGZdq{}t0\PYGZdq{}}\PYG{p}{:} \PYG{l+m+mi}{0}\PYG{p}{,} \PYG{n+nt}{\PYGZdq{}ch\PYGZdq{}}\PYG{p}{:} \PYG{l+s+s2}{\PYGZdq{}d0\PYGZdq{}}\PYG{p}{\PYGZcb{}]}
\PYG{p}{\PYGZcb{}}
\end{Verbatim}
\end{jsonexamplebox}
where
\begin{itemize}
\item \texttt{name}: Name of the OpenQASM command; either a gate name from the list of basis gates or \texttt{measure}.
\item \texttt{qubits}: Qubit(s) for which this sequence defines the OpenQASM command.
\item \texttt{sequence}: Experiment sequence of commands that defines the OpenQASM command in terms of pulses (see \S~\ref{sect:actionitems}). This sequence may involve commands on several channels. The start of the OpenQASM command is $t=0$, and the length of the OpenQASM command is the time until the end of the last pulse.
\end{itemize}
Since OpenQASM gates can have parameters, these are indicated in the \texttt{cmd\_def} as numbers replaced by the string ``PX'' where ``X'' is the parameter number (X starts at zero). Given a \texttt{cmd\_def}, a layer will be included in Qiskit to convert an OpenQASM circuit into the OpenPulse format.  The following is an example pulse library and \texttt{cmd\_def} for two-qubits. The pulse library is:
\begin{jsonexamplebox}{Two Qubit Pulse Library}
\begin{Verbatim}[commandchars=\\\{\}]
\PYG{p}{\PYGZob{}}
  \PYG{n+nt}{\PYGZdq{}pulse\PYGZus{}library\PYGZdq{}}\PYG{p}{:[}
   \PYG{p}{\PYGZob{}}\PYG{n+nt}{\PYGZdq{}name\PYGZdq{}}\PYG{p}{:} \PYG{l+s+s2}{\PYGZdq{}pulse0\PYGZdq{}}\PYG{p}{,}
    \PYG{n+nt}{\PYGZdq{}samples\PYGZdq{}}\PYG{p}{:} \PYG{p}{[[}\PYG{l+m+mf}{0.004}\PYG{p}{,}\PYG{l+m+mf}{0.009}\PYG{p}{],}
          \PYG{p}{[}\PYG{l+m+mf}{0.029}\PYG{p}{,}\PYG{l+m+mf}{0.05}\PYG{p}{],[}\PYG{l+m+mf}{0.135}\PYG{p}{,}\PYG{l+m+mf}{0.18}\PYG{p}{],[}\PYG{l+m+mf}{0.41}\PYG{p}{,}\PYG{l+m+mf}{0.365}\PYG{p}{],}
          \PYG{p}{[}\PYG{l+m+mf}{0.8}\PYG{p}{,}\PYG{l+m+mf}{0.355}\PYG{p}{],[}\PYG{l+m+mf}{1.0}\PYG{p}{,}\PYG{l+m+mf}{0.0}\PYG{p}{],[}\PYG{l+m+mf}{0.8}\PYG{p}{,}\PYG{l+m+mf}{\PYGZhy{}0.355}\PYG{p}{],}
          \PYG{p}{[}\PYG{l+m+mf}{0.41}\PYG{p}{,}\PYG{l+m+mf}{\PYGZhy{}0.365}\PYG{p}{],[}\PYG{l+m+mf}{0.135}\PYG{p}{,}\PYG{l+m+mf}{\PYGZhy{}0.18}\PYG{p}{],[}\PYG{l+m+mf}{0.029}\PYG{p}{,}\PYG{l+m+mf}{\PYGZhy{}0.05}\PYG{p}{],}
          \PYG{p}{[}\PYG{l+m+mf}{0.004}\PYG{p}{,}\PYG{l+m+mf}{\PYGZhy{}0.009}\PYG{p}{]]\PYGZcb{},}
   \PYG{p}{\PYGZob{}}\PYG{n+nt}{\PYGZdq{}name\PYGZdq{}}\PYG{p}{:} \PYG{l+s+s2}{\PYGZdq{}pulse1\PYGZdq{}}\PYG{p}{,}
    \PYG{n+nt}{\PYGZdq{}samples\PYGZdq{}}\PYG{p}{:} \PYG{p}{[[}\PYG{l+m+mf}{0.004}\PYG{p}{,}\PYG{l+m+mf}{0.009}\PYG{p}{],[}\PYG{l+m+mf}{0.029}\PYG{p}{,}\PYG{l+m+mf}{0.05}\PYG{p}{],[}\PYG{l+m+mf}{0.135}\PYG{p}{,}\PYG{l+m+mf}{0.18}\PYG{p}{],}
          \PYG{p}{[}\PYG{l+m+mf}{0.41}\PYG{p}{,}\PYG{l+m+mf}{0.365}\PYG{p}{],[}\PYG{l+m+mf}{0.8}\PYG{p}{,}\PYG{l+m+mf}{0.355}\PYG{p}{],[}\PYG{l+m+mf}{1.0}\PYG{p}{,}\PYG{l+m+mf}{0.0}\PYG{p}{],}
          \PYG{p}{[}\PYG{l+m+mf}{0.8}\PYG{p}{,}\PYG{l+m+mf}{\PYGZhy{}0.355}\PYG{p}{],[}\PYG{l+m+mf}{0.41}\PYG{p}{,}\PYG{l+m+mf}{\PYGZhy{}0.365}\PYG{p}{],[}\PYG{l+m+mf}{0.135}\PYG{p}{,}\PYG{l+m+mf}{\PYGZhy{}0.18}\PYG{p}{],}
          \PYG{p}{[}\PYG{l+m+mf}{0.029}\PYG{p}{,}\PYG{l+m+mf}{\PYGZhy{}0.05}\PYG{p}{],[}\PYG{l+m+mf}{0.004}\PYG{p}{,}\PYG{l+m+mf}{\PYGZhy{}0.009}\PYG{p}{]]\PYGZcb{},}
   \PYG{p}{\PYGZob{}}\PYG{n+nt}{\PYGZdq{}name\PYGZdq{}}\PYG{p}{:} \PYG{l+s+s2}{\PYGZdq{}gauss\PYGZus{}square\PYGZdq{}}\PYG{p}{,}
    \PYG{n+nt}{\PYGZdq{}samples\PYGZdq{}}\PYG{p}{:} \PYG{p}{[[}\PYG{l+m+mf}{0.05}\PYG{p}{,}\PYG{l+m+mf}{0.0}\PYG{p}{],[}\PYG{l+m+mf}{0.15}\PYG{p}{,}\PYG{l+m+mf}{0.0}\PYG{p}{],[}\PYG{l+m+mf}{0.175}\PYG{p}{,}\PYG{l+m+mf}{0.0}\PYG{p}{],}
          \PYG{p}{[}\PYG{l+m+mf}{0.2}\PYG{p}{,}\PYG{l+m+mf}{0.0}\PYG{p}{],[}\PYG{l+m+mf}{0.2}\PYG{p}{,}\PYG{l+m+mf}{0.0}\PYG{p}{],[}\PYG{l+m+mf}{0.2}\PYG{p}{,}\PYG{l+m+mf}{0.0}\PYG{p}{],[}\PYG{l+m+mf}{0.2}\PYG{p}{,}\PYG{l+m+mf}{0.0}\PYG{p}{],}
          \PYG{p}{[}\PYG{l+m+mf}{0.175}\PYG{p}{,}\PYG{l+m+mf}{0.0}\PYG{p}{],[}\PYG{l+m+mf}{0.15}\PYG{p}{,}\PYG{l+m+mf}{0.0}\PYG{p}{],[}\PYG{l+m+mf}{0.05}\PYG{p}{,}\PYG{l+m+mf}{0.0}\PYG{p}{]]\PYGZcb{},}
   \PYG{p}{\PYGZob{}}\PYG{n+nt}{\PYGZdq{}name\PYGZdq{}}\PYG{p}{:} \PYG{l+s+s2}{\PYGZdq{}square\PYGZus{}pulse\PYGZdq{}}\PYG{p}{,}
    \PYG{n+nt}{\PYGZdq{}samples\PYGZdq{}}\PYG{p}{:} \PYG{p}{[[}\PYG{l+m+mf}{0.1}\PYG{p}{,}\PYG{l+m+mf}{0.0}\PYG{p}{],[}\PYG{l+m+mf}{0.1}\PYG{p}{,}\PYG{l+m+mf}{0.0}\PYG{p}{],[}\PYG{l+m+mf}{0.1}\PYG{p}{,}\PYG{l+m+mf}{0.0}\PYG{p}{],}
          \PYG{p}{[}\PYG{l+m+mf}{0.1}\PYG{p}{,}\PYG{l+m+mf}{0.0}\PYG{p}{],[}\PYG{l+m+mf}{0.1}\PYG{p}{,}\PYG{l+m+mf}{0.0}\PYG{p}{],[}\PYG{l+m+mf}{0.1}\PYG{p}{,}\PYG{l+m+mf}{0.0}\PYG{p}{]]\PYGZcb{}}
  \PYG{p}{]}
\PYG{p}{\PYGZcb{}}
\end{Verbatim}
\end{jsonexamplebox}
and the \texttt{cmd\_def} (omitting the single qubit gates and measurement for Q1 for brevity) is:
\begin{jsonexamplebox}{Two Qubit Command Definition}
\begin{Verbatim}[commandchars=\\\{\}]
\PYG{p}{[}
\PYG{p}{\PYGZob{}}\PYG{n+nt}{\PYGZdq{}name\PYGZdq{}}\PYG{p}{:}\PYG{l+s+s2}{\PYGZdq{}u1\PYGZdq{}}\PYG{p}{,} \PYG{n+nt}{\PYGZdq{}qubits\PYGZdq{}}\PYG{p}{:} \PYG{p}{[}\PYG{l+m+mi}{0}\PYG{p}{],}
\PYG{n+nt}{\PYGZdq{}instructions\PYGZdq{}} \PYG{p}{:[\PYGZob{}}\PYG{n+nt}{\PYGZdq{}name\PYGZdq{}}\PYG{p}{:} \PYG{l+s+s2}{\PYGZdq{}fc\PYGZdq{}}\PYG{p}{,} \PYG{n+nt}{\PYGZdq{}phase\PYGZdq{}}\PYG{p}{:} \PYG{l+s+s2}{\PYGZdq{}p0\PYGZdq{}}\PYG{p}{,}
              \PYG{n+nt}{\PYGZdq{}t0\PYGZdq{}}\PYG{p}{:} \PYG{l+m+mi}{0}\PYG{p}{,} \PYG{n+nt}{\PYGZdq{}ch\PYGZdq{}}\PYG{p}{:} \PYG{l+s+s2}{\PYGZdq{}d0\PYGZdq{}}\PYG{p}{\PYGZcb{}]\PYGZcb{},}
\PYG{p}{\PYGZob{}}\PYG{n+nt}{\PYGZdq{}name\PYGZdq{}}\PYG{p}{:} \PYG{l+s+s2}{\PYGZdq{}u2\PYGZdq{}}\PYG{p}{,} \PYG{n+nt}{\PYGZdq{}qubits\PYGZdq{}}\PYG{p}{:} \PYG{p}{[}\PYG{l+m+mi}{0}\PYG{p}{],}
\PYG{n+nt}{\PYGZdq{}instructions\PYGZdq{}}\PYG{p}{:} \PYG{p}{[\PYGZob{}}\PYG{n+nt}{\PYGZdq{}name\PYGZdq{}}\PYG{p}{:} \PYG{l+s+s2}{\PYGZdq{}fc\PYGZdq{}}\PYG{p}{,} \PYG{n+nt}{\PYGZdq{}phase\PYGZdq{}}\PYG{p}{:} \PYG{l+s+s2}{\PYGZdq{}p1\PYGZdq{}}\PYG{p}{,}
              \PYG{n+nt}{\PYGZdq{}t0\PYGZdq{}}\PYG{p}{:} \PYG{l+m+mi}{0}\PYG{p}{,} \PYG{n+nt}{\PYGZdq{}ch\PYGZdq{}}\PYG{p}{:} \PYG{l+s+s2}{\PYGZdq{}d0\PYGZdq{}}\PYG{p}{\PYGZcb{},}
\PYG{p}{\PYGZob{}}\PYG{n+nt}{\PYGZdq{}name\PYGZdq{}}\PYG{p}{:} \PYG{l+s+s2}{\PYGZdq{}fc\PYGZdq{}}\PYG{p}{,} \PYG{n+nt}{\PYGZdq{}phase\PYGZdq{}}\PYG{p}{:} \PYG{l+m+mf}{1.5708}\PYG{p}{,} \PYG{n+nt}{\PYGZdq{}t0\PYGZdq{}}\PYG{p}{:} \PYG{l+m+mi}{0}\PYG{p}{,} \PYG{n+nt}{\PYGZdq{}ch\PYGZdq{}}\PYG{p}{:} \PYG{l+s+s2}{\PYGZdq{}d0\PYGZdq{}}\PYG{p}{\PYGZcb{},}
\PYG{p}{\PYGZob{}}\PYG{n+nt}{\PYGZdq{}name\PYGZdq{}}\PYG{p}{:} \PYG{l+s+s2}{\PYGZdq{}pulse0\PYGZdq{}}\PYG{p}{,} \PYG{n+nt}{\PYGZdq{}t0\PYGZdq{}}\PYG{p}{:} \PYG{l+m+mi}{0}\PYG{p}{,} \PYG{n+nt}{\PYGZdq{}ch\PYGZdq{}}\PYG{p}{:} \PYG{l+s+s2}{\PYGZdq{}d0\PYGZdq{}}\PYG{p}{\PYGZcb{},}
\PYG{p}{\PYGZob{}}\PYG{n+nt}{\PYGZdq{}name\PYGZdq{}}\PYG{p}{:} \PYG{l+s+s2}{\PYGZdq{}fc\PYGZdq{}}\PYG{p}{,} \PYG{n+nt}{\PYGZdq{}phase\PYGZdq{}}\PYG{p}{:} \PYG{l+m+mf}{\PYGZhy{}1.5708}\PYG{p}{,} \PYG{n+nt}{\PYGZdq{}t0\PYGZdq{}}\PYG{p}{:} \PYG{l+m+mi}{11}\PYG{p}{,} \PYG{n+nt}{\PYGZdq{}ch\PYGZdq{}}\PYG{p}{:} \PYG{l+s+s2}{\PYGZdq{}d0\PYGZdq{}}\PYG{p}{\PYGZcb{},}
\PYG{p}{\PYGZob{}}\PYG{n+nt}{\PYGZdq{}name\PYGZdq{}}\PYG{p}{:} \PYG{l+s+s2}{\PYGZdq{}fc\PYGZdq{}}\PYG{p}{,} \PYG{n+nt}{\PYGZdq{}phase\PYGZdq{}}\PYG{p}{:} \PYG{l+s+s2}{\PYGZdq{}p0\PYGZdq{}}\PYG{p}{,} \PYG{n+nt}{\PYGZdq{}t0\PYGZdq{}}\PYG{p}{:} \PYG{l+m+mi}{11}\PYG{p}{,} \PYG{n+nt}{\PYGZdq{}ch\PYGZdq{}}\PYG{p}{:} \PYG{l+s+s2}{\PYGZdq{}d0\PYGZdq{}}\PYG{p}{\PYGZcb{}]\PYGZcb{},}
\PYG{p}{\PYGZob{}}\PYG{n+nt}{\PYGZdq{}name\PYGZdq{}}\PYG{p}{:} \PYG{l+s+s2}{\PYGZdq{}u3\PYGZdq{}}\PYG{p}{,} \PYG{n+nt}{\PYGZdq{}qubits\PYGZdq{}}\PYG{p}{:} \PYG{p}{[}\PYG{l+m+mi}{0}\PYG{p}{],}
\PYG{n+nt}{\PYGZdq{}instructions\PYGZdq{}}\PYG{p}{:} \PYG{p}{[\PYGZob{}}\PYG{n+nt}{\PYGZdq{}name\PYGZdq{}}\PYG{p}{:} \PYG{l+s+s2}{\PYGZdq{}fc\PYGZdq{}}\PYG{p}{,} \PYG{n+nt}{\PYGZdq{}phase\PYGZdq{}}\PYG{p}{:} \PYG{l+s+s2}{\PYGZdq{}p2\PYGZdq{}}\PYG{p}{,}
              \PYG{n+nt}{\PYGZdq{}t0\PYGZdq{}}\PYG{p}{:} \PYG{l+m+mi}{0}\PYG{p}{,} \PYG{n+nt}{\PYGZdq{}ch\PYGZdq{}}\PYG{p}{:} \PYG{l+s+s2}{\PYGZdq{}d0\PYGZdq{}}\PYG{p}{\PYGZcb{},}
\PYG{p}{\PYGZob{}}\PYG{n+nt}{\PYGZdq{}name\PYGZdq{}}\PYG{p}{:} \PYG{l+s+s2}{\PYGZdq{}pulse0\PYGZdq{}}\PYG{p}{,} \PYG{n+nt}{\PYGZdq{}t0\PYGZdq{}}\PYG{p}{:} \PYG{l+m+mi}{0}\PYG{p}{,} \PYG{n+nt}{\PYGZdq{}ch\PYGZdq{}}\PYG{p}{:} \PYG{l+s+s2}{\PYGZdq{}d0\PYGZdq{}}\PYG{p}{\PYGZcb{},}
\PYG{p}{\PYGZob{}}\PYG{n+nt}{\PYGZdq{}name\PYGZdq{}}\PYG{p}{:} \PYG{l+s+s2}{\PYGZdq{}fc\PYGZdq{}}\PYG{p}{,} \PYG{n+nt}{\PYGZdq{}phase\PYGZdq{}}\PYG{p}{:} \PYG{l+s+s2}{\PYGZdq{}p0\PYGZdq{}}\PYG{p}{,} \PYG{n+nt}{\PYGZdq{}t0\PYGZdq{}}\PYG{p}{:} \PYG{l+m+mi}{11}\PYG{p}{,} \PYG{n+nt}{\PYGZdq{}ch\PYGZdq{}}\PYG{p}{:} \PYG{l+s+s2}{\PYGZdq{}d0\PYGZdq{}}\PYG{p}{\PYGZcb{},}
\PYG{p}{\PYGZob{}}\PYG{n+nt}{\PYGZdq{}name\PYGZdq{}}\PYG{p}{:} \PYG{l+s+s2}{\PYGZdq{}fc\PYGZdq{}}\PYG{p}{,} \PYG{n+nt}{\PYGZdq{}phase\PYGZdq{}}\PYG{p}{:} \PYG{l+m+mf}{3.14}\PYG{p}{,} \PYG{n+nt}{\PYGZdq{}t0\PYGZdq{}}\PYG{p}{:} \PYG{l+m+mi}{11}\PYG{p}{,} \PYG{n+nt}{\PYGZdq{}ch\PYGZdq{}}\PYG{p}{:} \PYG{l+s+s2}{\PYGZdq{}d0\PYGZdq{}}\PYG{p}{\PYGZcb{},}
\PYG{p}{\PYGZob{}}\PYG{n+nt}{\PYGZdq{}name\PYGZdq{}}\PYG{p}{:} \PYG{l+s+s2}{\PYGZdq{}pulse0\PYGZdq{}}\PYG{p}{,} \PYG{n+nt}{\PYGZdq{}t0\PYGZdq{}}\PYG{p}{:} \PYG{l+m+mi}{11}\PYG{p}{,} \PYG{n+nt}{\PYGZdq{}ch\PYGZdq{}}\PYG{p}{:} \PYG{l+s+s2}{\PYGZdq{}d0\PYGZdq{}}\PYG{p}{\PYGZcb{},}
\PYG{p}{\PYGZob{}}\PYG{n+nt}{\PYGZdq{}name\PYGZdq{}}\PYG{p}{:} \PYG{l+s+s2}{\PYGZdq{}fc\PYGZdq{}}\PYG{p}{,} \PYG{n+nt}{\PYGZdq{}phase\PYGZdq{}}\PYG{p}{:} \PYG{l+m+mf}{3.14}\PYG{p}{,} \PYG{n+nt}{\PYGZdq{}t0\PYGZdq{}}\PYG{p}{:} \PYG{l+m+mi}{22}\PYG{p}{,} \PYG{n+nt}{\PYGZdq{}ch\PYGZdq{}}\PYG{p}{:} \PYG{l+s+s2}{\PYGZdq{}d0\PYGZdq{}}\PYG{p}{\PYGZcb{},}
\PYG{p}{\PYGZob{}}\PYG{n+nt}{\PYGZdq{}name\PYGZdq{}}\PYG{p}{:} \PYG{l+s+s2}{\PYGZdq{}fc\PYGZdq{}}\PYG{p}{,} \PYG{n+nt}{\PYGZdq{}phase\PYGZdq{}}\PYG{p}{:} \PYG{l+s+s2}{\PYGZdq{}p1\PYGZdq{}}\PYG{p}{,} \PYG{n+nt}{\PYGZdq{}t0\PYGZdq{}}\PYG{p}{:} \PYG{l+m+mi}{22}\PYG{p}{,} \PYG{n+nt}{\PYGZdq{}ch\PYGZdq{}}\PYG{p}{:} \PYG{l+s+s2}{\PYGZdq{}d0\PYGZdq{}}\PYG{p}{\PYGZcb{}]\PYGZcb{},}
\PYG{p}{\PYGZob{}}\PYG{n+nt}{\PYGZdq{}name\PYGZdq{}}\PYG{p}{:} \PYG{l+s+s2}{\PYGZdq{}cx\PYGZdq{}}\PYG{p}{,} \PYG{n+nt}{\PYGZdq{}qubits\PYGZdq{}}\PYG{p}{:} \PYG{p}{[}\PYG{l+m+mi}{0}\PYG{p}{,}\PYG{l+m+mi}{1}\PYG{p}{],}
 \PYG{n+nt}{\PYGZdq{}instructions\PYGZdq{}}\PYG{p}{:} \PYG{p}{[\PYGZob{}}\PYG{n+nt}{\PYGZdq{}name\PYGZdq{}}\PYG{p}{:} \PYG{l+s+s2}{\PYGZdq{}pulse0\PYGZdq{}}\PYG{p}{,}
               \PYG{n+nt}{\PYGZdq{}t0\PYGZdq{}}\PYG{p}{:} \PYG{l+m+mi}{0}\PYG{p}{,} \PYG{n+nt}{\PYGZdq{}ch\PYGZdq{}}\PYG{p}{:} \PYG{l+s+s2}{\PYGZdq{}d0\PYGZdq{}}\PYG{p}{\PYGZcb{},}
\PYG{p}{\PYGZob{}}\PYG{n+nt}{\PYGZdq{}name\PYGZdq{}}\PYG{p}{:} \PYG{l+s+s2}{\PYGZdq{}pulse1\PYGZdq{}}\PYG{p}{,} \PYG{n+nt}{\PYGZdq{}t0\PYGZdq{}}\PYG{p}{:} \PYG{l+m+mi}{0}\PYG{p}{,} \PYG{n+nt}{\PYGZdq{}ch\PYGZdq{}}\PYG{p}{:} \PYG{l+s+s2}{\PYGZdq{}d1\PYGZdq{}}\PYG{p}{\PYGZcb{},}
\PYG{p}{\PYGZob{}}\PYG{n+nt}{\PYGZdq{}name\PYGZdq{}}\PYG{p}{:} \PYG{l+s+s2}{\PYGZdq{}pulse0\PYGZdq{}}\PYG{p}{,} \PYG{n+nt}{\PYGZdq{}t0\PYGZdq{}}\PYG{p}{:} \PYG{l+m+mi}{11}\PYG{p}{,} \PYG{n+nt}{\PYGZdq{}ch\PYGZdq{}}\PYG{p}{:} \PYG{l+s+s2}{\PYGZdq{}d0\PYGZdq{}}\PYG{p}{\PYGZcb{},}
\PYG{p}{\PYGZob{}}\PYG{n+nt}{\PYGZdq{}name\PYGZdq{}}\PYG{p}{:} \PYG{l+s+s2}{\PYGZdq{}fc\PYGZdq{}}\PYG{p}{,} \PYG{n+nt}{\PYGZdq{}phase\PYGZdq{}}\PYG{p}{:} \PYG{l+m+mf}{\PYGZhy{}4.71239}\PYG{p}{,} \PYG{n+nt}{\PYGZdq{}t0\PYGZdq{}}\PYG{p}{:} \PYG{l+m+mi}{22}\PYG{p}{,} \PYG{n+nt}{\PYGZdq{}ch\PYGZdq{}}\PYG{p}{:} \PYG{l+s+s2}{\PYGZdq{}d0\PYGZdq{}}\PYG{p}{\PYGZcb{},}
\PYG{p}{\PYGZob{}}\PYG{n+nt}{\PYGZdq{}name\PYGZdq{}}\PYG{p}{:} \PYG{l+s+s2}{\PYGZdq{}gauss\PYGZus{}square\PYGZdq{}}\PYG{p}{,} \PYG{n+nt}{\PYGZdq{}t0\PYGZdq{}}\PYG{p}{:} \PYG{l+m+mi}{22}\PYG{p}{,} \PYG{n+nt}{\PYGZdq{}ch\PYGZdq{}}\PYG{p}{:} \PYG{l+s+s2}{\PYGZdq{}u0\PYGZdq{}}\PYG{p}{\PYGZcb{},}
\PYG{p}{\PYGZob{}}\PYG{n+nt}{\PYGZdq{}name\PYGZdq{}}\PYG{p}{:} \PYG{l+s+s2}{\PYGZdq{}pulse0\PYGZdq{}}\PYG{p}{,} \PYG{n+nt}{\PYGZdq{}t0\PYGZdq{}}\PYG{p}{:} \PYG{l+m+mi}{32}\PYG{p}{,} \PYG{n+nt}{\PYGZdq{}ch\PYGZdq{}}\PYG{p}{:} \PYG{l+s+s2}{\PYGZdq{}d0\PYGZdq{}}\PYG{p}{\PYGZcb{},}
\PYG{p}{\PYGZob{}}\PYG{n+nt}{\PYGZdq{}name\PYGZdq{}}\PYG{p}{:} \PYG{l+s+s2}{\PYGZdq{}pulse0\PYGZdq{}}\PYG{p}{,} \PYG{n+nt}{\PYGZdq{}t0\PYGZdq{}}\PYG{p}{:} \PYG{l+m+mi}{43}\PYG{p}{,} \PYG{n+nt}{\PYGZdq{}ch\PYGZdq{}}\PYG{p}{:} \PYG{l+s+s2}{\PYGZdq{}d0\PYGZdq{}}\PYG{p}{\PYGZcb{},}
\PYG{p}{\PYGZob{}}\PYG{n+nt}{\PYGZdq{}name\PYGZdq{}}\PYG{p}{:} \PYG{l+s+s2}{\PYGZdq{}fc\PYGZdq{}}\PYG{p}{,} \PYG{n+nt}{\PYGZdq{}phase\PYGZdq{}}\PYG{p}{:} \PYG{l+m+mf}{3.14}\PYG{p}{,} \PYG{n+nt}{\PYGZdq{}t0\PYGZdq{}}\PYG{p}{:} \PYG{l+m+mi}{54}\PYG{p}{,} \PYG{n+nt}{\PYGZdq{}ch\PYGZdq{}}\PYG{p}{:} \PYG{l+s+s2}{\PYGZdq{}u0\PYGZdq{}}\PYG{p}{\PYGZcb{},}
\PYG{p}{\PYGZob{}}\PYG{n+nt}{\PYGZdq{}name\PYGZdq{}}\PYG{p}{:} \PYG{l+s+s2}{\PYGZdq{}gauss\PYGZus{}square\PYGZdq{}}\PYG{p}{,} \PYG{n+nt}{\PYGZdq{}t0\PYGZdq{}}\PYG{p}{:} \PYG{l+m+mi}{54}\PYG{p}{,} \PYG{n+nt}{\PYGZdq{}ch\PYGZdq{}}\PYG{p}{:} \PYG{l+s+s2}{\PYGZdq{}u0\PYGZdq{}}\PYG{p}{\PYGZcb{},}
\PYG{p}{\PYGZob{}}\PYG{n+nt}{\PYGZdq{}name\PYGZdq{}}\PYG{p}{:} \PYG{l+s+s2}{\PYGZdq{}fc\PYGZdq{}}\PYG{p}{,} \PYG{n+nt}{\PYGZdq{}phase\PYGZdq{}}\PYG{p}{:} \PYG{l+m+mf}{\PYGZhy{}3.14}\PYG{p}{,} \PYG{n+nt}{\PYGZdq{}t0\PYGZdq{}}\PYG{p}{:} \PYG{l+m+mi}{64}\PYG{p}{,} \PYG{n+nt}{\PYGZdq{}ch\PYGZdq{}}\PYG{p}{:} \PYG{l+s+s2}{\PYGZdq{}u0\PYGZdq{}}\PYG{p}{\PYGZcb{}]\PYGZcb{},}
\PYG{p}{\PYGZob{}}\PYG{n+nt}{\PYGZdq{}name\PYGZdq{}}\PYG{p}{:} \PYG{l+s+s2}{\PYGZdq{}measure\PYGZdq{}}\PYG{p}{,} \PYG{n+nt}{\PYGZdq{}qubits\PYGZdq{}}\PYG{p}{:} \PYG{p}{[}\PYG{l+m+mi}{0}\PYG{p}{],}
 \PYG{n+nt}{\PYGZdq{}instructions\PYGZdq{}}\PYG{p}{:} \PYG{p}{[\PYGZob{}}\PYG{n+nt}{\PYGZdq{}name\PYGZdq{}}\PYG{p}{:} \PYG{l+s+s2}{\PYGZdq{}square\PYGZus{}pulse\PYGZdq{}}\PYG{p}{,}
	 \PYG{n+nt}{\PYGZdq{}t0\PYGZdq{}}\PYG{p}{:} \PYG{l+m+mi}{0}\PYG{p}{,} \PYG{n+nt}{\PYGZdq{}ch\PYGZdq{}}\PYG{p}{:} \PYG{l+s+s2}{\PYGZdq{}m0\PYGZdq{}}\PYG{p}{\PYGZcb{},}
 \PYG{p}{\PYGZob{}}\PYG{n+nt}{\PYGZdq{}name\PYGZdq{}}\PYG{p}{:} \PYG{l+s+s2}{\PYGZdq{}acquire\PYGZdq{}}\PYG{p}{,} \PYG{n+nt}{\PYGZdq{}t0\PYGZdq{}}\PYG{p}{:} \PYG{l+m+mi}{3}\PYG{p}{,} \PYG{n+nt}{\PYGZdq{}duration\PYGZdq{}}\PYG{p}{:} \PYG{l+m+mi}{10}\PYG{p}{\PYGZcb{}]\PYGZcb{}}
\PYG{p}{]}
\end{Verbatim}
\end{jsonexamplebox}
The \texttt{acquire} command requires a \texttt{memory\_slot} key; this is assumed to come from the openQASM command being converted and will be appended by the conversion software. The two qubit gate (``cx'') illustrates how the \texttt{cmd\_def} can apply pulse sequences to multiple channels.

\subsubsection{Additional Configuration Settings in Qobj \label{sect:pulseQObjconfig}}

The \texttt{user\_config} data structure that is passed in through the \texttt{Qobj} (see \S~\ref{sect:rundict}) has the following additional settings for OpenPulse.
\begin{jsonexamplebox}{OpenPulse \texttt}
\begin{Verbatim}[commandchars=\\\{\}]
\PYG{p}{\PYGZob{}}
  \PYG{n+nt}{\PYGZdq{}meas\PYGZus{}level\PYGZdq{}}\PYG{p}{:} \PYG{l+m+mi}{1}\PYG{p}{,}
  \PYG{n+nt}{\PYGZdq{}pulse\PYGZus{}library\PYGZdq{}}\PYG{p}{:} \PYG{err}{pulse\PYGZus{}lib}\PYG{p}{,}
  \PYG{n+nt}{\PYGZdq{}memory\PYGZus{}slot\PYGZus{}size\PYGZdq{}}\PYG{p}{:} \PYG{l+m+mi}{100}\PYG{p}{,}
  \PYG{n+nt}{\PYGZdq{}meas\PYGZus{}return\PYGZdq{}}\PYG{p}{:} \PYG{l+s+s2}{\PYGZdq{}single\PYGZdq{}}\PYG{p}{,}
  \PYG{n+nt}{\PYGZdq{}qubit\PYGZus{}lo\PYGZus{}freq\PYGZdq{}}\PYG{p}{:} \PYG{p}{[}\PYG{l+m+mf}{5.034}\PYG{p}{,}\PYG{l+m+mf}{5.1242}\PYG{p}{,}\PYG{l+m+mf}{5.2353}\PYG{p}{],}
  \PYG{n+nt}{\PYGZdq{}meas\PYGZus{}lo\PYGZus{}freq\PYGZdq{}}\PYG{p}{:} \PYG{p}{[}\PYG{l+m+mf}{6.5324}\PYG{p}{,}\PYG{l+m+mf}{6.6745}\PYG{p}{,}\PYG{l+m+mf}{6.754}\PYG{p}{],}
  \PYG{n+nt}{\PYGZdq{}rep\PYGZus{}time\PYGZdq{}}\PYG{p}{:} \PYG{l+m+mi}{1000}
\PYG{p}{\PYGZcb{}}
\end{Verbatim}
\end{jsonexamplebox}
where
\begin{itemize}
\item \texttt{meas\_level}: Set the appropriate level of the measurement output as described in \S~\ref{sect:meas}
\item \texttt{pulse\_library}: \texttt{pulse\_lib} data structure defining the set of primitive pulses as described in \S~\ref{sect:pulselib}.
\item \texttt{memory\_slot\_size}: Size of each memory slot if the output is Level 0. The total number of memory slots ($r$) is defined in the general \texttt{Qobj} specification (\S~\ref{sect:rundict}).
\item \texttt{meas\_return}: Indicates the level of measurement information to return. \texttt{single} returns information from every shot of the experiment. \texttt{avg} returns the average measurement output (averaged over the number of shots). If the \texttt{meas\_level} is 2 then this is fixed to \texttt{single}.
\item \texttt{qubit\_lo\_freq}: List of frequencies for the qubit drive LO's (in GHz). Must be within \texttt{qubit\_lo\_range} given by \texttt{Backend.configuration}.
\item \texttt{meas\_lo\_freq}: List of frequencies for the measurement drive LO's (in GHz). Must be within \texttt{meas\_lo\_range} given by \texttt{Backend.configuration}.
\item \texttt{rep\_time}: Repetition time of the experiment in $\mu$s, ie. the delay between experiments will be \texttt{rep\_time}. Must be from the list provided by the device.
\end{itemize}

\subsection{Experiment Sequence Commands \label{sect:actionitems}}

The main part of the user/device interface, as defined in this document, is the specification of an experimental time sequence of commands. These can be either input pulses to drive the qubits, control lines or measurement stimuli (\S~\ref{sect:drivepulse}), frame changes (\S~\ref{sect:fc}), persistent value pulses (\S~\ref{sect:pv}) or measurement acquisitions (\S~\ref{sect:acq}). Here we specify these four different commands. Time sequences will be discussed in more detail in \S~\ref{sect:seq}. If the backend supports conditionals and the measurement level is set to 2 then boolean and copy functions (as described in \S~\ref{sect:qasmclassicalfunc}) are also allowed.

\subsubsection{Input Drive Pulse \label{sect:drivepulse}}

An input drive pulse (i.e. a pulse on channels $d$, $m$ or $U$) is specified with a structure of pre-defined pulses (pulse library \S~\ref{sect:pulselib}) at given times.  A drive pulse command is specified in an experiment sequence as an input pulse data structure which specifies the pulse name in the \texttt{pulse\_library}, the time of the pulse, and the pulse masking (if the measurement output is set to level 2, see \S~\ref{sect:meas}). Here is an example input-pulse command structure:
\begin{jsonexamplebox}{Input Pulse}
\begin{Verbatim}[commandchars=\\\{\}]
\PYG{p}{\PYGZob{}}
  \PYG{n+nt}{\PYGZdq{}name\PYGZdq{}}\PYG{p}{:} \PYG{l+s+s2}{\PYGZdq{}drag\PYGZus{}pulse\PYGZdq{}}\PYG{p}{,}
  \PYG{n+nt}{\PYGZdq{}t0\PYGZdq{}}\PYG{p}{:} \PYG{l+m+mi}{10}\PYG{p}{,}
  \PYG{n+nt}{\PYGZdq{}ch\PYGZdq{}}\PYG{p}{:} \PYG{l+s+s2}{\PYGZdq{}d0\PYGZdq{}}\PYG{p}{,}
  \PYG{n+nt}{\PYGZdq{}conditional\PYGZdq{}}\PYG{p}{:} \PYG{l+m+mi}{3}
\PYG{p}{\PYGZcb{}}
\end{Verbatim}
\end{jsonexamplebox}
where
\begin{itemize}
\item \texttt{name}: The name of the pulse in the pulse library (the name cannot be a reserved name ``fc'', ``pv'' or ``acquire'').
\item \texttt{t0}: The pulse start time in integer \texttt{dt} units (all experiments will start at the same $dt=0$ time).
\item \texttt{ch}: The channel to apply this pulse (one of the $d$, $m$ or $U$ channels).
\item \texttt{conditional} (optional): If conditionals are allowed and the measurement output is level 2 then apply the pulse if the given register (in this example register 3) is 1 (true). If left blank then the pulse has no conditional element (i.e. no feedback). By default this is blank.
\end{itemize}

\subsubsection{Frame Change \label{sect:fc}}

There is a special type of input pulse, implementing a phase advance of all subsequent pulses on that channel, i.e., each subsequent pulse is multiplied by $e^{-\iu\phi}$. The phase advances on each channel are independent. The conditional option for the drive pulse described above also applies to the frame change pulse. Here is an example pulse data structure for an FC pulse (with the conditional left blank):
\begin{jsonexamplebox}{Frame Change Pulse}
\begin{Verbatim}[commandchars=\\\{\}]
\PYG{p}{\PYGZob{}}
  \PYG{n+nt}{\PYGZdq{}name\PYGZdq{}}\PYG{p}{:} \PYG{l+s+s2}{\PYGZdq{}fc\PYGZdq{}}\PYG{p}{,}
  \PYG{n+nt}{\PYGZdq{}t0\PYGZdq{}}\PYG{p}{:} \PYG{l+m+mi}{10}\PYG{p}{,}
  \PYG{n+nt}{\PYGZdq{}ch\PYGZdq{}}\PYG{p}{:} \PYG{l+s+s2}{\PYGZdq{}d0\PYGZdq{}}\PYG{p}{,}
  \PYG{n+nt}{\PYGZdq{}phase\PYGZdq{}}\PYG{p}{:} \PYG{l+m+mf}{0.2}
\PYG{p}{\PYGZcb{}}
\end{Verbatim}
\end{jsonexamplebox}
where
\begin{itemize}
\item \texttt{name}: Reserved name for the frame change pulse (``fc'').
\item \texttt{t0}: Time at which the frame change will apply (the frame change is applied to all drive pulses after this time) in integer \texttt{dt} units.
\item \texttt{ch}: Channel to apply the frame change.
\item \texttt{phase}: Frame change phase in radians. The allowable precision is device specific.
\end{itemize}

\subsubsection{Persistent Value \label{sect:pv}}

The Persistent Value pulse is a special type of input pulse which holds the value until the time of the next pulse, including between experiments in the same run if allowed by the device. One use case for a persistent value pulse is to create a long square pulse. Note that after a \texttt{Qobj} is complete, all channels are set back to zero. Here is an example pulse data structure for a Persistent Value pulse:
\begin{jsonexamplebox}{Persistent Value Pulse}
\begin{Verbatim}[commandchars=\\\{\}]
\PYG{p}{\PYGZob{}}
  \PYG{n+nt}{\PYGZdq{}name\PYGZdq{}}\PYG{p}{:} \PYG{l+s+s2}{\PYGZdq{}pv\PYGZdq{}}\PYG{p}{,}
  \PYG{n+nt}{\PYGZdq{}t0\PYGZdq{}}\PYG{p}{:} \PYG{l+m+mi}{10}\PYG{p}{,}
  \PYG{n+nt}{\PYGZdq{}ch\PYGZdq{}}\PYG{p}{:} \PYG{l+s+s2}{\PYGZdq{}d0\PYGZdq{}}\PYG{p}{,}
  \PYG{n+nt}{\PYGZdq{}val\PYGZdq{}}\PYG{p}{:} \PYG{p}{[}\PYG{l+m+mf}{0.2}\PYG{p}{,}\PYG{l+m+mf}{\PYGZhy{}0.2}\PYG{p}{]}
\PYG{p}{\PYGZcb{}}
\end{Verbatim}
\end{jsonexamplebox}
where
\begin{itemize}
\item \texttt{name}: Reserved name for the persistent value pulse (``pv'').
\item \texttt{t0}: Time to apply the value in integer \texttt{dt} units.
\item \texttt{ch}: Channel which to apply this persistent value.
\item \texttt{val}: Complex value to apply, bounded by an absolute value of 1. The allowable precision is device specific.
\end{itemize}

\subsubsection{Acquisition Item \label{sect:acq}}

In this specification, a measurement is only acquired when indicated by an acquisition command, i.e., the measurement output is not set by the stimulus, although they should be coordinated. The acquisition command is a data structure that specifies the start time and duration of the measurement, which qubits to measure, which memory slot to store the measurement, which kernels to use (for Level 1), and which discriminators to use (for Level 2). For information on the levels see \S~\ref{sect:meas}. The acquisition data structure has the form:
\begin{jsonexamplebox}{Acquisition Command}
\begin{Verbatim}[commandchars=\\\{\}]
\PYG{p}{\PYGZob{}}
  \PYG{n+nt}{\PYGZdq{}name\PYGZdq{}}\PYG{p}{:} \PYG{l+s+s2}{\PYGZdq{}acquire\PYGZdq{}}\PYG{p}{,}
  \PYG{n+nt}{\PYGZdq{}t0\PYGZdq{}}\PYG{p}{:} \PYG{l+m+mi}{100}\PYG{p}{,}
  \PYG{n+nt}{\PYGZdq{}duration\PYGZdq{}}\PYG{p}{:} \PYG{l+m+mi}{7}\PYG{p}{,}
  \PYG{n+nt}{\PYGZdq{}qubits\PYGZdq{}}\PYG{p}{:} \PYG{p}{[}\PYG{l+m+mi}{0}\PYG{p}{,}\PYG{l+m+mi}{1}\PYG{p}{],}
  \PYG{n+nt}{\PYGZdq{}memory\PYGZus{}slot\PYGZdq{}}\PYG{p}{:} \PYG{p}{[}\PYG{l+m+mi}{0}\PYG{p}{,}\PYG{l+m+mi}{2}\PYG{p}{],}
  \PYG{n+nt}{\PYGZdq{}register\PYGZus{}slot\PYGZdq{}}\PYG{p}{:} \PYG{p}{[}\PYG{l+m+mi}{0}\PYG{p}{,}\PYG{l+m+mi}{1}\PYG{p}{],}
  \PYG{n+nt}{\PYGZdq{}kernels\PYGZdq{}}\PYG{p}{:} \PYG{p}{[\PYGZob{}}\PYG{n+nt}{\PYGZdq{}name\PYGZdq{}}\PYG{p}{:} \PYG{l+s+s2}{\PYGZdq{}boxcar\PYGZdq{}}\PYG{p}{,} \PYG{n+nt}{\PYGZdq{}params\PYGZdq{}}\PYG{p}{:} \PYG{p}{[]\PYGZcb{},}
              \PYG{p}{\PYGZob{}}\PYG{n+nt}{\PYGZdq{}name\PYGZdq{}}\PYG{p}{:} \PYG{l+s+s2}{\PYGZdq{}boxcar\PYGZdq{}}\PYG{p}{,} \PYG{n+nt}{\PYGZdq{}params\PYGZdq{}}\PYG{p}{:} \PYG{p}{[]\PYGZcb{}],}
  \PYG{n+nt}{\PYGZdq{}discriminators\PYGZdq{}}\PYG{p}{:} \PYG{p}{[\PYGZob{}}\PYG{n+nt}{\PYGZdq{}name\PYGZdq{}}\PYG{p}{:} \PYG{l+s+s2}{\PYGZdq{}max\PYGZus{}2Q\PYGZus{}fidelity\PYGZdq{}}\PYG{p}{,}
  \PYG{n+nt}{\PYGZdq{}params\PYGZdq{}}\PYG{p}{:} \PYG{p}{[]\PYGZcb{}]}
\PYG{p}{\PYGZcb{}}
\end{Verbatim}
\end{jsonexamplebox}

where
\begin{itemize}
\item \texttt{name}: Reserved name for the acquisition pulse (``acquire'').
\item \texttt{t0}, \texttt{duration}: Start and duration of that specific measurement in integer \texttt{dt} units. If the measurement output level is 0 (raw) then the data returned will be in \texttt{dtm} units (see \S~\ref{sect:pulseconfig} for the definition of \texttt{dt} and \texttt{dtm}).
\item \texttt{qubits}: List of the qubits to measure during this acquisition.
\item \texttt{memory\_slot}: List of the classical memory slots to store the measurement results. Must be the same length as the qubit list. The total number of memory slots is specified through the \texttt{Qobj} (see \S~\ref{sect:pulseQObjconfig}). Memory slot numbering starts at 0, and memory slots can be overwritten if they are specified twice in the same experiment (separate experiments have separate memories).
\item \texttt{register\_slot} (optional): List of the classical register slots to store the measurement results. Must be the same length as the qubit list. The total number of register slots is specified specified by the backend. This is only allowed if the backend supports conditionals \emph{and} the memory level is 2 (discriminated). Registers can only accept bits.
\item \texttt{kernels}: List of the data structures defining the measurement kernels to be used (from the list of available kernels) and set of parameters (if applicable) if the measurement level is 1 or 2. If this is left blank, then the default kernel is used. If the parameters are left blank, then the default parameters are used (see \S~\ref{sect:kernel}).
\item \texttt{discriminators}: Discriminators to be used (from the list of available discriminator) if the measurement level is 2. If left blank then a default discriminator is used (see \S~\ref{sect:kernel}).
\end{itemize}
The qubits are applied to the kernels and discriminators in the same order as listed in the \texttt{qubits} list. If only one kernel/discriminator is listed, then this is a multi-qubit kernel/discriminator function, and the data from all qubits will be sent jointly. Whether or not a kernel/discriminator can process multi-qubit data is information that has to be conveyed by the device in separate documentation. In the example listed above, the kernels are applied to each qubit, but the discriminator operates on both qubits. \\

If multiple acquisition commands are applied at identical times on identical qubits, this will acquire one set of data, but send the data through different kernel/discriminators, and to different memory slots as specified in the individual acquisition commands. However, this may not be supported by the device. Also, note that the device may limit the number of acquisitions per experiment and/or the time between acquisitions.

\subsubsection{Snapshot}

OpenPulse simulators may also support the snapshot as detailed in \S~\ref{sect:snapshot} except with an additional field $\texttt{t0}$ to indicate the time of the snapshot.

\subsection{Time Sequence Specification \label{sect:seq}}

In OpenPulse an experiment is specified as a sequence of the action items defined in \S~\ref{sect:actionitems}. Items are to be time ordered in the list. For example, a sequence for several qubit drive pulses (no feedback), a measurement stimulus, and acquisition would be the following,
\begin{jsonexamplebox}{Sequence of Action Items}
\begin{Verbatim}[commandchars=\\\{\}]
\PYG{p}{[\PYGZob{}}\PYG{n+nt}{\PYGZdq{}name\PYGZdq{}}\PYG{p}{:} \PYG{l+s+s2}{\PYGZdq{}pulse1\PYGZdq{}}\PYG{p}{,} \PYG{n+nt}{\PYGZdq{}t0\PYGZdq{}}\PYG{p}{:} \PYG{l+m+mi}{0}\PYG{p}{,} \PYG{n+nt}{\PYGZdq{}ch\PYGZdq{}}\PYG{p}{:} \PYG{l+s+s2}{\PYGZdq{}d0\PYGZdq{}}\PYG{p}{\PYGZcb{},}
\PYG{p}{\PYGZob{}}\PYG{n+nt}{\PYGZdq{}name\PYGZdq{}}\PYG{p}{:} \PYG{l+s+s2}{\PYGZdq{}fc\PYGZdq{}}\PYG{p}{,} \PYG{n+nt}{\PYGZdq{}t0\PYGZdq{}}\PYG{p}{:} \PYG{l+m+mi}{10}\PYG{p}{,} \PYG{n+nt}{\PYGZdq{}phase\PYGZdq{}}\PYG{p}{:} \PYG{l+m+mf}{0.1}\PYG{p}{,} \PYG{n+nt}{\PYGZdq{}ch\PYGZdq{}}\PYG{p}{:} \PYG{l+s+s2}{\PYGZdq{}d0\PYGZdq{}}\PYG{p}{\PYGZcb{},}
\PYG{p}{\PYGZob{}}\PYG{n+nt}{\PYGZdq{}name\PYGZdq{}}\PYG{p}{:} \PYG{l+s+s2}{\PYGZdq{}drag\PYGZus{}pulse\PYGZdq{}}\PYG{p}{,} \PYG{n+nt}{\PYGZdq{}t0\PYGZdq{}}\PYG{p}{:} \PYG{l+m+mi}{10}\PYG{p}{,} \PYG{n+nt}{\PYGZdq{}ch\PYGZdq{}}\PYG{p}{:} \PYG{l+s+s2}{\PYGZdq{}d0\PYGZdq{}}\PYG{p}{\PYGZcb{},}
\PYG{p}{\PYGZob{}}\PYG{n+nt}{\PYGZdq{}name\PYGZdq{}}\PYG{p}{:} \PYG{l+s+s2}{\PYGZdq{}square\PYGZus{}pulse\PYGZdq{}}\PYG{p}{,} \PYG{n+nt}{\PYGZdq{}t0\PYGZdq{}}\PYG{p}{:} \PYG{l+m+mi}{25}\PYG{p}{,} \PYG{n+nt}{\PYGZdq{}ch\PYGZdq{}}\PYG{p}{:} \PYG{l+s+s2}{\PYGZdq{}m0\PYGZdq{}}\PYG{p}{\PYGZcb{},}
\PYG{p}{\PYGZob{}}\PYG{n+nt}{\PYGZdq{}name\PYGZdq{}}\PYG{p}{:} \PYG{l+s+s2}{\PYGZdq{}acquire\PYGZdq{}}\PYG{p}{,} \PYG{n+nt}{\PYGZdq{}t0\PYGZdq{}}\PYG{p}{:} \PYG{l+m+mi}{25}\PYG{p}{,} \PYG{n+nt}{\PYGZdq{}duration\PYGZdq{}}\PYG{p}{:} \PYG{l+m+mi}{10}\PYG{p}{,}
 \PYG{n+nt}{\PYGZdq{}qubits\PYGZdq{}}\PYG{p}{:} \PYG{p}{[}\PYG{l+m+mi}{0}\PYG{p}{],} \PYG{n+nt}{\PYGZdq{}memory\PYGZus{}slot\PYGZdq{}}\PYG{p}{:} \PYG{p}{[}\PYG{l+m+mi}{0}\PYG{p}{]\PYGZcb{}]}
\end{Verbatim}
\end{jsonexamplebox}

Additionally, a sequence with two acquisitions and feedback based on the first can be written as,
\begin{jsonexamplebox}{Action Items with Feedback}
\begin{Verbatim}[commandchars=\\\{\}]
\PYG{p}{[}
\PYG{p}{\PYGZob{}}\PYG{n+nt}{\PYGZdq{}name\PYGZdq{}}\PYG{p}{:} \PYG{l+s+s2}{\PYGZdq{}square\PYGZus{}pulse\PYGZdq{}}\PYG{p}{,} \PYG{n+nt}{\PYGZdq{}t0\PYGZdq{}}\PYG{p}{:} \PYG{l+m+mi}{0}\PYG{p}{,} \PYG{n+nt}{\PYGZdq{}ch\PYGZdq{}}\PYG{p}{:} \PYG{l+s+s2}{\PYGZdq{}m0\PYGZdq{}}\PYG{p}{\PYGZcb{},}
\PYG{p}{\PYGZob{}}\PYG{n+nt}{\PYGZdq{}name\PYGZdq{}}\PYG{p}{:} \PYG{l+s+s2}{\PYGZdq{}acquire\PYGZdq{}}\PYG{p}{,} \PYG{n+nt}{\PYGZdq{}t0\PYGZdq{}}\PYG{p}{:} \PYG{l+m+mi}{0}\PYG{p}{,} \PYG{n+nt}{\PYGZdq{}duration\PYGZdq{}}\PYG{p}{:} \PYG{l+m+mi}{10}\PYG{p}{,}
  \PYG{n+nt}{\PYGZdq{}qubits\PYGZdq{}}\PYG{p}{:} \PYG{p}{[}\PYG{l+m+mi}{0}\PYG{p}{],} \PYG{n+nt}{\PYGZdq{}memory\PYGZus{}slot\PYGZdq{}}\PYG{p}{:} \PYG{p}{[}\PYG{l+m+mi}{0}\PYG{p}{],} \PYG{n+nt}{\PYGZdq{}register\PYGZus{}slot\PYGZdq{}}\PYG{p}{:} \PYG{p}{[}\PYG{l+m+mi}{0}\PYG{p}{]\PYGZcb{},}
\PYG{p}{\PYGZob{}}\PYG{n+nt}{\PYGZdq{}name\PYGZdq{}}\PYG{p}{:} \PYG{l+s+s2}{\PYGZdq{}pulse1\PYGZdq{}}\PYG{p}{,} \PYG{n+nt}{\PYGZdq{}t0\PYGZdq{}}\PYG{p}{:} \PYG{l+m+mi}{10}\PYG{p}{,} \PYG{n+nt}{\PYGZdq{}ch\PYGZdq{}}\PYG{p}{:} \PYG{l+s+s2}{\PYGZdq{}d0\PYGZdq{}}\PYG{p}{,}
  \PYG{n+nt}{\PYGZdq{}conditional\PYGZdq{}}\PYG{p}{:} \PYG{l+m+mi}{0}\PYG{p}{\PYGZcb{},}
\PYG{p}{\PYGZob{}}\PYG{n+nt}{\PYGZdq{}name\PYGZdq{}}\PYG{p}{:} \PYG{l+s+s2}{\PYGZdq{}square\PYGZus{}pulse\PYGZdq{}}\PYG{p}{,} \PYG{n+nt}{\PYGZdq{}t0\PYGZdq{}}\PYG{p}{:} \PYG{l+m+mi}{25}\PYG{p}{,} \PYG{n+nt}{\PYGZdq{}ch\PYGZdq{}}\PYG{p}{:} \PYG{l+s+s2}{\PYGZdq{}m0\PYGZdq{}}\PYG{p}{\PYGZcb{},}
\PYG{p}{\PYGZob{}}\PYG{n+nt}{\PYGZdq{}name\PYGZdq{}}\PYG{p}{:} \PYG{l+s+s2}{\PYGZdq{}acquire\PYGZdq{}}\PYG{p}{,} \PYG{n+nt}{\PYGZdq{}t0\PYGZdq{}}\PYG{p}{:} \PYG{l+m+mi}{25}\PYG{p}{,} \PYG{n+nt}{\PYGZdq{}duration\PYGZdq{}}\PYG{p}{:} \PYG{l+m+mi}{10}\PYG{p}{,}
  \PYG{n+nt}{\PYGZdq{}qubits\PYGZdq{}}\PYG{p}{:} \PYG{p}{[}\PYG{l+m+mi}{0}\PYG{p}{],} \PYG{n+nt}{\PYGZdq{}memory\PYGZus{}slot\PYGZdq{}}\PYG{p}{:} \PYG{p}{[}\PYG{l+m+mi}{1}\PYG{p}{]\PYGZcb{}}
\PYG{p}{]}
\end{Verbatim}
\end{jsonexamplebox}

More examples are given in the sample experiments section \S~\ref{sect:samples}. Between pulses on the same channel the output will be zero \emph{unless} a persistent value pulse is specified \S~\ref{sect:pv}, in which case the output will be the value of the persistent value pulse until another pulse is applied.

\subsection{Measurement Output \label{sect:meas}}

Measurements (instigated by an acquisition command in the sequence) are retrieved from the call to \texttt{Job.result()} (\S~\ref{sect:jobresult}). The type of data collected, and the level of processing depends on the measurement level selected. There are three levels of measurement output (devices do not need to support all three). In ``Level 0'', the output is the raw measurement pulse, in ``Level 1'' the output is a complex number (IQ) obtained after the application of a kernel and in  ``Level 2'' the output is the qubit state in the computational basis (0 or 1) comparable to the output from OpenQasm experiments. For Level 1 the user specifies a measurement kernel (from a list of available kernels), and for Level 2 the user also specifies a discriminator (from a list of available discriminators). These measurement options are summarized in Table~\ref{table:meas} and discussed in detail in the sections below. If the measurement is level 2 or the backend is a simulator returning either the statevector or unitary matrix, then the output is described in \S~\ref{sect:measdata}. Otherwise, the \texttt{result} data structure for measurement levels 0 and 1 has the following form,
\begin{jsonexamplebox}{Level 0 and 1 Result}
\begin{Verbatim}[commandchars=\\\{\}]
\PYG{p}{\PYGZob{}}
  \PYG{n+nt}{\PYGZdq{}memory\PYGZdq{}}\PYG{p}{:} \PYG{err}{data}
\PYG{p}{\PYGZcb{}}
\end{Verbatim}
\end{jsonexamplebox}
where
\begin{itemize}
\item \texttt{memory}: List of dimension $s \times r \times l$ where $s$ is the number of shots, $r$ is the number of memory slots, and $l$ is the slot size (1 for level 1, value of ``memory\_slot\_size'' for level 0). If the data is averaged then $s=1$.
\end{itemize}

\begin{table}
    \begin{tabular}{|c|p{0.2\textwidth}|p{0.3\textwidth}|p{0.2\textwidth}|}
    \hline Level & User Inputs & Limitations & Output \\ \hline \hline
        0 & N/A & High bandwidth, should only be used in averaging mode, no feedback & Time sequences of complex amplitudes \\ \hline
    1 & Kernel (from list) & No feedback & IQ values \\ \hline
    2 & Kernel and Discriminator (from list) & Single shot only & Qubit state \\ \hline
\end{tabular}
\caption{Summary of measurement output levels. \label{table:meas}}
\end{table}

\subsubsection{Measurement Output Level 0: Raw}

In output Level 0, the measurement output is a direct stream from the device during the acquisition period after mixing down with the measurement stimulus LO, i.e., the envelope of the measurement output. This output pulse should be similar to the input measurement pulse defined by the user, except for transformation due to the measurement process.  Data is a list of amplitude points spaced in equal \texttt{dtm} time intervals (see \S~\ref{sect:pulseconfig}). For example, in Level 0 average mode with two memory slots and each slot is three samples long,
\begin{jsonexamplebox}{Level 0 Average Mode Measurement Output}
\begin{Verbatim}[commandchars=\\\{\}]
\PYG{p}{\PYGZob{}}
  \PYG{n+nt}{\PYGZdq{}memory\PYGZdq{}}\PYG{p}{:} \PYG{p}{[[[}\PYG{l+m+mf}{0.1}\PYG{p}{,}\PYG{l+m+mf}{0.2}\PYG{p}{],[}\PYG{l+m+mf}{0.3}\PYG{p}{,}\PYG{l+m+mf}{\PYGZhy{}0.1}\PYG{p}{],[}\PYG{l+m+mf}{0.5}\PYG{p}{,}\PYG{l+m+mf}{0.8}\PYG{p}{]],}
             \PYG{p}{[[}\PYG{l+m+mf}{0.15}\PYG{p}{,}\PYG{l+m+mf}{0.7}\PYG{p}{],[}\PYG{l+m+mf}{0.13}\PYG{p}{,}\PYG{l+m+mf}{0.3}\PYG{p}{],[}\PYG{l+m+mf}{\PYGZhy{}0.5}\PYG{p}{,}\PYG{l+m+mf}{0.4}\PYG{p}{]]]}
\PYG{p}{\PYGZcb{}}
\end{Verbatim}
\end{jsonexamplebox}

In single shot mode (for two shots),
\begin{jsonexamplebox}{Level 0 Single Shot Measurement Output}
\begin{Verbatim}[commandchars=\\\{\}]
\PYG{p}{\PYGZob{}}
  \PYG{n+nt}{\PYGZdq{}memory\PYGZdq{}}\PYG{p}{:} \PYG{p}{[[[[}\PYG{l+m+mf}{0.1}\PYG{p}{,}\PYG{l+m+mf}{0.2}\PYG{p}{],[}\PYG{l+m+mf}{0.3}\PYG{p}{,}\PYG{l+m+mf}{\PYGZhy{}0.1}\PYG{p}{],[}\PYG{l+m+mf}{0.5}\PYG{p}{,}\PYG{l+m+mf}{0.8}\PYG{p}{]],}
              \PYG{p}{[[}\PYG{l+m+mf}{0.15}\PYG{p}{,}\PYG{l+m+mf}{0.7}\PYG{p}{],[}\PYG{l+m+mf}{0.13}\PYG{p}{,}\PYG{l+m+mf}{0.3}\PYG{p}{],[}\PYG{l+m+mf}{\PYGZhy{}0.5}\PYG{p}{,}\PYG{l+m+mf}{0.4}\PYG{p}{]]],}
             \PYG{p}{[[[}\PYG{l+m+mf}{0.1}\PYG{p}{,}\PYG{l+m+mf}{0.2}\PYG{p}{],[}\PYG{l+m+mf}{0.3}\PYG{p}{,}\PYG{l+m+mf}{\PYGZhy{}0.1}\PYG{p}{],[}\PYG{l+m+mf}{0.5}\PYG{p}{,}\PYG{l+m+mf}{0.8}\PYG{p}{]],}
              \PYG{p}{[[}\PYG{l+m+mf}{0.15}\PYG{p}{,}\PYG{l+m+mf}{0.7}\PYG{p}{],[}\PYG{l+m+mf}{0.13}\PYG{p}{,}\PYG{l+m+mf}{0.3}\PYG{p}{],[}\PYG{l+m+mf}{\PYGZhy{}0.5}\PYG{p}{,}\PYG{l+m+mf}{0.4}\PYG{p}{]]]]}
\PYG{p}{\PYGZcb{}}
\end{Verbatim}
\end{jsonexamplebox}

This output is very high bandwidth and memory, so should only be used sparingly and in averaging mode. One limitation of this level is that all measurements must be the same length, i.e., all memory slots must be the same length.

\subsubsection{Measurement Output Level 1: IQ Values}

Level 1 for the measurement output specification means that the user specifies a measurement kernel (from a list of available kernels specified by the device) or uses a default kernel. The kernel is a function that accepts the output measurement pulse (i.e. the Level 0 data) and converts it into a complex number (IQ Value). For more information see \S~\ref{sect:kernel}. Here is the averaged output data from memory with two slots,
\begin{jsonexamplebox}{Level 1 Average Measurement Output}
\begin{Verbatim}[commandchars=\\\{\}]
\PYG{p}{\PYGZob{}}
\PYG{n+nt}{\PYGZdq{}memory\PYGZdq{}}\PYG{p}{:} \PYG{p}{[[}\PYG{l+m+mf}{0.1}\PYG{p}{,}\PYG{l+m+mf}{0.2}\PYG{p}{],[}\PYG{l+m+mf}{\PYGZhy{}0.5}\PYG{p}{,}\PYG{l+m+mf}{0.4}\PYG{p}{]]}
\PYG{p}{\PYGZcb{}}
\end{Verbatim}
\end{jsonexamplebox}
In single shot mode (for two shots),
\begin{jsonexamplebox}{Level 1 Single Shot Measurement Output}
\begin{Verbatim}[commandchars=\\\{\}]
\PYG{p}{\PYGZob{}}
  \PYG{n+nt}{\PYGZdq{}memory\PYGZdq{}}\PYG{p}{:} \PYG{p}{[[[}\PYG{l+m+mf}{0.1}\PYG{p}{,}\PYG{l+m+mf}{0.2}\PYG{p}{],[}\PYG{l+m+mf}{\PYGZhy{}0.5}\PYG{p}{,}\PYG{l+m+mf}{0.4}\PYG{p}{]],}
             \PYG{p}{[[}\PYG{l+m+mf}{0.1}\PYG{p}{,}\PYG{l+m+mf}{0.2}\PYG{p}{],[}\PYG{l+m+mf}{\PYGZhy{}0.5}\PYG{p}{,}\PYG{l+m+mf}{0.4}\PYG{p}{]]]}
\PYG{p}{\PYGZcb{}}
\end{Verbatim}
\end{jsonexamplebox}

\subsubsection{Measurement Output Level 2: Qubit State}

In Level 2 measurement output, the user selects both a kernel and a discriminator from the list of available discriminators (see \S~\ref{sect:kernel}). The discriminator takes qubit measurements from the kernel and outputs the computational state based on a thresholding function. The output is given as a histogram of counts similar to an OpenQASM experiment (see \S~\ref{sect:gateresults})

\subsubsection{Measurement Feedback}

With level 2 measurement output, the user can also write to the registers and specify a mask on pulses based on the register values (see \S~\ref{sect:drivepulse}).

\section{Example Configurations for OpenPulse \label{sect:expulse}}

Here we provide some example \texttt{Backend} return calls for different devices with two qubits. We will not give sample pulse libraries and \texttt{cmd\_def} (assume these devices return those as empty). We also leave off the \texttt{basis\_gates}, \texttt{coupling\_map} and \texttt{gates} fields which must be provided by a backend, but which we omit for brevity.

\subsection{Fixed-Frequency Qubits Coupled via Fixed-Frequency Buses}

In this example, the device consists of two fixed-frequency qubits coupled to a single bus. There are two additional control lines locked to the other qubits LO to allow the user to perform a cross-resonance interaction \cite{sheldon:2016}. The two measurements are multiplexed. Each drive line has up to 1GHz bandwidth. All drive and measurement levels are allowed. One kernel and one discriminator are available. The boxcar kernel takes a ``boxcar'' average over the measurement output. The ``max\_1Q\_fidelity'' selects a thresholding line so as to maximize the one-qubit readout fidelity. The call to \texttt{Backend.configuration()} returns:
\begin{jsonexamplebox}{Fixed-Frequency Qubits Coupled via Fixed-Frequency Buses - Backend Configuration}
\begin{Verbatim}[commandchars=\\\{\}]
\PYG{p}{\PYGZob{}}
  \PYG{n+nt}{\PYGZdq{}backend\PYGZus{}name\PYGZdq{}}\PYG{p}{:} \PYG{l+s+s2}{\PYGZdq{}ibmqx1\PYGZdq{}}\PYG{p}{,}
  \PYG{n+nt}{\PYGZdq{}backend\PYGZus{}version\PYGZdq{}}\PYG{p}{:} \PYG{l+s+s2}{\PYGZdq{}1.1.5\PYGZdq{}}\PYG{p}{,}
  \PYG{n+nt}{\PYGZdq{}n\PYGZus{}qubits\PYGZdq{}}\PYG{p}{:} \PYG{l+m+mi}{2}\PYG{p}{,}
  \PYG{n+nt}{\PYGZdq{}local\PYGZdq{}}\PYG{p}{:} \PYG{k+kc}{false}\PYG{p}{,}
  \PYG{n+nt}{\PYGZdq{}simulator\PYGZdq{}}\PYG{p}{:} \PYG{k+kc}{false}\PYG{p}{,}
  \PYG{n+nt}{\PYGZdq{}conditional\PYGZdq{}}\PYG{p}{:} \PYG{k+kc}{false}\PYG{p}{,}
  \PYG{n+nt}{\PYGZdq{}open\PYGZus{}pulse\PYGZdq{}}\PYG{p}{:} \PYG{k+kc}{true}\PYG{p}{,}
  \PYG{n+nt}{\PYGZdq{}n\PYGZus{}uchannels\PYGZdq{}}\PYG{p}{:} \PYG{l+m+mi}{2}\PYG{p}{,}
  \PYG{n+nt}{\PYGZdq{}hamiltonian\PYGZdq{}}\PYG{p}{:} \PYG{err}{see\PYGZus{}below}\PYG{p}{,}
  \PYG{n+nt}{\PYGZdq{}u\PYGZus{}channel\PYGZus{}lo\PYGZdq{}}\PYG{p}{:} \PYG{p}{[[\PYGZob{}}\PYG{n+nt}{\PYGZdq{}q\PYGZdq{}}\PYG{p}{:} \PYG{l+m+mi}{0}\PYG{p}{,} \PYG{n+nt}{\PYGZdq{}scale\PYGZdq{}}\PYG{p}{:} \PYG{p}{[}\PYG{l+m+mi}{0}\PYG{p}{,}\PYG{l+m+mi}{1}\PYG{p}{]\PYGZcb{}],}
                   \PYG{p}{[\PYGZob{}}\PYG{n+nt}{\PYGZdq{}q\PYGZdq{}}\PYG{p}{:} \PYG{l+m+mi}{1}\PYG{p}{,} \PYG{n+nt}{\PYGZdq{}scale\PYGZdq{}}\PYG{p}{:} \PYG{p}{[}\PYG{l+m+mi}{1}\PYG{p}{,}\PYG{l+m+mi}{0}\PYG{p}{]\PYGZcb{}]],}
  \PYG{n+nt}{\PYGZdq{}meas\PYGZus{}levels\PYGZdq{}}\PYG{p}{:} \PYG{p}{[}\PYG{l+m+mi}{0}\PYG{p}{,}\PYG{l+m+mi}{1}\PYG{p}{,}\PYG{l+m+mi}{2}\PYG{p}{],}
  \PYG{n+nt}{\PYGZdq{}qubit\PYGZus{}lo\PYGZus{}range\PYGZdq{}}\PYG{p}{:} \PYG{p}{[[}\PYG{l+m+mf}{4.5}\PYG{p}{,}\PYG{l+m+mf}{5.5}\PYG{p}{],[}\PYG{l+m+mf}{4.5}\PYG{p}{,}\PYG{l+m+mf}{5.5}\PYG{p}{]],}
  \PYG{n+nt}{\PYGZdq{}meas\PYGZus{}lo\PYGZus{}range\PYGZdq{}}\PYG{p}{:} \PYG{p}{[[}\PYG{l+m+mf}{6.0}\PYG{p}{,}\PYG{l+m+mf}{7.0}\PYG{p}{],[}\PYG{l+m+mf}{6.0}\PYG{p}{,}\PYG{l+m+mf}{7.0}\PYG{p}{]],}
  \PYG{n+nt}{\PYGZdq{}dt\PYGZdq{}}\PYG{p}{:} \PYG{l+m+mf}{1.333}\PYG{p}{,}
  \PYG{n+nt}{\PYGZdq{}dtm\PYGZdq{}}\PYG{p}{:} \PYG{l+m+mi}{10}\PYG{p}{,}
  \PYG{n+nt}{\PYGZdq{}basis\PYGZus{}gates\PYGZdq{}}\PYG{p}{:} \PYG{p}{[}\PYG{l+s+s2}{\PYGZdq{}gate1\PYGZdq{}}\PYG{p}{,} \PYG{l+s+s2}{\PYGZdq{}gate2\PYGZdq{}}\PYG{p}{,}\PYG{err}{...}\PYG{p}{],}
  \PYG{n+nt}{\PYGZdq{}gates\PYGZdq{}}\PYG{p}{:} \PYG{p}{[}\PYG{err}{gate1}\PYG{p}{,} \PYG{err}{gate2}\PYG{p}{,}\PYG{err}{...}\PYG{p}{],}
  \PYG{n+nt}{\PYGZdq{}rep\PYGZus{}times\PYGZdq{}}\PYG{p}{:} \PYG{p}{[}\PYG{l+m+mi}{100}\PYG{p}{,}\PYG{l+m+mi}{250}\PYG{p}{,}\PYG{l+m+mi}{500}\PYG{p}{,}\PYG{l+m+mi}{1000}\PYG{p}{],}
  \PYG{n+nt}{\PYGZdq{}meas\PYGZus{}map\PYGZdq{}}\PYG{p}{:} \PYG{p}{[[}\PYG{l+m+mi}{0}\PYG{p}{,}\PYG{l+m+mi}{1}\PYG{p}{]],}
  \PYG{n+nt}{\PYGZdq{}channel\PYGZus{}bandwidth\PYGZdq{}}\PYG{p}{:} \PYG{p}{[[}\PYG{l+m+mf}{\PYGZhy{}0.5}\PYG{p}{,}\PYG{l+m+mf}{0.5}\PYG{p}{],[}\PYG{l+m+mf}{\PYGZhy{}0.5}\PYG{p}{,}\PYG{l+m+mf}{0.5}\PYG{p}{]],}
  \PYG{n+nt}{\PYGZdq{}meas\PYGZus{}kernels\PYGZdq{}}\PYG{p}{:} \PYG{p}{[}\PYG{l+s+s2}{\PYGZdq{}boxcar\PYGZdq{}}\PYG{p}{],}
  \PYG{n+nt}{\PYGZdq{}discriminators\PYGZdq{}}\PYG{p}{:} \PYG{p}{[}\PYG{l+s+s2}{\PYGZdq{}max\PYGZus{}1Q\PYGZus{}fidelity\PYGZdq{}}\PYG{p}{]}
\PYG{p}{\PYGZcb{}}
\end{Verbatim}
\end{jsonexamplebox}
and the call to \texttt{Backend.defaults()}
\begin{jsonexamplebox}{Fixed-Frequency Qubits Coupled via Fixed-Frequency Buses - Backend Defaults}
\begin{Verbatim}[commandchars=\\\{\}]
\PYG{p}{\PYGZob{}}
  \PYG{n+nt}{\PYGZdq{}qubit\PYGZus{}freq\PYGZus{}est\PYGZdq{}}\PYG{p}{:} \PYG{p}{[}\PYG{l+m+mf}{4.9}\PYG{p}{,}\PYG{l+m+mf}{5.0}\PYG{p}{],}
  \PYG{n+nt}{\PYGZdq{}meas\PYGZus{}freq\PYGZus{}est\PYGZdq{}}\PYG{p}{:} \PYG{p}{[}\PYG{l+m+mf}{6.5}\PYG{p}{,}\PYG{l+m+mf}{6.6}\PYG{p}{],}
  \PYG{n+nt}{\PYGZdq{}buffer\PYGZdq{}}\PYG{p}{:} \PYG{l+m+mi}{10}\PYG{p}{,}
  \PYG{n+nt}{\PYGZdq{}pulse\PYGZus{}library\PYGZdq{}}\PYG{p}{:} \PYG{p}{[],}
  \PYG{n+nt}{\PYGZdq{}cmd\PYGZus{}def\PYGZdq{}}\PYG{p}{:} \PYG{p}{[],}
  \PYG{n+nt}{\PYGZdq{}meas\PYGZus{}kernel\PYGZdq{}}\PYG{p}{:} \PYG{p}{\PYGZob{}}\PYG{n+nt}{\PYGZdq{}name\PYGZdq{}}\PYG{p}{:} \PYG{l+s+s2}{\PYGZdq{}boxcar\PYGZdq{}}\PYG{p}{,} \PYG{n+nt}{\PYGZdq{}params\PYGZdq{}}\PYG{p}{:} \PYG{p}{[]\PYGZcb{},}
  \PYG{n+nt}{\PYGZdq{}discriminator\PYGZdq{}}\PYG{p}{:} \PYG{p}{\PYGZob{}}\PYG{n+nt}{\PYGZdq{}name\PYGZdq{}}\PYG{p}{:} \PYG{l+s+s2}{\PYGZdq{}max\PYGZus{}1Q\PYGZus{}fidelity\PYGZdq{}}\PYG{p}{,} \PYG{n+nt}{\PYGZdq{}params\PYGZdq{}}\PYG{p}{:} \PYG{p}{[]\PYGZcb{}}
\PYG{p}{\PYGZcb{}}
\end{Verbatim}
\end{jsonexamplebox}

The Hamiltonian dictionary returned by the device is:
\begin{jsonexamplebox}{Fixed-Frequency Qubits Coupled via Fixed-Frequency Buses - Hamiltonian Dictionary}
\begin{Verbatim}[commandchars=\\\{\}]
\PYG{p}{\PYGZob{}}
  \PYG{n+nt}{\PYGZdq{}h\PYGZus{}latex\PYGZdq{}}\PYG{p}{:} \PYG{l+s+s2}{\PYGZdq{}}
\PYG{l+s+s2}{      \PYGZbs{}\PYGZbs{}sum\PYGZus{}\PYGZob{}i=0\PYGZcb{}\PYGZca{}\PYGZob{}1\PYGZcb{} (U\PYGZus{}i(t)+D\PYGZus{}i(t)) \PYGZbs{}\PYGZbs{}sigma\PYGZus{}i\PYGZca{}\PYGZob{}X\PYGZcb{} +}
\PYG{l+s+s2}{      \PYGZbs{}\PYGZbs{}sum\PYGZus{}\PYGZob{}i=0\PYGZcb{}\PYGZca{}\PYGZob{}1\PYGZcb{} 2\PYGZbs{}\PYGZbs{}pi \PYGZbs{}nu\PYGZus{}i (1\PYGZhy{}\PYGZbs{}\PYGZbs{}sigma\PYGZus{}i\PYGZca{}\PYGZob{}Z\PYGZcb{})/2 +}
\PYG{l+s+s2}{      \PYGZbs{}\PYGZbs{}omega\PYGZus{}B a\PYGZus{}B a\PYGZca{}\PYGZob{}\PYGZbs{}\PYGZbs{}dagger\PYGZcb{}\PYGZus{}B +}
\PYG{l+s+s2}{      \PYGZbs{}\PYGZbs{}sum\PYGZus{}\PYGZob{}i=0\PYGZcb{}\PYGZca{}\PYGZob{}1\PYGZcb{} g\PYGZus{}i \PYGZbs{}\PYGZbs{}sigma\PYGZus{}i\PYGZca{}\PYGZob{}X\PYGZcb{} (a\PYGZus{}B + a\PYGZus{}B\PYGZca{}\PYGZob{}\PYGZbs{}\PYGZbs{}dagger\PYGZcb{})\PYGZdq{}}\PYG{p}{,}
  \PYG{n+nt}{\PYGZdq{}h\PYGZus{}str\PYGZdq{}}\PYG{p}{:}  \PYG{p}{[}
      \PYG{l+s+s2}{\PYGZdq{}\PYGZus{}\PYGZus{}SUM[i,0,1,\PYGZus{}X\PYGZob{}i\PYGZcb{}\PYGZus{}*\PYGZus{}D\PYGZob{}i\PYGZcb{}\PYGZus{}+\PYGZus{}X\PYGZob{}i\PYGZcb{}\PYGZus{}*\PYGZus{}U\PYGZob{}i\PYGZcb{}\PYGZus{}]\PYGZdq{}}\PYG{p}{,}
      \PYG{l+s+s2}{\PYGZdq{}\PYGZus{}\PYGZus{}SUM[i,0,1,2*pi*\PYGZus{}v\PYGZob{}i\PYGZcb{}\PYGZus{}*\PYGZus{}O\PYGZob{}i\PYGZcb{}\PYGZus{}]\PYGZdq{}}\PYG{p}{,}
      \PYG{l+s+s2}{\PYGZdq{}\PYGZus{}+2*pi*\PYGZus{}wb\PYGZus{}*\PYGZus{}O\PYGZob{}i\PYGZcb{}\PYGZus{}\PYGZdq{}}\PYG{p}{,}
      \PYG{l+s+s2}{\PYGZdq{}\PYGZus{}\PYGZus{}SUM[i,0,1,\PYGZus{}g\PYGZob{}i\PYGZcb{}\PYGZus{}*\PYGZus{}X\PYGZob{}i\PYGZcb{}\PYGZus{}(\PYGZus{}a\PYGZus{}+\PYGZus{}A\PYGZus{})]\PYGZdq{}}\PYG{p}{],}
  \PYG{n+nt}{\PYGZdq{}vars\PYGZdq{}} \PYG{p}{:} \PYG{p}{\PYGZob{}}\PYG{n+nt}{\PYGZdq{}v0\PYGZdq{}}\PYG{p}{:} \PYG{l+m+mi}{5}\PYG{p}{,} \PYG{n+nt}{\PYGZdq{}v1\PYGZdq{}} \PYG{p}{:} \PYG{l+m+mf}{5.1}\PYG{p}{,}\PYG{n+nt}{\PYGZdq{}g0\PYGZdq{}}\PYG{p}{:} \PYG{l+m+mf}{0.1}\PYG{p}{,}
            \PYG{n+nt}{\PYGZdq{}g1\PYGZdq{}}\PYG{p}{:} \PYG{l+m+mf}{0.1}\PYG{p}{,} \PYG{n+nt}{\PYGZdq{}wb\PYGZdq{}}\PYG{p}{:} \PYG{l+m+mi}{6}\PYG{p}{\PYGZcb{}}
\PYG{p}{\PYGZcb{}}
\end{Verbatim}
\end{jsonexamplebox}

The latex corresponds to the Hamiltonian,
\begin{equation}
H = \sum_{i=0}^{1} (U_i(t)+D_i(t)) \sigma_i^{X} + \sum_{i=0}^{1} 2\pi \nu_i (1-\sigma_i^{Z})/2 + \omega_B a_B a^{\dagger}_B + \sum_{i=0}^{1} g_i \sigma_i^{X} (a_B + a_B^{\dagger})
\end{equation}
At the discretion of the device, the symbols may be replaced by estimates of these values.

\subsection{Tunable Qubits Coupled via Fixed-Frequency Buses}

In this example, the device consists of two tunable-frequency qubits coupled to a single bus. Each qubit has a flux control knob with 200~MHz bandwidth centered around DC since all the LO terms are zero. The two measurements are on separate lines. Each drive line has up to 1GHz bandwidth. There is a DC flux that is set by the device. The call to \texttt{Backend.defaults()} returns:
\begin{jsonexamplebox}{Tunable Qubits Coupled via Fixed-Frequency Buses - Backend Configuration}
\begin{Verbatim}[commandchars=\\\{\}]
\PYG{p}{\PYGZob{}}
  \PYG{n+nt}{\PYGZdq{}backend\PYGZus{}name\PYGZdq{}}\PYG{p}{:} \PYG{l+s+s2}{\PYGZdq{}ibmqx1\PYGZdq{}}\PYG{p}{,}
  \PYG{n+nt}{\PYGZdq{}backend\PYGZus{}version\PYGZdq{}}\PYG{p}{:} \PYG{l+s+s2}{\PYGZdq{}1.1.5\PYGZdq{}}\PYG{p}{,}
  \PYG{n+nt}{\PYGZdq{}n\PYGZus{}qubits\PYGZdq{}}\PYG{p}{:} \PYG{l+m+mi}{2}\PYG{p}{,}
  \PYG{n+nt}{\PYGZdq{}local\PYGZdq{}}\PYG{p}{:} \PYG{k+kc}{false}\PYG{p}{,}
  \PYG{n+nt}{\PYGZdq{}simulator\PYGZdq{}}\PYG{p}{:} \PYG{k+kc}{false}\PYG{p}{,}
  \PYG{n+nt}{\PYGZdq{}conditional\PYGZdq{}}\PYG{p}{:} \PYG{k+kc}{false}\PYG{p}{,}
  \PYG{n+nt}{\PYGZdq{}open\PYGZus{}pulse\PYGZdq{}}\PYG{p}{:} \PYG{k+kc}{true}\PYG{p}{,}
  \PYG{n+nt}{\PYGZdq{}n\PYGZus{}uchannels\PYGZdq{}}\PYG{p}{:} \PYG{l+m+mi}{2}\PYG{p}{,}
  \PYG{n+nt}{\PYGZdq{}hamiltonian\PYGZdq{}}\PYG{p}{:} \PYG{err}{see\PYGZus{}below}\PYG{p}{,}
  \PYG{n+nt}{\PYGZdq{}u\PYGZus{}channel\PYGZus{}lo\PYGZdq{}}\PYG{p}{:} \PYG{p}{[],}
  \PYG{n+nt}{\PYGZdq{}meas\PYGZus{}levels\PYGZdq{}}\PYG{p}{:} \PYG{p}{[}\PYG{l+m+mi}{0}\PYG{p}{,}\PYG{l+m+mi}{1}\PYG{p}{,}\PYG{l+m+mi}{2}\PYG{p}{],}
  \PYG{n+nt}{\PYGZdq{}qubit\PYGZus{}lo\PYGZus{}range\PYGZdq{}}\PYG{p}{:} \PYG{p}{[[}\PYG{l+m+mf}{4.5}\PYG{p}{,}\PYG{l+m+mf}{5.5}\PYG{p}{],[}\PYG{l+m+mf}{4.5}\PYG{p}{,}\PYG{l+m+mf}{5.5}\PYG{p}{]],}
  \PYG{n+nt}{\PYGZdq{}meas\PYGZus{}lo\PYGZus{}range\PYGZdq{}}\PYG{p}{:} \PYG{p}{[[}\PYG{l+m+mf}{6.0}\PYG{p}{,}\PYG{l+m+mf}{7.0}\PYG{p}{],[}\PYG{l+m+mf}{6.0}\PYG{p}{,}\PYG{l+m+mf}{7.0}\PYG{p}{]],}
  \PYG{n+nt}{\PYGZdq{}dt\PYGZdq{}}\PYG{p}{:} \PYG{l+m+mf}{1.333}\PYG{p}{,}
  \PYG{n+nt}{\PYGZdq{}dtm\PYGZdq{}}\PYG{p}{:} \PYG{l+m+mi}{10}\PYG{p}{,}
  \PYG{n+nt}{\PYGZdq{}basis\PYGZus{}gates\PYGZdq{}}\PYG{p}{:} \PYG{p}{[}\PYG{l+s+s2}{\PYGZdq{}gate1\PYGZdq{}}\PYG{p}{,} \PYG{l+s+s2}{\PYGZdq{}gate2\PYGZdq{}}\PYG{p}{,}\PYG{err}{...}\PYG{p}{],}
  \PYG{n+nt}{\PYGZdq{}gates\PYGZdq{}}\PYG{p}{:} \PYG{p}{[}\PYG{err}{gate1}\PYG{p}{,} \PYG{err}{gate2}\PYG{p}{,}\PYG{err}{...}\PYG{p}{],}
  \PYG{n+nt}{\PYGZdq{}rep\PYGZus{}times\PYGZdq{}}\PYG{p}{:} \PYG{p}{[}\PYG{l+m+mi}{100}\PYG{p}{,}\PYG{l+m+mi}{250}\PYG{p}{,}\PYG{l+m+mi}{500}\PYG{p}{,}\PYG{l+m+mi}{1000}\PYG{p}{],}
  \PYG{n+nt}{\PYGZdq{}meas\PYGZus{}map\PYGZdq{}}\PYG{p}{:} \PYG{p}{[[}\PYG{l+m+mi}{0}\PYG{p}{],[}\PYG{l+m+mi}{1}\PYG{p}{]],}
  \PYG{n+nt}{\PYGZdq{}channel\PYGZus{}bandwidth\PYGZdq{}}\PYG{p}{:} \PYG{p}{[[}\PYG{l+m+mf}{\PYGZhy{}0.5}\PYG{p}{,}\PYG{l+m+mf}{0.5}\PYG{p}{],[}\PYG{l+m+mf}{\PYGZhy{}0.5}\PYG{p}{,}\PYG{l+m+mf}{0.5}\PYG{p}{],}
                        \PYG{p}{[}\PYG{l+m+mf}{\PYGZhy{}0.1}\PYG{p}{,}\PYG{l+m+mf}{0.1}\PYG{p}{],[}\PYG{l+m+mf}{\PYGZhy{}0.1}\PYG{p}{,}\PYG{l+m+mf}{0.1}\PYG{p}{]],}
  \PYG{n+nt}{\PYGZdq{}meas\PYGZus{}kernels\PYGZdq{}}\PYG{p}{:} \PYG{p}{[}\PYG{l+s+s2}{\PYGZdq{}boxcar\PYGZdq{}}\PYG{p}{],}
  \PYG{n+nt}{\PYGZdq{}discriminators\PYGZdq{}}\PYG{p}{:} \PYG{p}{[}\PYG{l+s+s2}{\PYGZdq{}max\PYGZus{}1Q\PYGZus{}fidelity\PYGZdq{}}\PYG{p}{]}
\PYG{p}{\PYGZcb{}}
\end{Verbatim}
\end{jsonexamplebox}
and the call to \texttt{Backend.defaults()}
\begin{jsonexamplebox}{Tunable Qubits Coupled via Fixed-Frequency Buses - Backend Defaults}
\begin{Verbatim}[commandchars=\\\{\}]
\PYG{p}{\PYGZob{}}
  \PYG{n+nt}{\PYGZdq{}qubit\PYGZus{}freq\PYGZus{}est\PYGZdq{}}\PYG{p}{:} \PYG{p}{[}\PYG{l+m+mf}{4.9}\PYG{p}{,}\PYG{l+m+mf}{5.0}\PYG{p}{],}
  \PYG{n+nt}{\PYGZdq{}meas\PYGZus{}freq\PYGZus{}est\PYGZdq{}}\PYG{p}{:} \PYG{p}{[}\PYG{l+m+mf}{6.5}\PYG{p}{,}\PYG{l+m+mf}{6.6}\PYG{p}{],}
  \PYG{n+nt}{\PYGZdq{}buffer\PYGZdq{}}\PYG{p}{:} \PYG{l+m+mi}{10}\PYG{p}{,}
  \PYG{n+nt}{\PYGZdq{}pulse\PYGZus{}library\PYGZdq{}}\PYG{p}{:} \PYG{p}{[],}
  \PYG{n+nt}{\PYGZdq{}cmd\PYGZus{}def\PYGZdq{}}\PYG{p}{:} \PYG{p}{[],}
  \PYG{n+nt}{\PYGZdq{}meas\PYGZus{}kernel\PYGZdq{}}\PYG{p}{:} \PYG{p}{\PYGZob{}}\PYG{n+nt}{\PYGZdq{}name\PYGZdq{}}\PYG{p}{:} \PYG{l+s+s2}{\PYGZdq{}boxcar\PYGZdq{}}\PYG{p}{,} \PYG{n+nt}{\PYGZdq{}params\PYGZdq{}}\PYG{p}{:} \PYG{p}{[]\PYGZcb{},}
  \PYG{n+nt}{\PYGZdq{}discriminator\PYGZdq{}}\PYG{p}{:} \PYG{p}{\PYGZob{}}\PYG{n+nt}{\PYGZdq{}name\PYGZdq{}}\PYG{p}{:} \PYG{l+s+s2}{\PYGZdq{}max\PYGZus{}1Q\PYGZus{}fidelity\PYGZdq{}}\PYG{p}{,} \PYG{n+nt}{\PYGZdq{}params\PYGZdq{}}\PYG{p}{:} \PYG{p}{[]\PYGZcb{}}
\PYG{p}{\PYGZcb{}}
\end{Verbatim}
\end{jsonexamplebox}

The Hamiltonian returned by the device is:
\begin{eqnarray}
    H & = & \sum_{i=0}^{1} D_i(t) \sigma_i^{X} + \sum_{i=0}^{1} 2\pi \nu_i \sqrt{\left|\cos(\pi (dc_i + U_i(t))/\Phi_0)\right|} (1-\sigma_i^{Z})/2 + \nonumber \\
      & & \omega_B a_B a^{\dagger}_B + \sum_{i=0}^{1} g_i \sigma_i^{X} (a_B + a_B^{\dagger}) \\
    dc_0 & = & 0.2 \\
    dc_1 & = & -0.4
\end{eqnarray}
This Hamiltonian exposes that the device relies on existing DC flux biases which are set to the indicated values.

\subsection{Fixed-Frequency Qubits Coupled via Tunable Buses}

In this example, the device consists of two fixed-frequency qubits coupled to a single tunable bus. The bus has two flux control knobs, one to drive the difference frequency and one to drive the sum frequency. The two measurements are on separate lines. Each drive line has up to $1~\mathrm{GHz}$ bandwidth. The call to \texttt{Backend.configuration()} returns:
\begin{jsonexamplebox}{Fixed-Frequency Qubits Coupled via Tunable Buses - Backend Configuration}
\begin{Verbatim}[commandchars=\\\{\}]
\PYG{p}{\PYGZob{}}
  \PYG{n+nt}{\PYGZdq{}backend\PYGZus{}name\PYGZdq{}}\PYG{p}{:} \PYG{l+s+s2}{\PYGZdq{}ibmqx1\PYGZdq{}}\PYG{p}{,}
  \PYG{n+nt}{\PYGZdq{}backend\PYGZus{}version\PYGZdq{}}\PYG{p}{:} \PYG{l+s+s2}{\PYGZdq{}1.1.5\PYGZdq{}}\PYG{p}{,}
  \PYG{n+nt}{\PYGZdq{}n\PYGZus{}qubits\PYGZdq{}}\PYG{p}{:} \PYG{l+m+mi}{2}\PYG{p}{,}
  \PYG{n+nt}{\PYGZdq{}local\PYGZdq{}}\PYG{p}{:} \PYG{k+kc}{false}\PYG{p}{,}
  \PYG{n+nt}{\PYGZdq{}simulator\PYGZdq{}}\PYG{p}{:} \PYG{k+kc}{false}\PYG{p}{,}
  \PYG{n+nt}{\PYGZdq{}conditional\PYGZdq{}}\PYG{p}{:} \PYG{k+kc}{false}\PYG{p}{,}
  \PYG{n+nt}{\PYGZdq{}open\PYGZus{}pulse\PYGZdq{}}\PYG{p}{:} \PYG{k+kc}{true}\PYG{p}{,}
  \PYG{n+nt}{\PYGZdq{}n\PYGZus{}uchannels\PYGZdq{}}\PYG{p}{:} \PYG{l+m+mi}{2}\PYG{p}{,}
  \PYG{n+nt}{\PYGZdq{}hamiltonian\PYGZdq{}}\PYG{p}{:} \PYG{err}{see\PYGZus{}below}\PYG{p}{,}
  \PYG{n+nt}{\PYGZdq{}meas\PYGZus{}levels\PYGZdq{}}\PYG{p}{:} \PYG{p}{[}\PYG{l+m+mi}{0}\PYG{p}{,}\PYG{l+m+mi}{1}\PYG{p}{,}\PYG{l+m+mi}{2}\PYG{p}{],}
  \PYG{n+nt}{\PYGZdq{}u\PYGZus{}channel\PYGZus{}lo\PYGZdq{}}\PYG{p}{:} \PYG{p}{[[\PYGZob{}}\PYG{n+nt}{\PYGZdq{}q\PYGZdq{}}\PYG{p}{:} \PYG{l+m+mi}{0}\PYG{p}{,} \PYG{n+nt}{\PYGZdq{}scale\PYGZdq{}}\PYG{p}{:} \PYG{p}{[}\PYG{l+m+mi}{1}\PYG{p}{,}\PYG{l+m+mi}{0}\PYG{p}{]\PYGZcb{},}
                    \PYG{p}{\PYGZob{}}\PYG{n+nt}{\PYGZdq{}q\PYGZdq{}}\PYG{p}{:} \PYG{l+m+mi}{1}\PYG{p}{,} \PYG{n+nt}{\PYGZdq{}scale\PYGZdq{}}\PYG{p}{:} \PYG{p}{[}\PYG{l+m+mi}{\PYGZhy{}1}\PYG{p}{,}\PYG{l+m+mi}{0}\PYG{p}{]\PYGZcb{}],}
  		\PYG{p}{[\PYGZob{}}\PYG{n+nt}{\PYGZdq{}q\PYGZdq{}}\PYG{p}{:} \PYG{l+m+mi}{0}\PYG{p}{,} \PYG{n+nt}{\PYGZdq{}scale\PYGZdq{}}\PYG{p}{:} \PYG{p}{[}\PYG{l+m+mi}{1}\PYG{p}{,}\PYG{l+m+mi}{0}\PYG{p}{]\PYGZcb{},}
       \PYG{p}{\PYGZob{}}\PYG{n+nt}{\PYGZdq{}q\PYGZdq{}}\PYG{p}{:} \PYG{l+m+mi}{1}\PYG{p}{,} \PYG{n+nt}{\PYGZdq{}scale\PYGZdq{}}\PYG{p}{:} \PYG{p}{[}\PYG{l+m+mi}{1}\PYG{p}{,}\PYG{l+m+mi}{0}\PYG{p}{]\PYGZcb{}]],}
  \PYG{n+nt}{\PYGZdq{}qubit\PYGZus{}lo\PYGZus{}range\PYGZdq{}}\PYG{p}{:} \PYG{p}{[[}\PYG{l+m+mf}{4.5}\PYG{p}{,}\PYG{l+m+mf}{5.5}\PYG{p}{],[}\PYG{l+m+mf}{4.5}\PYG{p}{,}\PYG{l+m+mf}{5.5}\PYG{p}{]],}
  \PYG{n+nt}{\PYGZdq{}meas\PYGZus{}lo\PYGZus{}range\PYGZdq{}}\PYG{p}{:} \PYG{p}{[[}\PYG{l+m+mf}{6.0}\PYG{p}{,}\PYG{l+m+mf}{7.0}\PYG{p}{],[}\PYG{l+m+mf}{6.0}\PYG{p}{,}\PYG{l+m+mf}{7.0}\PYG{p}{]],}
  \PYG{n+nt}{\PYGZdq{}dt\PYGZdq{}}\PYG{p}{:} \PYG{l+m+mf}{1.333}\PYG{p}{,}
  \PYG{n+nt}{\PYGZdq{}dtm\PYGZdq{}}\PYG{p}{:} \PYG{l+m+mi}{10}\PYG{p}{,}
  \PYG{n+nt}{\PYGZdq{}basis\PYGZus{}gates\PYGZdq{}}\PYG{p}{:} \PYG{p}{[}\PYG{l+s+s2}{\PYGZdq{}gate1\PYGZdq{}}\PYG{p}{,} \PYG{l+s+s2}{\PYGZdq{}gate2\PYGZdq{}}\PYG{p}{,}\PYG{err}{...}\PYG{p}{],}
  \PYG{n+nt}{\PYGZdq{}gates\PYGZdq{}}\PYG{p}{:} \PYG{p}{[}\PYG{err}{gate1}\PYG{p}{,} \PYG{err}{gate2}\PYG{p}{,}\PYG{err}{...}\PYG{p}{],}
  \PYG{n+nt}{\PYGZdq{}rep\PYGZus{}times\PYGZdq{}}\PYG{p}{:} \PYG{p}{[}\PYG{l+m+mi}{100}\PYG{p}{,}\PYG{l+m+mi}{250}\PYG{p}{,}\PYG{l+m+mi}{500}\PYG{p}{,}\PYG{l+m+mi}{1000}\PYG{p}{],}
  \PYG{n+nt}{\PYGZdq{}meas\PYGZus{}map\PYGZdq{}}\PYG{p}{:} \PYG{p}{[[}\PYG{l+m+mi}{0}\PYG{p}{],[}\PYG{l+m+mi}{1}\PYG{p}{]],}
  \PYG{n+nt}{\PYGZdq{}channel\PYGZus{}bandwidth\PYGZdq{}}\PYG{p}{:} \PYG{p}{[[}\PYG{l+m+mf}{\PYGZhy{}0.5}\PYG{p}{,}\PYG{l+m+mf}{0.5}\PYG{p}{],[}\PYG{l+m+mf}{\PYGZhy{}0.5}\PYG{p}{,}\PYG{l+m+mf}{0.5}\PYG{p}{],}
                        \PYG{p}{[}\PYG{l+m+mf}{\PYGZhy{}0.1}\PYG{p}{,}\PYG{l+m+mf}{0.1}\PYG{p}{],[}\PYG{l+m+mf}{\PYGZhy{}0.1}\PYG{p}{,}\PYG{l+m+mf}{0.1}\PYG{p}{]],}
  \PYG{n+nt}{\PYGZdq{}meas\PYGZus{}kernels\PYGZdq{}}\PYG{p}{:} \PYG{p}{[}\PYG{l+s+s2}{\PYGZdq{}boxcar\PYGZdq{}}\PYG{p}{],}
  \PYG{n+nt}{\PYGZdq{}discriminators\PYGZdq{}}\PYG{p}{:} \PYG{p}{[}\PYG{l+s+s2}{\PYGZdq{}max\PYGZus{}1Q\PYGZus{}fidelity\PYGZdq{}}\PYG{p}{]}
\PYG{p}{\PYGZcb{}}
\end{Verbatim}
\end{jsonexamplebox}
and the call to \texttt{Backend.defaults()}
\begin{jsonexamplebox}{Fixed-Frequency Qubits Coupled via Tunable Buses - Backend Defaults}
\begin{Verbatim}[commandchars=\\\{\}]
\PYG{p}{\PYGZob{}}
  \PYG{n+nt}{\PYGZdq{}qubit\PYGZus{}freq\PYGZus{}est\PYGZdq{}}\PYG{p}{:} \PYG{p}{[}\PYG{l+m+mf}{4.9}\PYG{p}{,}\PYG{l+m+mf}{5.0}\PYG{p}{],}
  \PYG{n+nt}{\PYGZdq{}meas\PYGZus{}freq\PYGZus{}est\PYGZdq{}}\PYG{p}{:} \PYG{p}{[}\PYG{l+m+mf}{6.5}\PYG{p}{,}\PYG{l+m+mf}{6.6}\PYG{p}{],}
  \PYG{n+nt}{\PYGZdq{}buffer\PYGZdq{}}\PYG{p}{:} \PYG{l+m+mi}{10}\PYG{p}{,}
  \PYG{n+nt}{\PYGZdq{}pulse\PYGZus{}library\PYGZdq{}}\PYG{p}{:} \PYG{p}{[],}
  \PYG{n+nt}{\PYGZdq{}cmd\PYGZus{}def\PYGZdq{}}\PYG{p}{:} \PYG{p}{[],}
  \PYG{n+nt}{\PYGZdq{}meas\PYGZus{}kernel\PYGZdq{}}\PYG{p}{:} \PYG{p}{\PYGZob{}}\PYG{n+nt}{\PYGZdq{}name\PYGZdq{}}\PYG{p}{:} \PYG{l+s+s2}{\PYGZdq{}boxcar\PYGZdq{}}\PYG{p}{,} \PYG{n+nt}{\PYGZdq{}params\PYGZdq{}}\PYG{p}{:} \PYG{p}{[]\PYGZcb{},}
  \PYG{n+nt}{\PYGZdq{}discriminator\PYGZdq{}}\PYG{p}{:} \PYG{p}{\PYGZob{}}\PYG{n+nt}{\PYGZdq{}name\PYGZdq{}}\PYG{p}{:} \PYG{l+s+s2}{\PYGZdq{}max\PYGZus{}1Q\PYGZus{}fidelity\PYGZdq{}}\PYG{p}{,} \PYG{n+nt}{\PYGZdq{}params\PYGZdq{}}\PYG{p}{:} \PYG{p}{[]\PYGZcb{}}
\PYG{p}{\PYGZcb{}}
\end{Verbatim}
\end{jsonexamplebox}

The Hamiltonian returned by the device is:
\begin{eqnarray}
H & = & \sum_{i=0}^{1} D_i(t) \sigma_i^{X} + \sum_{i=0}^{1} 2\pi \nu_i  (1-\sigma_i^{Z})/2 + \nonumber \\
     & & \omega_B \sqrt{\left|\cos(\pi (U_0(t)+U_1(t)+dc_0(t))/\Phi_0)\right|} a_B a^{\dagger}_B \nonumber \\
     & & + \sum_{i=0}^{1} g_i \sigma_i^{X} (a_B + a_B^{\dagger}) \\
dc_0 & = & 0.1
\end{eqnarray}

\subsection{Ion Trap}

In this example, the device consists of a three-qubit Yb ion trap (this example is based on Ref~\cite{debnath:2016}). Measurement of the ion system is made with fluorescence detection so only output measurement level 2 is supplied (i.e. 0 and 1 detection). There is no real concept of a kernel so the available kernel will just be listed as ``default''. There is also no adjustment of the measurement frequency (the frequency of the fluorescence beam). The timescales for ion traps are scaled up by about a factor of $100-1000$ over SC qubits. Here we assume (as in the reference) addressable Raman beams for each qubit. Although not put into this example, one could also consider DC channels to control the trap electrodes. The call to \texttt{Backend.configuration()} returns:
\begin{jsonexamplebox}{Ion Trap - Backend Configuration}
\begin{Verbatim}[commandchars=\\\{\}]
\PYG{p}{\PYGZob{}}
  \PYG{n+nt}{\PYGZdq{}backend\PYGZus{}name\PYGZdq{}}\PYG{p}{:} \PYG{l+s+s2}{\PYGZdq{}ibmqx10\PYGZdq{}}\PYG{p}{,}
  \PYG{n+nt}{\PYGZdq{}backend\PYGZus{}version\PYGZdq{}}\PYG{p}{:} \PYG{l+s+s2}{\PYGZdq{}1.1.5\PYGZdq{}}\PYG{p}{,}
  \PYG{n+nt}{\PYGZdq{}n\PYGZus{}qubits\PYGZdq{}}\PYG{p}{:} \PYG{l+m+mi}{3}\PYG{p}{,}
  \PYG{n+nt}{\PYGZdq{}local\PYGZdq{}}\PYG{p}{:} \PYG{k+kc}{false}\PYG{p}{,}
  \PYG{n+nt}{\PYGZdq{}simulator\PYGZdq{}}\PYG{p}{:} \PYG{k+kc}{false}\PYG{p}{,}
  \PYG{n+nt}{\PYGZdq{}conditional\PYGZdq{}}\PYG{p}{:} \PYG{k+kc}{false}\PYG{p}{,}
  \PYG{n+nt}{\PYGZdq{}open\PYGZus{}pulse\PYGZdq{}}\PYG{p}{:} \PYG{k+kc}{true}\PYG{p}{,}
  \PYG{n+nt}{\PYGZdq{}n\PYGZus{}uchannels\PYGZdq{}}\PYG{p}{:} \PYG{l+m+mi}{0}\PYG{p}{,}
  \PYG{n+nt}{\PYGZdq{}hamiltonian\PYGZdq{}}\PYG{p}{:} \PYG{err}{see\PYGZus{}below}\PYG{p}{,}
  \PYG{n+nt}{\PYGZdq{}u\PYGZus{}channel\PYGZus{}lo\PYGZdq{}}\PYG{p}{:} \PYG{p}{[],}
  \PYG{n+nt}{\PYGZdq{}meas\PYGZus{}levels\PYGZdq{}}\PYG{p}{:} \PYG{p}{[}\PYG{l+m+mi}{2}\PYG{p}{],}
  \PYG{n+nt}{\PYGZdq{}qubit\PYGZus{}lo\PYGZus{}range\PYGZdq{}}\PYG{p}{:} \PYG{p}{[[}\PYG{l+m+mf}{12.6}\PYG{p}{,}\PYG{l+m+mf}{12.7}\PYG{p}{],[}\PYG{l+m+mf}{12.6}\PYG{p}{,}\PYG{l+m+mf}{12.7}\PYG{p}{],[}\PYG{l+m+mf}{12.6}\PYG{p}{,}\PYG{l+m+mf}{12.7}\PYG{p}{]],}
  \PYG{n+nt}{\PYGZdq{}meas\PYGZus{}lo\PYGZus{}range\PYGZdq{}}\PYG{p}{:} \PYG{p}{[[}\PYG{l+m+mi}{811280}\PYG{p}{,}\PYG{l+m+mi}{811280}\PYG{p}{],[}\PYG{l+m+mi}{811280}\PYG{p}{,}\PYG{l+m+mi}{811280}\PYG{p}{],}
                    \PYG{p}{[}\PYG{l+m+mi}{811280}\PYG{p}{,}\PYG{l+m+mi}{811280}\PYG{p}{]],}
  \PYG{n+nt}{\PYGZdq{}dt\PYGZdq{}}\PYG{p}{:} \PYG{l+m+mf}{10.5}\PYG{p}{,}
  \PYG{n+nt}{\PYGZdq{}dtm\PYGZdq{}}\PYG{p}{:} \PYG{l+m+mf}{10.5}\PYG{p}{,}
  \PYG{n+nt}{\PYGZdq{}rep\PYGZus{}times\PYGZdq{}}\PYG{p}{:} \PYG{p}{[}\PYG{l+m+mi}{10000}\PYG{p}{],}
  \PYG{n+nt}{\PYGZdq{}meas\PYGZus{}map\PYGZdq{}}\PYG{p}{:} \PYG{p}{[[}\PYG{l+m+mi}{0}\PYG{p}{],[}\PYG{l+m+mi}{1}\PYG{p}{],[}\PYG{l+m+mi}{2}\PYG{p}{]],}
  \PYG{n+nt}{\PYGZdq{}basis\PYGZus{}gates\PYGZdq{}}\PYG{p}{:} \PYG{p}{[}\PYG{l+s+s2}{\PYGZdq{}gate1\PYGZdq{}}\PYG{p}{,} \PYG{l+s+s2}{\PYGZdq{}gate2\PYGZdq{}}\PYG{p}{,}\PYG{err}{...}\PYG{p}{],}
  \PYG{n+nt}{\PYGZdq{}gates\PYGZdq{}}\PYG{p}{:} \PYG{p}{[}\PYG{err}{gate1}\PYG{p}{,} \PYG{err}{gate2}\PYG{p}{,}\PYG{err}{...}\PYG{p}{],}
  \PYG{n+nt}{\PYGZdq{}channel\PYGZus{}bandwidth\PYGZdq{}}\PYG{p}{:} \PYG{p}{[[}\PYG{l+m+mf}{\PYGZhy{}0.1}\PYG{p}{,}\PYG{l+m+mf}{0.1}\PYG{p}{],[}\PYG{l+m+mf}{\PYGZhy{}0.1}\PYG{p}{,}\PYG{l+m+mf}{0.1}\PYG{p}{],[}\PYG{l+m+mf}{\PYGZhy{}0.1}\PYG{p}{,}\PYG{l+m+mf}{0.1}\PYG{p}{]],}
  \PYG{n+nt}{\PYGZdq{}meas\PYGZus{}kernels\PYGZdq{}}\PYG{p}{:} \PYG{p}{[}\PYG{l+s+s2}{\PYGZdq{}default\PYGZdq{}}\PYG{p}{],}
  \PYG{n+nt}{\PYGZdq{}discriminators\PYGZdq{}}\PYG{p}{:} \PYG{p}{[}\PYG{l+s+s2}{\PYGZdq{}max\PYGZus{}1Q\PYGZus{}fidelity\PYGZdq{}}\PYG{p}{]}
\PYG{p}{\PYGZcb{}}
\end{Verbatim}
\end{jsonexamplebox}
and the call to \texttt{Backend.defaults()}
\begin{jsonexamplebox}{Ion Trap - Backend Defaults}
\begin{Verbatim}[commandchars=\\\{\}]
\PYG{p}{\PYGZob{}}
  \PYG{n+nt}{\PYGZdq{}qubit\PYGZus{}freq\PYGZus{}est\PYGZdq{}}\PYG{p}{:} \PYG{p}{[}\PYG{l+m+mf}{12.642}\PYG{p}{,}\PYG{l+m+mf}{12.642}\PYG{p}{,}\PYG{l+m+mf}{12.642}\PYG{p}{],}
  \PYG{n+nt}{\PYGZdq{}meas\PYGZus{}freq\PYGZus{}est\PYGZdq{}}\PYG{p}{:} \PYG{p}{[}\PYG{l+m+mi}{811280}\PYG{p}{,}\PYG{l+m+mi}{811280}\PYG{p}{,}\PYG{l+m+mi}{811280}\PYG{p}{],}
  \PYG{n+nt}{\PYGZdq{}buffer\PYGZdq{}}\PYG{p}{:} \PYG{l+m+mi}{10}\PYG{p}{,}
  \PYG{n+nt}{\PYGZdq{}pulse\PYGZus{}library\PYGZdq{}}\PYG{p}{:} \PYG{p}{[],}
  \PYG{n+nt}{\PYGZdq{}cmd\PYGZus{}def\PYGZdq{}}\PYG{p}{:} \PYG{p}{[],}
  \PYG{n+nt}{\PYGZdq{}meas\PYGZus{}kernel\PYGZdq{}}\PYG{p}{:} \PYG{p}{\PYGZob{}}\PYG{n+nt}{\PYGZdq{}name\PYGZdq{}}\PYG{p}{:} \PYG{l+s+s2}{\PYGZdq{}default\PYGZdq{}}\PYG{p}{,} \PYG{n+nt}{\PYGZdq{}params\PYGZdq{}}\PYG{p}{:} \PYG{p}{[]\PYGZcb{},}
  \PYG{n+nt}{\PYGZdq{}discriminator\PYGZdq{}}\PYG{p}{:} \PYG{p}{\PYGZob{}}\PYG{n+nt}{\PYGZdq{}name\PYGZdq{}}\PYG{p}{:} \PYG{l+s+s2}{\PYGZdq{}max\PYGZus{}1Q\PYGZus{}fidelity\PYGZdq{}}\PYG{p}{,} \PYG{n+nt}{\PYGZdq{}params\PYGZdq{}}\PYG{p}{:} \PYG{p}{[]\PYGZcb{}}
\PYG{p}{\PYGZcb{}}
\end{Verbatim}
\end{jsonexamplebox}

The Hamiltonian returned by the device is:
\begin{eqnarray}
    H & = & \sum_{i=0}^{2} D_i(t) \sigma_i^{X} + \sum_{i=0}^{2} 2\pi \nu_i  (1-\sigma_i^{Z})/2 + \nonumber \\
      & & \sum_{i=0}^{2}\sum_{p} D_i(t) \eta_{p}^{i} \sigma_i^{X} (\hat{a}_p + \hat{a}_p^{\dagger}) + \sum_{p} 2\pi 3.01~\textrm{MHz}~\hat{a}_{p}^{\dagger} \hat{a}_p
\end{eqnarray}
which describes three ions coupled by the Raman beam to the motional trap modes ``$p$''. With this information a Molmer-Sorensen gate can be constructed.

\subsection{NMR}

In this example, the device consists of a three-qubit liquid NMR system. The drive lines here represent different LO frequencies, but there is only one driving coil. Similarly, the different measurement channels represent a single receiver coil with different mix down LO frequencies. The measurement drive is a filter applied to the receiver signal. For NMR there is a large global field; this value would have to be conveyed outside the specification. All the timescales are greatly increased. The call to \texttt{Backend.configuration()} returns:
\begin{jsonexamplebox}{NMR - Backend Configuration}
\begin{Verbatim}[commandchars=\\\{\}]
\PYG{p}{\PYGZob{}}
  \PYG{n+nt}{\PYGZdq{}backend\PYGZus{}name\PYGZdq{}}\PYG{p}{:} \PYG{l+s+s2}{\PYGZdq{}ibmqx1\PYGZdq{}}\PYG{p}{,}
  \PYG{n+nt}{\PYGZdq{}backend\PYGZus{}version\PYGZdq{}}\PYG{p}{:} \PYG{l+s+s2}{\PYGZdq{}1.1.5\PYGZdq{}}\PYG{p}{,}
  \PYG{n+nt}{\PYGZdq{}n\PYGZus{}qubits\PYGZdq{}}\PYG{p}{:} \PYG{l+m+mi}{3}\PYG{p}{,}
  \PYG{n+nt}{\PYGZdq{}local\PYGZdq{}}\PYG{p}{:} \PYG{k+kc}{false}\PYG{p}{,}
  \PYG{n+nt}{\PYGZdq{}simulator\PYGZdq{}}\PYG{p}{:} \PYG{k+kc}{false}\PYG{p}{,}
  \PYG{n+nt}{\PYGZdq{}conditional\PYGZdq{}}\PYG{p}{:} \PYG{k+kc}{false}\PYG{p}{,}
  \PYG{n+nt}{\PYGZdq{}open\PYGZus{}pulse\PYGZdq{}}\PYG{p}{:} \PYG{k+kc}{true}\PYG{p}{,}
  \PYG{n+nt}{\PYGZdq{}n\PYGZus{}uchannels\PYGZdq{}}\PYG{p}{:} \PYG{l+m+mi}{0}\PYG{p}{,}
  \PYG{n+nt}{\PYGZdq{}hamiltonian\PYGZdq{}}\PYG{p}{:} \PYG{err}{see\PYGZus{}below}\PYG{p}{,}
  \PYG{n+nt}{\PYGZdq{}u\PYGZus{}channel\PYGZus{}lo\PYGZdq{}}\PYG{p}{:} \PYG{p}{[],}
  \PYG{n+nt}{\PYGZdq{}meas\PYGZus{}levels\PYGZdq{}}\PYG{p}{:} \PYG{p}{[}\PYG{l+m+mi}{0}\PYG{p}{,}\PYG{l+m+mi}{1}\PYG{p}{,}\PYG{l+m+mi}{2}\PYG{p}{],}
  \PYG{n+nt}{\PYGZdq{}qubit\PYGZus{}lo\PYGZus{}range\PYGZdq{}}\PYG{p}{:} \PYG{p}{[[}\PYG{l+m+mf}{0.05}\PYG{p}{,}\PYG{l+m+mf}{0.3}\PYG{p}{],[}\PYG{l+m+mf}{0.05}\PYG{p}{,}\PYG{l+m+mf}{0.3}\PYG{p}{],[}\PYG{l+m+mf}{0.05}\PYG{p}{,}\PYG{l+m+mf}{0.3}\PYG{p}{]],}
  \PYG{n+nt}{\PYGZdq{}meas\PYGZus{}lo\PYGZus{}range\PYGZdq{}}\PYG{p}{:} \PYG{p}{[[}\PYG{l+m+mf}{0.05}\PYG{p}{,}\PYG{l+m+mf}{0.3}\PYG{p}{],[}\PYG{l+m+mf}{0.05}\PYG{p}{,}\PYG{l+m+mf}{0.3}\PYG{p}{],[}\PYG{l+m+mf}{0.05}\PYG{p}{,}\PYG{l+m+mf}{0.4}\PYG{p}{]],}
  \PYG{n+nt}{\PYGZdq{}dt\PYGZdq{}}\PYG{p}{:} \PYG{l+m+mi}{1000}\PYG{p}{,}
  \PYG{n+nt}{\PYGZdq{}dtm\PYGZdq{}}\PYG{p}{:} \PYG{l+m+mi}{100}\PYG{p}{,}
  \PYG{n+nt}{\PYGZdq{}rep\PYGZus{}times\PYGZdq{}}\PYG{p}{:} \PYG{p}{[}\PYG{l+m+mi}{1000000}\PYG{p}{],}
  \PYG{n+nt}{\PYGZdq{}meas\PYGZus{}map\PYGZdq{}}\PYG{p}{:} \PYG{p}{[[}\PYG{l+m+mi}{0}\PYG{p}{,}\PYG{l+m+mi}{1}\PYG{p}{,}\PYG{l+m+mi}{2}\PYG{p}{]],}
  \PYG{n+nt}{\PYGZdq{}basis\PYGZus{}gates\PYGZdq{}}\PYG{p}{:} \PYG{p}{[}\PYG{l+s+s2}{\PYGZdq{}gate1\PYGZdq{}}\PYG{p}{,} \PYG{l+s+s2}{\PYGZdq{}gate2\PYGZdq{}}\PYG{p}{,}\PYG{err}{...}\PYG{p}{],}
  \PYG{n+nt}{\PYGZdq{}gates\PYGZdq{}}\PYG{p}{:} \PYG{p}{[}\PYG{err}{gate1}\PYG{p}{,} \PYG{err}{gate2}\PYG{p}{,}\PYG{err}{...}\PYG{p}{],}
  \PYG{n+nt}{\PYGZdq{}channel\PYGZus{}bandwidth\PYGZdq{}}\PYG{p}{:} \PYG{p}{[[}\PYG{l+m+mf}{\PYGZhy{}0.01}\PYG{p}{,}\PYG{l+m+mf}{0.01}\PYG{p}{],[}\PYG{l+m+mf}{\PYGZhy{}0.01}\PYG{p}{,}\PYG{l+m+mf}{0.01}\PYG{p}{],}
                        \PYG{p}{[}\PYG{l+m+mf}{\PYGZhy{}0.01}\PYG{p}{,}\PYG{l+m+mf}{0.01}\PYG{p}{]],}
  \PYG{n+nt}{\PYGZdq{}meas\PYGZus{}kernels\PYGZdq{}}\PYG{p}{:} \PYG{p}{[}\PYG{l+s+s2}{\PYGZdq{}boxcar\PYGZdq{}}\PYG{p}{],}
  \PYG{n+nt}{\PYGZdq{}discriminators\PYGZdq{}}\PYG{p}{:} \PYG{p}{[}\PYG{l+s+s2}{\PYGZdq{}max\PYGZus{}1Q\PYGZus{}fidelity\PYGZdq{}}\PYG{p}{]}
\PYG{p}{\PYGZcb{}}
\end{Verbatim}
\end{jsonexamplebox}
and the call to \texttt{Backend.defaults()}
\begin{jsonexamplebox}{NMR - Backend Defaults}
\begin{Verbatim}[commandchars=\\\{\}]
\PYG{p}{\PYGZob{}}
  \PYG{n+nt}{\PYGZdq{}qubit\PYGZus{}freq\PYGZus{}est\PYGZdq{}}\PYG{p}{:} \PYG{p}{[}\PYG{l+m+mf}{0.1}\PYG{p}{,}\PYG{l+m+mf}{0.09}\PYG{p}{,}\PYG{l+m+mf}{0.12}\PYG{p}{],}
  \PYG{n+nt}{\PYGZdq{}meas\PYGZus{}freq\PYGZus{}est\PYGZdq{}}\PYG{p}{:} \PYG{p}{[}\PYG{l+m+mf}{0.1}\PYG{p}{,}\PYG{l+m+mf}{0.09}\PYG{p}{,}\PYG{l+m+mf}{0.12}\PYG{p}{],}
  \PYG{n+nt}{\PYGZdq{}buffer\PYGZdq{}}\PYG{p}{:} \PYG{l+m+mi}{10}\PYG{p}{,}
  \PYG{n+nt}{\PYGZdq{}pulse\PYGZus{}library\PYGZdq{}}\PYG{p}{:} \PYG{p}{[],}
  \PYG{n+nt}{\PYGZdq{}cmd\PYGZus{}def\PYGZdq{}}\PYG{p}{:} \PYG{p}{[],}
  \PYG{n+nt}{\PYGZdq{}meas\PYGZus{}kernel\PYGZdq{}}\PYG{p}{:} \PYG{p}{\PYGZob{}}\PYG{n+nt}{\PYGZdq{}name\PYGZdq{}}\PYG{p}{:} \PYG{l+s+s2}{\PYGZdq{}boxcar\PYGZdq{}}\PYG{p}{,} \PYG{n+nt}{\PYGZdq{}params\PYGZdq{}}\PYG{p}{:} \PYG{p}{[]\PYGZcb{},}
  \PYG{n+nt}{\PYGZdq{}discriminator\PYGZdq{}}\PYG{p}{:} \PYG{p}{\PYGZob{}}\PYG{n+nt}{\PYGZdq{}name\PYGZdq{}}\PYG{p}{:} \PYG{l+s+s2}{\PYGZdq{}max\PYGZus{}1Q\PYGZus{}fidelity\PYGZdq{}}\PYG{p}{,} \PYG{n+nt}{\PYGZdq{}params\PYGZdq{}}\PYG{p}{:} \PYG{p}{[]\PYGZcb{}}
\PYG{p}{\PYGZcb{}}
\end{Verbatim}
\end{jsonexamplebox}

The Hamiltonian returned by the device is:
\begin{eqnarray}
    H & = & \sum_{i=0}^{2} D_i(t) \sigma_i^{X} + \sum_{i=0}^{2} 2\pi \nu_i (1+0.02(10.0-dc_0)) (1-\sigma_i^{Z})/2 + \nonumber \\
      & & \sum_{ij} J_{ij} \sigma_i^Z \sigma_j^Z \\
    dc_0 & = & 10
\end{eqnarray}

\section{Sample Pulse Libraries for OpenPulse \label{sect:samplelibs}}

\subsection{Q-CTRL Black Opal}

\begin{figure}[ht]
	\label{fig:blackopal}
	\centering
\includegraphics[width=\textwidth]{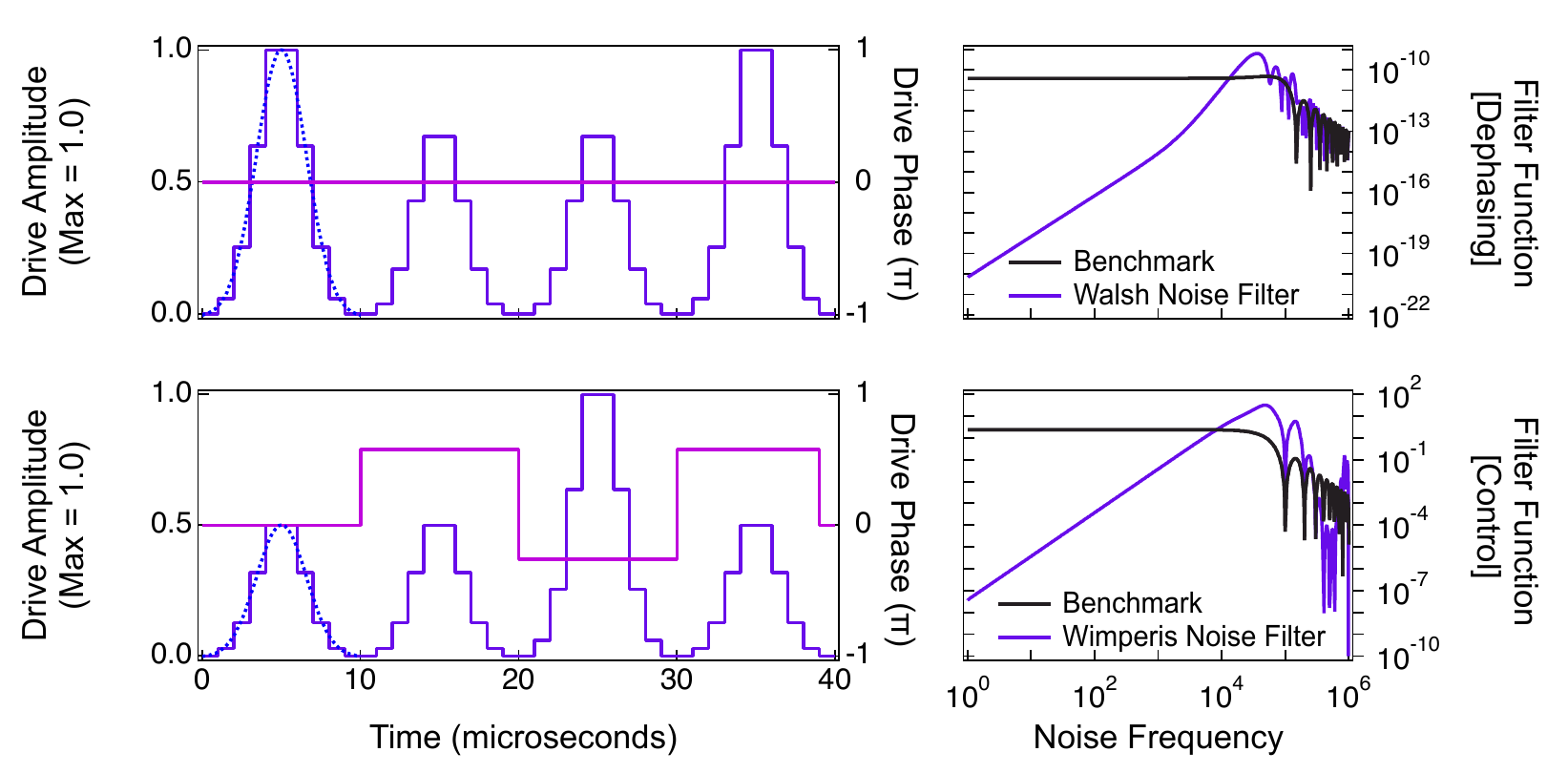}
	\caption{\textbf{Example pulse formats crafted using Black Opal.} Left panels indicate the drive amplitude and phase for given controls; a Gaussian profile is overlaid on each sampled control as a guide to the eye.  In both examples here we have elected to construct our pulse waveform using fixed segment durations rather than a fixed drive amplitude.  The right-hand panels show the relevant filter functions, calculated using the Black Opal package, for the corresponding control operations.  In order to present the filter functions in laboratory units of frequency the controls are scaled to have an arbitrarily selected duration of 40 $\mu$s.
}
\end{figure}

Here we include a sample pulse library from our partners at \href{https://q-ctrl.com}{Q-CTRL} that demonstrates some of the unique pulse shapes that can be enabled by OpenPulse. These are created using the \href{https://q-ctrl.com/products/}{Black Opal} package which can natively output library entries in the appropriate JSON format. The samples are shown in the JSON below and illustrated in Fig.~\ref{fig:blackopal}

\begin{jsonexamplebox}{Black Opal Pulse Library}
\begin{Verbatim}[commandchars=\\\{\}]
\PYG{p}{[}
 \PYG{p}{\PYGZob{}}\PYG{n+nt}{\PYGZdq{}name\PYGZdq{}}\PYG{p}{:} \PYG{l+s+s2}{\PYGZdq{}walsh\PYGZus{}gaussian\PYGZdq{}}\PYG{p}{,}  \PYG{n+nt}{\PYGZdq{}samples\PYGZdq{}}\PYG{p}{:} \PYG{p}{[[}\PYG{l+m+mf}{0.0}\PYG{p}{,} \PYG{l+m+mf}{0.0}\PYG{p}{],} 
 \PYG{p}{[}\PYG{l+m+mf}{0.013434}\PYG{p}{,} \PYG{l+m+mf}{0.0}\PYG{p}{],} \PYG{p}{[}\PYG{l+m+mf}{0.058597}\PYG{p}{,} \PYG{l+m+mf}{0.0}\PYG{p}{],} \PYG{p}{[}\PYG{l+m+mf}{0.146446}\PYG{p}{,} \PYG{l+m+mf}{0.0}\PYG{p}{],} 
 \PYG{p}{[}\PYG{l+m+mf}{0.229930}\PYG{p}{,} \PYG{l+m+mf}{0.0}\PYG{p}{],} \PYG{p}{[}\PYG{l+m+mf}{0.229930}\PYG{p}{,} \PYG{l+m+mf}{0.0}\PYG{p}{],} \PYG{p}{[}\PYG{l+m+mf}{0.1464464}\PYG{p}{,} \PYG{l+m+mf}{0.0}\PYG{p}{],} 
 \PYG{p}{[}\PYG{l+m+mf}{0.058597}\PYG{p}{,} \PYG{l+m+mf}{0.0}\PYG{p}{],} \PYG{p}{[}\PYG{l+m+mf}{0.013434}\PYG{p}{,} \PYG{l+m+mf}{0.0}\PYG{p}{],} \PYG{p}{[}\PYG{l+m+mf}{0.0}\PYG{p}{,} \PYG{l+m+mf}{0.0}\PYG{p}{],} \PYG{p}{[}\PYG{l+m+mf}{0.0}\PYG{p}{,} \PYG{l+m+mf}{0.0}\PYG{p}{],} 
 \PYG{p}{[}\PYG{l+m+mf}{0.009035}\PYG{p}{,} \PYG{l+m+mf}{0.0}\PYG{p}{],} \PYG{p}{[}\PYG{l+m+mf}{0.039411}\PYG{p}{,} \PYG{l+m+mf}{0.0}\PYG{p}{],} \PYG{p}{[}\PYG{l+m+mf}{0.098498}\PYG{p}{,} \PYG{l+m+mf}{0.0}\PYG{p}{],} 
 \PYG{p}{[}\PYG{l+m+mf}{0.154648}\PYG{p}{,} \PYG{l+m+mf}{0.0}\PYG{p}{],} \PYG{p}{[}\PYG{l+m+mf}{0.154648}\PYG{p}{,} \PYG{l+m+mf}{0.0}\PYG{p}{],} \PYG{p}{[}\PYG{l+m+mf}{0.098498}\PYG{p}{,} \PYG{l+m+mf}{0.0}\PYG{p}{],} 
 \PYG{p}{[}\PYG{l+m+mf}{0.039411}\PYG{p}{,} \PYG{l+m+mf}{0.0}\PYG{p}{],} \PYG{p}{[}\PYG{l+m+mf}{0.009035}\PYG{p}{,} \PYG{l+m+mf}{0.0}\PYG{p}{],} \PYG{p}{[}\PYG{l+m+mf}{0.0}\PYG{p}{,} \PYG{l+m+mf}{0.0}\PYG{p}{],} \PYG{p}{[}\PYG{l+m+mf}{0.0}\PYG{p}{,} \PYG{l+m+mf}{0.0}\PYG{p}{],} 
 \PYG{p}{[}\PYG{l+m+mf}{0.009035}\PYG{p}{,} \PYG{l+m+mf}{0.0}\PYG{p}{],} \PYG{p}{[}\PYG{l+m+mf}{0.039411}\PYG{p}{,} \PYG{l+m+mf}{0.0}\PYG{p}{],} \PYG{p}{[}\PYG{l+m+mf}{0.098498}\PYG{p}{,} \PYG{l+m+mf}{0.0}\PYG{p}{],} 
 \PYG{p}{[}\PYG{l+m+mf}{0.154648}\PYG{p}{,} \PYG{l+m+mf}{0.0}\PYG{p}{],} \PYG{p}{[}\PYG{l+m+mf}{0.154648}\PYG{p}{,} \PYG{l+m+mf}{0.0}\PYG{p}{],} \PYG{p}{[}\PYG{l+m+mf}{0.098498}\PYG{p}{,} \PYG{l+m+mf}{0.0}\PYG{p}{],} 
 \PYG{p}{[}\PYG{l+m+mf}{0.039411}\PYG{p}{,} \PYG{l+m+mf}{0.0}\PYG{p}{],} \PYG{p}{[}\PYG{l+m+mf}{0.009035}\PYG{p}{,} \PYG{l+m+mf}{0.0}\PYG{p}{],} \PYG{p}{[}\PYG{l+m+mf}{0.0}\PYG{p}{,} \PYG{l+m+mf}{0.0}\PYG{p}{],} \PYG{p}{[}\PYG{l+m+mf}{0.0}\PYG{p}{,} \PYG{l+m+mf}{0.0}\PYG{p}{],}
 \PYG{p}{[}\PYG{l+m+mf}{0.013434}\PYG{p}{,} \PYG{l+m+mf}{0.0}\PYG{p}{],} \PYG{p}{[}\PYG{l+m+mf}{0.058597}\PYG{p}{,} \PYG{l+m+mf}{0.0}\PYG{p}{],} \PYG{p}{[}\PYG{l+m+mf}{0.146446}\PYG{p}{,} \PYG{l+m+mf}{0.0}\PYG{p}{],} 
 \PYG{p}{[}\PYG{l+m+mf}{0.229930}\PYG{p}{,} \PYG{l+m+mf}{0.0}\PYG{p}{],} \PYG{p}{[}\PYG{l+m+mf}{0.229930}\PYG{p}{,} \PYG{l+m+mf}{0.0}\PYG{p}{],} \PYG{p}{[}\PYG{l+m+mf}{0.146446}\PYG{p}{,} \PYG{l+m+mf}{0.0}\PYG{p}{],} 
 \PYG{p}{[}\PYG{l+m+mf}{0.058597}\PYG{p}{,} \PYG{l+m+mf}{0.0}\PYG{p}{],} \PYG{p}{[}\PYG{l+m+mf}{0.013434}\PYG{p}{,} \PYG{l+m+mf}{0.0}\PYG{p}{],} \PYG{p}{[}\PYG{l+m+mf}{0.0}\PYG{p}{,} \PYG{l+m+mf}{0.0}\PYG{p}{]]\PYGZcb{},}
 \PYG{p}{\PYGZob{}}\PYG{n+nt}{\PYGZdq{}name\PYGZdq{}}\PYG{p}{:} \PYG{l+s+s2}{\PYGZdq{}BB1\PYGZus{}gaussian\PYGZdq{}}\PYG{p}{,}  \PYG{n+nt}{\PYGZdq{}samples\PYGZdq{}}\PYG{p}{:} \PYG{p}{[[}\PYG{l+m+mf}{0.0}\PYG{p}{,} \PYG{l+m+mf}{0.0}\PYG{p}{],}
 \PYG{p}{[}\PYG{l+m+mf}{0.014980}\PYG{p}{,} \PYG{l+m+mf}{0.0}\PYG{p}{],} \PYG{p}{[}\PYG{l+m+mf}{0.065339}\PYG{p}{,} \PYG{l+m+mf}{0.0}\PYG{p}{],} \PYG{p}{[}\PYG{l+m+mf}{0.163296}\PYG{p}{,} \PYG{l+m+mf}{0.0}\PYG{p}{],}
 \PYG{p}{[}\PYG{l+m+mf}{0.256385}\PYG{p}{,} \PYG{l+m+mf}{0.0}\PYG{p}{],} \PYG{p}{[}\PYG{l+m+mf}{0.256385}\PYG{p}{,} \PYG{l+m+mf}{0.0}\PYG{p}{],} \PYG{p}{[}\PYG{l+m+mf}{0.163296}\PYG{p}{,} \PYG{l+m+mf}{0.0}\PYG{p}{],}
 \PYG{p}{[}\PYG{l+m+mf}{0.065339}\PYG{p}{,} \PYG{l+m+mf}{0.0}\PYG{p}{],} \PYG{p}{[}\PYG{l+m+mf}{0.014980}\PYG{p}{,} \PYG{l+m+mf}{0.0}\PYG{p}{],} \PYG{p}{[}\PYG{l+m+mf}{0.0}\PYG{p}{,} \PYG{l+m+mf}{0.0}\PYG{p}{],} \PYG{p}{[}\PYG{l+m+mf}{0.0}\PYG{p}{,} \PYG{l+m+mf}{0.0}\PYG{p}{],} 
 \PYG{p}{[}\PYG{l+m+mf}{\PYGZhy{}0.003745}\PYG{p}{,} \PYG{l+m+mf}{0.014504}\PYG{p}{],} \PYG{p}{[}\PYG{l+m+mf}{\PYGZhy{}0.016335}\PYG{p}{,} \PYG{l+m+mf}{0.063264}\PYG{p}{],}
 \PYG{p}{[}\PYG{l+m+mf}{\PYGZhy{}0.040824}\PYG{p}{,} \PYG{l+m+mf}{0.158111}\PYG{p}{],} \PYG{p}{[}\PYG{l+m+mf}{\PYGZhy{}0.064096}\PYG{p}{,} \PYG{l+m+mf}{0.248244}\PYG{p}{],} 
 \PYG{p}{[}\PYG{l+m+mf}{\PYGZhy{}0.064096}\PYG{p}{,} \PYG{l+m+mf}{0.248244}\PYG{p}{],} \PYG{p}{[}\PYG{l+m+mf}{\PYGZhy{}0.040824}\PYG{p}{,} \PYG{l+m+mf}{0.158111}\PYG{p}{],}
 \PYG{p}{[}\PYG{l+m+mf}{\PYGZhy{}0.016335}\PYG{p}{,} \PYG{l+m+mf}{0.063264}\PYG{p}{],} \PYG{p}{[}\PYG{l+m+mf}{\PYGZhy{}0.003745}\PYG{p}{,} \PYG{l+m+mf}{0.014504}\PYG{p}{],} \PYG{p}{[}\PYG{l+m+mf}{0.0}\PYG{p}{,} \PYG{l+m+mf}{0.0}\PYG{p}{],}
 \PYG{p}{[}\PYG{l+m+mf}{0.0}\PYG{p}{,} \PYG{l+m+mf}{0.0}\PYG{p}{],} \PYG{p}{[}\PYG{l+m+mf}{0.020597}\PYG{p}{,} \PYG{l+m+mf}{\PYGZhy{}0.021756}\PYG{p}{],} \PYG{p}{[}\PYG{l+m+mf}{0.089841}\PYG{p}{,} \PYG{l+m+mf}{\PYGZhy{}0.094896}\PYG{p}{],}
 \PYG{p}{[}\PYG{l+m+mf}{0.224532}\PYG{p}{,} \PYG{l+m+mf}{\PYGZhy{}0.237166}\PYG{p}{],} \PYG{p}{[}\PYG{l+m+mf}{0.352530}\PYG{p}{,} \PYG{l+m+mf}{\PYGZhy{}0.372366}\PYG{p}{],} 
 \PYG{p}{[}\PYG{l+m+mf}{0.352530}\PYG{p}{,} \PYG{l+m+mf}{\PYGZhy{}0.372366}\PYG{p}{],} \PYG{p}{[}\PYG{l+m+mf}{0.224532}\PYG{p}{,} \PYG{l+m+mf}{\PYGZhy{}0.237166}\PYG{p}{],}
 \PYG{p}{[}\PYG{l+m+mf}{0.089841}\PYG{p}{,} \PYG{l+m+mf}{\PYGZhy{}0.094896}\PYG{p}{],} \PYG{p}{[}\PYG{l+m+mf}{0.020597}\PYG{p}{,} \PYG{l+m+mf}{\PYGZhy{}0.021756}\PYG{p}{],} \PYG{p}{[}\PYG{l+m+mf}{0.0}\PYG{p}{,} \PYG{l+m+mf}{0.0}\PYG{p}{],}
 \PYG{p}{[}\PYG{l+m+mf}{0.0}\PYG{p}{,} \PYG{l+m+mf}{0.0}\PYG{p}{],} \PYG{p}{[}\PYG{l+m+mf}{\PYGZhy{}0.003745}\PYG{p}{,} \PYG{l+m+mf}{0.014504}\PYG{p}{],} \PYG{p}{[}\PYG{l+m+mf}{\PYGZhy{}0.016335}\PYG{p}{,} \PYG{l+m+mf}{0.063264}\PYG{p}{],} 
 \PYG{p}{[}\PYG{l+m+mf}{\PYGZhy{}0.040824}\PYG{p}{,} \PYG{l+m+mf}{0.158111}\PYG{p}{],} \PYG{p}{[}\PYG{l+m+mf}{\PYGZhy{}0.064096}\PYG{p}{,} \PYG{l+m+mf}{0.248244}\PYG{p}{],}
 \PYG{p}{[}\PYG{l+m+mf}{\PYGZhy{}0.064096}\PYG{p}{,} \PYG{l+m+mf}{0.248244}\PYG{p}{],} \PYG{p}{[}\PYG{l+m+mf}{\PYGZhy{}0.040824}\PYG{p}{,} \PYG{l+m+mf}{0.158111}\PYG{p}{],} 
 \PYG{p}{[}\PYG{l+m+mf}{\PYGZhy{}0.016335}\PYG{p}{,} \PYG{l+m+mf}{0.063264}\PYG{p}{],} \PYG{p}{[}\PYG{l+m+mf}{\PYGZhy{}0.003745}\PYG{p}{,} \PYG{l+m+mf}{0.014504}\PYG{p}{],} \PYG{p}{[}\PYG{l+m+mf}{0.0}\PYG{p}{,} \PYG{l+m+mf}{0.0}\PYG{p}{]]\PYGZcb{}}
\PYG{p}{]}
\end{Verbatim}
\end{jsonexamplebox}

In our first example (``walsh\_gaussian'' pulse in the library, top panel of Fig.~\ref{fig:blackopal}) we implement a control which produces a net $X$ operation (a rotation by angle $\pi$ about $\hat{x}$) in a manner that is robust against dephasing noise processes which can cause both overrotation and phase errors during a driven operation.  This control is synthesized using the basis of Walsh functions, and is implemented as a series of Gaussian \emph{segments} commonly employed in superconducting processors.  The specific control waveform is optimized for the discretized Gaussian segments employed here. Each segment is in turn defined by multiple \emph{samples} used to define the relevant control envelope; the sample duration is set by the device time unit $dt$.  In both examples presented here we have constrained the number of samples in order to keep the JSON files easily readable.  As this particular control is performed exclusively using rotations about the $\hat{x}$ axis, all samples in the JSON file are real valued. \\

In a second example (``BB1\_gaussian'' pulse in the library, bottom panel of Fig.~\ref{fig:blackopal}) we implement a control which again produces a net $X$ operation, but this time provides robustness against fluctuations in the strength of the driving field or pulse timing errors.  This is a composite pulse derived from the Wimperis family of controls, but is again structured to incorporate Gaussian pulse segments.  Here, due to the varying rotation axes (indicated by the varying drive phase) each sample is formatted as a complex number representing the amplitude of the driving field around the $\hat{x}$ and $\hat{y}$ axes.  \\

The performance of these two controls is captured through the filter function which is a heuristic measure of noise susceptibility. Low values of the filter function indicate noise suppression, and are presented in Fig.~\ref{fig:blackopal}.  For further discussion, alternative control selections, and detailed documentation see the Q-CTRL \href{https://docs.q-ctrl.com/}{Technical Documentation}.  

\section{Sample Experiments for OpenPulse \label{sect:samples}}

In this section, we detail sample experiments. In the Rabi experiment example, we include the different measurement levels to detail how those work. In the other examples, we leave off the actual measurements for brevity.

\subsection{Rabi Oscillation}

This experiment is a simple Rabi oscillation accomplished by changing the drive amplitude. For brevity we consider only three amplitude points $(0,0.5,1.0)$ where we assume that amplitude 1.0 is a $\pi$-pulse. A call to \texttt{Backend.configuration()} returns:

\begin{jsonexamplebox}{Rabi Oscillation - Backend Configuration}
\begin{Verbatim}[commandchars=\\\{\}]
\PYG{p}{\PYGZob{}}
  \PYG{n+nt}{\PYGZdq{}backend\PYGZus{}name\PYGZdq{}}\PYG{p}{:} \PYG{l+s+s2}{\PYGZdq{}ibmqx1\PYGZdq{}}\PYG{p}{,}
  \PYG{n+nt}{\PYGZdq{}backend\PYGZus{}version\PYGZdq{}}\PYG{p}{:} \PYG{l+s+s2}{\PYGZdq{}1.1.5\PYGZdq{}}\PYG{p}{,}
  \PYG{n+nt}{\PYGZdq{}n\PYGZus{}qubits\PYGZdq{}}\PYG{p}{:} \PYG{l+m+mi}{1}\PYG{p}{,}
  \PYG{n+nt}{\PYGZdq{}local\PYGZdq{}}\PYG{p}{:} \PYG{k+kc}{false}\PYG{p}{,}
  \PYG{n+nt}{\PYGZdq{}simulator\PYGZdq{}}\PYG{p}{:} \PYG{k+kc}{false}\PYG{p}{,}
  \PYG{n+nt}{\PYGZdq{}conditional\PYGZdq{}}\PYG{p}{:} \PYG{k+kc}{false}\PYG{p}{,}
  \PYG{n+nt}{\PYGZdq{}open\PYGZus{}pulse\PYGZdq{}}\PYG{p}{:} \PYG{k+kc}{true}\PYG{p}{,}
  \PYG{n+nt}{\PYGZdq{}n\PYGZus{}uchannels\PYGZdq{}}\PYG{p}{:} \PYG{l+m+mi}{0}\PYG{p}{,}
  \PYG{n+nt}{\PYGZdq{}hamiltonian\PYGZdq{}}\PYG{p}{:} \PYG{err}{see\PYGZus{}below}\PYG{p}{,}
  \PYG{n+nt}{\PYGZdq{}u\PYGZus{}channel\PYGZus{}lo\PYGZdq{}}\PYG{p}{:} \PYG{p}{[],}
  \PYG{n+nt}{\PYGZdq{}meas\PYGZus{}levels\PYGZdq{}}\PYG{p}{:} \PYG{p}{[}\PYG{l+m+mi}{0}\PYG{p}{,}\PYG{l+m+mi}{1}\PYG{p}{,}\PYG{l+m+mi}{2}\PYG{p}{],}
  \PYG{n+nt}{\PYGZdq{}qubit\PYGZus{}lo\PYGZus{}range\PYGZdq{}}\PYG{p}{:} \PYG{p}{[[}\PYG{l+m+mf}{4.9}\PYG{p}{,}\PYG{l+m+mf}{5.1}\PYG{p}{]],}
  \PYG{n+nt}{\PYGZdq{}meas\PYGZus{}lo\PYGZus{}range\PYGZdq{}}\PYG{p}{:} \PYG{p}{[[}\PYG{l+m+mf}{6.0}\PYG{p}{,}\PYG{l+m+mf}{7.0}\PYG{p}{]],}
  \PYG{n+nt}{\PYGZdq{}dt\PYGZdq{}}\PYG{p}{:} \PYG{l+m+mf}{0.83333}\PYG{p}{,}
  \PYG{n+nt}{\PYGZdq{}dtm\PYGZdq{}}\PYG{p}{:} \PYG{l+m+mf}{0.83333}\PYG{p}{,}
  \PYG{n+nt}{\PYGZdq{}rep\PYGZus{}times\PYGZdq{}}\PYG{p}{:} \PYG{p}{[}\PYG{l+m+mi}{100}\PYG{p}{,}\PYG{l+m+mi}{250}\PYG{p}{,}\PYG{l+m+mi}{500}\PYG{p}{,}\PYG{l+m+mi}{1000}\PYG{p}{],}
  \PYG{n+nt}{\PYGZdq{}meas\PYGZus{}map\PYGZdq{}}\PYG{p}{:} \PYG{p}{[[}\PYG{l+m+mi}{0}\PYG{p}{]],}
  \PYG{n+nt}{\PYGZdq{}basis\PYGZus{}gates\PYGZdq{}}\PYG{p}{:} \PYG{p}{[}\PYG{l+s+s2}{\PYGZdq{}gate1\PYGZdq{}}\PYG{p}{,} \PYG{l+s+s2}{\PYGZdq{}gate2\PYGZdq{}}\PYG{p}{,}\PYG{err}{...}\PYG{p}{],}
  \PYG{n+nt}{\PYGZdq{}gates\PYGZdq{}}\PYG{p}{:} \PYG{p}{[}\PYG{err}{gate1}\PYG{p}{,} \PYG{err}{gate2}\PYG{p}{,}\PYG{err}{...}\PYG{p}{],}
  \PYG{n+nt}{\PYGZdq{}channel\PYGZus{}bandwidth\PYGZdq{}}\PYG{p}{:} \PYG{p}{[[}\PYG{l+m+mf}{\PYGZhy{}0.5}\PYG{p}{,}\PYG{l+m+mf}{0.5}\PYG{p}{],[}\PYG{l+m+mf}{\PYGZhy{}0.5}\PYG{p}{,}\PYG{l+m+mf}{0.5}\PYG{p}{]],}
  \PYG{n+nt}{\PYGZdq{}meas\PYGZus{}kernels\PYGZdq{}}\PYG{p}{:} \PYG{p}{[}\PYG{l+s+s2}{\PYGZdq{}default\PYGZdq{}}\PYG{p}{],}
  \PYG{n+nt}{\PYGZdq{}discriminators\PYGZdq{}}\PYG{p}{:} \PYG{p}{[}\PYG{l+s+s2}{\PYGZdq{}max\PYGZus{}1Q\PYGZus{}fidelity\PYGZdq{}}\PYG{p}{]}
\PYG{p}{\PYGZcb{}}
\end{Verbatim}
\end{jsonexamplebox}
and the call to \texttt{Backend.defaults()}
\begin{jsonexamplebox}{Rabi Oscillation - Backend Defaults }
\begin{Verbatim}[commandchars=\\\{\}]
\PYG{p}{\PYGZob{}}
  \PYG{n+nt}{\PYGZdq{}qubit\PYGZus{}freq\PYGZus{}est\PYGZdq{}}\PYG{p}{:} \PYG{p}{[}\PYG{l+m+mf}{5.0}\PYG{p}{],}
  \PYG{n+nt}{\PYGZdq{}meas\PYGZus{}freq\PYGZus{}est\PYGZdq{}}\PYG{p}{:} \PYG{p}{[}\PYG{l+m+mf}{6.5}\PYG{p}{],}
  \PYG{n+nt}{\PYGZdq{}buffer\PYGZdq{}}\PYG{p}{:} \PYG{l+m+mi}{10}\PYG{p}{,}
  \PYG{n+nt}{\PYGZdq{}pulse\PYGZus{}library\PYGZdq{}}\PYG{p}{:} \PYG{p}{[],}
  \PYG{n+nt}{\PYGZdq{}cmd\PYGZus{}def\PYGZdq{}}\PYG{p}{:} \PYG{p}{[],}
  \PYG{n+nt}{\PYGZdq{}meas\PYGZus{}kernel\PYGZdq{}}\PYG{p}{:} \PYG{p}{\PYGZob{}}\PYG{n+nt}{\PYGZdq{}name\PYGZdq{}}\PYG{p}{:} \PYG{l+s+s2}{\PYGZdq{}boxcar\PYGZdq{}}\PYG{p}{,} \PYG{n+nt}{\PYGZdq{}params\PYGZdq{}}\PYG{p}{:} \PYG{p}{[]\PYGZcb{},}
  \PYG{n+nt}{\PYGZdq{}discriminator\PYGZdq{}}\PYG{p}{:} \PYG{p}{\PYGZob{}}\PYG{n+nt}{\PYGZdq{}name\PYGZdq{}}\PYG{p}{:} \PYG{l+s+s2}{\PYGZdq{}max\PYGZus{}1Q\PYGZus{}fidelity\PYGZdq{}}\PYG{p}{,} \PYG{n+nt}{\PYGZdq{}params\PYGZdq{}}\PYG{p}{:} \PYG{p}{[]\PYGZcb{}}
\PYG{p}{\PYGZcb{}}
\end{Verbatim}
\end{jsonexamplebox}

The Hamiltonian returned by the device is:
\begin{equation}
    H = D_0(t) \sigma_0^{X} + 2\pi \nu_0 (1-\sigma_0^{Z})/2
\end{equation}

For this example the device returns an empty \texttt{pulse\_library} and \texttt{cmd\_def}.

\subsubsection{Pulse Library and Time Sequence}

The pulse library for this experiment is,
\begin{jsonexamplebox}{Rabi Oscillation - Pulse Library}
\begin{Verbatim}[commandchars=\\\{\}]
\PYG{p}{\PYGZob{}}
  \PYG{n+nt}{\PYGZdq{}pulse\PYGZus{}library\PYGZdq{}} \PYG{p}{:} \PYG{p}{[}
   \PYG{p}{\PYGZob{}}\PYG{n+nt}{\PYGZdq{}name\PYGZdq{}}\PYG{p}{:} \PYG{l+s+s2}{\PYGZdq{}pulse1\PYGZdq{}}\PYG{p}{,} \PYG{n+nt}{\PYGZdq{}samples\PYGZdq{}}\PYG{p}{:} \PYG{p}{[[}\PYG{l+m+mf}{0.002}\PYG{p}{,}\PYG{l+m+mf}{0.0}\PYG{p}{],}
      \PYG{p}{[}\PYG{l+m+mf}{0.015}\PYG{p}{,}\PYG{l+m+mf}{0.0}\PYG{p}{],[}\PYG{l+m+mf}{0.065}\PYG{p}{,}\PYG{l+m+mf}{0.0}\PYG{p}{],[}\PYG{l+m+mf}{0.2}\PYG{p}{,}\PYG{l+m+mf}{0.0}\PYG{p}{],[}\PYG{l+m+mf}{0.4}\PYG{p}{,}\PYG{l+m+mf}{0.0}\PYG{p}{],}
      \PYG{p}{[}\PYG{l+m+mf}{0.5}\PYG{p}{,}\PYG{l+m+mf}{0.0}\PYG{p}{],[}\PYG{l+m+mf}{0.4}\PYG{p}{,}\PYG{l+m+mf}{0.0}\PYG{p}{],[}\PYG{l+m+mf}{0.2}\PYG{p}{,}\PYG{l+m+mf}{0.0}\PYG{p}{],[}\PYG{l+m+mf}{0.065}\PYG{p}{,}\PYG{l+m+mf}{\PYGZhy{}0.0}\PYG{p}{],}
      \PYG{p}{[}\PYG{l+m+mf}{0.015}\PYG{p}{,}\PYG{l+m+mf}{0.0}\PYG{p}{],[}\PYG{l+m+mf}{0.002}\PYG{p}{,}\PYG{l+m+mf}{0.0}\PYG{p}{]]\PYGZcb{},}
   \PYG{p}{\PYGZob{}}\PYG{n+nt}{\PYGZdq{}name\PYGZdq{}}\PYG{p}{:} \PYG{l+s+s2}{\PYGZdq{}pulse2\PYGZdq{}}\PYG{p}{,} \PYG{n+nt}{\PYGZdq{}samples\PYGZdq{}}\PYG{p}{:} \PYG{p}{[[}\PYG{l+m+mf}{0.004}\PYG{p}{,}\PYG{l+m+mf}{0.0}\PYG{p}{],}
      \PYG{p}{[}\PYG{l+m+mf}{0.029}\PYG{p}{,}\PYG{l+m+mf}{0.0}\PYG{p}{],[}\PYG{l+m+mf}{0.135}\PYG{p}{,}\PYG{l+m+mf}{0.0}\PYG{p}{],[}\PYG{l+m+mf}{0.41}\PYG{p}{,}\PYG{l+m+mf}{0.0}\PYG{p}{],[}\PYG{l+m+mf}{0.8}\PYG{p}{,}\PYG{l+m+mf}{0.0}\PYG{p}{],}
      \PYG{p}{[}\PYG{l+m+mf}{1.0}\PYG{p}{,}\PYG{l+m+mf}{0.0}\PYG{p}{],[}\PYG{l+m+mf}{0.8}\PYG{p}{,}\PYG{l+m+mf}{0.0}\PYG{p}{],[}\PYG{l+m+mf}{0.41}\PYG{p}{,}\PYG{l+m+mf}{0.0}\PYG{p}{],[}\PYG{l+m+mf}{0.135}\PYG{p}{,}\PYG{l+m+mf}{0.0}\PYG{p}{],}
      \PYG{p}{[}\PYG{l+m+mf}{0.029}\PYG{p}{,}\PYG{l+m+mf}{0.0}\PYG{p}{],[}\PYG{l+m+mf}{0.004}\PYG{p}{,}\PYG{l+m+mf}{0.0}\PYG{p}{]]\PYGZcb{},}
   \PYG{p}{\PYGZob{}}\PYG{n+nt}{\PYGZdq{}name\PYGZdq{}}\PYG{p}{:} \PYG{l+s+s2}{\PYGZdq{}square\PYGZus{}pulse\PYGZdq{}}\PYG{p}{,} \PYG{n+nt}{\PYGZdq{}samples\PYGZdq{}}\PYG{p}{:} \PYG{p}{[[}\PYG{l+m+mf}{0.1}\PYG{p}{,}\PYG{l+m+mf}{0.0}\PYG{p}{],}
      \PYG{p}{[}\PYG{l+m+mf}{0.1}\PYG{p}{,}\PYG{l+m+mf}{0.0}\PYG{p}{],[}\PYG{l+m+mf}{0.1}\PYG{p}{,}\PYG{l+m+mf}{0.0}\PYG{p}{],[}\PYG{l+m+mf}{0.1}\PYG{p}{,}\PYG{l+m+mf}{0.0}\PYG{p}{],[}\PYG{l+m+mf}{0.1}\PYG{p}{,}\PYG{l+m+mf}{0.0}\PYG{p}{],}
      \PYG{p}{[}\PYG{l+m+mf}{0.1}\PYG{p}{,}\PYG{l+m+mf}{0.0}\PYG{p}{]]\PYGZcb{}]}
\PYG{p}{\PYGZcb{}}
\end{Verbatim}
\end{jsonexamplebox}
which is two Gaussian pulses (one of peak amplitude $1.0$ and the other of peak amplitude $0.5$) and then a square pulse for measurement. The dictionaries for the different sequences of the experiment are,
\begin{jsonexamplebox}{Rabi Oscillation - Time Sequence}
\begin{Verbatim}[commandchars=\\\{\}]
\PYG{p}{[}
\PYG{p}{\PYGZob{}}\PYG{n+nt}{\PYGZdq{}header\PYGZdq{}}\PYG{p}{:} \PYG{p}{\PYGZob{}}\PYG{n+nt}{\PYGZdq{}name\PYGZdq{}}\PYG{p}{:} \PYG{l+s+s2}{\PYGZdq{}Amplitude 0\PYGZdq{}}\PYG{p}{\PYGZcb{},}
  \PYG{n+nt}{\PYGZdq{}instructions\PYGZdq{}}\PYG{p}{:[}
       \PYG{p}{\PYGZob{}}\PYG{n+nt}{\PYGZdq{}name\PYGZdq{}}\PYG{p}{:} \PYG{l+s+s2}{\PYGZdq{}square\PYGZus{}pulse\PYGZdq{}}\PYG{p}{,} \PYG{n+nt}{\PYGZdq{}t0\PYGZdq{}}\PYG{p}{:} \PYG{l+m+mi}{12}\PYG{p}{,} \PYG{n+nt}{\PYGZdq{}ch\PYGZdq{}}\PYG{p}{:} \PYG{l+s+s2}{\PYGZdq{}m0\PYGZdq{}}\PYG{p}{\PYGZcb{},}
       \PYG{p}{\PYGZob{}}\PYG{n+nt}{\PYGZdq{}name\PYGZdq{}}\PYG{p}{:} \PYG{l+s+s2}{\PYGZdq{}acquire\PYGZdq{}}\PYG{p}{,} \PYG{n+nt}{\PYGZdq{}t0\PYGZdq{}}\PYG{p}{:} \PYG{l+m+mi}{12}\PYG{p}{,} \PYG{n+nt}{\PYGZdq{}duration\PYGZdq{}}\PYG{p}{:} \PYG{l+m+mi}{6}\PYG{p}{,}
        \PYG{n+nt}{\PYGZdq{}qubits\PYGZdq{}}\PYG{p}{:} \PYG{p}{[}\PYG{l+m+mi}{0}\PYG{p}{],} \PYG{n+nt}{\PYGZdq{}memory\PYGZus{}slot\PYGZdq{}}\PYG{p}{:} \PYG{p}{[}\PYG{l+m+mi}{0}\PYG{p}{]\PYGZcb{}]\PYGZcb{},}
\PYG{p}{\PYGZob{}}\PYG{n+nt}{\PYGZdq{}header\PYGZdq{}}\PYG{p}{:} \PYG{p}{\PYGZob{}}\PYG{n+nt}{\PYGZdq{}name\PYGZdq{}}\PYG{p}{:} \PYG{l+s+s2}{\PYGZdq{}Amplitude 0.5\PYGZdq{}}\PYG{p}{\PYGZcb{},}
  \PYG{n+nt}{\PYGZdq{}instructions\PYGZdq{}}\PYG{p}{:[\PYGZob{}}\PYG{n+nt}{\PYGZdq{}name\PYGZdq{}}\PYG{p}{:} \PYG{l+s+s2}{\PYGZdq{}pulse1\PYGZdq{}}\PYG{p}{,} \PYG{n+nt}{\PYGZdq{}t0\PYGZdq{}}\PYG{p}{:} \PYG{l+m+mi}{0}\PYG{p}{,} \PYG{n+nt}{\PYGZdq{}ch\PYGZdq{}}\PYG{p}{:} \PYG{l+s+s2}{\PYGZdq{}d0\PYGZdq{}}\PYG{p}{\PYGZcb{},}
       \PYG{p}{\PYGZob{}}\PYG{n+nt}{\PYGZdq{}name\PYGZdq{}}\PYG{p}{:} \PYG{l+s+s2}{\PYGZdq{}square\PYGZus{}pulse\PYGZdq{}}\PYG{p}{,} \PYG{n+nt}{\PYGZdq{}t0\PYGZdq{}}\PYG{p}{:} \PYG{l+m+mi}{12}\PYG{p}{,} \PYG{n+nt}{\PYGZdq{}ch\PYGZdq{}}\PYG{p}{:} \PYG{l+s+s2}{\PYGZdq{}m0\PYGZdq{}}\PYG{p}{\PYGZcb{},}
       \PYG{p}{\PYGZob{}}\PYG{n+nt}{\PYGZdq{}name\PYGZdq{}}\PYG{p}{:} \PYG{l+s+s2}{\PYGZdq{}acquire\PYGZdq{}}\PYG{p}{,} \PYG{n+nt}{\PYGZdq{}t0\PYGZdq{}}\PYG{p}{:} \PYG{l+m+mi}{12}\PYG{p}{,} \PYG{n+nt}{\PYGZdq{}duration\PYGZdq{}}\PYG{p}{:} \PYG{l+m+mi}{6}\PYG{p}{,}
        \PYG{n+nt}{\PYGZdq{}qubits\PYGZdq{}}\PYG{p}{:} \PYG{p}{[}\PYG{l+m+mi}{0}\PYG{p}{],} \PYG{n+nt}{\PYGZdq{}memory\PYGZus{}slot\PYGZdq{}}\PYG{p}{:} \PYG{p}{[}\PYG{l+m+mi}{0}\PYG{p}{]\PYGZcb{}]\PYGZcb{},}
\PYG{p}{\PYGZob{}}\PYG{n+nt}{\PYGZdq{}header\PYGZdq{}}\PYG{p}{:} \PYG{p}{\PYGZob{}}\PYG{n+nt}{\PYGZdq{}name\PYGZdq{}}\PYG{p}{:} \PYG{l+s+s2}{\PYGZdq{}Amplitude 1.0\PYGZdq{}}\PYG{p}{\PYGZcb{},}
  \PYG{n+nt}{\PYGZdq{}instructions\PYGZdq{}}\PYG{p}{:} \PYG{p}{[\PYGZob{}}\PYG{n+nt}{\PYGZdq{}name\PYGZdq{}}\PYG{p}{:} \PYG{l+s+s2}{\PYGZdq{}pulse2\PYGZdq{}}\PYG{p}{,} \PYG{n+nt}{\PYGZdq{}t0\PYGZdq{}}\PYG{p}{:} \PYG{l+m+mi}{0}\PYG{p}{,} \PYG{n+nt}{\PYGZdq{}ch\PYGZdq{}}\PYG{p}{:} \PYG{l+s+s2}{\PYGZdq{}d0\PYGZdq{}}\PYG{p}{\PYGZcb{},}
       \PYG{p}{\PYGZob{}}\PYG{n+nt}{\PYGZdq{}name\PYGZdq{}}\PYG{p}{:} \PYG{l+s+s2}{\PYGZdq{}square\PYGZus{}pulse\PYGZdq{}}\PYG{p}{,} \PYG{n+nt}{\PYGZdq{}t0\PYGZdq{}}\PYG{p}{:} \PYG{l+m+mi}{12}\PYG{p}{,} \PYG{n+nt}{\PYGZdq{}ch\PYGZdq{}}\PYG{p}{:} \PYG{l+s+s2}{\PYGZdq{}m0\PYGZdq{}}\PYG{p}{\PYGZcb{},}
       \PYG{p}{\PYGZob{}}\PYG{n+nt}{\PYGZdq{}name\PYGZdq{}}\PYG{p}{:} \PYG{l+s+s2}{\PYGZdq{}acquire\PYGZdq{}}\PYG{p}{,} \PYG{n+nt}{\PYGZdq{}t0\PYGZdq{}}\PYG{p}{:} \PYG{l+m+mi}{12}\PYG{p}{,} \PYG{n+nt}{\PYGZdq{}duration\PYGZdq{}}\PYG{p}{:} \PYG{l+m+mi}{6}\PYG{p}{,}
        \PYG{n+nt}{\PYGZdq{}qubits\PYGZdq{}}\PYG{p}{:} \PYG{p}{[}\PYG{l+m+mi}{0}\PYG{p}{],} \PYG{n+nt}{\PYGZdq{}memory\PYGZus{}slot\PYGZdq{}}\PYG{p}{:} \PYG{p}{[}\PYG{l+m+mi}{0}\PYG{p}{]\PYGZcb{}]\PYGZcb{}}
\PYG{p}{]}
\end{Verbatim}
\end{jsonexamplebox}
where the acquisition commands do not specify the kernel or discriminator so the defaults are used.

\subsubsection{Level 0 Measurement}

Here we will run the experiment and retrieve the measurements as a level 0 averaged measurement. The run dictionary is,
\begin{jsonexamplebox}{Rabi Oscillation - Level 0 Measurement}
\begin{Verbatim}[commandchars=\\\{\}]
\PYG{p}{\PYGZob{}}
  \PYG{n+nt}{\PYGZdq{}qobj\PYGZus{}id\PYGZdq{}}\PYG{p}{:} \PYG{l+s+s2}{\PYGZdq{}Qobj\PYGZus{}sample\PYGZus{}test\PYGZus{}0726\PYGZdq{}}\PYG{p}{,}
  \PYG{n+nt}{\PYGZdq{}schema\PYGZus{}version\PYGZdq{}}\PYG{p}{:} \PYG{l+s+s2}{\PYGZdq{}1.0.0\PYGZdq{}}\PYG{p}{,}
  \PYG{n+nt}{\PYGZdq{}type\PYGZdq{}}\PYG{p}{:} \PYG{l+s+s2}{\PYGZdq{}PULSE\PYGZdq{}}\PYG{p}{,}
  \PYG{n+nt}{\PYGZdq{}header\PYGZdq{}}\PYG{p}{:} \PYG{p}{\PYGZob{}\PYGZcb{},}
	\PYG{n+nt}{\PYGZdq{}experiments\PYGZdq{}}\PYG{p}{:} \PYG{p}{[}\PYG{err}{exp1}\PYG{p}{,} \PYG{err}{exp2}\PYG{p}{,} \PYG{err}{exp3}\PYG{p}{],}
	\PYG{n+nt}{\PYGZdq{}config\PYGZdq{}}\PYG{p}{:} \PYG{p}{\PYGZob{}}
    \PYG{n+nt}{\PYGZdq{}meas\PYGZus{}level\PYGZdq{}}\PYG{p}{:} \PYG{l+m+mi}{0}\PYG{p}{,}
    \PYG{n+nt}{\PYGZdq{}pulse\PYGZus{}library\PYGZdq{}}\PYG{p}{:} \PYG{err}{pulse\PYGZus{}lib}\PYG{p}{,}
    \PYG{n+nt}{\PYGZdq{}memory\PYGZus{}slots\PYGZdq{}}\PYG{p}{:} \PYG{l+m+mi}{1}\PYG{p}{,}
  	\PYG{n+nt}{\PYGZdq{}memory\PYGZus{}slot\PYGZus{}size\PYGZdq{}}\PYG{p}{:} \PYG{l+m+mi}{6}\PYG{p}{,}
  	\PYG{n+nt}{\PYGZdq{}meas\PYGZus{}return\PYGZdq{}}\PYG{p}{:} \PYG{l+s+s2}{\PYGZdq{}avg\PYGZdq{}}\PYG{p}{,}
  	\PYG{n+nt}{\PYGZdq{}qubit\PYGZus{}lo\PYGZus{}freq\PYGZdq{}}\PYG{p}{:} \PYG{p}{[}\PYG{l+m+mf}{5.0}\PYG{p}{],}
  	\PYG{n+nt}{\PYGZdq{}meas\PYGZus{}lo\PYGZus{}freq\PYGZdq{}}\PYG{p}{:} \PYG{p}{[}\PYG{l+m+mf}{6.5}\PYG{p}{],}
  	\PYG{n+nt}{\PYGZdq{}rep\PYGZus{}time\PYGZdq{}}\PYG{p}{:} \PYG{l+m+mi}{1000}\PYG{p}{,}
  	\PYG{n+nt}{\PYGZdq{}shots\PYGZdq{}}\PYG{p}{:} \PYG{l+m+mi}{5}
  \PYG{p}{\PYGZcb{}}
\PYG{p}{\PYGZcb{}}
\end{Verbatim}
\end{jsonexamplebox}

We start the experiment with \texttt{job=Backend.run(Qobj)}. We then call \texttt{Job.result()} to get the measurement output. For simplicity we assume that the measurement acquisition is the measurement stimulus pulse with a state dependent phase (0 phase for qubit in 0 and $\pi/2$ for a qubit in 1). The measurement dictionary is:
\begin{jsonexamplebox}{Rabi Oscillation - Level 0 Measurement Result}
\begin{Verbatim}[commandchars=\\\{\}]
\PYG{p}{\PYGZob{}}
  \PYG{n+nt}{\PYGZdq{}backend\PYGZus{}name\PYGZdq{}}\PYG{p}{:} \PYG{l+s+s2}{\PYGZdq{}ibmqx1\PYGZdq{}}\PYG{p}{,}
  \PYG{n+nt}{\PYGZdq{}backend\PYGZus{}version\PYGZdq{}}\PYG{p}{:} \PYG{l+s+s2}{\PYGZdq{}1.1.5\PYGZdq{}}\PYG{p}{,}
  \PYG{n+nt}{\PYGZdq{}job\PYGZus{}id\PYGZdq{}}\PYG{p}{:} \PYG{l+s+s2}{\PYGZdq{}XYFSKO!123\PYGZdq{}}\PYG{p}{,}
  \PYG{n+nt}{\PYGZdq{}qobj\PYGZus{}id\PYGZdq{}}\PYG{p}{:} \PYG{l+s+s2}{\PYGZdq{}Qobj\PYGZus{}sample\PYGZus{}test\PYGZus{}0726\PYGZdq{}}\PYG{p}{,}
  \PYG{n+nt}{\PYGZdq{}date\PYGZdq{}}\PYG{p}{:} \PYG{l+s+s2}{\PYGZdq{}2018\PYGZhy{}04\PYGZhy{}02 15:00:00Z\PYGZdq{}}\PYG{p}{,}
  \PYG{n+nt}{\PYGZdq{}header\PYGZdq{}}\PYG{p}{:} \PYG{p}{\PYGZob{}\PYGZcb{},}
  \PYG{n+nt}{\PYGZdq{}success\PYGZdq{}}\PYG{p}{:} \PYG{k+kc}{true}\PYG{p}{,}
  \PYG{n+nt}{\PYGZdq{}results\PYGZdq{}}\PYG{p}{:} \PYG{p}{[}
     \PYG{p}{\PYGZob{}}\PYG{n+nt}{\PYGZdq{}shots\PYGZdq{}}\PYG{p}{:} \PYG{l+m+mi}{5}\PYG{p}{,}
      \PYG{n+nt}{\PYGZdq{}success\PYGZdq{}}\PYG{p}{:} \PYG{k+kc}{true}\PYG{p}{,}
		  \PYG{n+nt}{\PYGZdq{}status\PYGZdq{}}\PYG{p}{:} \PYG{l+s+s2}{\PYGZdq{}DONE\PYGZdq{}}\PYG{p}{,}
		  \PYG{n+nt}{\PYGZdq{}header\PYGZdq{}}\PYG{p}{:} \PYG{p}{\PYGZob{}}\PYG{n+nt}{\PYGZdq{}name\PYGZdq{}}\PYG{p}{:} \PYG{l+s+s2}{\PYGZdq{}Amplitude 0\PYGZdq{}}\PYG{p}{\PYGZcb{},}
      \PYG{n+nt}{\PYGZdq{}meas\PYGZus{}return\PYGZdq{}}\PYG{p}{:} \PYG{l+s+s2}{\PYGZdq{}avg\PYGZdq{}}\PYG{p}{,}
		  \PYG{n+nt}{\PYGZdq{}data\PYGZdq{}}\PYG{p}{:} \PYG{p}{\PYGZob{}}\PYG{n+nt}{\PYGZdq{}memory\PYGZdq{}}\PYG{p}{:[}
                \PYG{p}{[[}\PYG{l+m+mf}{0.1}\PYG{p}{,}\PYG{l+m+mf}{0.0}\PYG{p}{],[}\PYG{l+m+mf}{0.1}\PYG{p}{,}\PYG{l+m+mf}{0.0}\PYG{p}{],}
                 \PYG{p}{[}\PYG{l+m+mf}{0.1}\PYG{p}{,}\PYG{l+m+mf}{0.0}\PYG{p}{],[}\PYG{l+m+mf}{0.1}\PYG{p}{,}\PYG{l+m+mf}{0.0}\PYG{p}{],}
                 \PYG{p}{[}\PYG{l+m+mf}{0.1}\PYG{p}{,}\PYG{l+m+mf}{0.0}\PYG{p}{],[}\PYG{l+m+mf}{0.1}\PYG{p}{,}\PYG{l+m+mf}{0.0}\PYG{p}{]]]\PYGZcb{}\PYGZcb{},}
	   \PYG{p}{\PYGZob{}}\PYG{n+nt}{\PYGZdq{}shots\PYGZdq{}}\PYG{p}{:} \PYG{l+m+mi}{5}\PYG{p}{,}
      \PYG{n+nt}{\PYGZdq{}success\PYGZdq{}}\PYG{p}{:} \PYG{k+kc}{true}\PYG{p}{,}
      \PYG{n+nt}{\PYGZdq{}status\PYGZdq{}}\PYG{p}{:} \PYG{l+s+s2}{\PYGZdq{}DONE\PYGZdq{}}\PYG{p}{,}
		  \PYG{n+nt}{\PYGZdq{}header\PYGZdq{}}\PYG{p}{:} \PYG{p}{\PYGZob{}}\PYG{n+nt}{\PYGZdq{}name\PYGZdq{}}\PYG{p}{:} \PYG{l+s+s2}{\PYGZdq{}Amplitude 0.5\PYGZdq{}}\PYG{p}{\PYGZcb{},}
      \PYG{n+nt}{\PYGZdq{}meas\PYGZus{}return\PYGZdq{}}\PYG{p}{:} \PYG{l+s+s2}{\PYGZdq{}avg\PYGZdq{}}\PYG{p}{,}
	    \PYG{n+nt}{\PYGZdq{}data\PYGZdq{}}\PYG{p}{:} \PYG{p}{\PYGZob{}}\PYG{n+nt}{\PYGZdq{}memory\PYGZdq{}}\PYG{p}{:} \PYG{p}{[}
              \PYG{p}{[[}\PYG{l+m+mf}{0.07}\PYG{p}{,}\PYG{l+m+mf}{0.07}\PYG{p}{],[}\PYG{l+m+mf}{0.07}\PYG{p}{,}\PYG{l+m+mf}{0.07}\PYG{p}{],}
               \PYG{p}{[}\PYG{l+m+mf}{0.07}\PYG{p}{,}\PYG{l+m+mf}{0.07}\PYG{p}{],[}\PYG{l+m+mf}{0.07}\PYG{p}{,}\PYG{l+m+mf}{0.07}\PYG{p}{],}
               \PYG{p}{[}\PYG{l+m+mf}{0.07}\PYG{p}{,}\PYG{l+m+mf}{0.07}\PYG{p}{],[}\PYG{l+m+mf}{0.07}\PYG{p}{,}\PYG{l+m+mf}{0.07}\PYG{p}{]]]\PYGZcb{}\PYGZcb{},}
	   \PYG{p}{\PYGZob{}}\PYG{n+nt}{\PYGZdq{}shots\PYGZdq{}}\PYG{p}{:} \PYG{l+m+mi}{5}\PYG{p}{,}
      \PYG{n+nt}{\PYGZdq{}success\PYGZdq{}}\PYG{p}{:} \PYG{k+kc}{true}\PYG{p}{,}
      \PYG{n+nt}{\PYGZdq{}status\PYGZdq{}}\PYG{p}{:} \PYG{l+s+s2}{\PYGZdq{}DONE\PYGZdq{}}\PYG{p}{,}
		  \PYG{n+nt}{\PYGZdq{}header\PYGZdq{}}\PYG{p}{:} \PYG{p}{\PYGZob{}}\PYG{n+nt}{\PYGZdq{}name\PYGZdq{}}\PYG{p}{:} \PYG{l+s+s2}{\PYGZdq{}Amplitude 1.0\PYGZdq{}}\PYG{p}{\PYGZcb{},}
      \PYG{n+nt}{\PYGZdq{}meas\PYGZus{}return\PYGZdq{}}\PYG{p}{:} \PYG{l+s+s2}{\PYGZdq{}avg\PYGZdq{}}\PYG{p}{,}
	    \PYG{n+nt}{\PYGZdq{}data\PYGZdq{}}\PYG{p}{:} \PYG{p}{\PYGZob{}}\PYG{n+nt}{\PYGZdq{}memory\PYGZdq{}}\PYG{p}{:} \PYG{p}{[}
              \PYG{p}{[[}\PYG{l+m+mf}{0.0}\PYG{p}{,}\PYG{l+m+mf}{0.1}\PYG{p}{],[}\PYG{l+m+mf}{0.0}\PYG{p}{,}\PYG{l+m+mf}{0.1}\PYG{p}{],}
               \PYG{p}{[}\PYG{l+m+mf}{0.0}\PYG{p}{,}\PYG{l+m+mf}{0.1}\PYG{p}{],[}\PYG{l+m+mf}{0.0}\PYG{p}{,}\PYG{l+m+mf}{0.1}\PYG{p}{],}
               \PYG{p}{[}\PYG{l+m+mf}{0.0}\PYG{p}{,}\PYG{l+m+mf}{0.1}\PYG{p}{],[}\PYG{l+m+mf}{0.0}\PYG{p}{,}\PYG{l+m+mf}{0.1}\PYG{p}{]]]\PYGZcb{}\PYGZcb{}]}
  \PYG{p}{\PYGZcb{}}
\end{Verbatim}
\end{jsonexamplebox}

\subsubsection{Level 1 Measurement}

Here we will run the experiment and retrieve the measurements as a level 1 single shot measurement. The run dictionary is,
\begin{jsonexamplebox}{Rabi Oscillation - Level 1 Measurement}
\begin{Verbatim}[commandchars=\\\{\}]
\PYG{p}{\PYGZob{}}
	\PYG{n+nt}{\PYGZdq{}qobj\PYGZus{}id\PYGZdq{}}\PYG{p}{:} \PYG{l+s+s2}{\PYGZdq{}Qobj\PYGZus{}sample\PYGZus{}test\PYGZus{}0726\PYGZdq{}}\PYG{p}{,}
	\PYG{n+nt}{\PYGZdq{}schema\PYGZus{}version\PYGZdq{}}\PYG{p}{:} \PYG{l+s+s2}{\PYGZdq{}1.0.0\PYGZdq{}}\PYG{p}{,}
	\PYG{n+nt}{\PYGZdq{}type\PYGZdq{}}\PYG{p}{:} \PYG{l+s+s2}{\PYGZdq{}PULSE\PYGZdq{}}\PYG{p}{,}
	\PYG{n+nt}{\PYGZdq{}header\PYGZdq{}}\PYG{p}{:} \PYG{p}{\PYGZob{}\PYGZcb{},}
	\PYG{n+nt}{\PYGZdq{}experiments\PYGZdq{}}\PYG{p}{:} \PYG{p}{[}\PYG{err}{exp1}\PYG{p}{,} \PYG{err}{exp2}\PYG{p}{,} \PYG{err}{exp3}\PYG{p}{],}
	\PYG{n+nt}{\PYGZdq{}config\PYGZdq{}}\PYG{p}{:} \PYG{p}{\PYGZob{}}
		\PYG{n+nt}{\PYGZdq{}meas\PYGZus{}level\PYGZdq{}}\PYG{p}{:} \PYG{l+m+mi}{1}\PYG{p}{,}
	  \PYG{n+nt}{\PYGZdq{}pulse\PYGZus{}library\PYGZdq{}}\PYG{p}{:} \PYG{err}{pulse\PYGZus{}lib}\PYG{p}{,}
		\PYG{n+nt}{\PYGZdq{}memory\PYGZus{}slots\PYGZdq{}}\PYG{p}{:} \PYG{l+m+mi}{1}\PYG{p}{,}
		\PYG{n+nt}{\PYGZdq{}memory\PYGZus{}slot\PYGZus{}size\PYGZdq{}}\PYG{p}{:} \PYG{l+m+mi}{6}\PYG{p}{,}
		\PYG{n+nt}{\PYGZdq{}meas\PYGZus{}return\PYGZdq{}}\PYG{p}{:} \PYG{l+s+s2}{\PYGZdq{}single\PYGZdq{}}\PYG{p}{,}
		\PYG{n+nt}{\PYGZdq{}qubit\PYGZus{}lo\PYGZus{}freq\PYGZdq{}}\PYG{p}{:} \PYG{p}{[}\PYG{l+m+mf}{5.0}\PYG{p}{],}
		\PYG{n+nt}{\PYGZdq{}meas\PYGZus{}lo\PYGZus{}freq\PYGZdq{}}\PYG{p}{:} \PYG{p}{[}\PYG{l+m+mf}{6.5}\PYG{p}{],}
		\PYG{n+nt}{\PYGZdq{}rep\PYGZus{}time\PYGZdq{}}\PYG{p}{:} \PYG{l+m+mi}{1000}\PYG{p}{,}
		\PYG{n+nt}{\PYGZdq{}shots\PYGZdq{}}\PYG{p}{:} \PYG{l+m+mi}{5}
						\PYG{p}{\PYGZcb{}}
\PYG{p}{\PYGZcb{}}
\end{Verbatim}
\end{jsonexamplebox}

We start the experiment with \texttt{job=Backend.run(Qobj)}. We then call \texttt{Job.result} to get the measurement output. For simplicity we assume that the measurement acquisition is the measurement stimulus pulse with a state dependent phase (0 phase for qubit in 0 and $\pi/2$ for a qubit in 1). The measurement dictionary is:
\begin{jsonexamplebox}{Rabi Oscillation - Level 1 Measurement Results}
\begin{Verbatim}[commandchars=\\\{\}]
\PYG{p}{\PYGZob{}}
\PYG{n+nt}{\PYGZdq{}backend\PYGZus{}name\PYGZdq{}}\PYG{p}{:} \PYG{l+s+s2}{\PYGZdq{}ibmqx1\PYGZdq{}}\PYG{p}{,}
\PYG{n+nt}{\PYGZdq{}backend\PYGZus{}version\PYGZdq{}}\PYG{p}{:} \PYG{l+s+s2}{\PYGZdq{}1.1.5\PYGZdq{}}\PYG{p}{,}
\PYG{n+nt}{\PYGZdq{}job\PYGZus{}id\PYGZdq{}}\PYG{p}{:} \PYG{l+s+s2}{\PYGZdq{}XYFSKO!123\PYGZdq{}}\PYG{p}{,}
\PYG{n+nt}{\PYGZdq{}qobj\PYGZus{}id\PYGZdq{}}\PYG{p}{:} \PYG{l+s+s2}{\PYGZdq{}Qobj\PYGZus{}sample\PYGZus{}test\PYGZus{}0726\PYGZdq{}}\PYG{p}{,}
\PYG{n+nt}{\PYGZdq{}date\PYGZdq{}}\PYG{p}{:} \PYG{l+s+s2}{\PYGZdq{}2018\PYGZhy{}04\PYGZhy{}02 15:00:00Z\PYGZdq{}}\PYG{p}{,}
\PYG{n+nt}{\PYGZdq{}header\PYGZdq{}}\PYG{p}{:} \PYG{p}{\PYGZob{}\PYGZcb{},}
\PYG{n+nt}{\PYGZdq{}success\PYGZdq{}}\PYG{p}{:} \PYG{k+kc}{true}\PYG{p}{,}
\PYG{n+nt}{\PYGZdq{}results\PYGZdq{}}\PYG{p}{:} \PYG{p}{[}
  \PYG{p}{\PYGZob{}}\PYG{n+nt}{\PYGZdq{}shots\PYGZdq{}}\PYG{p}{:} \PYG{l+m+mi}{5}\PYG{p}{,}
   \PYG{n+nt}{\PYGZdq{}success\PYGZdq{}}\PYG{p}{:} \PYG{k+kc}{true}\PYG{p}{,}
   \PYG{n+nt}{\PYGZdq{}status\PYGZdq{}}\PYG{p}{:} \PYG{l+s+s2}{\PYGZdq{}DONE\PYGZdq{}}\PYG{p}{,}
	 \PYG{n+nt}{\PYGZdq{}header\PYGZdq{}}\PYG{p}{:} \PYG{p}{\PYGZob{}}\PYG{n+nt}{\PYGZdq{}name\PYGZdq{}}\PYG{p}{:} \PYG{l+s+s2}{\PYGZdq{}Amplitude 0\PYGZdq{}}\PYG{p}{\PYGZcb{},}
	 \PYG{n+nt}{\PYGZdq{}data\PYGZdq{}}\PYG{p}{:} \PYG{p}{\PYGZob{}}
     \PYG{n+nt}{\PYGZdq{}memory\PYGZdq{}}\PYG{p}{:} \PYG{p}{[[[}\PYG{l+m+mf}{0.1}\PYG{p}{,}\PYG{l+m+mf}{0.05}\PYG{p}{]],}
                \PYG{p}{[[}\PYG{l+m+mf}{0.11}\PYG{p}{,}\PYG{l+m+mf}{\PYGZhy{}0.05}\PYG{p}{]],[[}\PYG{l+m+mf}{0.09}\PYG{p}{,}\PYG{l+m+mf}{0.02}\PYG{p}{]],}
                \PYG{p}{[[}\PYG{l+m+mf}{0.095}\PYG{p}{,}\PYG{l+m+mf}{0.01}\PYG{p}{]],[[}\PYG{l+m+mf}{0.105}\PYG{p}{,}\PYG{l+m+mf}{\PYGZhy{}0.03}\PYG{p}{]]]\PYGZcb{}}
        \PYG{p}{\PYGZcb{},}
  \PYG{p}{\PYGZob{}}\PYG{n+nt}{\PYGZdq{}shots\PYGZdq{}}\PYG{p}{:} \PYG{l+m+mi}{5}\PYG{p}{,}
   \PYG{n+nt}{\PYGZdq{}success\PYGZdq{}}\PYG{p}{:} \PYG{k+kc}{true}\PYG{p}{,}
   \PYG{n+nt}{\PYGZdq{}status\PYGZdq{}}\PYG{p}{:} \PYG{l+s+s2}{\PYGZdq{}DONE\PYGZdq{}}\PYG{p}{,}
	 \PYG{n+nt}{\PYGZdq{}header\PYGZdq{}}\PYG{p}{:} \PYG{p}{\PYGZob{}}\PYG{n+nt}{\PYGZdq{}name\PYGZdq{}}\PYG{p}{:} \PYG{l+s+s2}{\PYGZdq{}Amplitude 0\PYGZdq{}}\PYG{p}{\PYGZcb{},}
	 \PYG{n+nt}{\PYGZdq{}data\PYGZdq{}}\PYG{p}{:} \PYG{p}{\PYGZob{}}
     \PYG{n+nt}{\PYGZdq{}memory\PYGZdq{}}\PYG{p}{:} \PYG{p}{[[[}\PYG{l+m+mf}{0.08}\PYG{p}{,}\PYG{l+m+mf}{0.075}\PYG{p}{]],}
                \PYG{p}{[[}\PYG{l+m+mf}{0.06}\PYG{p}{,}\PYG{l+m+mf}{0.075}\PYG{p}{]],[[}\PYG{l+m+mf}{0.07}\PYG{p}{,}\PYG{l+m+mf}{0.06}\PYG{p}{]],}
                \PYG{p}{[[}\PYG{l+m+mf}{0.05}\PYG{p}{,}\PYG{l+m+mf}{0.08}\PYG{p}{]],[[}\PYG{l+m+mf}{0.09}\PYG{p}{,}\PYG{l+m+mf}{0.05}\PYG{p}{]]]\PYGZcb{}\PYGZcb{},}
  \PYG{p}{\PYGZob{}}\PYG{n+nt}{\PYGZdq{}shots\PYGZdq{}}\PYG{p}{:} \PYG{l+m+mi}{5}\PYG{p}{,}
   \PYG{n+nt}{\PYGZdq{}success\PYGZdq{}}\PYG{p}{:} \PYG{k+kc}{true}\PYG{p}{,}
   \PYG{n+nt}{\PYGZdq{}status\PYGZdq{}}\PYG{p}{:} \PYG{l+s+s2}{\PYGZdq{}DONE\PYGZdq{}}\PYG{p}{,}
	 \PYG{n+nt}{\PYGZdq{}header\PYGZdq{}}\PYG{p}{:} \PYG{p}{\PYGZob{}}\PYG{n+nt}{\PYGZdq{}name\PYGZdq{}}\PYG{p}{:} \PYG{l+s+s2}{\PYGZdq{}Amplitude 0\PYGZdq{}}\PYG{p}{\PYGZcb{},}
	 \PYG{n+nt}{\PYGZdq{}data\PYGZdq{}}\PYG{p}{:} \PYG{p}{\PYGZob{}}
     \PYG{n+nt}{\PYGZdq{}memory\PYGZdq{}}\PYG{p}{:} \PYG{p}{[[[}\PYG{l+m+mf}{0.01}\PYG{p}{,}\PYG{l+m+mf}{0.11}\PYG{p}{]],}
                \PYG{p}{[[}\PYG{l+m+mf}{0.01}\PYG{p}{,}\PYG{l+m+mf}{0.08}\PYG{p}{]],[[}\PYG{l+m+mf}{0.01}\PYG{p}{,}\PYG{l+m+mf}{0.09}\PYG{p}{]],}
                \PYG{p}{[[}\PYG{l+m+mf}{\PYGZhy{}0.03}\PYG{p}{,}\PYG{l+m+mf}{0.12}\PYG{p}{]],[[}\PYG{l+m+mf}{0.0}\PYG{p}{,}\PYG{l+m+mf}{0.1}\PYG{p}{]]]\PYGZcb{}\PYGZcb{}}
       \PYG{p}{]}
\PYG{p}{\PYGZcb{}}
\end{Verbatim}
\end{jsonexamplebox}

\subsubsection{Level 2 Measurement}

Here we will run the experiment and retrieve the measurements as a level 2 measurement. The run dictionary is,

\begin{jsonexamplebox}{Rabi Oscillation - Level 2 Measurement}
\begin{Verbatim}[commandchars=\\\{\}]
\PYG{p}{\PYGZob{}}
	\PYG{n+nt}{\PYGZdq{}qobj\PYGZus{}id\PYGZdq{}}\PYG{p}{:} \PYG{l+s+s2}{\PYGZdq{}Qobj\PYGZus{}sample\PYGZus{}test\PYGZus{}0726\PYGZdq{}}\PYG{p}{,}
	\PYG{n+nt}{\PYGZdq{}schema\PYGZus{}version\PYGZdq{}}\PYG{p}{:} \PYG{l+s+s2}{\PYGZdq{}1.0.0\PYGZdq{}}\PYG{p}{,}
	\PYG{n+nt}{\PYGZdq{}type\PYGZdq{}}\PYG{p}{:} \PYG{l+s+s2}{\PYGZdq{}PULSE\PYGZdq{}}\PYG{p}{,}
	\PYG{n+nt}{\PYGZdq{}header\PYGZdq{}}\PYG{p}{:} \PYG{p}{\PYGZob{}\PYGZcb{},}
	\PYG{n+nt}{\PYGZdq{}experiments\PYGZdq{}}\PYG{p}{:} \PYG{p}{[}\PYG{err}{exp1}\PYG{p}{,} \PYG{err}{exp2}\PYG{p}{,} \PYG{err}{exp3}\PYG{p}{],}
	\PYG{n+nt}{\PYGZdq{}config\PYGZdq{}}\PYG{p}{:} \PYG{p}{\PYGZob{}}
		\PYG{n+nt}{\PYGZdq{}meas\PYGZus{}level\PYGZdq{}}\PYG{p}{:} \PYG{l+m+mi}{2}\PYG{p}{,}
	  \PYG{n+nt}{\PYGZdq{}pulse\PYGZus{}library\PYGZdq{}}\PYG{p}{:} \PYG{err}{pulse\PYGZus{}lib}\PYG{p}{,}
		\PYG{n+nt}{\PYGZdq{}memory\PYGZus{}slots\PYGZdq{}}\PYG{p}{:} \PYG{l+m+mi}{1}\PYG{p}{,}
		\PYG{n+nt}{\PYGZdq{}memory\PYGZus{}slot\PYGZus{}size\PYGZdq{}}\PYG{p}{:} \PYG{l+m+mi}{6}\PYG{p}{,}
		\PYG{n+nt}{\PYGZdq{}meas\PYGZus{}return\PYGZdq{}}\PYG{p}{:} \PYG{l+s+s2}{\PYGZdq{}single\PYGZdq{}}\PYG{p}{,}
		\PYG{n+nt}{\PYGZdq{}qubit\PYGZus{}lo\PYGZus{}freq\PYGZdq{}}\PYG{p}{:} \PYG{p}{[}\PYG{l+m+mf}{5.0}\PYG{p}{],}
		\PYG{n+nt}{\PYGZdq{}meas\PYGZus{}lo\PYGZus{}freq\PYGZdq{}}\PYG{p}{:} \PYG{p}{[}\PYG{l+m+mf}{6.5}\PYG{p}{],}
		\PYG{n+nt}{\PYGZdq{}rep\PYGZus{}time\PYGZdq{}}\PYG{p}{:} \PYG{l+m+mi}{1000}\PYG{p}{,}
		\PYG{n+nt}{\PYGZdq{}shots\PYGZdq{}}\PYG{p}{:} \PYG{l+m+mi}{5}

	\PYG{p}{\PYGZcb{}}
\PYG{p}{\PYGZcb{}}
\end{Verbatim}
\end{jsonexamplebox}

We start the experiment with \texttt{job=Backend.run(Qobj)}. We then call \texttt{Job.result()} to get the measurement output. The measurement dictionary is:
\begin{jsonexamplebox}{Rabi Oscillation - Level 2 Measurement Results}
\begin{Verbatim}[commandchars=\\\{\}]
\PYG{p}{\PYGZob{}}
\PYG{n+nt}{\PYGZdq{}backend\PYGZus{}name\PYGZdq{}}\PYG{p}{:} \PYG{l+s+s2}{\PYGZdq{}ibmqx1\PYGZdq{}}\PYG{p}{,}
\PYG{n+nt}{\PYGZdq{}backend\PYGZus{}version\PYGZdq{}}\PYG{p}{:} \PYG{l+s+s2}{\PYGZdq{}1.1.5\PYGZdq{}}\PYG{p}{,}
\PYG{n+nt}{\PYGZdq{}job\PYGZus{}id\PYGZdq{}}\PYG{p}{:} \PYG{l+s+s2}{\PYGZdq{}XYJSXO!123\PYGZdq{}}\PYG{p}{,}
\PYG{n+nt}{\PYGZdq{}qobj\PYGZus{}id\PYGZdq{}}\PYG{p}{:} \PYG{l+s+s2}{\PYGZdq{}Qobj\PYGZus{}sample\PYGZus{}test\PYGZus{}0726\PYGZdq{}}\PYG{p}{,}
\PYG{n+nt}{\PYGZdq{}success\PYGZdq{}}\PYG{p}{:} \PYG{k+kc}{true}\PYG{p}{,}
\PYG{n+nt}{\PYGZdq{}header\PYGZdq{}}\PYG{p}{:\PYGZob{}\PYGZcb{},}
\PYG{n+nt}{\PYGZdq{}results\PYGZdq{}}\PYG{p}{:} \PYG{p}{[}
    \PYG{p}{\PYGZob{}}\PYG{n+nt}{\PYGZdq{}shots\PYGZdq{}}\PYG{p}{:} \PYG{l+m+mi}{5}\PYG{p}{,}
     \PYG{n+nt}{\PYGZdq{}success\PYGZdq{}}\PYG{p}{:} \PYG{k+kc}{true}\PYG{p}{,}
     \PYG{n+nt}{\PYGZdq{}status\PYGZdq{}}\PYG{p}{:} \PYG{l+s+s2}{\PYGZdq{}DONE\PYGZdq{}}\PYG{p}{,}
     \PYG{n+nt}{\PYGZdq{}data\PYGZdq{}}\PYG{p}{:} \PYG{p}{\PYGZob{}}\PYG{n+nt}{\PYGZdq{}counts\PYGZdq{}}\PYG{p}{:} \PYG{p}{\PYGZob{}}\PYG{n+nt}{\PYGZdq{}0x0\PYGZdq{}}\PYG{p}{:} \PYG{l+m+mi}{5}\PYG{p}{\PYGZcb{},}
              \PYG{n+nt}{\PYGZdq{}memory\PYGZdq{}}\PYG{p}{:} \PYG{p}{[}\PYG{l+s+s2}{\PYGZdq{}0x0\PYGZdq{}}\PYG{p}{,}\PYG{l+s+s2}{\PYGZdq{}0x0\PYGZdq{}}\PYG{p}{,}\PYG{l+s+s2}{\PYGZdq{}0x0\PYGZdq{}}\PYG{p}{,}\PYG{l+s+s2}{\PYGZdq{}0x0\PYGZdq{}}\PYG{p}{,}\PYG{l+s+s2}{\PYGZdq{}0x0\PYGZdq{}}\PYG{p}{]\PYGZcb{}\PYGZcb{},}
    \PYG{p}{\PYGZob{}}\PYG{n+nt}{\PYGZdq{}shots\PYGZdq{}}\PYG{p}{:} \PYG{l+m+mi}{5}\PYG{p}{,}
     \PYG{n+nt}{\PYGZdq{}success\PYGZdq{}}\PYG{p}{:} \PYG{k+kc}{true}\PYG{p}{,}
     \PYG{n+nt}{\PYGZdq{}status\PYGZdq{}}\PYG{p}{:} \PYG{l+s+s2}{\PYGZdq{}DONE\PYGZdq{}}\PYG{p}{,}
     \PYG{n+nt}{\PYGZdq{}data\PYGZdq{}}\PYG{p}{:} \PYG{p}{\PYGZob{}}\PYG{n+nt}{\PYGZdq{}counts\PYGZdq{}}\PYG{p}{:} \PYG{p}{\PYGZob{}}\PYG{n+nt}{\PYGZdq{}0x0\PYGZdq{}}\PYG{p}{:} \PYG{l+m+mi}{2}\PYG{p}{,} \PYG{n+nt}{\PYGZdq{}0x1\PYGZdq{}}\PYG{p}{:} \PYG{l+m+mi}{3}\PYG{p}{\PYGZcb{},}
              \PYG{n+nt}{\PYGZdq{}memory\PYGZdq{}}\PYG{p}{:} \PYG{p}{[}\PYG{l+s+s2}{\PYGZdq{}0x0\PYGZdq{}}\PYG{p}{,}\PYG{l+s+s2}{\PYGZdq{}0x1\PYGZdq{}}\PYG{p}{,}\PYG{l+s+s2}{\PYGZdq{}0x1\PYGZdq{}}\PYG{p}{,}\PYG{l+s+s2}{\PYGZdq{}0x0\PYGZdq{}}\PYG{p}{,}\PYG{l+s+s2}{\PYGZdq{}0x1\PYGZdq{}}\PYG{p}{]\PYGZcb{}\PYGZcb{},}
    \PYG{p}{\PYGZob{}}\PYG{n+nt}{\PYGZdq{}shots\PYGZdq{}}\PYG{p}{:} \PYG{l+m+mi}{5}\PYG{p}{,}
     \PYG{n+nt}{\PYGZdq{}success\PYGZdq{}}\PYG{p}{:} \PYG{k+kc}{true}\PYG{p}{,}
     \PYG{n+nt}{\PYGZdq{}status\PYGZdq{}}\PYG{p}{:} \PYG{l+s+s2}{\PYGZdq{}DONE\PYGZdq{}}\PYG{p}{,}
     \PYG{n+nt}{\PYGZdq{}data\PYGZdq{}}\PYG{p}{:} \PYG{p}{\PYGZob{}}\PYG{n+nt}{\PYGZdq{}counts\PYGZdq{}}\PYG{p}{:} \PYG{p}{\PYGZob{}}\PYG{n+nt}{\PYGZdq{}0x1\PYGZdq{}}\PYG{p}{:} \PYG{l+m+mi}{5}\PYG{p}{\PYGZcb{},}
              \PYG{n+nt}{\PYGZdq{}memory\PYGZdq{}}\PYG{p}{:} \PYG{p}{[}\PYG{l+s+s2}{\PYGZdq{}0x1\PYGZdq{}}\PYG{p}{,}\PYG{l+s+s2}{\PYGZdq{}0x1\PYGZdq{}}\PYG{p}{,}\PYG{l+s+s2}{\PYGZdq{}0x1\PYGZdq{}}\PYG{p}{,}\PYG{l+s+s2}{\PYGZdq{}0x1\PYGZdq{}}\PYG{p}{,}\PYG{l+s+s2}{\PYGZdq{}0x1\PYGZdq{}}\PYG{p}{]\PYGZcb{}\PYGZcb{}}
           \PYG{p}{]}
\PYG{p}{\PYGZcb{}}
\end{Verbatim}
\end{jsonexamplebox}

\subsection{T1}

Here will give the Qobj for a T1 experiment with the measurement pulse aligned between experiments. Assume the pulse library as given in the Rabi experiment and that ``pulse2'' is a $\pi$ pulse. The different experiment dictionaries are:
\begin{jsonexamplebox}{T1 - Experiments}
\begin{Verbatim}[commandchars=\\\{\}]
\PYG{p}{[}
\PYG{p}{\PYGZob{}}\PYG{n+nt}{\PYGZdq{}header\PYGZdq{}}\PYG{p}{:} \PYG{p}{\PYGZob{}}\PYG{n+nt}{\PYGZdq{}name\PYGZdq{}}\PYG{p}{:} \PYG{l+s+s2}{\PYGZdq{}G Cal\PYGZdq{}}\PYG{p}{\PYGZcb{},} \PYG{n+nt}{\PYGZdq{}instructions\PYGZdq{}}\PYG{p}{:}
  \PYG{p}{[\PYGZob{}}\PYG{n+nt}{\PYGZdq{}name\PYGZdq{}}\PYG{p}{:} \PYG{l+s+s2}{\PYGZdq{}square\PYGZus{}pulse\PYGZdq{}}\PYG{p}{,} \PYG{n+nt}{\PYGZdq{}t0\PYGZdq{}}\PYG{p}{:} \PYG{l+m+mi}{200}\PYG{p}{,} \PYG{n+nt}{\PYGZdq{}ch\PYGZdq{}}\PYG{p}{:} \PYG{l+s+s2}{\PYGZdq{}m0\PYGZdq{}}\PYG{p}{\PYGZcb{},}
   \PYG{p}{\PYGZob{}}\PYG{n+nt}{\PYGZdq{}name\PYGZdq{}}\PYG{p}{:} \PYG{l+s+s2}{\PYGZdq{}acquire\PYGZdq{}}\PYG{p}{,} \PYG{n+nt}{\PYGZdq{}t0\PYGZdq{}}\PYG{p}{:} \PYG{l+m+mi}{200}\PYG{p}{,} \PYG{n+nt}{\PYGZdq{}duration\PYGZdq{}}\PYG{p}{:} \PYG{l+m+mi}{6}\PYG{p}{,}
                    \PYG{n+nt}{\PYGZdq{}qubits\PYGZdq{}}\PYG{p}{:} \PYG{p}{[}\PYG{l+m+mi}{0}\PYG{p}{],} \PYG{n+nt}{\PYGZdq{}memory\PYGZus{}slot\PYGZdq{}}\PYG{p}{:} \PYG{p}{[}\PYG{l+m+mi}{0}\PYG{p}{]\PYGZcb{}]\PYGZcb{},}
\PYG{p}{\PYGZob{}}\PYG{n+nt}{\PYGZdq{}header\PYGZdq{}}\PYG{p}{:} \PYG{p}{\PYGZob{}}\PYG{n+nt}{\PYGZdq{}name\PYGZdq{}}\PYG{p}{:} \PYG{l+s+s2}{\PYGZdq{}E Cal\PYGZdq{}}\PYG{p}{\PYGZcb{},} \PYG{n+nt}{\PYGZdq{}instructions\PYGZdq{}}\PYG{p}{:}
  \PYG{p}{[\PYGZob{}}\PYG{n+nt}{\PYGZdq{}name\PYGZdq{}}\PYG{p}{:} \PYG{l+s+s2}{\PYGZdq{}pulse2\PYGZdq{}}\PYG{p}{,} \PYG{n+nt}{\PYGZdq{}t0\PYGZdq{}}\PYG{p}{:} \PYG{l+m+mi}{190}\PYG{p}{,} \PYG{n+nt}{\PYGZdq{}ch\PYGZdq{}}\PYG{p}{:} \PYG{l+s+s2}{\PYGZdq{}d0\PYGZdq{}}\PYG{p}{\PYGZcb{},}
   \PYG{p}{\PYGZob{}}\PYG{n+nt}{\PYGZdq{}name\PYGZdq{}}\PYG{p}{:} \PYG{l+s+s2}{\PYGZdq{}square\PYGZus{}pulse\PYGZdq{}}\PYG{p}{,} \PYG{n+nt}{\PYGZdq{}t0\PYGZdq{}}\PYG{p}{:} \PYG{l+m+mi}{200}\PYG{p}{,} \PYG{n+nt}{\PYGZdq{}ch\PYGZdq{}}\PYG{p}{:} \PYG{l+s+s2}{\PYGZdq{}m0\PYGZdq{}}\PYG{p}{\PYGZcb{},}
   \PYG{p}{\PYGZob{}}\PYG{n+nt}{\PYGZdq{}name\PYGZdq{}}\PYG{p}{:} \PYG{l+s+s2}{\PYGZdq{}acquire\PYGZdq{}}\PYG{p}{,} \PYG{n+nt}{\PYGZdq{}t0\PYGZdq{}}\PYG{p}{:} \PYG{l+m+mi}{200}\PYG{p}{,} \PYG{n+nt}{\PYGZdq{}duration\PYGZdq{}}\PYG{p}{:} \PYG{l+m+mi}{6}\PYG{p}{,}
                    \PYG{n+nt}{\PYGZdq{}qubits\PYGZdq{}}\PYG{p}{:} \PYG{p}{[}\PYG{l+m+mi}{0}\PYG{p}{],} \PYG{n+nt}{\PYGZdq{}memory\PYGZus{}slot\PYGZdq{}}\PYG{p}{:} \PYG{p}{[}\PYG{l+m+mi}{0}\PYG{p}{]\PYGZcb{}]\PYGZcb{},}
\PYG{p}{\PYGZob{}}\PYG{n+nt}{\PYGZdq{}header\PYGZdq{}}\PYG{p}{:} \PYG{p}{\PYGZob{}}\PYG{n+nt}{\PYGZdq{}name\PYGZdq{}}\PYG{p}{:} \PYG{l+s+s2}{\PYGZdq{}Wait 1\PYGZdq{}}\PYG{p}{\PYGZcb{},} \PYG{n+nt}{\PYGZdq{}instructions\PYGZdq{}}\PYG{p}{:}
  \PYG{p}{[\PYGZob{}}\PYG{n+nt}{\PYGZdq{}name\PYGZdq{}}\PYG{p}{:} \PYG{l+s+s2}{\PYGZdq{}pulse2\PYGZdq{}}\PYG{p}{,} \PYG{n+nt}{\PYGZdq{}t0\PYGZdq{}}\PYG{p}{:} \PYG{l+m+mi}{190}\PYG{p}{,} \PYG{n+nt}{\PYGZdq{}ch\PYGZdq{}}\PYG{p}{:} \PYG{l+s+s2}{\PYGZdq{}d0\PYGZdq{}}\PYG{p}{\PYGZcb{},}
   \PYG{p}{\PYGZob{}}\PYG{n+nt}{\PYGZdq{}name\PYGZdq{}}\PYG{p}{:} \PYG{l+s+s2}{\PYGZdq{}square\PYGZus{}pulse\PYGZdq{}}\PYG{p}{,} \PYG{n+nt}{\PYGZdq{}t0\PYGZdq{}}\PYG{p}{:} \PYG{l+m+mi}{200}\PYG{p}{,} \PYG{n+nt}{\PYGZdq{}ch\PYGZdq{}}\PYG{p}{:} \PYG{l+s+s2}{\PYGZdq{}m0\PYGZdq{}}\PYG{p}{\PYGZcb{},}
   \PYG{p}{\PYGZob{}}\PYG{n+nt}{\PYGZdq{}name\PYGZdq{}}\PYG{p}{:} \PYG{l+s+s2}{\PYGZdq{}acquire\PYGZdq{}}\PYG{p}{,} \PYG{n+nt}{\PYGZdq{}t0\PYGZdq{}}\PYG{p}{:} \PYG{l+m+mi}{200}\PYG{p}{,} \PYG{n+nt}{\PYGZdq{}duration\PYGZdq{}}\PYG{p}{:} \PYG{l+m+mi}{6}\PYG{p}{,}
                    \PYG{n+nt}{\PYGZdq{}qubits\PYGZdq{}}\PYG{p}{:} \PYG{p}{[}\PYG{l+m+mi}{0}\PYG{p}{],} \PYG{n+nt}{\PYGZdq{}memory\PYGZus{}slot\PYGZdq{}}\PYG{p}{:} \PYG{p}{[}\PYG{l+m+mi}{0}\PYG{p}{]\PYGZcb{}]\PYGZcb{},}
\PYG{p}{\PYGZob{}}\PYG{n+nt}{\PYGZdq{}header\PYGZdq{}}\PYG{p}{:} \PYG{p}{\PYGZob{}}\PYG{n+nt}{\PYGZdq{}name\PYGZdq{}}\PYG{p}{:} \PYG{l+s+s2}{\PYGZdq{}Wait 2\PYGZdq{}}\PYG{p}{\PYGZcb{},} \PYG{n+nt}{\PYGZdq{}instructions\PYGZdq{}}\PYG{p}{:}
  \PYG{p}{[\PYGZob{}}\PYG{n+nt}{\PYGZdq{}name\PYGZdq{}}\PYG{p}{:} \PYG{l+s+s2}{\PYGZdq{}pulse2\PYGZdq{}}\PYG{p}{,} \PYG{n+nt}{\PYGZdq{}t0\PYGZdq{}}\PYG{p}{:} \PYG{l+m+mi}{160}\PYG{p}{,} \PYG{n+nt}{\PYGZdq{}ch\PYGZdq{}}\PYG{p}{:} \PYG{l+s+s2}{\PYGZdq{}d0\PYGZdq{}}\PYG{p}{\PYGZcb{},}
   \PYG{p}{\PYGZob{}}\PYG{n+nt}{\PYGZdq{}name\PYGZdq{}}\PYG{p}{:} \PYG{l+s+s2}{\PYGZdq{}square\PYGZus{}pulse\PYGZdq{}}\PYG{p}{,} \PYG{n+nt}{\PYGZdq{}t0\PYGZdq{}}\PYG{p}{:} \PYG{l+m+mi}{200}\PYG{p}{,} \PYG{n+nt}{\PYGZdq{}ch\PYGZdq{}}\PYG{p}{:} \PYG{l+s+s2}{\PYGZdq{}m0\PYGZdq{}}\PYG{p}{\PYGZcb{},}
   \PYG{p}{\PYGZob{}}\PYG{n+nt}{\PYGZdq{}name\PYGZdq{}}\PYG{p}{:} \PYG{l+s+s2}{\PYGZdq{}acquire\PYGZdq{}}\PYG{p}{,} \PYG{n+nt}{\PYGZdq{}t0\PYGZdq{}}\PYG{p}{:} \PYG{l+m+mi}{200}\PYG{p}{,} \PYG{n+nt}{\PYGZdq{}duration\PYGZdq{}}\PYG{p}{:} \PYG{l+m+mi}{6}\PYG{p}{,}
                    \PYG{n+nt}{\PYGZdq{}qubits\PYGZdq{}}\PYG{p}{:} \PYG{p}{[}\PYG{l+m+mi}{0}\PYG{p}{],} \PYG{n+nt}{\PYGZdq{}memory\PYGZus{}slot\PYGZdq{}}\PYG{p}{:} \PYG{p}{[}\PYG{l+m+mi}{0}\PYG{p}{]\PYGZcb{}]\PYGZcb{},}
\PYG{p}{\PYGZob{}}\PYG{n+nt}{\PYGZdq{}header\PYGZdq{}}\PYG{p}{:} \PYG{p}{\PYGZob{}}\PYG{n+nt}{\PYGZdq{}name\PYGZdq{}}\PYG{p}{:} \PYG{l+s+s2}{\PYGZdq{}Wait 3\PYGZdq{}}\PYG{p}{\PYGZcb{},} \PYG{n+nt}{\PYGZdq{}instructions\PYGZdq{}}\PYG{p}{:}
  \PYG{p}{[\PYGZob{}}\PYG{n+nt}{\PYGZdq{}name\PYGZdq{}}\PYG{p}{:} \PYG{l+s+s2}{\PYGZdq{}pulse2\PYGZdq{}}\PYG{p}{,} \PYG{n+nt}{\PYGZdq{}t0\PYGZdq{}}\PYG{p}{:} \PYG{l+m+mi}{130}\PYG{p}{,} \PYG{n+nt}{\PYGZdq{}ch\PYGZdq{}}\PYG{p}{:} \PYG{l+s+s2}{\PYGZdq{}d0\PYGZdq{}}\PYG{p}{\PYGZcb{},}
   \PYG{p}{\PYGZob{}}\PYG{n+nt}{\PYGZdq{}name\PYGZdq{}}\PYG{p}{:} \PYG{l+s+s2}{\PYGZdq{}square\PYGZus{}pulse\PYGZdq{}}\PYG{p}{,} \PYG{n+nt}{\PYGZdq{}t0\PYGZdq{}}\PYG{p}{:} \PYG{l+m+mi}{200}\PYG{p}{,} \PYG{n+nt}{\PYGZdq{}ch\PYGZdq{}}\PYG{p}{:} \PYG{l+s+s2}{\PYGZdq{}m0\PYGZdq{}}\PYG{p}{\PYGZcb{},}
   \PYG{p}{\PYGZob{}}\PYG{n+nt}{\PYGZdq{}name\PYGZdq{}}\PYG{p}{:} \PYG{l+s+s2}{\PYGZdq{}acquire\PYGZdq{}}\PYG{p}{,} \PYG{n+nt}{\PYGZdq{}t0\PYGZdq{}}\PYG{p}{:} \PYG{l+m+mi}{200}\PYG{p}{,} \PYG{n+nt}{\PYGZdq{}duration\PYGZdq{}}\PYG{p}{:} \PYG{l+m+mi}{6}\PYG{p}{,}
                    \PYG{n+nt}{\PYGZdq{}qubits\PYGZdq{}}\PYG{p}{:} \PYG{p}{[}\PYG{l+m+mi}{0}\PYG{p}{],} \PYG{n+nt}{\PYGZdq{}memory\PYGZus{}slot\PYGZdq{}}\PYG{p}{:} \PYG{p}{[}\PYG{l+m+mi}{0}\PYG{p}{]\PYGZcb{}]\PYGZcb{},}
\PYG{p}{\PYGZob{}}\PYG{n+nt}{\PYGZdq{}header\PYGZdq{}}\PYG{p}{:} \PYG{p}{\PYGZob{}}\PYG{n+nt}{\PYGZdq{}name\PYGZdq{}}\PYG{p}{:} \PYG{l+s+s2}{\PYGZdq{}Wait 4\PYGZdq{}}\PYG{p}{\PYGZcb{},} \PYG{n+nt}{\PYGZdq{}instructions\PYGZdq{}}\PYG{p}{:}
  \PYG{p}{[\PYGZob{}}\PYG{n+nt}{\PYGZdq{}name\PYGZdq{}}\PYG{p}{:} \PYG{l+s+s2}{\PYGZdq{}pulse2\PYGZdq{}}\PYG{p}{,} \PYG{n+nt}{\PYGZdq{}t0\PYGZdq{}}\PYG{p}{:} \PYG{l+m+mi}{100}\PYG{p}{,} \PYG{n+nt}{\PYGZdq{}ch\PYGZdq{}}\PYG{p}{:} \PYG{l+s+s2}{\PYGZdq{}d0\PYGZdq{}}\PYG{p}{\PYGZcb{},}
   \PYG{p}{\PYGZob{}}\PYG{n+nt}{\PYGZdq{}name\PYGZdq{}}\PYG{p}{:} \PYG{l+s+s2}{\PYGZdq{}square\PYGZus{}pulse\PYGZdq{}}\PYG{p}{,} \PYG{n+nt}{\PYGZdq{}t0\PYGZdq{}}\PYG{p}{:} \PYG{l+m+mi}{200}\PYG{p}{,} \PYG{n+nt}{\PYGZdq{}ch\PYGZdq{}}\PYG{p}{:} \PYG{l+s+s2}{\PYGZdq{}m0\PYGZdq{}}\PYG{p}{\PYGZcb{},}
   \PYG{p}{\PYGZob{}}\PYG{n+nt}{\PYGZdq{}name\PYGZdq{}}\PYG{p}{:} \PYG{l+s+s2}{\PYGZdq{}acquire\PYGZdq{}}\PYG{p}{,} \PYG{n+nt}{\PYGZdq{}t0\PYGZdq{}}\PYG{p}{:} \PYG{l+m+mi}{200}\PYG{p}{,} \PYG{n+nt}{\PYGZdq{}duration\PYGZdq{}}\PYG{p}{:} \PYG{l+m+mi}{6}\PYG{p}{,}
                    \PYG{n+nt}{\PYGZdq{}qubits\PYGZdq{}}\PYG{p}{:} \PYG{p}{[}\PYG{l+m+mi}{0}\PYG{p}{],} \PYG{n+nt}{\PYGZdq{}memory\PYGZus{}slot\PYGZdq{}}\PYG{p}{:} \PYG{p}{[}\PYG{l+m+mi}{0}\PYG{p}{]\PYGZcb{}]\PYGZcb{},}
\PYG{p}{\PYGZob{}}\PYG{n+nt}{\PYGZdq{}header\PYGZdq{}}\PYG{p}{:} \PYG{p}{\PYGZob{}}\PYG{n+nt}{\PYGZdq{}name\PYGZdq{}}\PYG{p}{:} \PYG{l+s+s2}{\PYGZdq{}Wait 5\PYGZdq{}}\PYG{p}{\PYGZcb{},} \PYG{n+nt}{\PYGZdq{}instructions\PYGZdq{}}\PYG{p}{:}
  \PYG{p}{[\PYGZob{}}\PYG{n+nt}{\PYGZdq{}name\PYGZdq{}}\PYG{p}{:} \PYG{l+s+s2}{\PYGZdq{}pulse2\PYGZdq{}}\PYG{p}{,} \PYG{n+nt}{\PYGZdq{}t0\PYGZdq{}}\PYG{p}{:} \PYG{l+m+mi}{60}\PYG{p}{,} \PYG{n+nt}{\PYGZdq{}ch\PYGZdq{}}\PYG{p}{:} \PYG{l+s+s2}{\PYGZdq{}d0\PYGZdq{}}\PYG{p}{\PYGZcb{},}
   \PYG{p}{\PYGZob{}}\PYG{n+nt}{\PYGZdq{}name\PYGZdq{}}\PYG{p}{:} \PYG{l+s+s2}{\PYGZdq{}square\PYGZus{}pulse\PYGZdq{}}\PYG{p}{,} \PYG{n+nt}{\PYGZdq{}t0\PYGZdq{}}\PYG{p}{:} \PYG{l+m+mi}{200}\PYG{p}{,} \PYG{n+nt}{\PYGZdq{}ch\PYGZdq{}}\PYG{p}{:} \PYG{l+s+s2}{\PYGZdq{}m0\PYGZdq{}}\PYG{p}{\PYGZcb{},}
   \PYG{p}{\PYGZob{}}\PYG{n+nt}{\PYGZdq{}name\PYGZdq{}}\PYG{p}{:} \PYG{l+s+s2}{\PYGZdq{}acquire\PYGZdq{}}\PYG{p}{,} \PYG{n+nt}{\PYGZdq{}t0\PYGZdq{}}\PYG{p}{:} \PYG{l+m+mi}{200}\PYG{p}{,} \PYG{n+nt}{\PYGZdq{}duration\PYGZdq{}}\PYG{p}{:} \PYG{l+m+mi}{6}\PYG{p}{,}
                    \PYG{n+nt}{\PYGZdq{}qubits\PYGZdq{}}\PYG{p}{:} \PYG{p}{[}\PYG{l+m+mi}{0}\PYG{p}{],} \PYG{n+nt}{\PYGZdq{}memory\PYGZus{}slot\PYGZdq{}}\PYG{p}{:} \PYG{p}{[}\PYG{l+m+mi}{0}\PYG{p}{]\PYGZcb{}]\PYGZcb{}}
\PYG{p}{]}
\end{Verbatim}
\end{jsonexamplebox}
where the acquisition commands do not specify the kernel or discriminator so the defaults are used.
Here we will run the experiment and retrieve the measurements as a level 1 averaged measurement. The run dictionary is,
\begin{jsonexamplebox}{T1 - Level 1 Measurement}
\begin{Verbatim}[commandchars=\\\{\}]
\PYG{p}{\PYGZob{}}
	\PYG{n+nt}{\PYGZdq{}qobj\PYGZus{}id\PYGZdq{}}\PYG{p}{:} \PYG{l+s+s2}{\PYGZdq{}t1\PYGZus{}exp\PYGZus{}07312018\PYGZdq{}}\PYG{p}{,}
	\PYG{n+nt}{\PYGZdq{}schema\PYGZus{}version\PYGZdq{}}\PYG{p}{:} \PYG{l+s+s2}{\PYGZdq{}1.0.0\PYGZdq{}}\PYG{p}{,}
	\PYG{n+nt}{\PYGZdq{}type\PYGZdq{}}\PYG{p}{:} \PYG{l+s+s2}{\PYGZdq{}PULSE\PYGZdq{}}\PYG{p}{,}
	\PYG{n+nt}{\PYGZdq{}header\PYGZdq{}}\PYG{p}{:} \PYG{p}{\PYGZob{}\PYGZcb{},}
	\PYG{n+nt}{\PYGZdq{}experiments\PYGZdq{}}\PYG{p}{:} \PYG{p}{[}\PYG{err}{cal1}\PYG{p}{,} \PYG{err}{cal2}\PYG{p}{,} \PYG{err}{exp1}\PYG{p}{,}\PYG{err}{...}\PYG{p}{],}
	\PYG{n+nt}{\PYGZdq{}config\PYGZdq{}}\PYG{p}{:} \PYG{p}{\PYGZob{}}
		\PYG{n+nt}{\PYGZdq{}meas\PYGZus{}level\PYGZdq{}}\PYG{p}{:} \PYG{l+m+mi}{1}\PYG{p}{,}
	  \PYG{n+nt}{\PYGZdq{}pulse\PYGZus{}library\PYGZdq{}}\PYG{p}{:} \PYG{err}{pulse\PYGZus{}lib}\PYG{p}{,}
		\PYG{n+nt}{\PYGZdq{}memory\PYGZus{}slots\PYGZdq{}}\PYG{p}{:} \PYG{l+m+mi}{1}\PYG{p}{,}
		\PYG{n+nt}{\PYGZdq{}memory\PYGZus{}slot\PYGZus{}size\PYGZdq{}}\PYG{p}{:} \PYG{l+m+mi}{6}\PYG{p}{,}
		\PYG{n+nt}{\PYGZdq{}meas\PYGZus{}return\PYGZdq{}}\PYG{p}{:} \PYG{l+s+s2}{\PYGZdq{}avg\PYGZdq{}}\PYG{p}{,}
		\PYG{n+nt}{\PYGZdq{}qubit\PYGZus{}lo\PYGZus{}freq\PYGZdq{}}\PYG{p}{:} \PYG{p}{[}\PYG{l+m+mf}{5.0}\PYG{p}{],}
		\PYG{n+nt}{\PYGZdq{}meas\PYGZus{}lo\PYGZus{}freq\PYGZdq{}}\PYG{p}{:} \PYG{p}{[}\PYG{l+m+mf}{6.5}\PYG{p}{],}
		\PYG{n+nt}{\PYGZdq{}rep\PYGZus{}time\PYGZdq{}}\PYG{p}{:} \PYG{l+m+mi}{1000}\PYG{p}{,}
		\PYG{n+nt}{\PYGZdq{}shots\PYGZdq{}}\PYG{p}{:} \PYG{l+m+mi}{5}
						\PYG{p}{\PYGZcb{}}
\PYG{p}{\PYGZcb{}}
\end{Verbatim}
\end{jsonexamplebox}

\renewcommand{\tocbibname}{References}
\bibliographystyle{apsrev4-1}
\bibliography{openquantum}

\end{document}